\documentclass{article}

\usepackage{arxiv}

\usepackage[utf8]{inputenc} 
\usepackage[T1]{fontenc}    
\usepackage{hyperref}       
\usepackage{url}            
\usepackage{booktabs}       
\usepackage{amsfonts}       
\usepackage{nicefrac}       
\usepackage{microtype}      
\usepackage{lipsum}

\usepackage[displaymath]{lineno}
\usepackage{hyperref}

\usepackage{graphicx,pstricks}
\usepackage{graphics}
\usepackage{epsfig}
\usepackage{amssymb}
\usepackage{amsmath, amsthm}
\usepackage{amsfonts}
\usepackage{tabu}
\usepackage{multirow}
\usepackage{subfig}
\usepackage{caption}
\usepackage[ruled,vlined]{algorithm2e}

\newcolumntype{P}[1]{>{\centering\arraybackslash}p{#1}} 
\newcommand{\norm}[1]{\left\lVert #1 \right\rVert}

\title{A new engineering theory describing oblique free surface impact by flexible plates}

\author{
  Wensi Wu\\
  Civil \& Environmental Engineering\\
  Cornell University\\
  Ithaca, NY 14850 \\
  \texttt{ww382@cornell.edu} \\
    \And
    \textbf{Christopher Earls}\\
    Center for Applied Mathematics\\
  Civil \& Environmental Engineering\\
  Cornell University\\
  Ithaca, NY 14850 \\
  \texttt{earls@cornell.edu} \\
}

\begin{document}
\maketitle

\begin{abstract}
Consideration of slamming loads within the structural design of planning hulls is of critical importance in ensuring adequate structural performance. However, because of the intricacy in the interplay between complex fluid flows and nonlinear structural deformations that accompany slamming, a general engineering theory for slamming has yet to be uncovered, and so design relies on specialized theories. In this paper, we propose one such theory for a design case that has, until now, eluded a proper description. In pursuit of this theory, we employ a specialized implicit, partitioned \textit{fluid-structural interaction (FSI)} simulation approach, to study the underlying physical mechanisms accompanying the oblique impact of a flexible plate during water entry. In the present work, we first present validation results from flexible plate water entry experiments. Subsequent to validation, we carry out a series of numerical analyses to characterize the impact force and plate out-of-plane deformations. Finally, we use our FSI solver to study the mechanistic evolution of fluid flows and elastic plate deformations that occur during slamming. Based on these observations, we propose a novel, but simple engineering theory for flexible plates obliquely impacting the water free surface (\textit{e.g.,} high speed porpoising water craft reentry).
\end{abstract}

\keywords{Fluid-structure interaction \and Volume of Fluid \and Oblique slamming impact \and Added Mass Approximation \and Engineering theory}


\section{Introduction}

In planning watercraft, sailing in head seas at high speeds, the outer hull plating frequently experiences violent impacts with the water free surface. These violent impacts are known as \textit{slamming loads}. Slamming loads are characterized as instantaneous, acute, and concentrated pressure impulses acting on hull plating and underlying structural components. This type of extreme loading occurs at small spatiotemporal scales and can induce high strain rates and nonlinear behaviors within the structural system -- in extreme cases, slamming may result in serious ship accidents, with accompanying loss of life \cite{Kapsenberg2011}. Indeed, the American Bureau of Shipping, (ABS), in its HSNC 2007 guide for building and classing naval craft, has called out slamming loads as the single most important consideration when proportioning hull scantlings in high speed watercraft. That said, a proper consideration of slamming loads in the design of planning hulls is of critical importance for safe and satisfactory performance. 

\subsection{Literature review}
Theoretical study of hydrodynamic slamming phenomenology was pioneered by von Karman in his 1929 report focused on the impact of seaplane floats during landing. In that study, von Karman approximated the seaplane float as a rigid wedge-shaped body, and subsequently developed a mathematical framework (based on the conservation of momentum) to estimate the hydrodynamic impact force experienced by the structure of the float \cite{karman1929}.  A few years after that seminal work, Herbert Wagner proposed an improvement to von Karman's theory. He included the so-called \textit{splash-up} effect in the force calculation, and derived a new mathematical model on the basis of \textit{potential flow theory} to approximate the impact force \cite{Wagner1932} acting on the wetted surface of a wedge. Wagner's model was further extended by many researchers in an effort to explain the various aspects of slamming phenomenology. These studies include, to name a few, three-dimensional impact on various geometries during water entry \cite{Howison1991}, structural elastic deflections resulting from wave loads \cite{Newman1994}, the hydroelastic effects on elastic plates and wedges subjected to wetdeck slamming \cite{Faltinsen1997, Faltinsen1997[2], Faltinsen1999}, \textit{etc}; many mathematical and analytical models have been developed to better understand the underlying hydrodynamic and hydroelastic impacts centering around slamming \cite{ Dobrovolskaya1969, Divitiis2002, Korobkin2004, oliver2007, taghipour2008, Faltinsen2008}.

Accompanying the development of slamming theories, experimental investigations of various slamming contexts have been undertaken in an effort to validate and refine the proposed theories. For example, the importance of hydroelastic effects in wedge drop testing was discussed in \cite{Faltinsen1997}, water slamming loads and the effects of air entrapment for flat plate slamming was reported in \cite{Huera2011}, experimental studies of flat plate subjected to high horizontal speed was reported in \cite{Iafrati2015}, and the oblique impact of flexible plates was studied in \cite{Wang2016, Wang2019, Wang2020}. Although experimental data concerning slamming impacts have helped in accelerating the development of slamming theories, the existing experimental findings only represent a subset of slamming scenarios that are encountered within design practice. The need for considering many specimen geometries and loading contexts creates a combinatorial ``explosion" within the space of possible experiments. The highly complex nature of slamming involves the interaction of complex fluid flows and nonlinear structural response that occurs at a confined spatiotemporal scale in order to properly characterize the underlying phenomenology; this further complicates experimental investigation. As a result of all the foregoing, we pursue a program of investigation involving \textit{experimentally validated} computer simulations, as a stand-in for large scale experimentation, as we pursue a useful engineering theory suitable for design.  

Considerable work has already been undertaken to uncover useful numerical strategies for analyzing slamming impacts: some of these approaches include water entry of a wedge with large deadrise angle using a nonlinear boundary element method \cite{zhao1993}, numerical simulation of the vertical impact of a two dimensional rigid plate onto water surface via boundary value problem formulation \cite{Iafrati2004, Iafrati2008}, and a state-space models using a Newmark $\beta$ time integration scheme for analyzing a flexible barge \cite{Taghipour2007}. While the previous numerical studies mentioned provide valuable insights into slamming, the fluid and structure domains are solved explicitly within these numerical frameworks (\textit{i.e.,} the slamming problems are treated as one-way coupling FSI problems where the structural bodies are assumed rigid). Since structural deformations have significant effects on the fluid dynamics and vice versa \cite{Korobkin2006}, a one-way coupling strategy undermines the importance of the interdependent nature between the structural and fluid responses in the context of water entry and exit problems: a vital element within the current work. As a result, \textit{implicit} fluid-structure interaction (FSI) analysis is adopted herein; to accurately assess the hydrodynamic and hydroelastic effects accompanying slamming. Exemplars of the FSI formulations for analyzing the hydrodynamics and hydroelastics response in the context of slamming can  be found in \cite{Lu2000, Walhorn2005, Korobkin2008, Paik2009, Tallec2001, Wall2007, Piro2013}.

\subsection{Contribution}
The present work describes a state-of-the-art, partitioned, implicit FSI simulation tool built from open source components and experimentally validated for the oblique, hydroelastic plate context that is salient for high speed vessels operating within seaways. We use subsequent high fidelity simulations to afford unprecedented insight into the physical mechanisms at work within this important engineering context, so that we may propose a useful engineering theory to support design. 

\subsection{Paper structure}
In the present work, we aim to study the slamming phenomenon in unprecedented detail, through an implicit partitioned FSI framework. Specifically, we first validate the impact forces and the out-of-plane deflections of flexible plates using the experimental investigations reported in \cite{Wang2020}. Subsequently, we perform a series of complementary simulations for this slamming context, to elucidate the missing details from experimental results. Finally, we furnish a new engineering theory for slamming with engineering insights gleaned from high fidelity FSI simulations.

The outline of this manuscript is as follows: Section \ref{Goveqns} and \ref{solver} provide the governing equations for the FSI coupled system along with the FSI coupling framework adopted in the present work; Section \ref{slamming} provides in-depth information regarding the experimental and computational details concerning the flexible plate slamming experiments, as well as FSI validation results;  Section \ref{applications} presents slamming results from additional FSI simulations germane to the slamming context we study; Section \ref{theory} discusses the derivation of the proposed engineering theory; and Section \ref{conclusion} concludes the work with a summary of findings.

\section{Mathematical formulations}\label{Goveqns}

We now introduce the mathematical formulations that govern the implicitly coupled FSI system employed in the present work. These mathematical models include the governing equations for the fluid and structural domains, along with mathematical formulations for free surface tracking, and imposition of transmission conditions along the FSI interface, as well as the algorithmic approach to implicit coupling within the FSI system.  

\subsection{Governing equations for incompressible flows}

The fluid domain within the present work is modeled as a homogenous, incompressible Newtonian fluid governed by the Navier-Stokes equation; compressibility of the air is assumed negligible in the present work given that no significant air entrapments were observed in oblique slamming impact \cite{Huera2011, Hicks2010, Khabakhpasheva2020} with angle of attack greater than $5^\circ$. The Navier-Stokes and the continuity equation are expressed in vectorial form in Equation \ref{NSeqn} 

\begin{equation} \label{NSeqn}
\begin{gathered}
	\rho^f[\frac{\partial {\textbf{u}^f}}{\partial t}+ (\textbf{u}^f \cdot \nabla)\textbf{u}^f] =\rho^f g-\nabla{\textbf{p}^f}+ \mu^f \nabla^2 \textbf{u}^f, \\
	\nabla \cdot \textbf{u}^f = 0.
\end{gathered}
\end{equation}

\noindent where $\rho^f$ is the fluid density, $\mathbf{u}^f$ represents the fluid velocity, $g$ is the gravitational body force, $\mathbf{p}^f$ is the fluid pressure.

\subsubsection{Free surface tracking method}

\begin{figure}[h]
\centering
\includegraphics[width=0.8\textwidth]{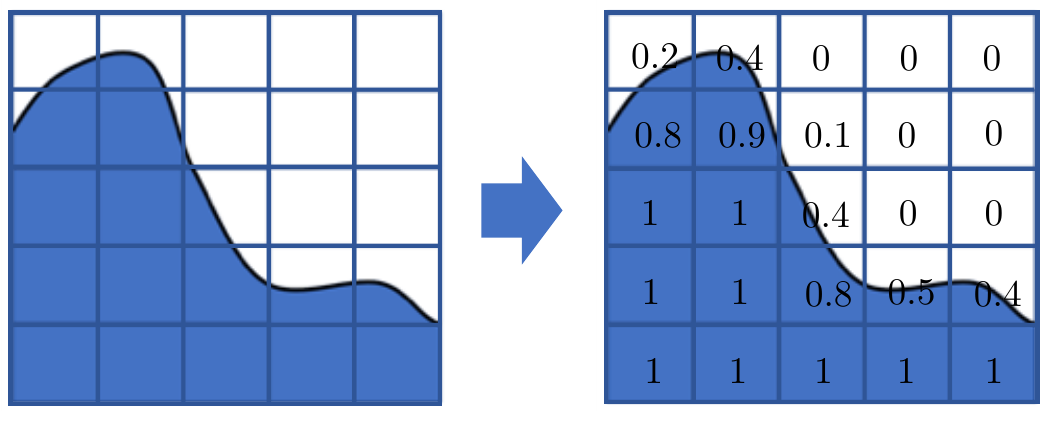}
\caption{A representation of the VOF implementation is shown. The cells with blue color represents water and white color represent air. Each cell has an assigned phase fraction value, $\alpha$, which is used to identify the amount of water within each cell.}\label{voffig}
\end{figure}

Within the present FSI context, it is vitally important to correctly track the free surface: the interface between the air phase and water phase within the fluid domain. In order to efficiently resolve and capture the free surface motion, the \textit{volume-of-fluid (VOF)} method \cite{hirt1981} is used to model the transient fluid dynamic flow of the two immiscible fluids in the present work. The VOF method is a popular interface tracking method in marine application because of its ability to capture the water free surface when undergoing large deformations (\emph{e.g.,} breaking waves) \cite{pedersen2017}. A visual representation of the VOF methodology is shown in Figure \ref{voffig}. The control volumes within the fluid domain are characterized using so-called  $\alpha$ values, which are used to quantify the volume fraction of the water and air phase within the volume cells used to discretize the fluid domain, as well as to track the motion of the free surface (\textit{i.e.,} the layer between air and water). The $\alpha$ values used in this work are defined in Equation \ref{phase_var} and Figure 1 displays representative values. 

\begin{equation} \label {phase_var}
    \alpha = 
	\begin{cases}
   		  		1		& \text{water},\\
  			  	0              & \text{air}, \\
				0 < \alpha < 1 & \text{fluid interface}.
	\end{cases}
\end{equation}

Within the VOF framework, the air and water phases are each treated as individual continua, and the fluid properties of the air and water are interpolated through a phase fraction variable, $\alpha$, along the interface, as shown in Equation \ref{vof_eqn}

\begin{equation} \label {vof_eqn}
\begin{gathered} 
\rho = \alpha \rho_w + (1 - \alpha) \rho_a, \\
\mu = \alpha \mu_w + (1 - \alpha) \mu_a,
\end{gathered}
\end{equation}

\noindent where subscript $w$ denotes the fluid properties of water and the subscript $a$ denotes the fluid properties of air. The continuity equation for the phase fraction $\alpha$ is expressed as

\begin{equation} \label {conser_alpha}
\begin{gathered} 
\frac{\partial \alpha}{\partial t} + \nabla \cdot (\alpha\mathbf{u}^f)+\nabla \cdot (\alpha(1-\alpha)\mathbf{u}^f_r) = 0,
\end{gathered}
\end{equation}
where $\mathbf{u}^f_r$ is the relative velocity between water and air (\textit{i.e.,} the artificial compression velocity \cite{berberovi2009}). It is critical to maintain a sharp interface within the computational domain, in order to maintain appropriate simulation resolution. Since the sharpness in the vicinity of the interface can no longer be guaranteed if the term $\nabla \cdot (\alpha\mathbf{u}^f)$ is diffusive, an artificial compressive convective term, $\nabla \cdot (\alpha(1-\alpha)\mathbf{u}^f_r)$, is added to the continuity equation, in order to enhance interface sharpness. 
\subsection{Governing equations for isotropic linear elastic materials}
The structural domain in the present work is described within a Lagrangian frame of reference, wherein the momentum equation appears as

\begin{equation} \label{structure_momentum}
	\rho^s[ \frac{\partial d^s_i}{\partial t}+ (\textbf{d}^s \cdot \nabla)d^s_i]=\rho^s g_i +\frac{\partial \sigma^s_{ij}}{\partial x_j},\\
\end{equation}
\noindent where $\rho^s$ is the structure density, $\bf{d^s}$ represents the structural displacement vector, $g$ is the gravitational body force, $\sigma^s$ is the Cauchy stress tensor, and $x^f$ represents the rectangular Cartesian spatial coordinates of the structural response variables, and $t$ is the temporal variable. 

Consider an isotropic linear elastic material model

\begin{equation} \label{constitutive_model}
\begin{gathered}
\sigma^s_{ij} = 2 \mu^s \epsilon^s_{ij}+ \lambda \epsilon^s_{kk} \delta_{ij}, \\
\epsilon^s_{ij} = \frac{1}{2} (\frac{\partial d_i^s}{\partial x^s_j}+\frac{\partial d^s_j}{\partial x^s_i}), 
\end{gathered}
\end{equation}
where $\epsilon^s$ is the infinitesimal strain tensor, $\lambda$ and $\mu^s$ are the Lam\'e constants. ($\mu^s$ is also the shear modulus.) The conservation of momentum equation governing within the structural domain may be written as 

\begin{equation} \label{structure_momentum2}
\rho^s[ \frac{(\partial d^2_i)^s}{\partial^2 t}+ (\frac{\partial d_i^s}{\partial t} \cdot \nabla)\frac{\partial d_i^s}{\partial t}]=\rho^s g_i + \mu^s(\frac{\partial d_i^s}{\partial x^s_j}+\frac{\partial d^s_j}{\partial x^s_i})+\lambda\frac{\partial d^s_k}{\partial x^s_k}.\\
\end{equation}
The advection term of the momentum equation vanishes in a Lagrangian framework; thus the vectorial form of the momentum equation is expressed as

\begin{equation} \label{structure_momentum3}
\rho^s \frac{\partial^2 \textbf{d}^s}{\partial t^2}= \rho^s \textbf{g}+\mu^s \nabla \textbf{d}^s+(\lambda+\mu^s)(\nabla \cdot \textbf{d}^s).\\
\end{equation}

\subsection{FSI interface transmission conditions}
In contrast with an explicit FSI coupling scheme, where equilibrium along the fluid-structure interface is not strictly enforced, the following transmission conditions in Equation \ref{eq_interface} are imposed on the fluid-structure interface within the implicit FSI framework adopted herein, to ensure no-slip boundary condition for the fluid flow, as well as force equilibrium along the moving boundary. 

\begin{equation} \label{eq_interface}
\begin{gathered}
\mathbf{u}^f = \frac{\partial \mathbf{d}^s}{\partial t}, \\
\mathbf{\sigma}^f \cdot \mathbf{n}^f = \mathbf{\sigma}^s \cdot \mathbf{n}^s, 
\end{gathered}
\end{equation}
where $\mathbf{n}^f$ is the unit outward normal vector w.r.t the fluid domain, and $\mathbf{n}^s$ is the unit outward normal vector w.r.t the structural domain. 

\subsection{Mathematical approach of the FSI coupled system}
In the realm of FSI, the computational mesh in the fluid domain needs to track to the interface deformations accompanying the structural deformation, in order to ensure numerical stability. Given that the fluid and structural domains are formulating in different mathematical frameworks (Lagrangian \textit{v.s.} Eulerian), a unified approach that incorporates the two frames of reference, into a single system context, is needed. The present work adopts the Arbitrary Lagrangian Eulerian (ALE) framework \cite{donea2004, hirt1974} to combine the best features of both the Lagrangian and Eulerian descriptions, while allowing for FSI analyses involving large structural deformations. In the ALE formulation, a field variable $\mathbf{u}^m$, associated with the velocity of mesh points is introduced to the convective term in the Navier-Stokes governing equation, in order to capture the mesh motion effects. This additional variable provides flexibility for the fluid computational mesh to move in a Lagrangian fashion, or to be fixed in an Eulerian manner, or to deform in some combination of the two. This moving grid methodology greatly improves the efficiency in FSI simulations by reducing mesh distortions resulting from large deformation in a purely Lagrangian perspective, meanwhile, offering higher resolution in the solution than that afforded by a purely Eulerian viewpoint.

The Navier-Stokes equation formulated in an ALE frame of reference is expressed in Equation \ref{eq_fluid} 

\begin{equation} \label{eq_fluid}
\begin{gathered}
	\rho^f[\frac{\partial {\mathbf{u}^f}}{\partial t}+ ((\mathbf{u}^f-\mathbf{u}^m) \cdot \nabla)\mathbf{u}^f] =\rho^f g-\nabla{p}+ \mu^f \nabla^2 \mathbf{u}^f,
\end{gathered}
\end{equation}
where $\mathbf{u}^f $ and $\mathbf{u}^m $ denote the velocity of the fluid particle and fluid mesh respectively. In the ALE framework above, the Navier-Stokes equation reduces to the Eulerian expression when $\mathbf{u}^m = 0$ and Lagrangian if $\mathbf{u}^m = \mathbf{u}^f$.

In the FSI context, the mesh motion propagates in a way such that the region closest to the FSI interface stays Lagrangian while allowing an Eulerian description to govern within the region away from the deforming FSI interface. In order to avoid spurious field values arising from artificial mass and momentum sources, as a result of the moving grid, the Space Conservation Scheme (SCL) \cite{demirdzic1988} or the Geometric Conservation Law (GCL) \cite{thomas1979} needs to be satisfied in the ALE scheme, to preserve a stable solution; especially in nonlinear flows \cite{farhat2001}. In other words, the SCL or GCL must be satisfied to ensure that the ALE discretization reproduces exactly a uniform flow, as if it were using a fixed grid. To do so, the velocity of the mesh motion must be computed for all first or second order time accurate methods (\textit{i.e.,} implicit Euler, Crank-Nicholson, \textit{etc.}) to ensure the condition 

\begin{equation} \label{GCL}
\dot{\bf{u}}^m = \frac{\bf{u^m_{n+1}}-\bf{u^m_n}}{\Delta{t}}
\end{equation}
is satisfied \cite{farhat2001}. In Equation \ref{GCL}, $\dot{\bf{u}}^m$ denotes the mesh acceleration and the subscript, $n$, denotes a time instance within the deformation evolution. In the present work, a vertex-based Laplacian motion diffuser \cite{Jasak2004} is chosen to determine $\mathbf{u^m}$ (\textit{i.e.,} mesh diffusivity) in Equation \ref{eq_fluid}. Detailed descriptions of the Laplacian motion diffuser is discussed in Section \ref{fluid_solver}.

\section{Partitioned FSI solver}\label{solver}

\begin{figure}[h]
\centering
\includegraphics[width=0.8\textwidth]{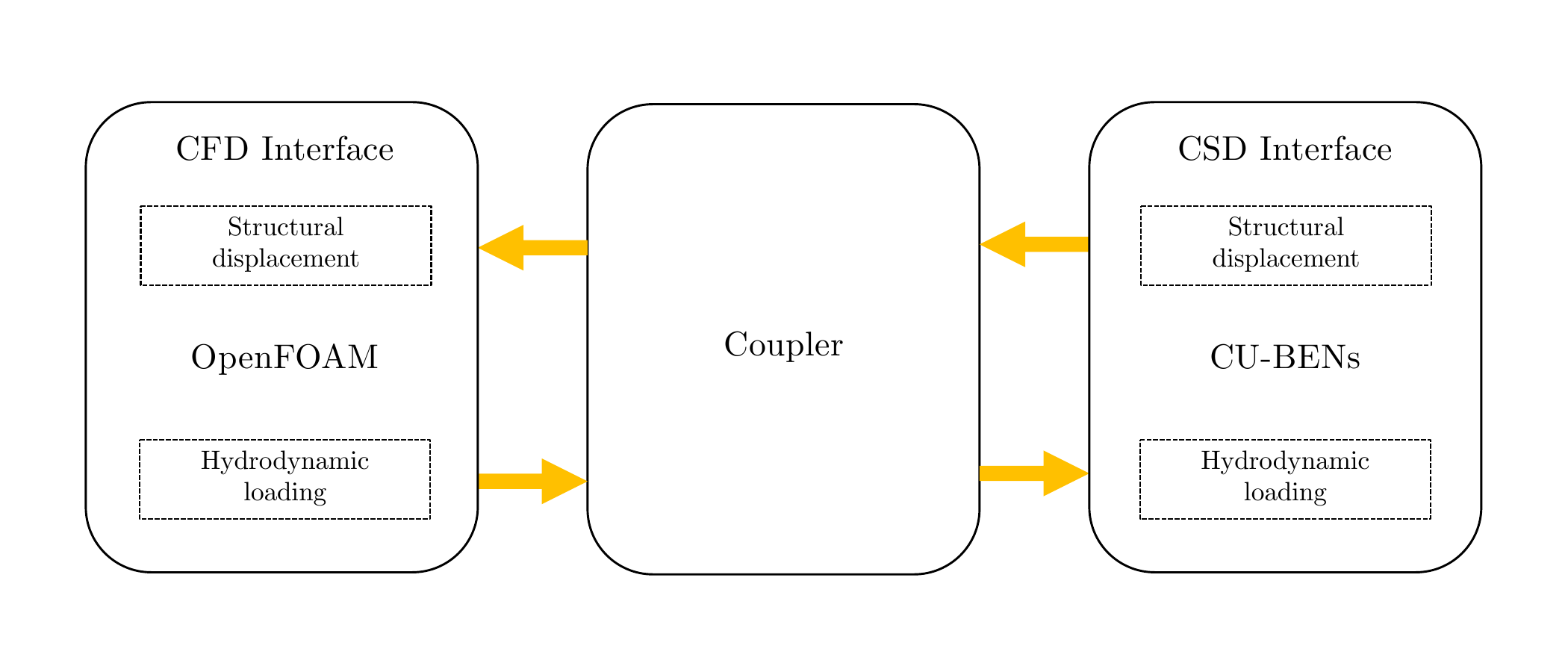}
\caption{Implicit partitioned FSI coupling framework with OpenFOAM as the CFD solver, CU-BENs as the CSD solver, and a Coupler that facilitates all necessary communication between the two solvers to ensure the fluid-structure interface conforms to the transmission conditions.}\label{coupling_framework}
\end{figure}

We adopt an implicit, partitioned FSI approach to carry out the FSI analyses in this work. Herein, the fluid and structural systems within the partitioned FSI framework are treated with two open source transient dynamics libraries (OpenFOAM 1.6-ext \cite{jasak2007, weller1998, jasak1996, openfoam} and CU-BENs \cite{wu2020}). OpenFOAM 1.6-ext is a computational fluid dynamics (CFD) finite volume solver written in C++, while, CU-BENs is a computational structural dynamics (CSD) finite element solver written in C. To ensure the two solvers communicate with each other properly, a coupling library written in Python, denoted as the ``\textit{Coupler}'' \cite{Miller2010}, is used to facilitate information transfer between the two solvers. Figure \ref{coupling_framework} provides a schematic representation of the implicit partitioned FSI solver coupling framework in the present work. A flow chart summarizing these steps appears as Figure 2. 

We will now discuss additional specialized details pertaining to the FSI modeling framework, adopted herein. 

\subsection{Structural solver}

The structural behavior within the present work is modeled using an open source finite element solver, CU-BENs \cite{wu2020}. CU-BENs exploits higher order 3D structural elements (\emph{i.e.,} truss, beam, and Discrete Kirchhoff triangular shell elements) in linear and nonlinear transient dynamic analyses. Details on the functionalities and implementations within CU-BENs can be found in \cite{wu2020}. Within the structural domain, geometric nonlinear transient analyses (as opposed to modal analyses in a few of the related work \cite{Paik2009, Piro2013}) were carried out to assess the local deformation on aluminum plates during slamming impact. In addition, Newton-Raphson subiteration was called within each time step to account for the nonlinear behavior within the structural system.  

\subsubsection{Structural elements}
Discrete Kirchhoff triangular (DKT) shell are used as the structural elements in this work \cite{bathe1981, earls2016}. The DKT shell element formulation considers a three node triangular element whose tangent stiffness matrix, $\bf{K_T}$ comprises the superimposition of the plane stress membrane stiffness, $\bf{K_m}$, plate bending stiffness, $\bf{K_b}$, and in-plane rotational stiffness, $\bf{K_\theta}$

\begin{equation} \label{DKT_stiffness}
\begin{gathered}
\bf{K_T} = \bf{K_m} + \bf{K_b} + \bf{K_\theta}, \\
\end{gathered}
\end{equation}
which takes the following matrix form 

\begin{equation} \label{DKT_stiffness2}
\begin{gathered}
\bf{K_T} = 
\begin{bmatrix}
\bf{K_m} & 0 & 0 \\
0 & \bf{K_b} & 0 \\
0 & 0 & \bf{K_\theta}
\end{bmatrix}.
\end{gathered}
\end{equation}
In Equation \ref{DKT_stiffness}, the plate stress membrane stiffness assumes a constant plane stress, the plate bending stiffness assumes homogeneous isotropic material properties, and the in-plane rotational stiffness assumes the value of $10^{-14}$ times the smallest bending stiffness (a value selected to remove the in-plane rotational singularity from the stiffness matrix in the case where the local element coordinates happen to coincide with the global coordinates). These highly efficient shell elements provide satisfactory performance in the small deflection regime, while also ensuring satisfactory performance in the large deformation regime. Since the present work concerns the deformation of flexible plates subjected to various intensities of slamming impacts, this response regime flexibility in DKT shell elements, is especially favorable when it comes to accurately approximating the plate deformations. Interested readers can refer to \cite{earls2016} for extensive derivations and discussions on the plate bending stiffness, plane stress membrane stiffness, and in-plane rotational stiffness.

\subsubsection{Implicit time integration scheme}
Numerical stability is often time a concern in FSI analyses. In particular, numerical artifacts arising from coupling two distinct spatial discretization schemes (finite element and finite volume methods) in combination with the presence of \textit{added mass effects}\cite{Causin2005, Forster2006, Forster2007} greatly amplify the spurious high frequency modes within the FSI system; thus, lead to potentially serious numerical instabilities. The generalized-$\alpha$ method appears to be effective in improving numerical stability within the partitioned FSI context, and therefore is used to facilitate time advancement in the structural domain. The generalized-$\alpha$  method was developed by Chung and Hulbert \cite{chung1993} to ameliorate spatially unresolved high frequency responses by introducing maximal numerical dissipation in high frequency structural modes, while, safeguarding solution accuracy in the low frequency structural modes. The generalized-$\alpha$ formulation is expressed as followed,

\begin{equation} \label{EOM_gen}
\mathbf{M_s}\ddot{\mathbf{d}}_{n+1-\alpha_m}^{t+\Delta t}+(1-\alpha_f)\mathbf{K_s\Delta d}=\mathbf{R}_{n+1-\alpha_f}^{t+\Delta t},
\end{equation}
where

\begin{equation} \label{EOM_gen1}
\begin{gathered}
\ddot{\mathbf{d}}_{n+1-\alpha_m}^{t+\Delta t}=(1-\alpha_m) \ddot{\mathbf{d}}_{n+1}^{t+\Delta t}+\alpha_m \ddot{\mathbf{d}}_n^{t+\Delta t}, \\ 
\end{gathered}
\end{equation}
and 

\begin{equation}  \label{EOM_gen2}
\mathbf{R}_{n+1-\alpha_f}^{t+\Delta t}=(1-\alpha_f ) \mathbf{R}_{n+1}^{t+\Delta t} +\alpha_f \mathbf{R}_{n}^{t+\Delta t}.
\end{equation}

In the equations above, $\mathbf{M_s}$ is the mass matrix, $\mathbf{K_s}$ is the stiffness matrix, $\ddot{\mathbf{d}}$ is the acceleration vector, $\mathbf{d}$ is the displacement vector, and $\mathbf{R}$ is the residual force vector.  

Integration constants $\alpha_f$ and $\alpha_m$ in Equation \ref{EOM_gen} to \ref{EOM_gen2} are expressed as

\begin{equation} \label{EOM_gen3}
\alpha_f=  \frac{\rho_\infty}{\rho_\infty+1}  \hspace{1cm} \text{and} \hspace{1cm} \alpha_m= \frac{2\rho_\infty - 1}{\rho_\infty+1},
\end{equation}
where $\rho_\infty$ is the spectral radius -- a measure of numerical dissipation in the high frequency limit. $\rho_\infty=1$  indicates no numerical dissipation and the algorithm recovers the standard Newmark implicit scheme, while, $\rho_\infty=0$ indicates ``asymptotic annihilation": meaning, the high-frequency response is entirely damped out after one time step. In the present work, we set $\rho_\infty$ to 

\begin{equation}
\begin{gathered}
  \rho_\infty  = 
\begin{cases}
   0.4,   & \text{if } W_n \geq 0.75~m/s, \\
    0.2,  & \text{otherwise},
\end{cases}
\end{gathered}
\end{equation}
where $W_n$ represents the normal slamming impact velocity. The maximum number of iterations for each time step is set to 100.

\subsection{Fluid solver} \label{fluid_solver}
The fluid responses in the present work are modeled using an open source finite volume solver, OpenFOAM 1.6-ext \cite{jasak2007}. OpenFOAM includes an extensive range of specialized solvers designed for specific continuum mechanics problems involving chemical reactions, turbulence, electromagnetics, multiphase, \textit{etc} \cite{jasak2007, weller1998, jasak1996, openfoam}. In particular, the \textit{interDyMFoam} solver is used as the multiphase fluid dynamics solver adopted in this work. 

\subsubsection{Mesh motion solver}
In contrast with the structural domain, where the structure is discretized into DKT shell elements, the fluid domain within the FSI system is discretized into hexahedral control volumes/cells. Since the DKT shell is a vertex-based element, while the hexahedral cell is a cell-centered volume, this gives rise to complications, if one wishes to transfer field variables between the structural and fluid systems (this constitutes an important aspect of FSI analyses). Due to the need for communication between FEM and FVM meshes through the FSI transmission conditions, the FEM decomposition library \cite{jasak2007, jasak2009} within OpenFOAM 1.6-ext is particularly attractive in terms of both communicating with finite element meshes and tracking fluid mesh motions. And therefore the FEM decomposition framework is adopted in the present work to preserve mesh quality and mesh motion boundedness within the FSI solutions.

The mesh motion and diffusivity in the fluid domain is managed and maintained by a Laplacian operator \cite{Jasak2004, jasak2009, kassiotis2008}. In particular, a mesh diffusion variable, $\gamma^f$, is introduced in the Laplace operator, to update the locations of mesh points. In this work, a linear distance-based method is chosen (\emph{i.e.,} $\gamma^f = \frac{1}{y}$ where $y$ is the distance between the cell center and the closest moving boundary). 

\subsubsection{Pressure-velocity coupling}
Due to the mutual linear dependency between the velocity and pressure fields in the incompressible Navier-Stokes system, inter-equation coupling between the velocity and pressure fields are required. The PIMPLE \cite{passalacqua2011, habchi2013, robertson2015} procedure is adopted in the present work for computing the pressure-velocity fields within the fluid domain. The PIMPLE algorithm adopts the procedure of PISO \cite{issa1986, jasak1996} algorithm and solves each time step as a steady-state problem, by introducing under-relaxation to the pressure and velocity fields. The inclusion of under-relaxation amplifies the influence of owner cells -- which increases the dominance of the diagonal in the system -- thus leads to better convergence in the numerical solutions. In this work, we define the maximum PIMPLE iterations to be 4. 

\subsubsection{Numerical schemes and iterative solvers}
OpenFOAM offers numerous discretization schemes \cite{openfoamuser}; Table \ref{NS_discretization_schemes} includes a list of the temporal and spatial discretization schemes used to estimate the incompressible Navier-Stokes equation in this work. In particular, a surface normal corrected scheme (a scheme which is used to ensure surface orthogonality) is applied to explicitly correct for non-orthogonality between two adjacent fluid cell faces, as to maintain a second-order accuracy in the solution. Non-orthogonality in the fluid mesh is determined by the angle between the vector that connects the cell centroids of two neighboring control volumes and the vector normal to the shared surface. It is imperative to keep non-orthogonality under $70^{\circ}$  to maintain a stable solution \cite{openfoamuser}.

\begin{table}[h!]
\centering
\caption{The temporal and spatial discretization schemes adopted in the present work for solving the incompressible Naiver-Stokes equation.} 
\label{NS_discretization_schemes}
\resizebox{\textwidth}{!}{\begin{tabular}{p{0.4\linewidth}p{0.2\linewidth}p{0.4\linewidth}}
 \hline
\textbf{Operators}                 		         & \textbf{Terms}                   		& \textbf{Numerical schemes}               \\
\hline
Time                                				&  $\frac{d}{dt} $      				& Implicit Euler                                       \\
\multirow{3}{*}{Gradient  }                        	& $\nabla \mathbf{u}^f$               	&  CellLimited Gauss linear 1.0       \\
 								& $\nabla \mathbf{p}^f$              	&  CellLimited Gauss linear 1.0     \\
								& $\nabla \alpha$               		&  CellLimited Gauss linear 1.0       \\
\multirow{3}{*}{Divergence  }                     	& $\nabla \cdot (\rho \mathbf{u}^f\mathbf{u}^f)$           & Gauss linearUpwind \\
								& $\nabla \cdot (\mathbf{u}^f_{r} \alpha)$         & Gauss interfaceCompression\\
								& $\nabla \cdot (\mathbf{u}^f \alpha)$       	    & Gauss vanLeer \cite{vanLeer1974} \\		
Laplacian                        				 & $\nabla^2$          				& Gauss linear corrected                \\
Surface normal gradient  				& $\nabla^{\perp}$    				& Corrected                                     \\
Interpolation                     				&                   					& Linear  \\
\hline
\end{tabular}}
\end{table}

The iterative solvers used in this work for each fluid variable are specified in Table \ref{iterative_solver}. Each iterative solver can be accompanied by a preconditioner and smoother, to improve numeric stability and solution smoothness. For more details on the iterative solvers, one may refer to \cite{axelsson1994, Frerziger2002, jasak2007_2, Moukalled2016} and the OpenFOAM User Manual \cite{openfoamuser}. In the present work, we set the convergence tolerances for pressure, velocity, mesh motion to $10^{-7}$; under-relaxation factors of 0.6 and 0.7 are applied the pressure and velocity fields respectively.

\begin{table}[h!]
\centering
\caption{The computational methods used in this work and associated with each of the fluid variables.}
\label{iterative_solver}
\resizebox{\textwidth}{!}{\begin{tabular}{p{0.25\linewidth}p{0.75\linewidth}}
 \hline
		                  	& \textbf{Iterative solvers}                   						 \\
\hline
$\mathbf{p}^f$                      & Geometric Agglomerated Algebraic MultiGrid (GAMG)  with Diagonal Incomplete Cholesky smoother \\                
$\mathbf{p}^f$ correction      & Pre-conditioned conjugate gradient (PCG) with GAMG preconditioner and DICGaussSeidel smoother \\
$\mathbf{p}^f$ final                     		& GAMG \\	
$\mathbf{u}^f$                     			& Gauss Seidel smooth solver \\			
$\alpha$				& Preconditioned bi-conjugate gradient with Diagonal incomplete LU preconditioner \\
FEM mesh motion            & Conjugate Gradient with Cholesky preconditioner	\\
\hline
\end{tabular}}
\end{table}

\subsection{Coupler}

In terms of communications between the fluid and structural solvers, we use the \textit{Coupler} (which was originally developed by the Navel Surface Warfare Center Carderock Division \cite{Miller2010} and later modified by the Authors, for use in the present work), to facilitate information passing along the fluid-structure interface. The grid-to-grid mapping and mesh updating is managed using a nonconforming mesh projecting technique \cite{Farhat1998}. Within the Coupler, inverse distance weighting interpolation (IDW) \cite{shepard1968} is used to convert hydrodynamic pressure, located at the center of each fluid patch (one face of the finite volume), along the interface, into point loads, before they are projected onto the structural mesh. While the fixed-point under-relaxation approach (also known as the Aitken's relaxation method \cite{Wall2007}) was a popular coupling algorithm in partitioned FSI analysis \cite{Piro2013, Campbell2011} due to its simplicity, it is rather computational expensive compared to other coupling schemes such as the iterative Quasi-Newton method (IQN-ILS) and Interface-GMRES \cite{degroote2009,degroote2010_2}. As such, the IQN-ILS method is used to ensure solution convergence on the fluid-structure interface within each FSI time step. Detailed descriptions of the algorithm and functionality within the Coupler can be found in \cite{wu2020_2}. 

\section{Experimental validation} \label{slamming}
In this section, we discuss validation results for our proposed partitioned FSI modeling framework. In particular, we compare our FSI simulation results against experimental work detailed in \cite{Wang2020}. The organization of this section is as follow: 1) we discuss the experimental setup of the oblique free surface impact experiments; 2) we provide details of the FSI computational models used in the present work; and 3) we present mesh convergence and FSI validation results.

\subsection{Experiment configuration}
\begin{figure}[h!]
\centering
\includegraphics[width=1\textwidth]{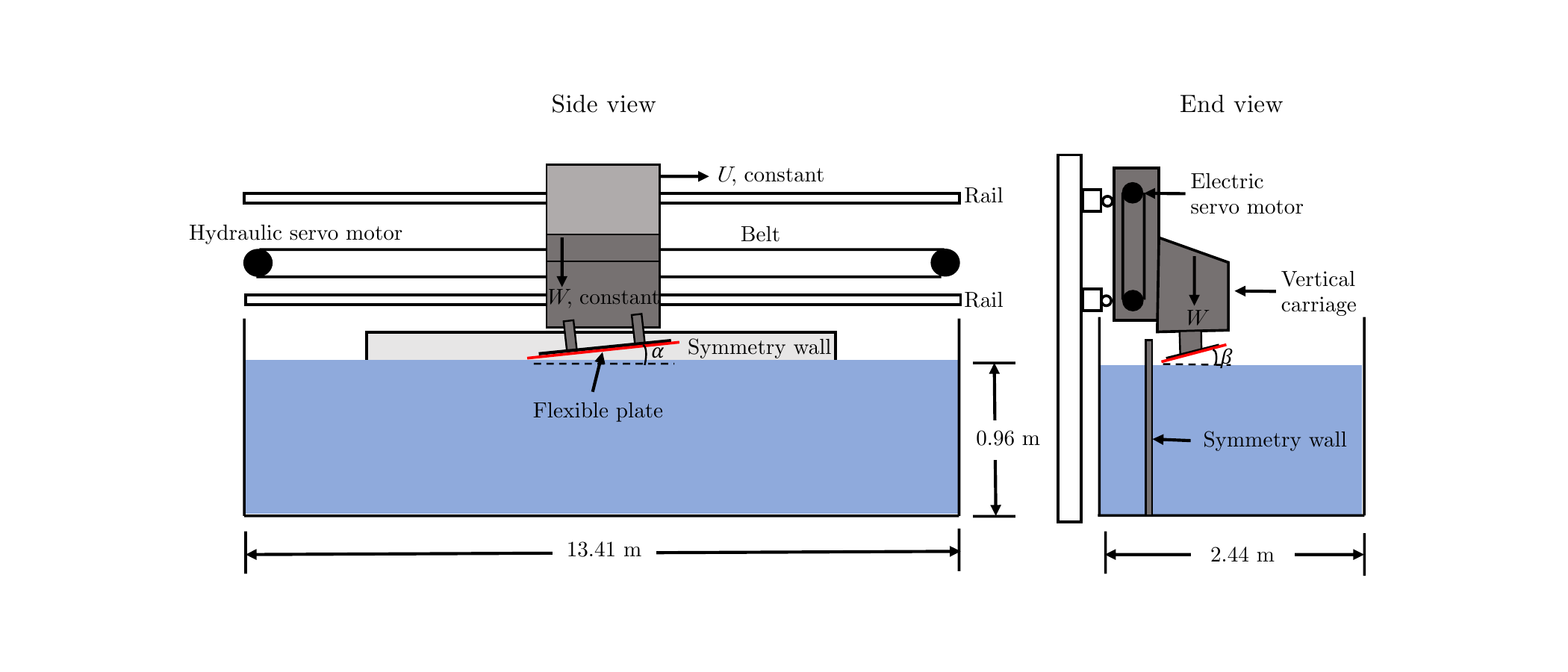}
\caption{The side view and end view of the open channel testing facility at which the flat plate slamming experiments are conducted \cite{Wang2020}.}\label{tank}
\end{figure}

The flexible plate oblique impact experiments are conducted at the University of Maryland in an open channel filled with water. The water channel measures 13.41 m long, 2.44 m wide, and 0.96 m deep, as shown in Figure \ref{tank}. An aluminum plate is mounted to a vertical carriage, which is attached to an electric servo motor. The plate mounting instrument offers flexibility for the plate to be adjusted to the desired angle of attack and deadrise angle (denoted as $\alpha$ and $\beta$ respectively). The flexible plates in this work are oriented at $10^{\circ}$ angle of attack ($\alpha=10^{\circ}$) and zero deadrise ($\beta=0^{\circ}$). The horizontal motion of the plate is driven by a horizontal carriage attached to two hydraulic servo motors, located at opposite ends of the tank. Both vertical and horizontal carriages are controlled by a computer-based feedback system that is connected to sensors, so that it precisely drives the position of the horizontal and vertical carriages. This system is used to control the impact velocities with which the plates strike in the water free surface (\textit{i.e.,} the impact velocities). 

\begin{figure}[h!]
\centering
\captionsetup[subfigure]{justification=centering}
\subfloat[Plate mounting system]{\scalebox{1}{\includegraphics[width=0.5\textwidth]{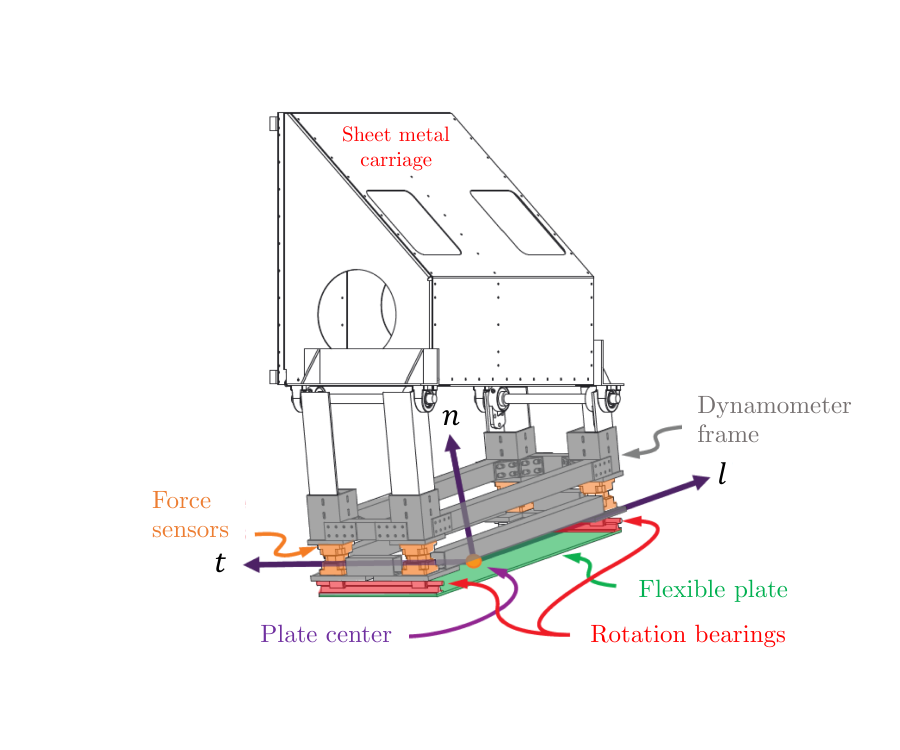}}}
\subfloat[Displacement gauge locations ]{\scalebox{1}{\includegraphics[width=0.5\textwidth]{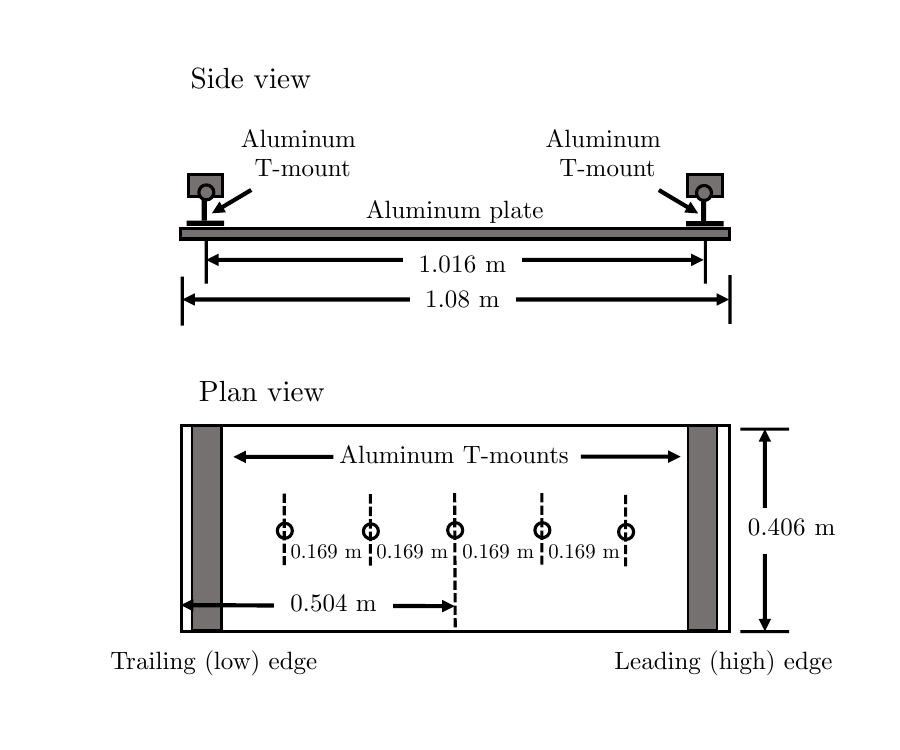}}}
\caption{Locations of the force sensors and displacement gauges for tracking the hydrodynamics and hydroelastics responses of the flexible plate slamming experiments \cite{Wang2020}.} \label{plate}
\end{figure}

The total impact force acting on the plate during the slamming process is recorded by force sensors that are attached at the four corners of the plate, as displayed in Figure \ref{plate}a. The rotation bearings shown in Figure \ref{plate}b (very stiff rails made of aluminum act as rigid stiffeners along the width of the plate) is modeled as rigid offset pinned condition in our FSI models (in order to properly capture the subtle kinematic effect). The deflection of the plate is measured by the five displacement gauges attached along the centerline of the plate, each 169 mm apart, as shown in Figure \ref{plate}b. 

\subsection{Numerical models}

\subsubsection{Fluid domain}

\begin{figure}[h!]
\centering
\includegraphics[width=1\textwidth]{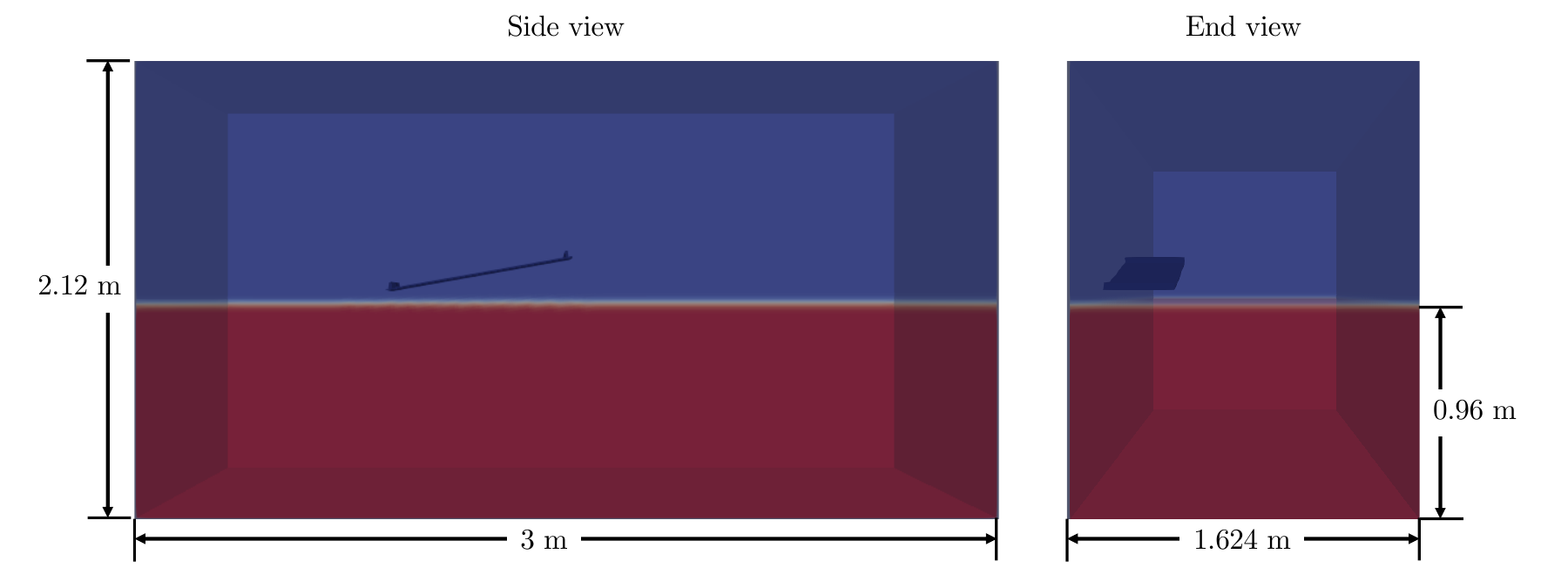}
\caption{The side view and end view of the open water channel fluid domain model; blue represents air, while red represents water.}\label{fsi_model}
\end{figure}

The fluid domain discretization is generated using a three-dimensional finite element generator, Gmsh \cite{Christophe2009}. Then, we convert the Gmsh model to a hexahedral mesh for OpenFOAM by issuing the \textit{gmshToFoam} command in OpenFOAM. The fluid model considers a truncated geometry of the open water channel to ease computational demands, while still capturing all salient physics. The water channel is modeled as a rectangular box measuring 3 m long, 1.624 m wide, and 2.12 m tall as presented in Figure \ref{fsi_model}. The rectangular box is partially filled with water (in red) that is 0.96 m deep. A ``placeholder" (\textit{i.e.,} ``cut-out" within fluid domain) for the flexible plate structural model is oriented at a $10^{\circ}$ angle of attack, and situated at 0.1 m above the water free surface (the white layer in between blue and red). 

We assume wall boundary condition (\textit{i.e.,} no slip boundary for velocity and zero gradient for pressure) on all boundaries enclosing the fluid domain, except the top plane and the left plane, looking from the end view, which we assume to be a standard atmosphere and symmetry wall boundary condition, respectively. The surfaces  of the placeholder, enclosing the flexible plate model, assume moving wall velocity and fixed flux pressure conditions. The densities of the water and air are taken as $\rho_w = 1000~kg/m^3$  and $\rho_a = 1~kg/m^3$ respectively. Furthermore, the kinematic viscosities of the water and air are given values of $\nu_w  = 1 \times 10^{-6}~m/s^2$ and $\nu_a  = 1.48 \times 10^{-5}~m/s^2$, respectively. 

\subsubsection{Structural domain}
\begin{figure}[h!]
\centering
\includegraphics[width=0.8\textwidth]{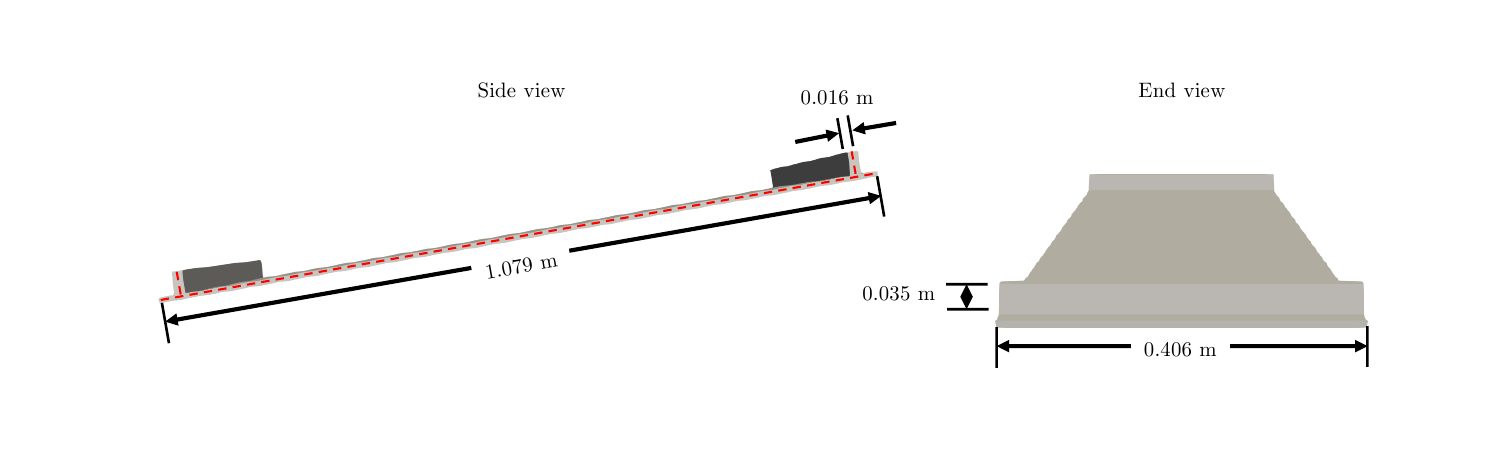}
\caption{The side view and end view of the flexible plate model; the dotted red line represents the finite element model of the flexible plate specimen (all other components correspond to the modeled experimental fixtures).}\label{plate_model}
\end{figure}

The structural model of the flexible plate is shown in Figure \ref{plate_model}. The plate is 1.079 m long and 0.406 m wide. There are two rectangular beams on two ends of the plate, to represent the high strength aluminum rails shown in Figure \ref{plate}b. The rectangular beam used to model the rails measures 0.016 m wide and 0.035 m tall. Both the flexible plate and the rigid rails are discretized using DKT shell elements: the latter being used to create the rigid offset pinned condition mentioned previously. As a result of using DKT shells, the thickness of the structure is treated as a shell cross-sectional property; thus, we place the shell elements at the mid-plane of the physical plate model (represented by the red dotted lines) to properly capture the geometry of the validation experiments. A pinned-pinned boundary condition is applied to the top ends of the rectangular beams. The flexible plate has Young's modulus, $E^s_p = 68.9~GPa$, Poisson's ratio, $\nu^s_p = 0.35$, and mass density, $\rho^s_p = 2700~kg/m^3$. The rectangular beams have Young's modulus, $E^s_b = 689~GPa$, Poisson's ratio, $\nu^s_b = 0.35$, and mass density, $\rho^s_b = 8050~kg/m^3$. We assign a high Young's modulus to the beams to ensure rigidity of the boundary rails. A summary of the engineering properties of the FSI model is provided in Table \ref{slamming_props}.

\begin{table}[h]
\centering
\caption{Engineering properties of the FSI model for flexible plate slamming tests.}\label{slamming_props} 
\resizebox{\textwidth}{!}{\begin{tabular}{l l l l}
\hline
 Fluid Properties & & Structural Properties & \\
  \hline
  Water density, $\rho_w$   & $10^3~\text{kg/m}^3$       & Plate density, $\rho_p^s$ & $2700~\text{kg/m}^3$  \\
Air  density, $\rho_a$        & $1~\text{kg/m}^3$            & Beam density, $\rho_b^s$ & $8050~\text{kg/m}^3$  \\
Water viscosity, $\mu_w$ & $10^{-6}~\text{m/s}^2$                          & Plate Young`s modulus, $ E_p^s$ & 68.9 GPa\\
 Air viscosity, $\mu_a$      & $1.48 \times 10^{-5}~\text{m/s}^2$                                               & Beam Young`s modulus, $ E_b^s$ & 689 GPa\\
                                            &                                                    & Plate Poisson ratio, $\nu_p^s$ & 0.35\\
                                           &                          & Plate Poisson ratio, $\nu_b^s$ & 0.35\\
  \hline
\end{tabular}}
\end{table} 

\subsection{FSI model validation}
In the present work, we perform a total of 24 simulated slamming impact tests on very flexible, moderately deformable, and approximately rigid aluminum plates, each characterized by plate thickness $0.00635~m$, $0.00795~m$, and $0.0127~m$ respectively. (thicker plate implies more resistance to deformation when subjected to bending.) \cite{Wang2020} The rigidity of the plate is expressed with the usual flexural rigidity, as 

\begin{equation} \label{flexural}
\begin{gathered}
D = \frac{E^s_ph^3}{12(1-(\nu_p^s)^2)},
\end{gathered}
\end{equation}
where $h$ is the plate thickness. The flexural rigidity of the three plates considered, in the order of increasing thickness, are $1675~Pa\cdot m^3$,  $3288~Pa\cdot m^3$, and $13403~Pa\cdot m^3$ respectively.

\begin{figure}[h!]
\centering
\includegraphics[width=1\textwidth]{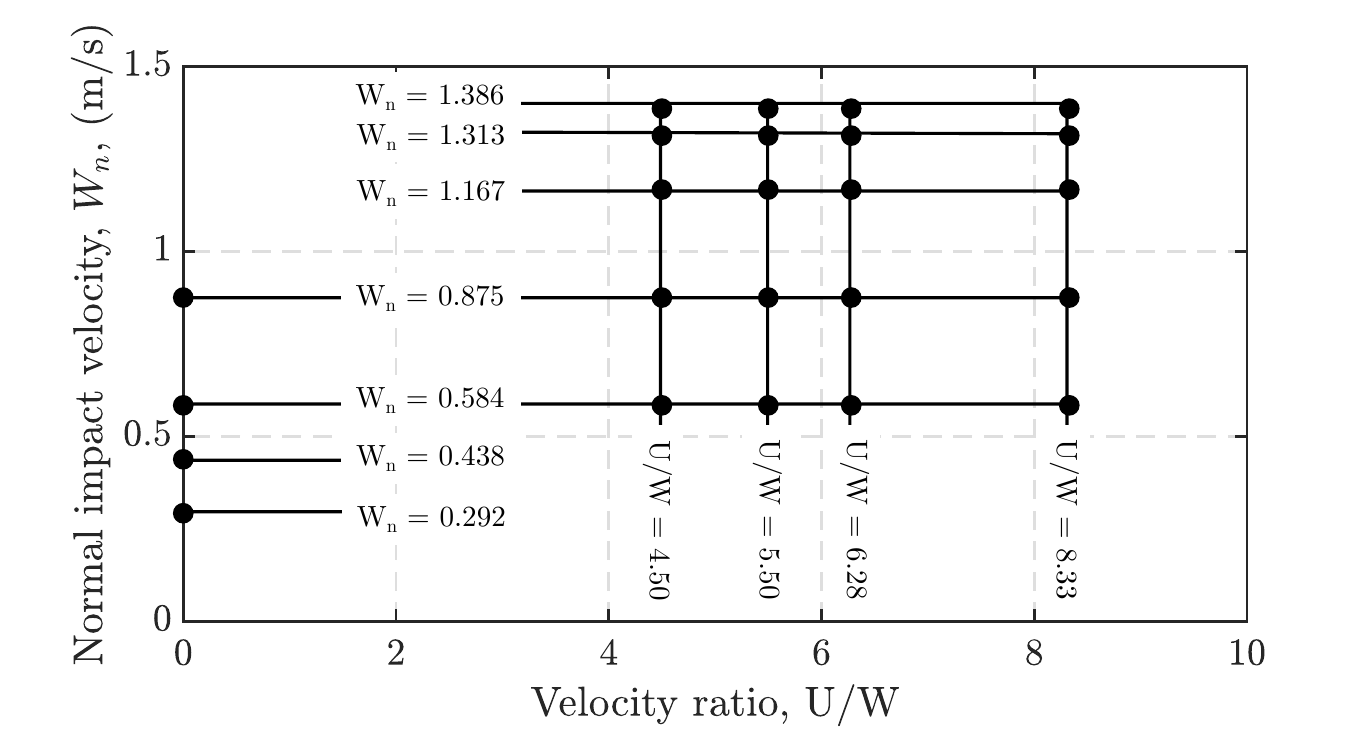}
\caption{The U/W ratios and normal impact velocities of the slamming tests from \cite{Wang2020} that are considered in the present work.}\label{test_cases}
\end{figure}

In the slamming test, the plate is subjected to a constant impact velocity in the direction normal to the plate surface (the impact velocity deviates by a maximum of $1.35\%$ in the experiment as a result of the closed-loop control system setup). The normal impact velocity can be be divided into two components, namely, the forward and downward velocities. The various slamming velocity conditions are shown in Figure \ref{test_cases}. (Laminar model is used in all FSI simulations in the present work). Therein, $U$ and $W$ denote horizontal and downward impact velocities, respectively. $W_n$ denotes impact velocities orthogonal to the plate surface, expressed as $W_n = Usin(\alpha)+Wcos(\alpha)$, where $\alpha$ is the angle of attack (in this work, $\alpha = 10^{\circ}$). In each test, we compute the displacements at the plate center and the total slamming impact force normal to the plate surface. Subsequently, we compare our FSI model results against experimental data.

All numerical simulations are performed on the Centennial high performance computing system, located at the U.S. Army Research Laboratory (ARL) DoD Supercomputing Resource Center (DSRC). Each compute node contains 40 Intel Xeon E5-2698v4 Broadwell cores, with a 2.2GHz clock speed and 128GB RAM per node. All FSI simulations are run in serial on one core (the large plate deformation in some slamming cases leads to inconsistent fluid patch area between adjacent cells among processors -- this causes simulation failures when the FSI analyses are run in parallel). 

\subsubsection{Mesh convergence tests}
\begin{figure}[h!]
\centering
\includegraphics[width=0.75\textwidth]{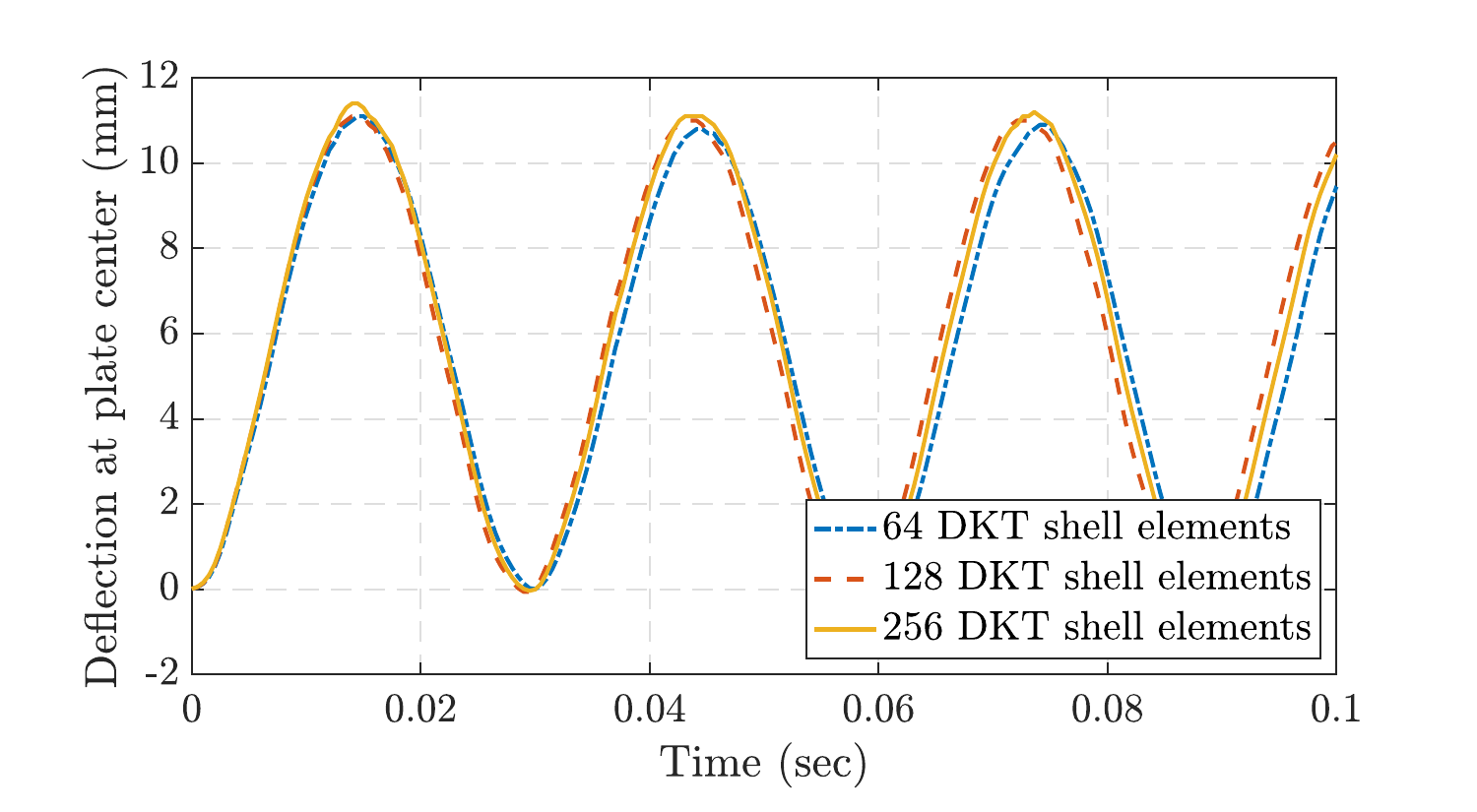}
\caption{Normal displacements at plate center of the flexible plate model using 64, 128, and 256 DKT shell elements for time $t = \left[0, 0.1\right]~sec$.}\label{csd_conv}
\end{figure}

In order to arrive at the most suitable FSI model for validation, we perform mesh convergence tests to identify a suitable mesh sizes in both the structural and fluid domains. We begin with a structural dynamics mesh convergence study. A uniformly distributed load, $P_n = 5000~N/m^2$, is applied perpendicularly to the surface of the moderately deformable plate (plate thickness 0.00795~m) for $0.1~sec$ with $\Delta t = 0.0005~sec$. The structural model is discretized into 64, 128, and 256 DKT shell elements for the mesh refinement study. The normal displacements at the plate center is obtained using the standard Newmark implicit scheme. Based on the results shown in Figure \ref{csd_conv}, we observe that 64 DKT shell elements slightly underpredicts the displacements, while displacement results with 128 and 256 DKT shell elements are virtually identical. Since the 256 DKT shell model does not provide much improvement in displacement predictions, compared to the 128 DKT shell model, we use 128 DKT shell model for the FSI validation tests. 

\begin{figure}[h!]
\centering
\subfloat[]{\scalebox{0.5}{\includegraphics[width=1\textwidth]{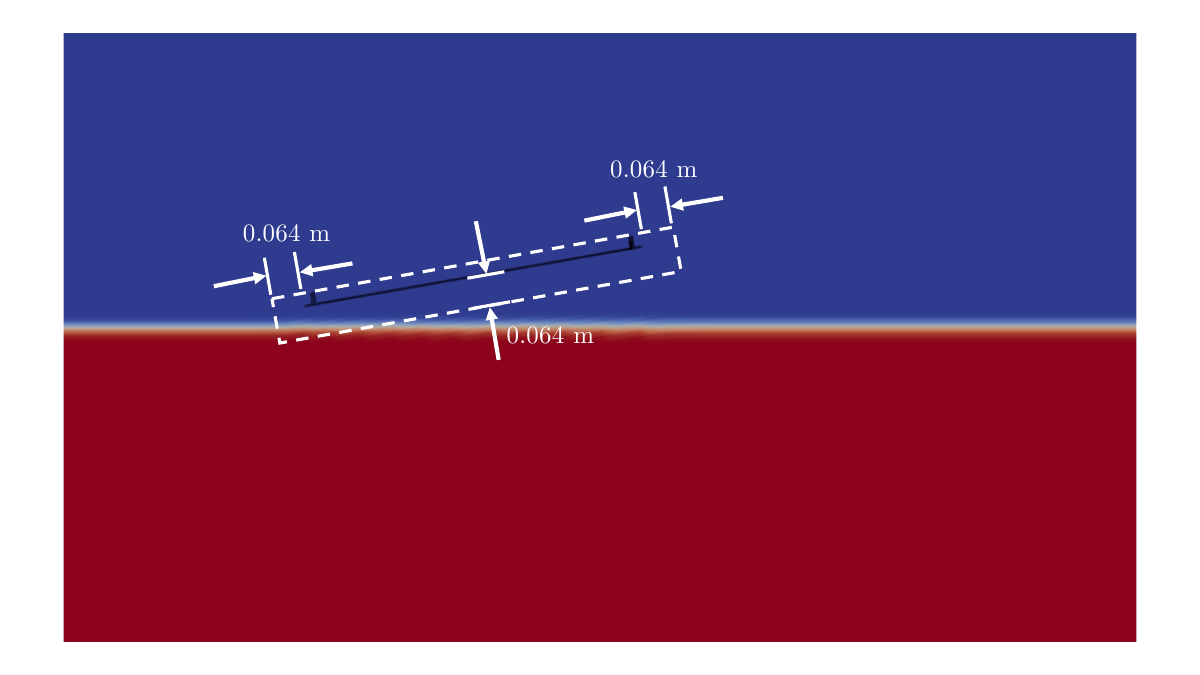}}}
\subfloat[]{\scalebox{0.5}{\includegraphics[width=1\textwidth]{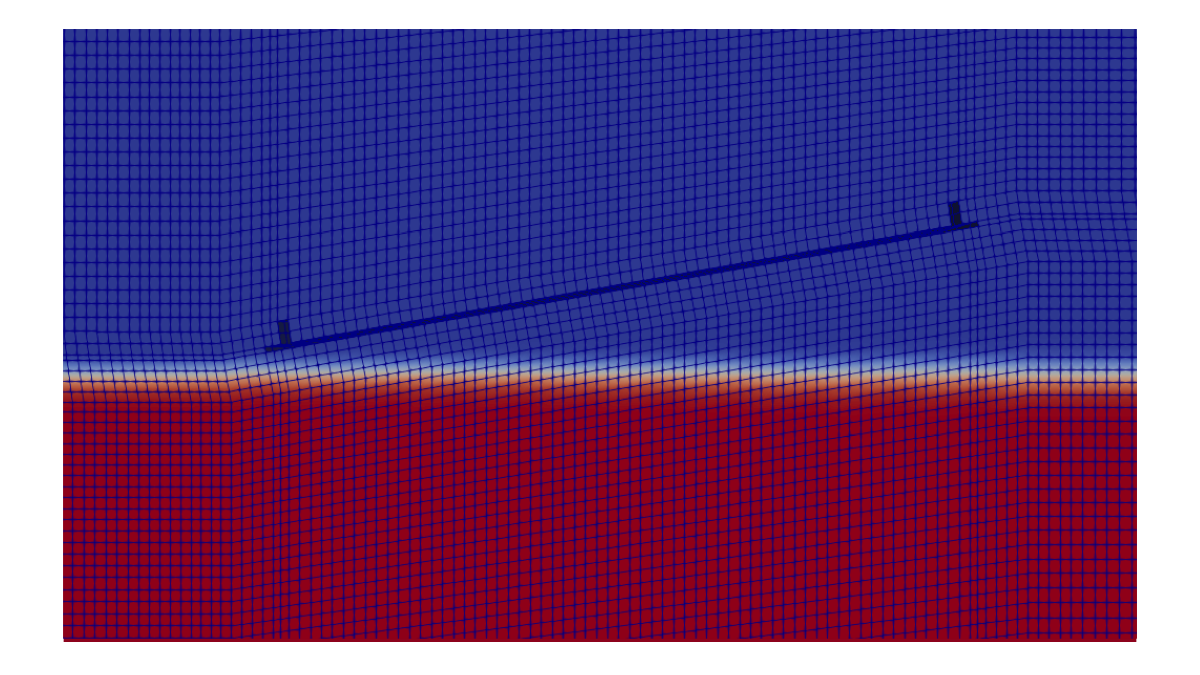}}}
\caption{A snapshot of the FSI simulation. (a) The region enclosed by the dotted white line is subjected to grid refinement to identify a suitable mesh size for the FSI slamming analysis. The dimension of each fluid cell within the refinement box is: $dx = 0.018~m$, $dy = 0.068~m$, and $dz = 0.015~m$. (b) The fluid mesh resolution of the medium mesh in the region near the plate.}\label{cfd_mesh}
\end{figure}

After we obtain a satisfactory structural model, we focus on the fluid domain, and the OpenFOAM mesh discretization that offers the best FSI results in both the plate displacement and total normal impact force. We focus on refining the mesh in the region bounded by the dotted black line in Figure \ref{cfd_mesh}. (the physical behavior within the refinement box is most relevant to the hydrodynamics and hydroelastics histories of the slamming test.) Furthermore, the mesh size away from the refinement box is adjusted correspondingly, based on appropriate aspect ratios (\textit{i.e.,} the mesh is proportionally coarsened w.r.t it's distance from the plate). The fluid mesh within the refinement box is discretized into $348$, $3120$, and $21218$ fluid cells respectively, for the mesh refinement study. 

\begin{figure}[h!]
\centering
\subfloat[]{\scalebox{0.7}{\includegraphics[width=1\textwidth]{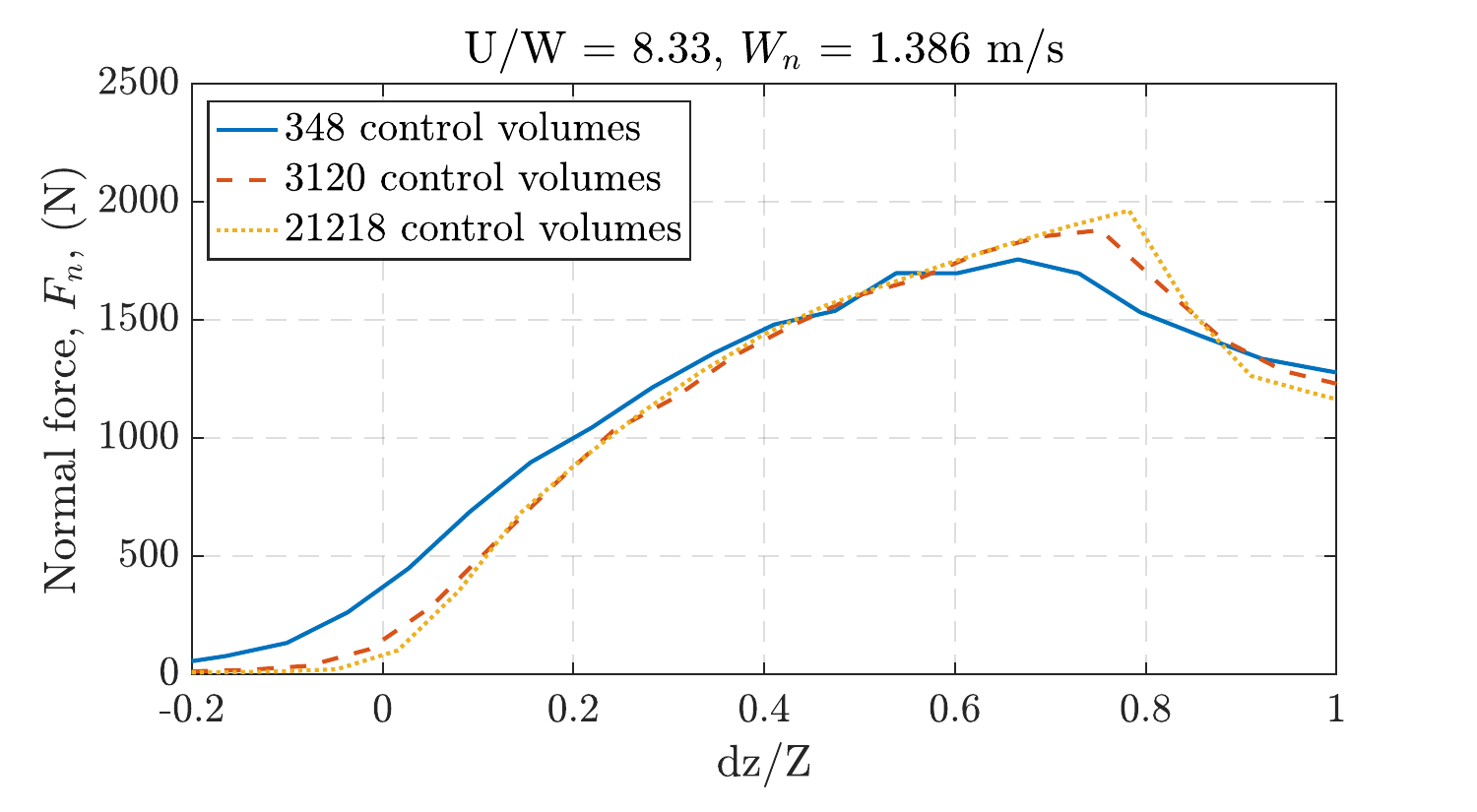}}}
\subfloat[]{\scalebox{0.3}{\includegraphics[width=1\textwidth]{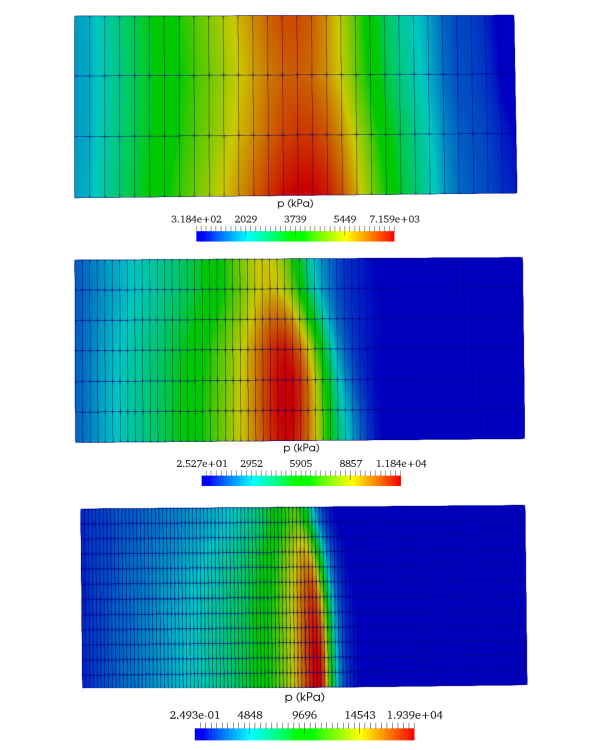}}}
\caption{(a) Total normal impact forces of the fluid model with 348, 3120, 21218 control volumes within the refinement box; (b) the pressure profile on all three plates with varying mesh densities.} \label{cfd_conv}
\end{figure}

We perform the mesh refinement study on the fluid model using a slamming condition of $U/W = 8.33$ and $W_n = 1.386~m/s$. The plate is assumed rigid in this case. The total normal impact force acting on the plate, for the three different refinements, are shown in Figure \ref{cfd_conv}. In Figure \ref{cfd_conv}, $dz/Z \in [-0.2 , 0]$ suggests the plate is in air -- the plate is reaching toward the water free surface as $dz/Z$ increases. Positive $dz$ represents the total vertical distance the plate has traveled in water, and $Z$ represents the total vertical distance from the lower to the upper edge of the plate. In other words, $dz/Z \in [0, 1]$ represents the fraction of the plate submerged within the water.  Since the free surface in the present work is modeled by linearly interpolating the fluid properties between the air and water using a VOF formulation, the non-zero fluid density of the free surface introduces additional force to the plate as it approaches the free surface. As such, non-zero forces are observed for zero submergence in the range closes to $dz/Z = 0$ in Figure \ref{cfd_conv}. However, as the mesh is refined, the artificial forces at the free surface are reduced. 

It is observed that all three meshes produce somewhat similar force histories. Figure \ref{cfd_conv}b shows the pressure profile along the plate for all three meshes -- the finest mesh shows excellent performance in resolving the sharp pressure peak. The CPU times for the three meshes, with increasing level of refinement, are 1071~sec, 9543~sec, and 61702~sec, respectively. Given that the medium mesh performs nearly as well as the fine mesh but requires much less computational time (which is a limitation in the present work) and that the total impact force is more relevant to the present work, the fluid mesh with 3120 control volumes within the refinement box is chosen for carrying out all FSI analyses.  

 \subsubsection{Validation results} \label{validation_results}
\begin{figure}[h!]
\centering
\captionsetup[subfigure]{justification=centering}
\subfloat[]{\scalebox{0.5}{\includegraphics[width=1\textwidth]{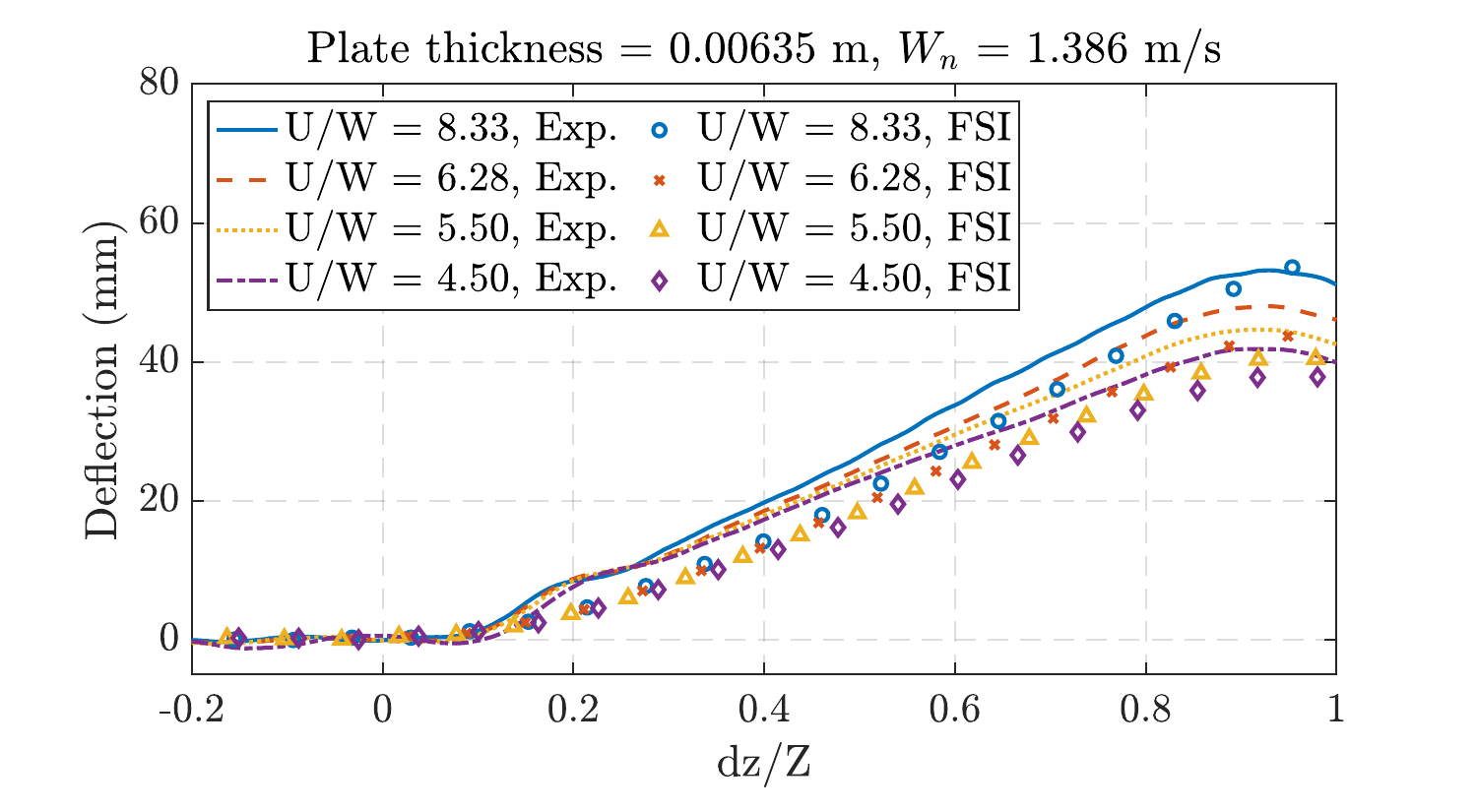}}}
\subfloat[]{\scalebox{0.5}{\includegraphics[width=1\textwidth]{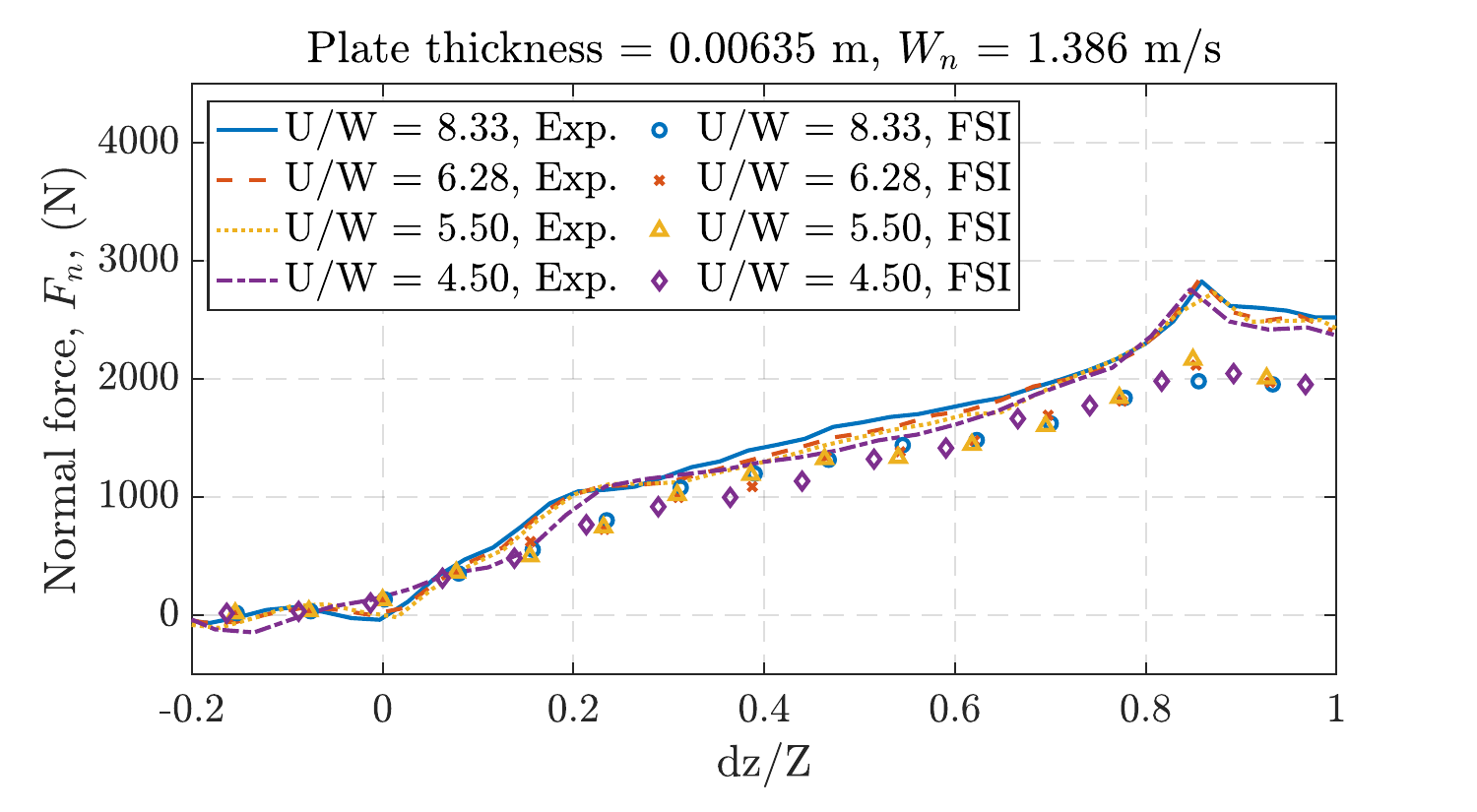}}} \\
\subfloat[]{\scalebox{0.5}{\includegraphics[width=1\textwidth]{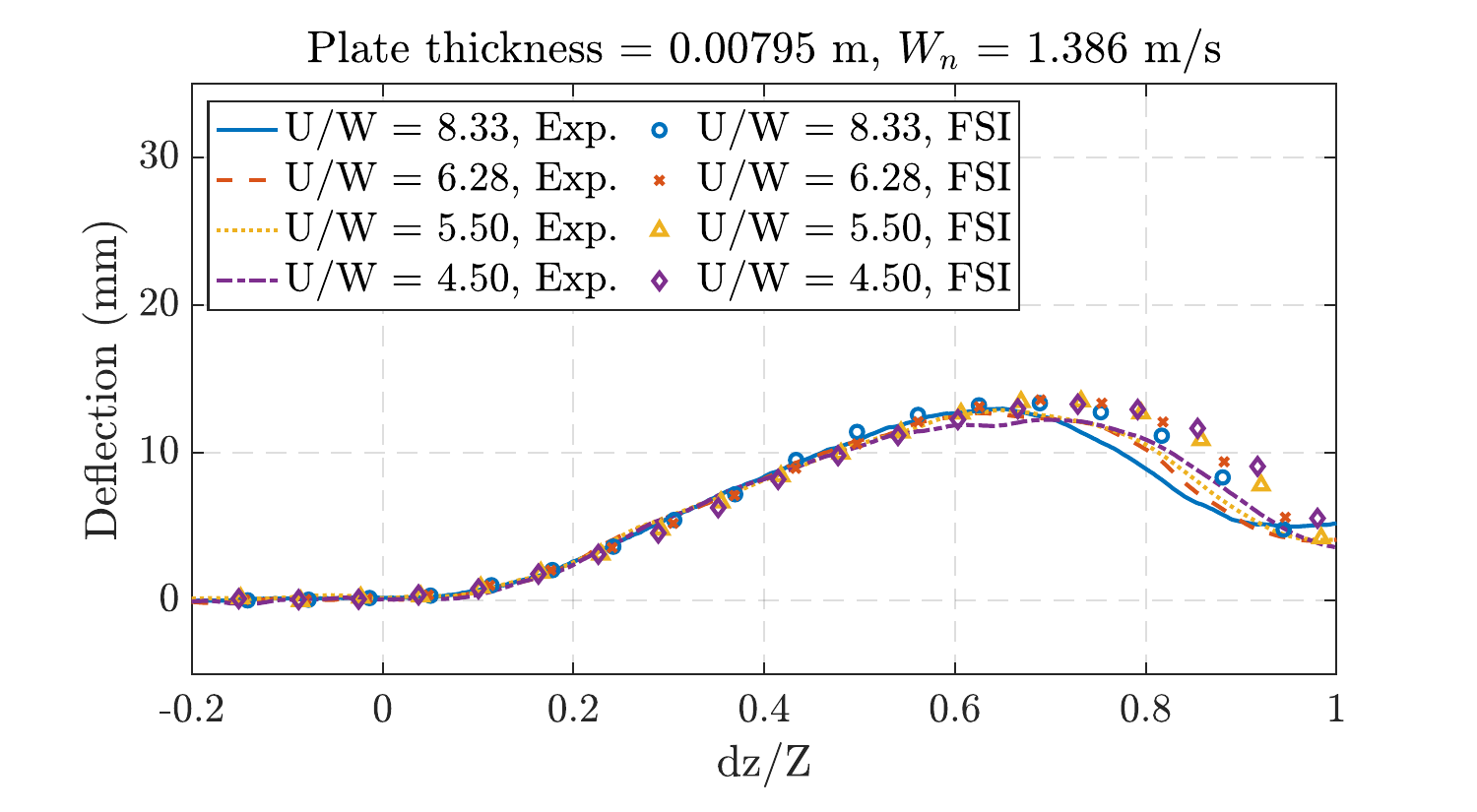}}}
\subfloat[]{\scalebox{0.5}{\includegraphics[width=1\textwidth]{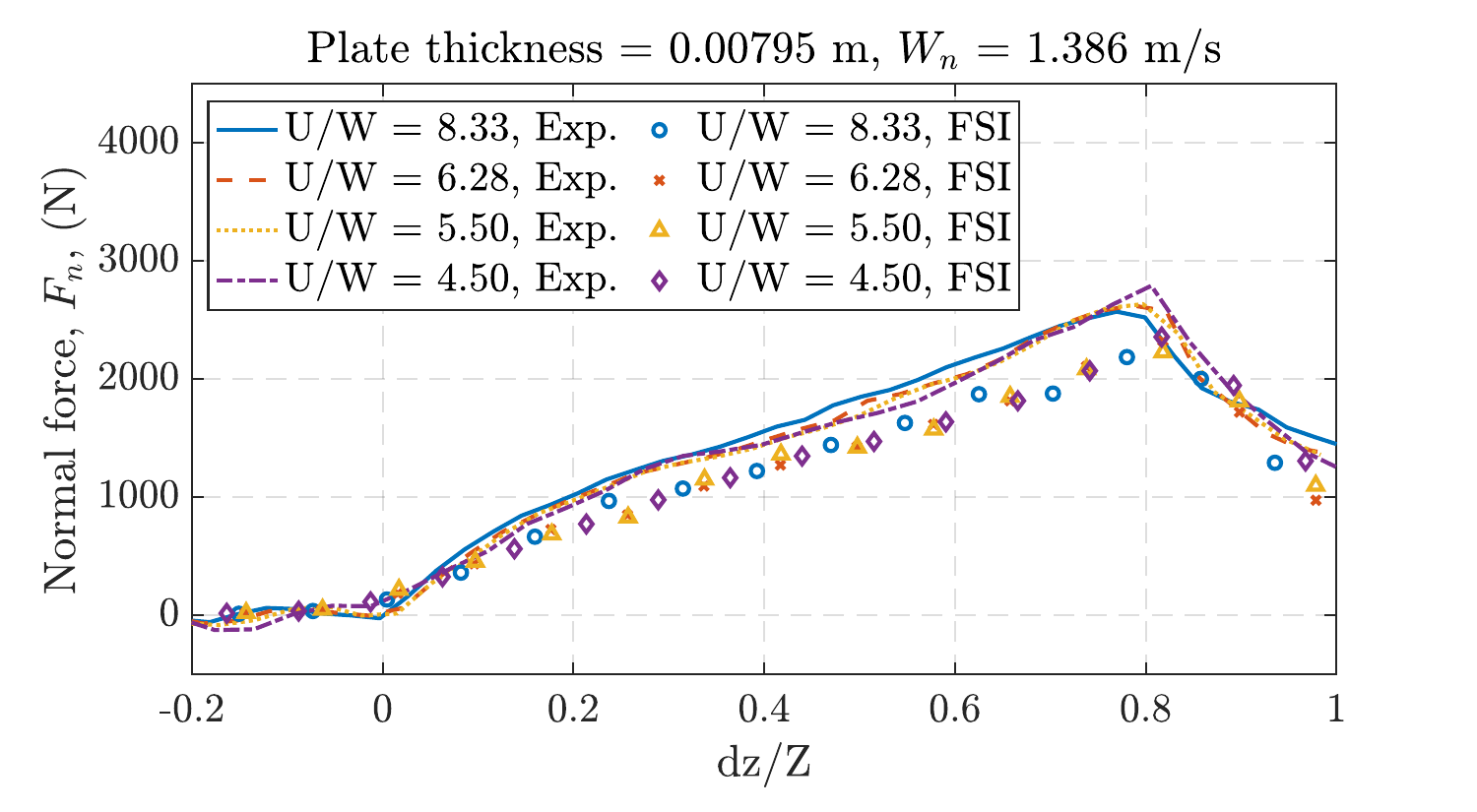}}}\\
\subfloat[]{\scalebox{0.5}{\includegraphics[width=1\textwidth]{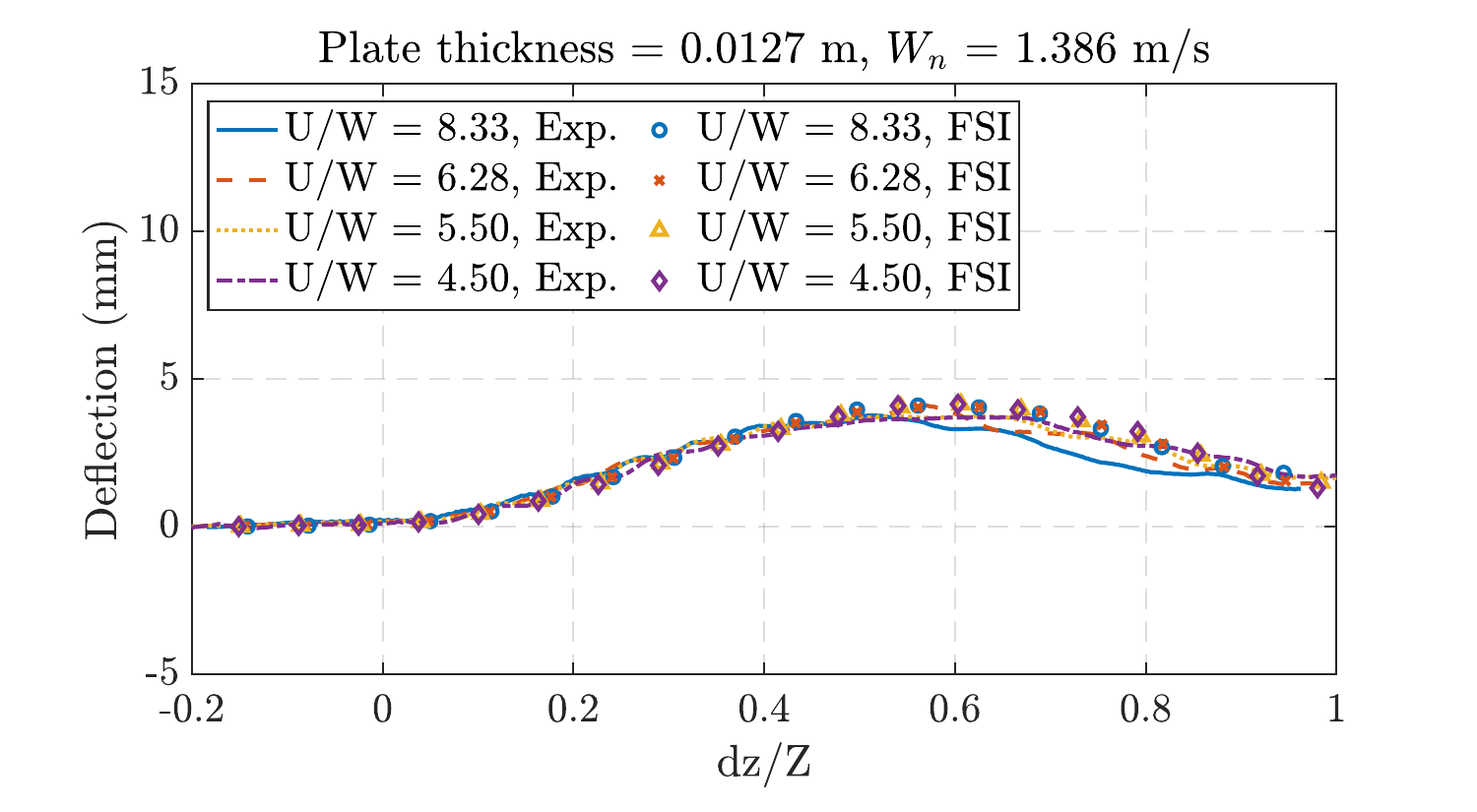}}} 
\subfloat[]{\scalebox{0.5}{\includegraphics[width=1\textwidth]{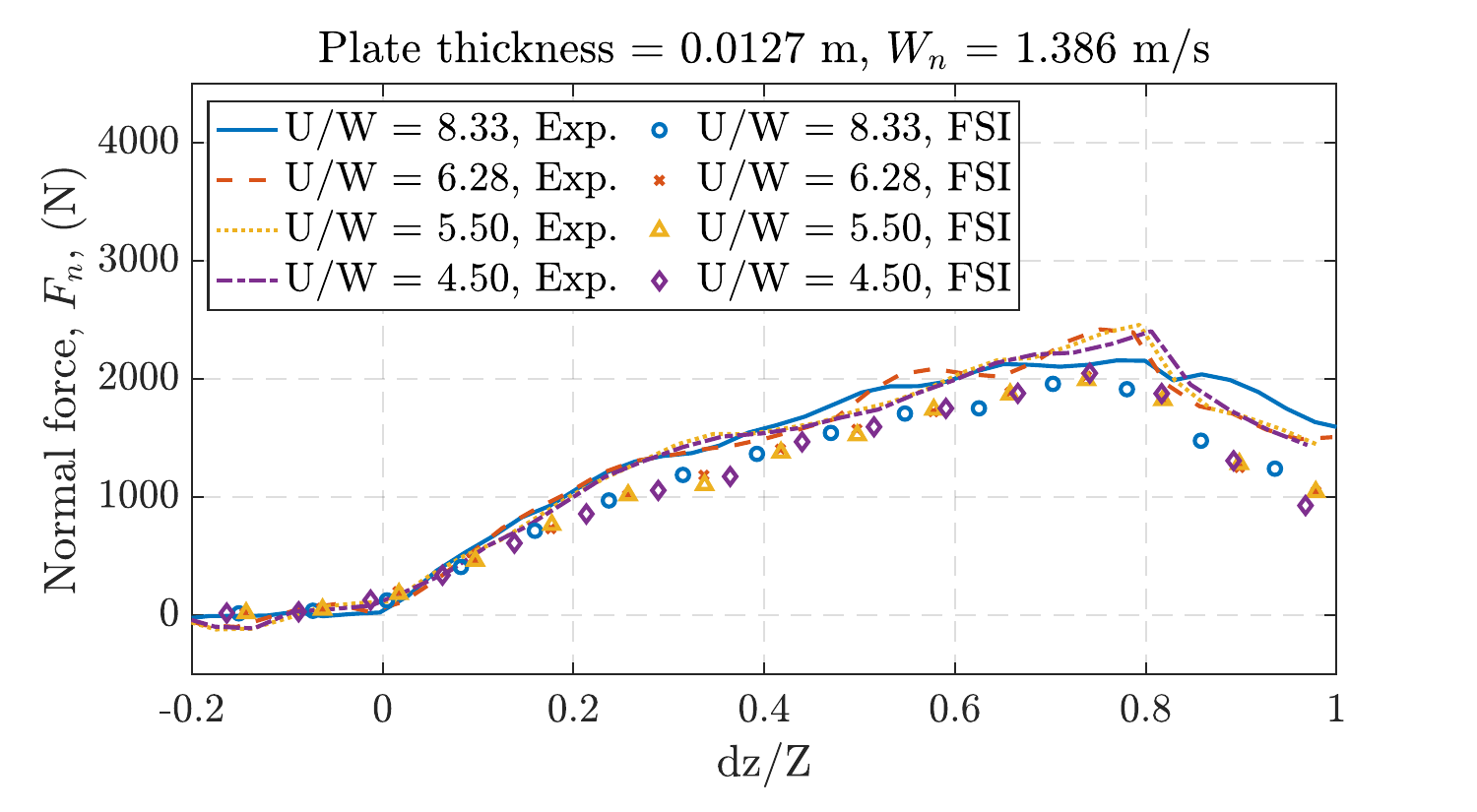}}}
\caption{Plate deflection and normal impact force histories of the highly flexible (with plate thickness 0.00635~m), moderately deformable (with plate thickness 0.00795~m), and nearly rigid (with plate thickness 0.0127~m) plates subjected to slamming impacts with $W_n = 1.386$~m/s.} \label{fr0p426}
\end{figure}

In the following, we will present three sets of validation results. First, we present validation results for oblique slamming scenarios with $W_n = 1.386~m/s$ and $U/W = 8.33, 6.28, 5.50,$ and $4.50$; higher $U/W$ ratio indicates higher forward velocity ($U$) but lower downward velocity ($W$). The $U$ and $W$ of the slamming tests in this validation set are listed in Table \ref{val_test1}. Displacement and normal force responses for plate thickness 0.00635~m, 0.00795~m, and 0.0127~m are shown in Figure \ref{fr0p426}. In Figure \ref{fr0p426}a, c, and e, we observe that thinner plates have higher deflections at the plate center. This observation agrees with our intuition. (as the plate gets thinner, it has lower flexural rigidity, and therefore it admits larger deformation.) Figure \ref{fr0p426}c and \ref{fr0p426}e imply that plate deformation is independent of $U/W$ ratios. This means for plates with thickness 0.00795~m, or larger (\textit{i.e.,} flexural rigidity $3288~Pa\cdot m^3$ or higher), forward impact velocity has negligible influence on plate deformations, as long as the normal impact velocity is fixed. However, the observation mentioned does not apply to the case with plate thickness of 0.00635~m. For such highly flexible plates, higher $U/W$ leads to larger plate deformation, even if the plates are subjected to the same normal impact. 

\begin{table}[h]
\centering
\caption{Forward and downward velocities of slamming tests with $W_n = 1.386$~m/s. $F_r$ denotes the Froude number expressed as $F_r = \frac{W_n}{\sqrt{gL}}$, where $g$ is gravity and $L$ is the length of the plate.} \label{val_test1} 
\begin{tabular}{ccccc} 
\hline
$W_n$ (m/s)          & U/W         & U (m/s)              & W (m/s)    & $F_r$ \\
  \hline      
  1.386         &  8.33        & 4.75                          & 0.57       &    0.426 \\
  1.386          &  6.28        & 4.19                          & 0.668     &    0.426 \\
  1.386           &  5.50        & 3.93                          & 0.715     &    0.426 \\
  1.386          &  4.50        & 3.53                          & 0.785      &     0.426 \\ 
  \hline
\end{tabular}
\end{table} 

In Figure \ref{fr0p426}b, d, and f, it is observed that the normal impact force histories resulting from the slamming impacts are almost identical between plate thickness 0.00795~m and 0.0127~m; these two plates experience linear force loading within the interval $dz/Z = [0, 0.8]$ and forces unloading within $dz/Z = [0.8, 1]$. For plate thickness 0.00635~m, a slightly higher peak impact force is observed, along with a delay in force unloading. Also, in contrast to the stiffer plates, the 0.00635 plate experiences nonlinear force loading within the range $dz/Z = [0, 0.85]$, and the impact force levels off afterward. Investigations into the inconsistency in structural behaviors with regard to plate rigidity are discussed in Section \ref{applications}, wherein we identify the plate thickness at which the $U/W$ ratio becomes important to plate deformations and normal impact force behaviors. 

The $L_2$ relative error norms in impact forces and plate deformations of the first set of FSI validation tests are presented in Table \ref{val1_error}; these norms are defined as 

\begin{equation} \label{L2}
L_2 \,\, \text{relative error norm} = \sqrt{\frac{\sum_{i=0}^{N}(S_{exp, i}-S_{fsi, i})^2}{\sum_{i=0}^{N}S_{exp, i}^2}}
\end{equation}
where $S_{exp, i}$ represents the $i^{th}$ index of the experiment solution vector, $S_{fsi, i}$ represents the $i^{th}$ index of the FSI solution vector, and $N$ represents the total number of sampled points. The $L_2$ error norms in impact forces among the three plates fall within the interval $\left[0.146, 0.201\right]$, with a median of 0.164 and a mean of 0.165. Likewise, the $L_2$ error norms in displacements among the three plates occur within the interval $\left[0.104, 0.222\right]$, with a median of 0.152 and a mean of 0.149. The $L_2$ error norms are consistent in this validation set and are only a small fraction of the typical impact forces and plate deflections; thus indicating favorable agreements between the models and experiments. 
 
 \begin{table}[h!]
\centering
\caption{$L_2$ relative error norms in slamming impact forces and plate deformations from slamming tests with $W_n = 1.386$~m/s.} \label{val1_error} 
\resizebox{\textwidth}{!}{\begin{tabular}{ccccccc}
 \hline  
 & \multicolumn{3}{c}{ $L_2$ error, impact force } &  \multicolumn{3}{c}{ $L_2$ error, displacement} \\
Plate thickness (m)  &  0.00635 & 0.00795   & 0.0127 & 0.00635 & 0.00795   & 0.0127 \\
 \hline   
U/W 8.33   &  0.201        &  0.170     & 0.161       & 0.132       & 0.149        & 0.222  \\
U/W 6.28   &  0.178        & 0.149      & 0.159        & 0.155       &  0.159       & 0.128 \\
U/W 5.50   &   0.171       &  0.146        & 0.160              & 0.158       &  0.146            &  0.104    \\ 
U/W 4.50   &   0.167       & 0.150           & 0.156           & 0.162       & 0.160           & 0.113  \\ 
  \hline
\end{tabular}}
\end{table} 

\begin{figure}[h!]
\centering
\captionsetup[subfigure]{justification=centering}
\subfloat[]{\scalebox{0.5}{\includegraphics[width=1\textwidth]{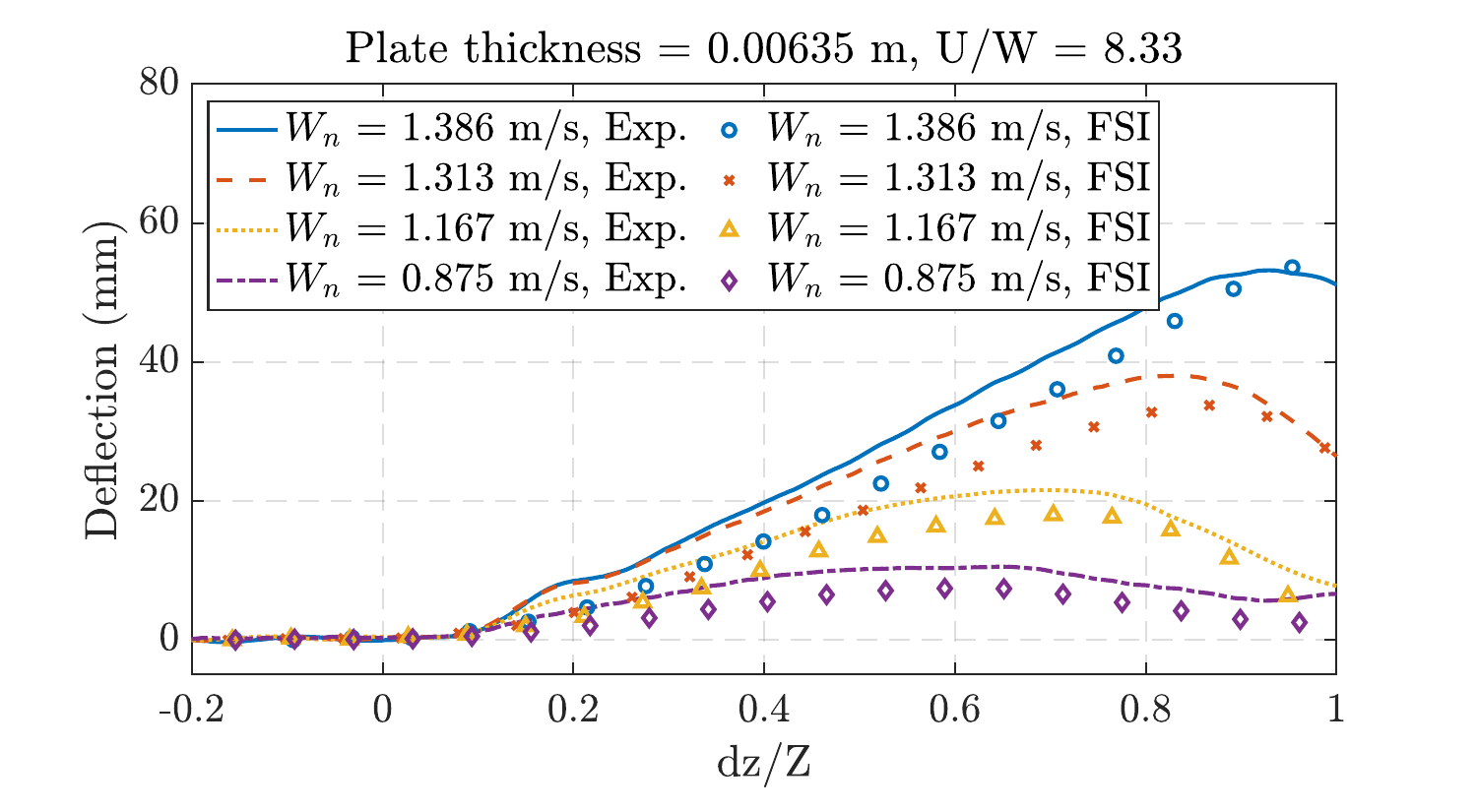}}}
\subfloat[]{\scalebox{0.5}{\includegraphics[width=1\textwidth]{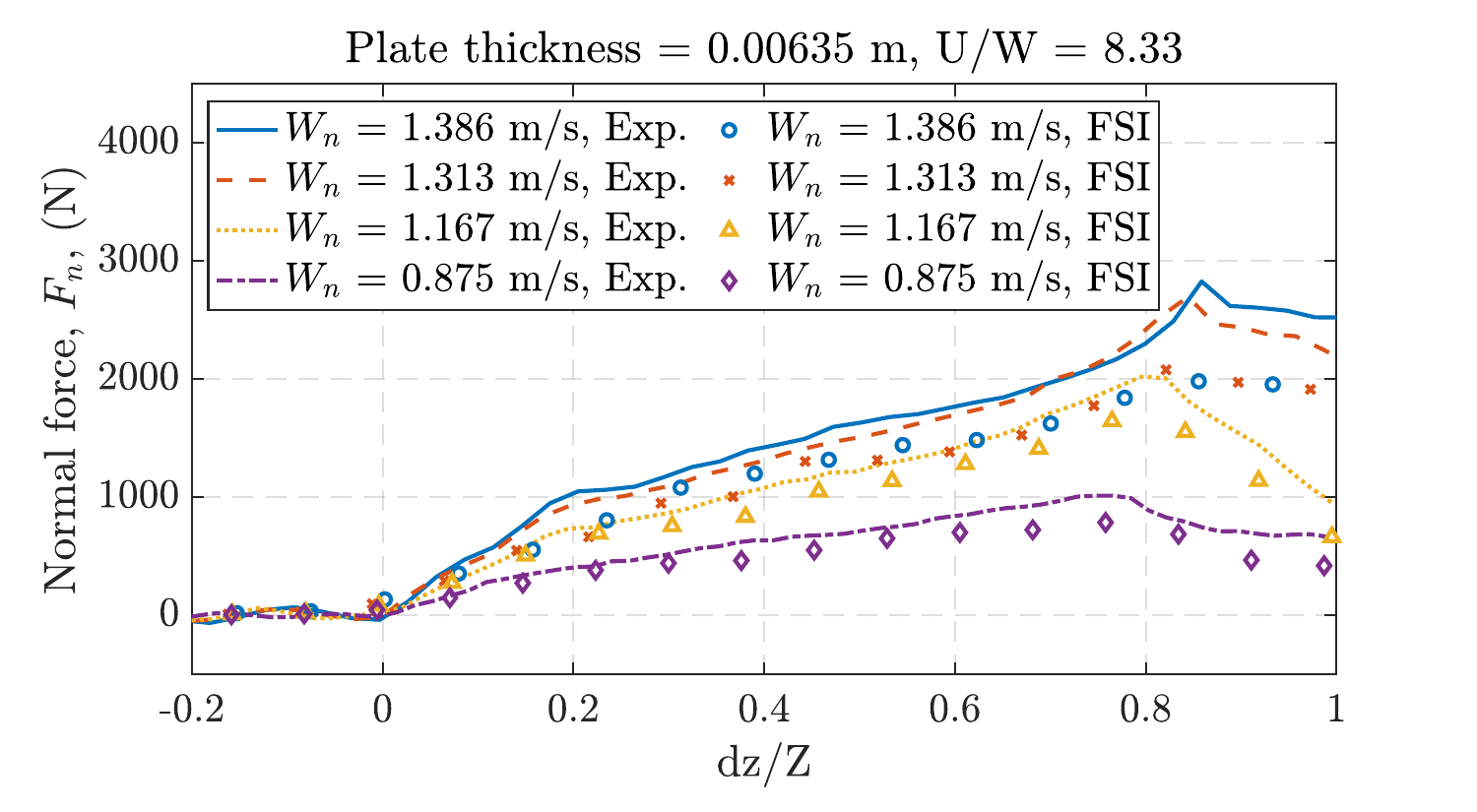}}} \\
\subfloat[]{\scalebox{0.5}{\includegraphics[width=1\textwidth]{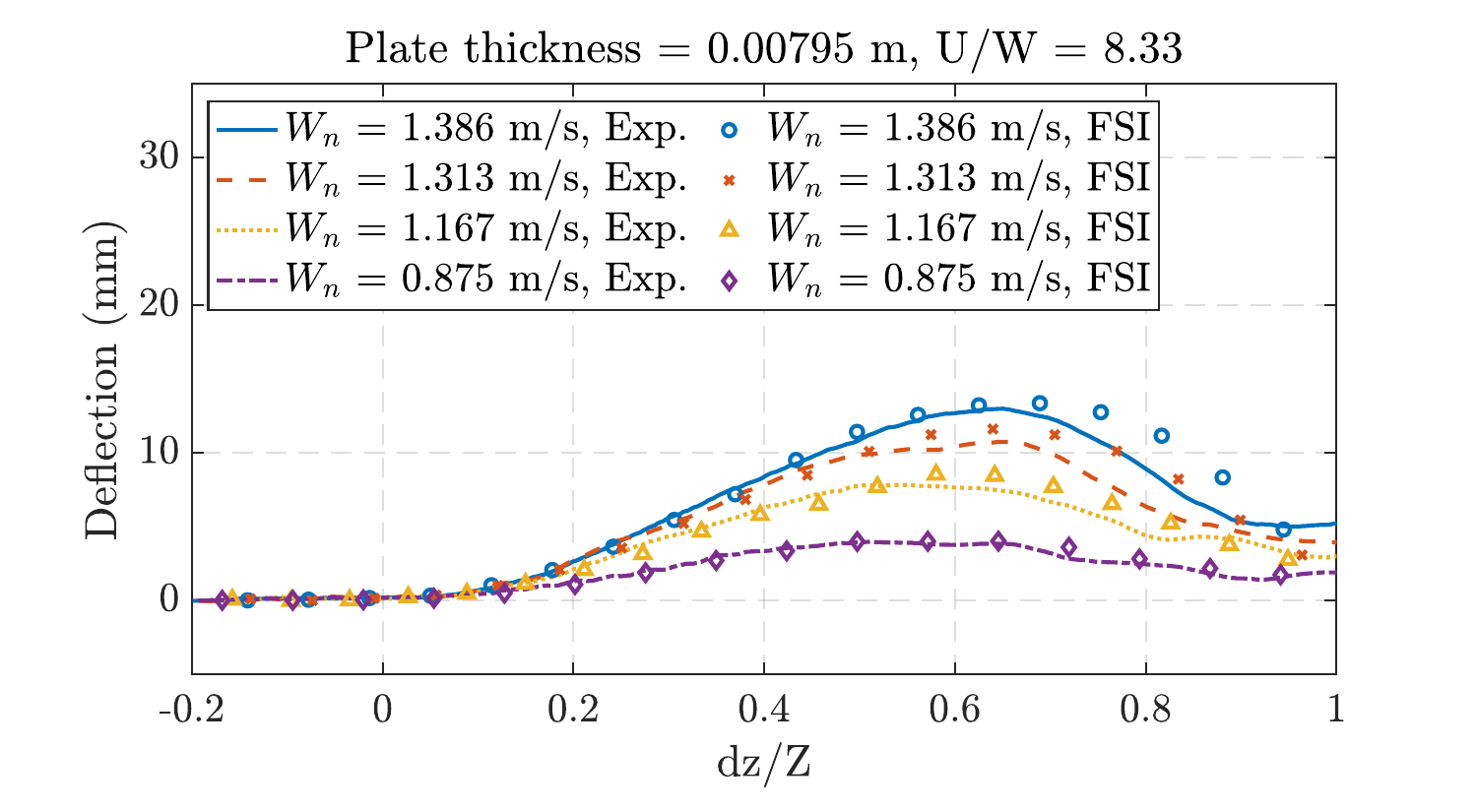}}}
\subfloat[]{\scalebox{0.5}{\includegraphics[width=1\textwidth]{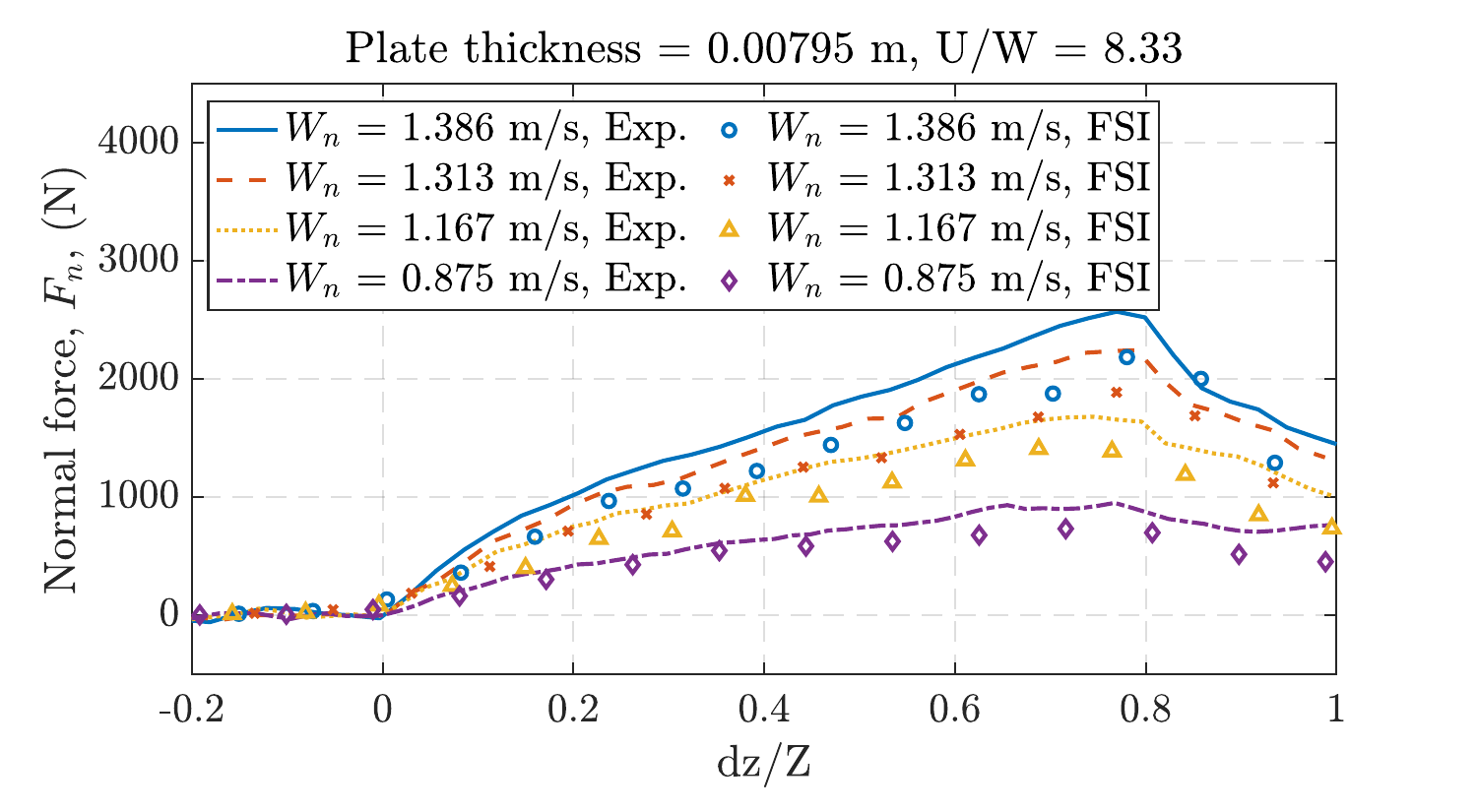}}} \\
\subfloat[]{\scalebox{0.5}{\includegraphics[width=1\textwidth]{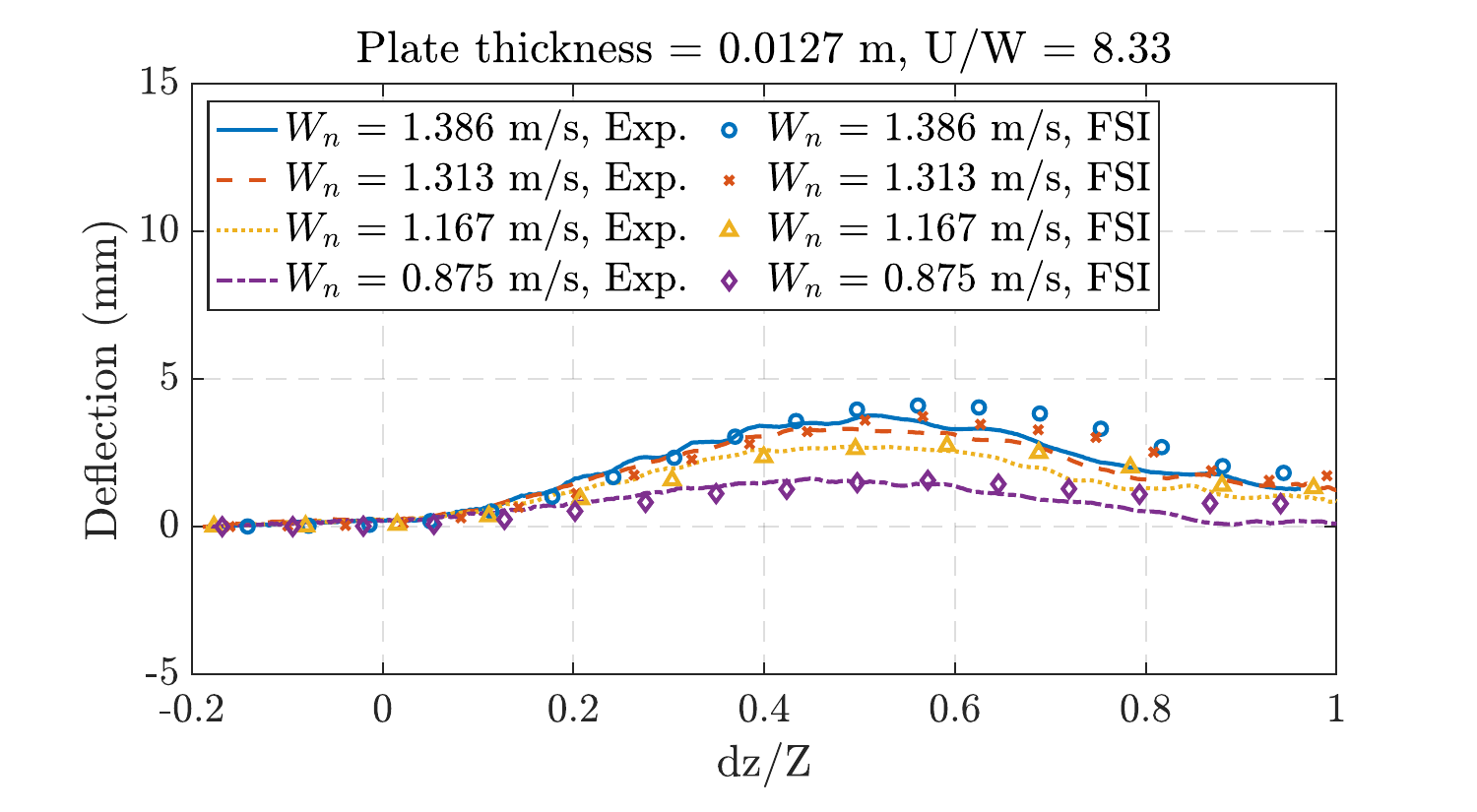}}} 
\subfloat[]{\scalebox{0.5}{\includegraphics[width=1\textwidth]{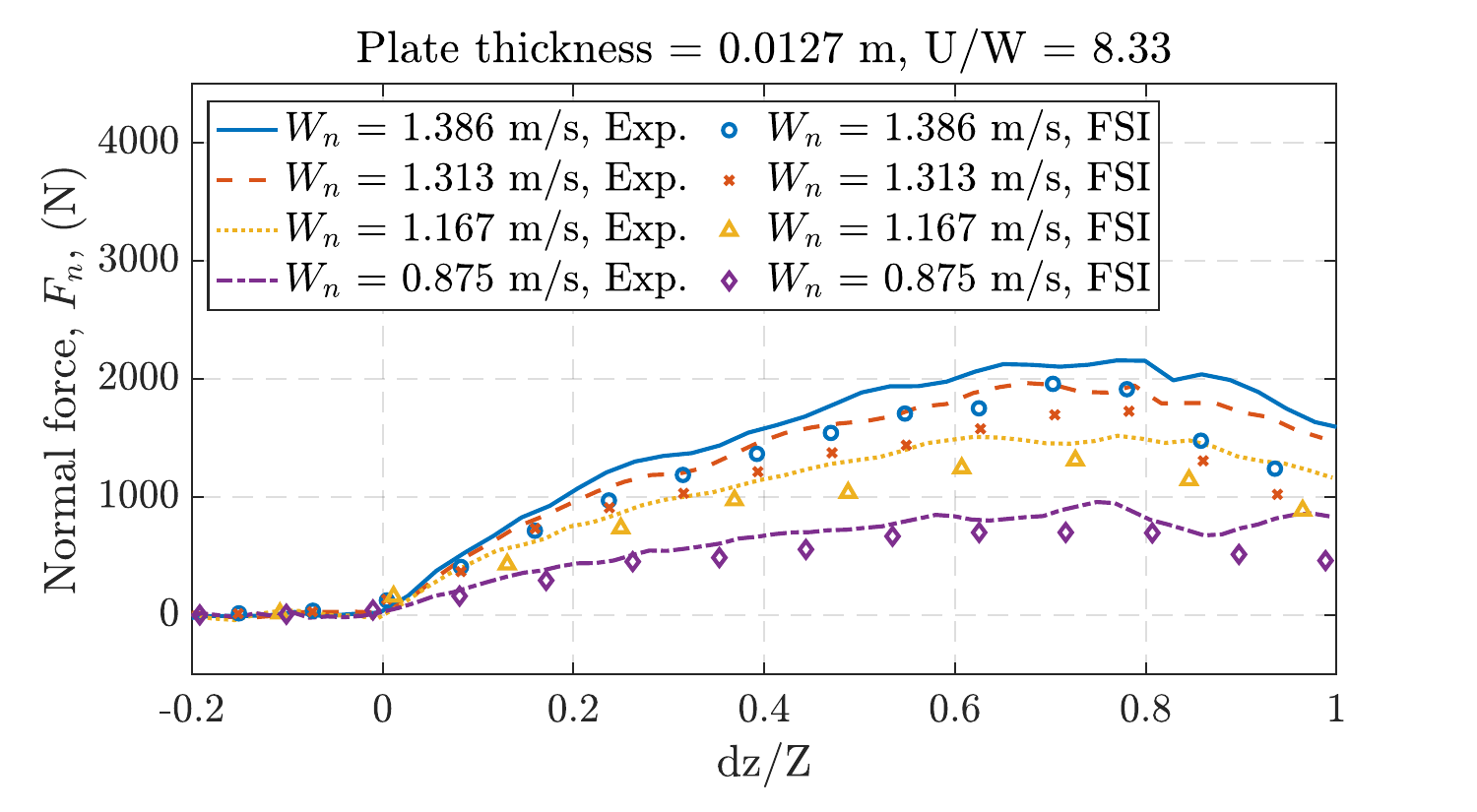}}}
\caption{Plate deflection and normal impact force histories of the highly flexible (with plate thickness 0.00635~m), moderately deformable (with plate thickness 0.00795~m), and nearly rigid (with plate thickness 0.0127~m) plates subjected to slamming impacts with $U/W = 8.33$.} \label{UW8p33}
\end{figure}

In the second set of FSI validation tests, we present validation results of slamming cases with $U/W = 8.33$ and $W_n = 1.386, 1.313, 1.167,$ and $0.875$~m/s; higher $W_n$ indicates higher $U$ and $W$. The $U$ and $W$ of the slamming tests are listed in Table \ref{val_test2}. In Figure \ref{UW8p33}a, c, and e, we observe that plate deflection increases with increasing $W_n$ and decreasing plate thickness. This observation is reasonable and agrees with our intuition. In Figure \ref{UW8p33}b, d, and f, once again, we observe nearly identical force histories in 0.00795~m and 0.0127~m plates, wherein the plates experience bilinear force loading-unloading behaviors. In Figure \ref{UW8p33}b, we observe that the 0.00635~m plate experiences nonlinear force loading in $W_n = 1.386$~m/s and $W_n = 1.313$~m/s, while in $W_n = 1.167$~m/s and $W_n = 0.875$~m/s, the 0.00635~m plate follows a similar bilinear force pattern as the two stiffer plates. These diverging structural behaviors in Figure \ref{UW8p33}b suggest that the plate's oblique slamming response is related to its rigidity and the intensity of impact.
 
\begin{table}[h!]
\centering
\caption{Forward and downward velocities of slamming tests with $U/W = 8.33$.} \label{val_test2} 
\begin{tabular}{ccccc} 
\hline
$W_n$ (m/s)         & U/W         & U (m/s)                     & W (m/s)  & $F_r$ \\
  \hline      
  1.386         &  8.33        & 4.75                          & 0.57   &    0.426 \\
  1.313         &  8.33        & 4.5                            & 0.54   &   0.404   \\
  1.167        &  8.33        & 4                               & 0.48    &    0.359   \\
  0.875         &  8.33        & 3                               & 0.36    &    0.269  \\ 
  \hline
\end{tabular}
\end{table} 

The $L_2$ relative error norms in impact forces and plate deformations of the second set of FSI validation tests are presented in Table \ref{val2_error}. The $L_2$ error norms in impact forces among the three plates fall within the interval $\left[0.161, 0.226\right]$, with a median of 0.180 and a mean of 0.187. The $L_2$ error norms in displacements among the three plates occur within the interval $\left[0.104, 0.379\right]$, with a median of 0.1885 and a mean of 0.205. The second validation set yields higher $L_2$ error norms, compared to the first FSI validation set. In Table \ref{val2_error}, we observe that, in general, the $L_2$ error norm increases with decreasing $W_n$. Interestingly, we observe in Figure \ref{UW8p33} that the discrepancies between the FSI and experimental results are somewhat consistent among all four slamming tests. Because lower $W_n$ yields lower impact forces and plate deformations, any modeling errors become more significant, as a percentage of response, in these cases. 
 
 \begin{table}[h!]
\centering
\caption{$L_2$ relative error norms in slamming impact forces and plate deformations from slamming tests with $U/W = 8.33$.} \label{val2_error} 
\resizebox{\textwidth}{!}{\begin{tabular}{ccccccc}
 \hline  
 & \multicolumn{3}{c}{ $L_2$ error, impact force } &  \multicolumn{3}{c}{ $L_2$ error, displacement} \\
Plate thickness (m)  &  0.00635 & 0.00795   & 0.0127 & 0.00635 & 0.00795   & 0.0127 \\
 \hline   
$W_n$ 1.386  &  0.201        &  0.170     & 0.161       & 0.132       & 0.149        & 0.222  \\
$W_n$ 1.313    &  0.177        & 0.182      & 0.171        & 0.182       &  0.168       & 0.205 \\
$W_n$ 1.167   &   0.166       &  0.194        & 0.178              & 0.228     &  0.124           &  0.195    \\ 
$W_n$ 0.875    &   0.201       & 0.226           & 0.219          & 0.379       & 0.104          & 0.377  \\ 
  \hline
\end{tabular}}
\end{table} 

\begin{figure}[h!]
\centering
\captionsetup[subfigure]{justification=centering}
\subfloat[]{\scalebox{0.5}{\includegraphics[width=1\textwidth]{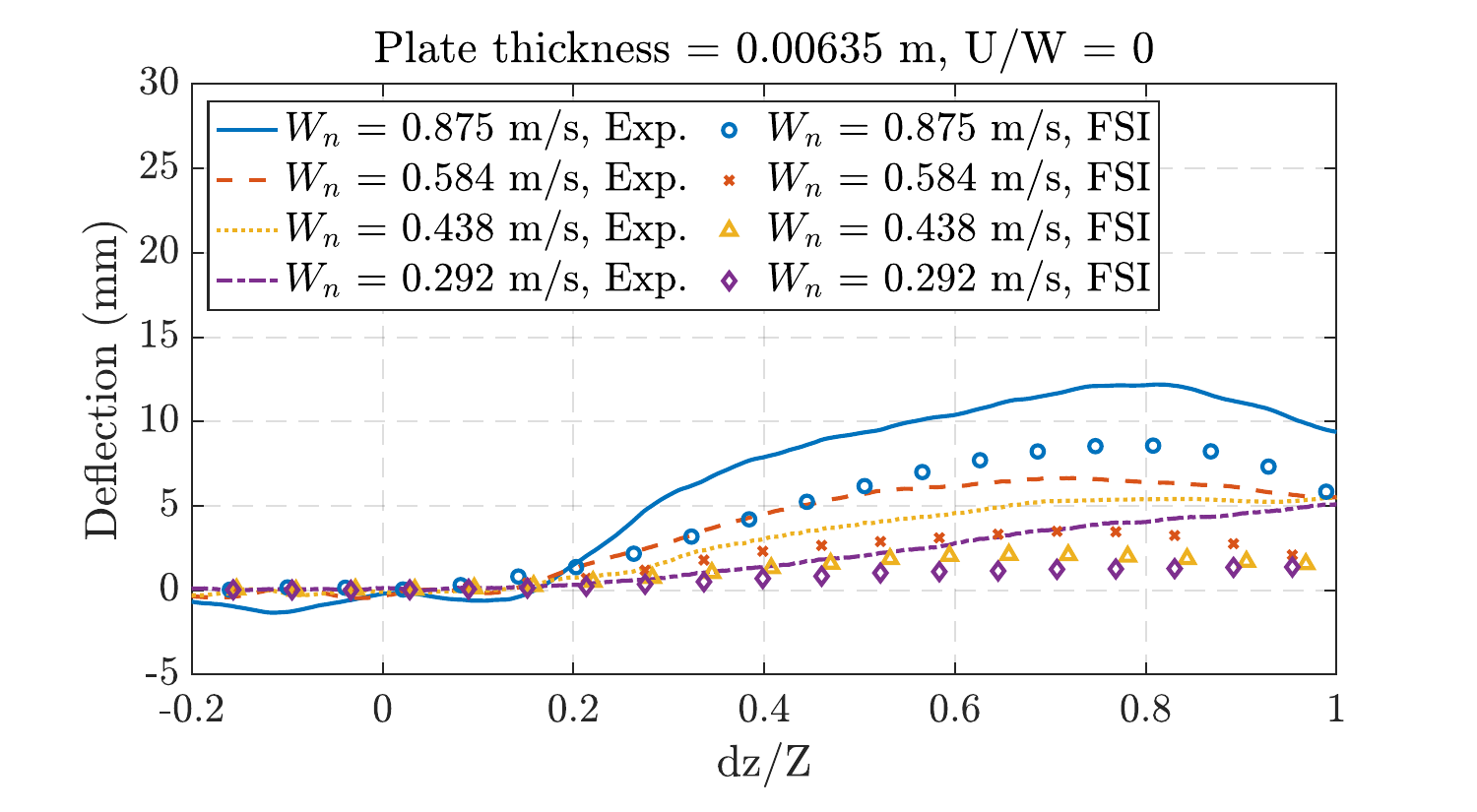}}}
\subfloat[]{\scalebox{0.5}{\includegraphics[width=1\textwidth]{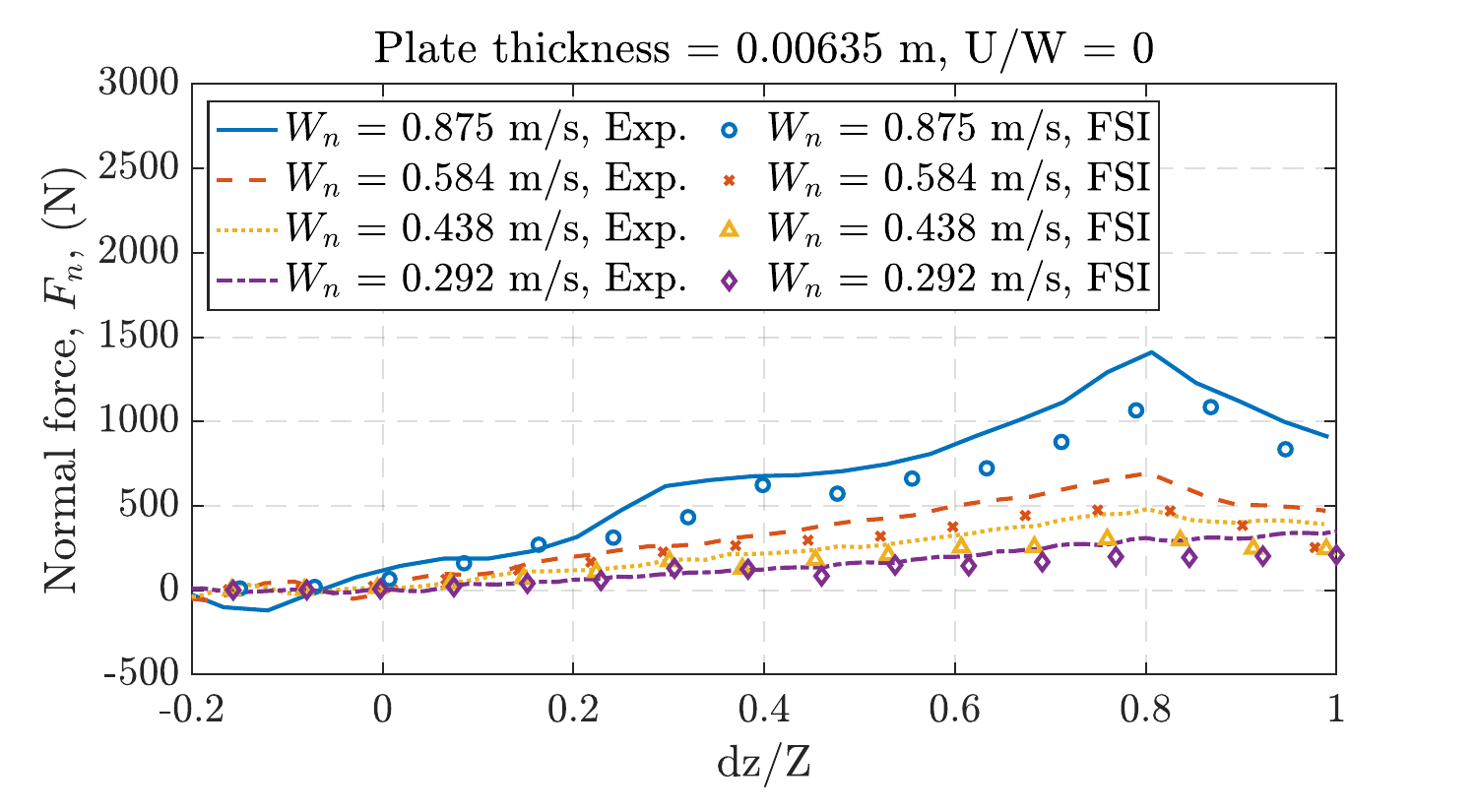}}} \\
\subfloat[]{\scalebox{0.5}{\includegraphics[width=1\textwidth]{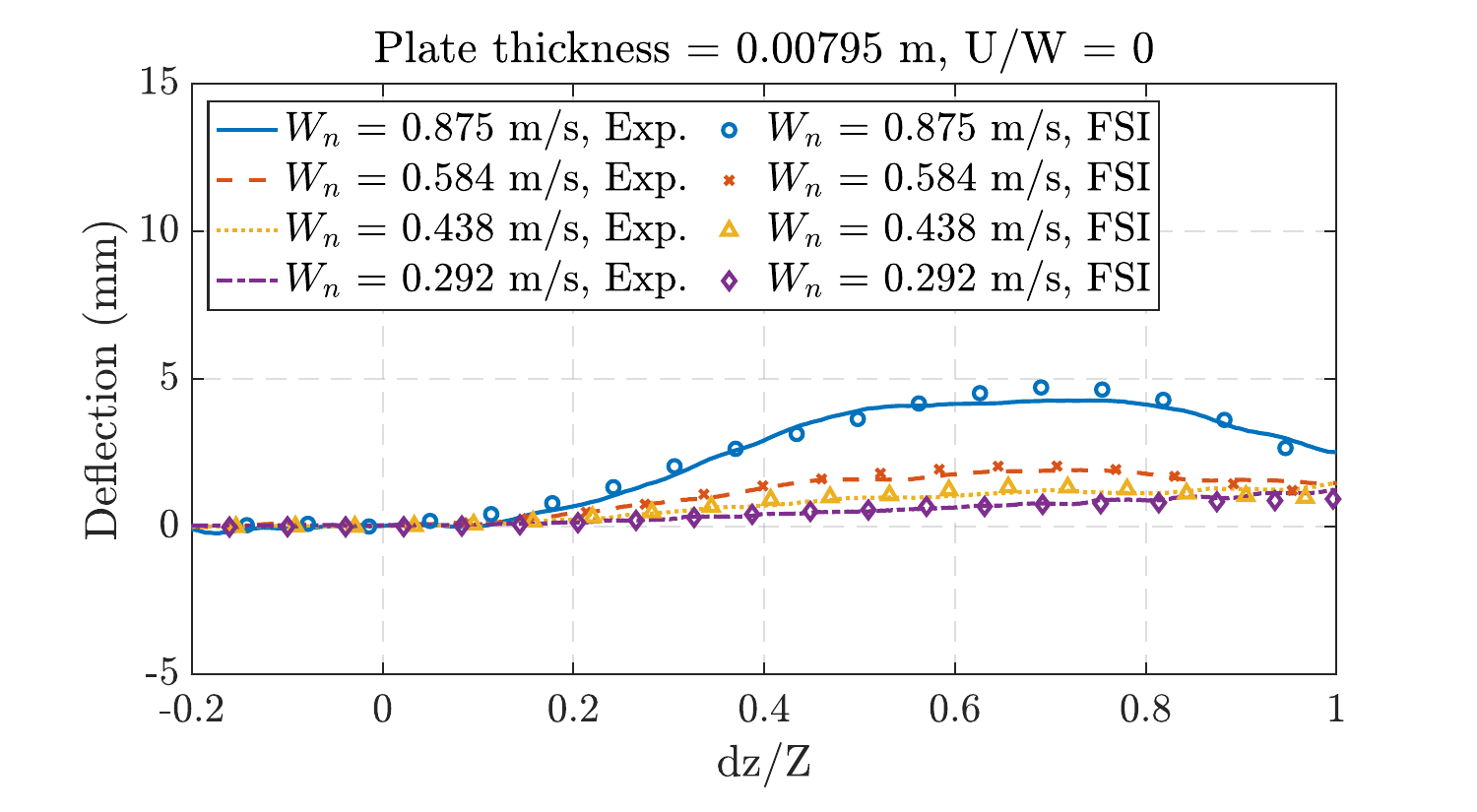}}}
\subfloat[]{\scalebox{0.5}{\includegraphics[width=1\textwidth]{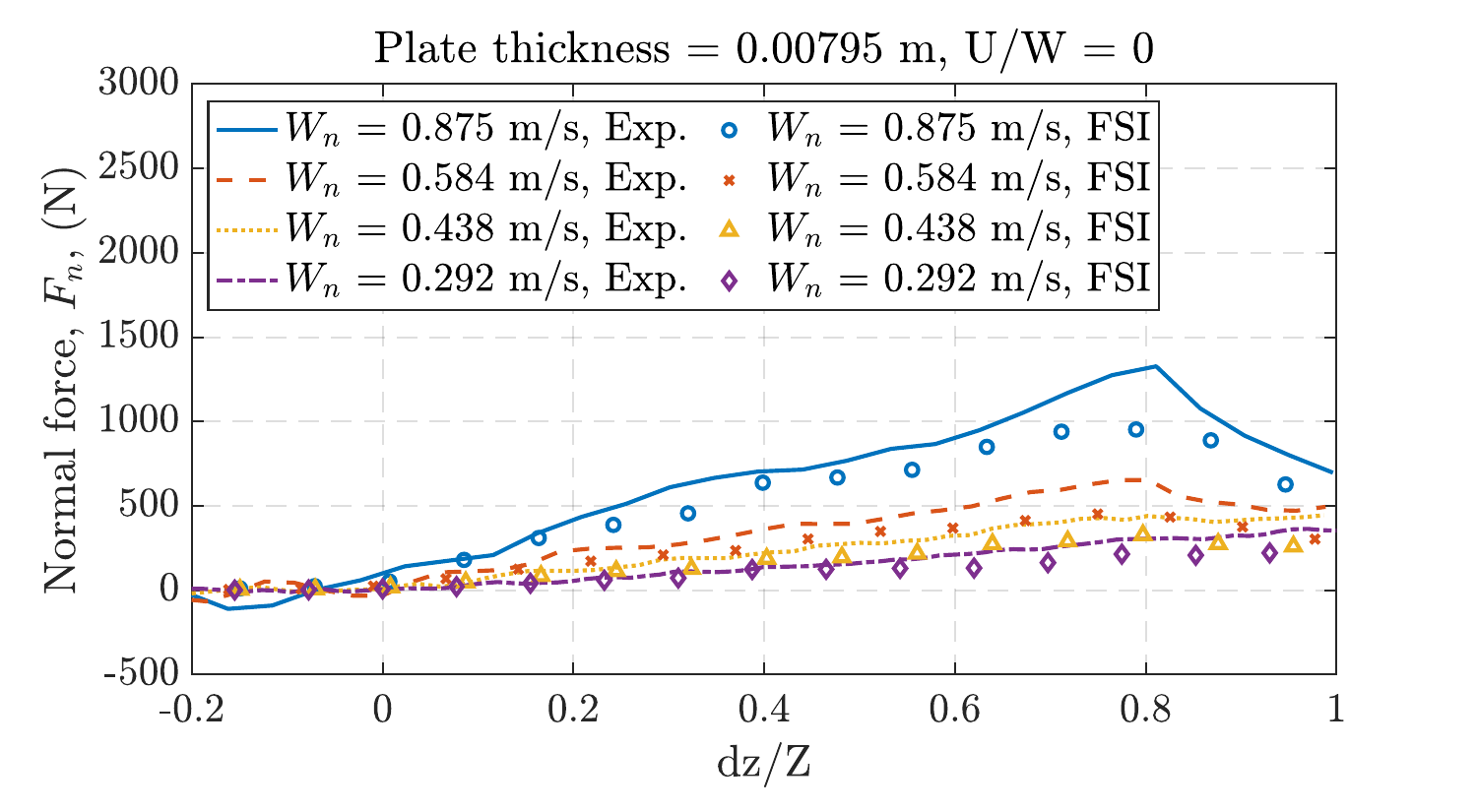}}} \\
\subfloat[]{\scalebox{0.5}{\includegraphics[width=1\textwidth]{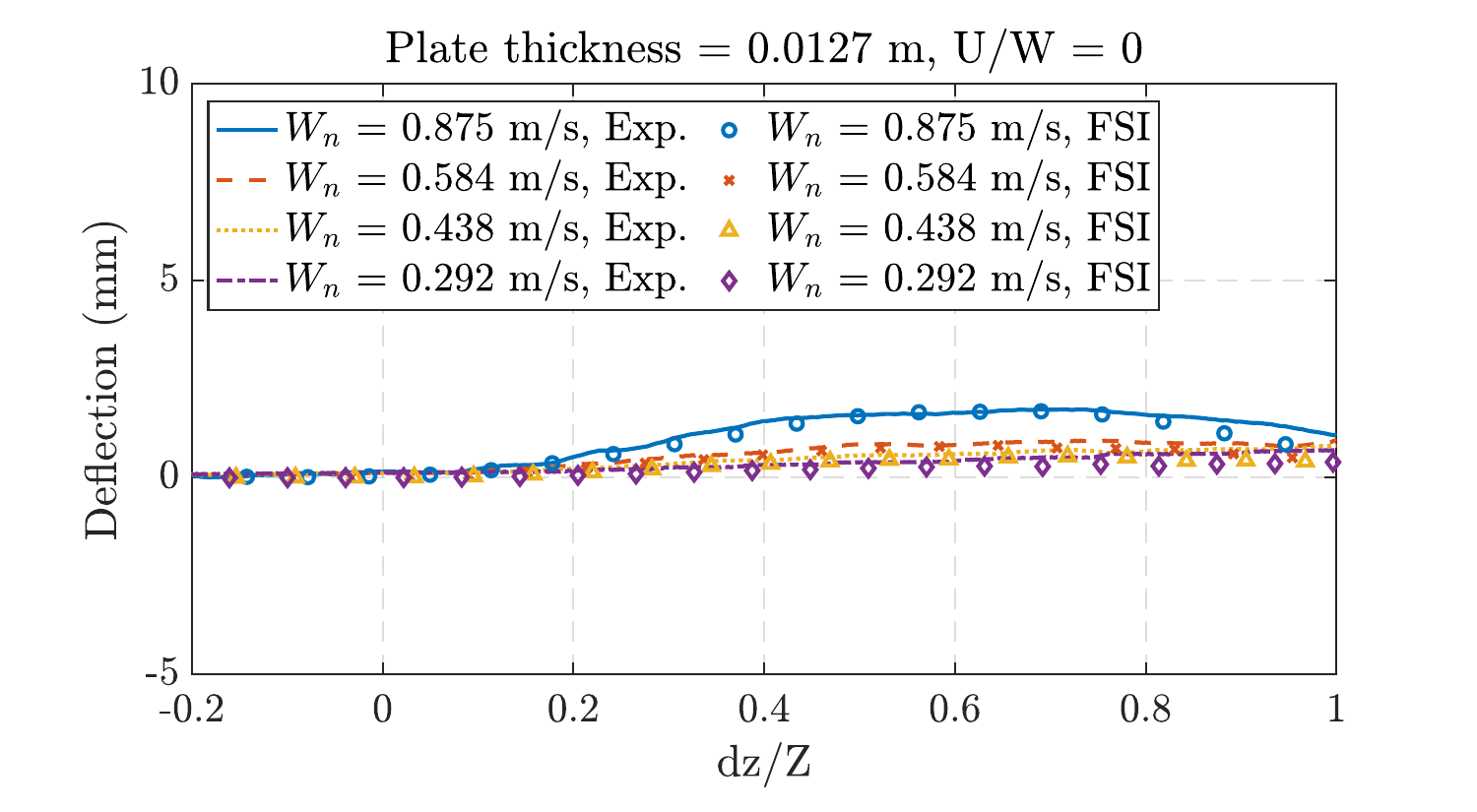}}}
\subfloat[]{\scalebox{0.5}{\includegraphics[width=1\textwidth]{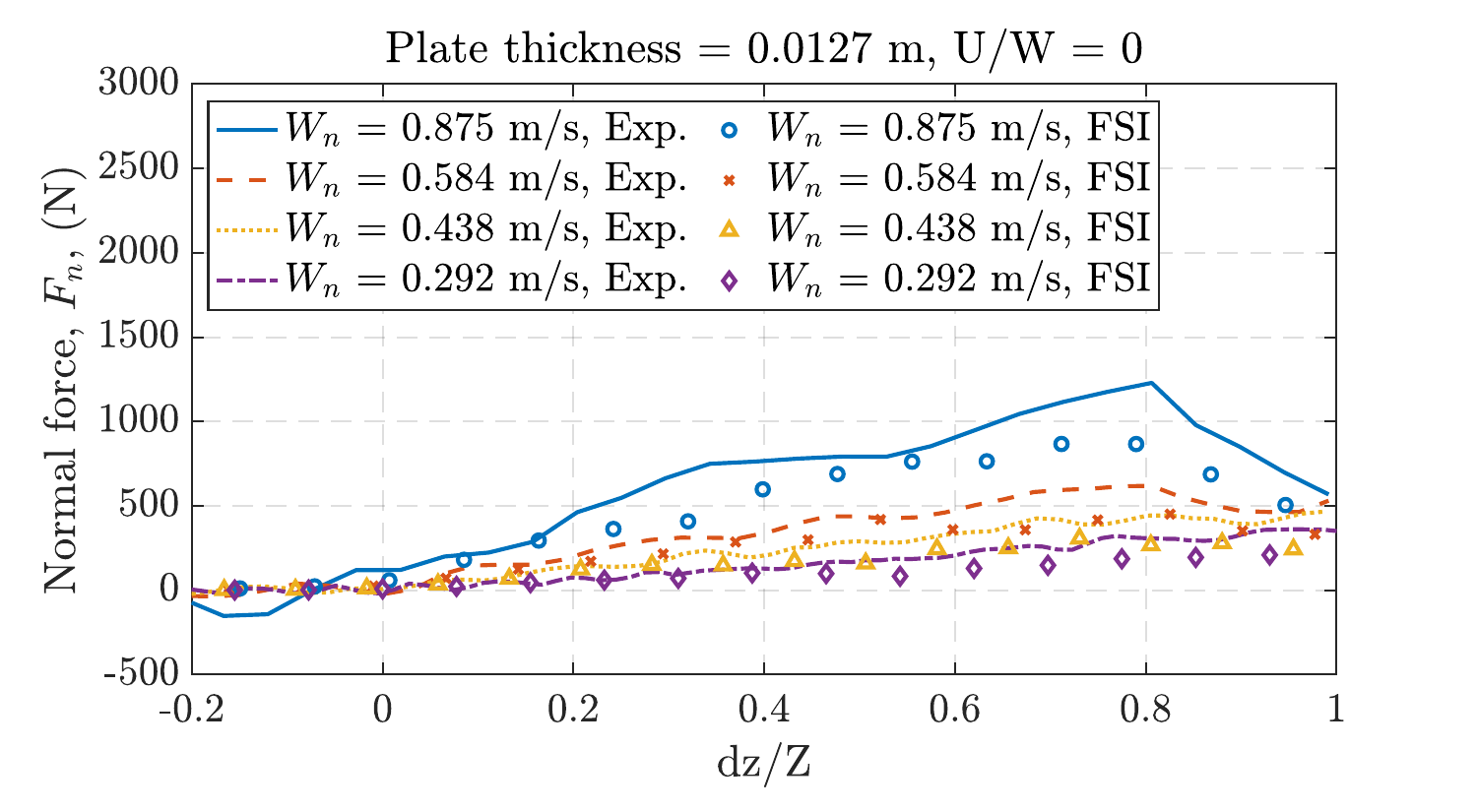}}} 
\caption{Plate deflection and normal impact force histories of the highly flexible (with plate thickness 0.00635~m), moderately deformable (with plate thickness 0.00795~m), and nearly rigid (with plate thickness 0.0127~m) plates subjected to slamming impacts with $U/W = 0$.} \label{UW0}
\end{figure}

In the last set of FSI validation tests, we present validation results for slamming tests with $U/W = 0 $ and $W_n = 0.875, 0.584, 0.438,$ and $0.292$~m/s. These cases are subjected to $W$ only; higher $W_n$ indicates higher $W$. The $U$ and $W$ of the slamming tests are listed in Table \ref{val_test3}. Displacement and normal force responses for plate thickness 0.00635~m, 0.00795~m, and 0.0127~m are shown in Figure \ref{UW0}. Similarly, we observe that the higher impact velocity and smaller plate thickness lead to higher plate deformation. In Figure \ref{UW0}b, d, and f, we observe bilinear force behaviors in $W_n = 0.875$~m/s and $W_n = 0.584$~m/s, while the forces are monotonically increasing in $W_n = 0.438$~m/s and $W_n = 0.292$~m/s.

\begin{table}[h!]
\centering
\caption{Forward and downward velocities of slamming tests with $U/W = 0$.} \label{val_test3} 
\begin{tabular}{ccccc} 
\hline
$W_n$ (m/s)     & U/W         & U (m/s)                     & W (m/s)  &  $F_r$  \\
  \hline      
  0.875         &  0       & 0                       & 0.889        &  0.269   \\
  0.584         &  0        & 0                            & 0.593   & 0.179  \\
  0.438         &  0        & 0                             & 0.445   & 0.135 \\
  0.292          &   0        & 0                               & 0.296   & 0.090   \\ 
  \hline
\end{tabular}
\end{table} 

The $L_2$ relative error norms in impact forces and plate deformations of this last set of FSI validation tests are presented in Table \ref{val3_error}. The $L_2$ error norms in impact forces among the three plates fall within the interval $\left[0.189, 0.348\right]$, with a median of 0.292 and a mean of 0.277. The $L_2$ error norms in displacements among the three plates occur within the interval $\left[0.091, 0.670\right]$, with a median of 0.270 and a mean of 0.313. Similar to the second set of FSI validation tests, it is observed that lower $W_n$ leads to higher $L_2$ errors. 

\begin{table}[h!]
\centering
\caption{$L_2$ relative error norms in slamming impact forces and plate deformations from slamming tests with $U/W = 0$.} \label{val3_error} 
\resizebox{\textwidth}{!}{\begin{tabular}{ccccccc}
 \hline  
 & \multicolumn{3}{c}{ $L_2$ error, impact force } &  \multicolumn{3}{c}{ $L_2$ error, displacement} \\
Plate thickness (m)  &  0.00635 & 0.00795   & 0.0127 & 0.00635 & 0.00795   & 0.0127 \\
 \hline   
$W_n$ 0.875   &  0.189        &  0.215        & 0.208             & 0.331       & 0.091            & 0.136  \\
$W_n$ 0.584    &  0.252        & 0.274        & 0.253            & 0.511        &  0.102             & 0.221 \\
$W_n$ 0.438    &   0.310       &  0.309       & 0.319             & 0.621       &  0.162            &  0.318    \\ 
$W_n$ 0.292    &   0.318       & 0.329       & 0.348             & 0.670        & 0.151             & 0.446  \\ 
  \hline
\end{tabular}}
\end{table} 

\begin{figure}[h!]
\centering
\captionsetup[subfigure]{justification=centering}
\subfloat[dz/Z = 0.7]{\scalebox{0.25}{\includegraphics[width=1\textwidth]{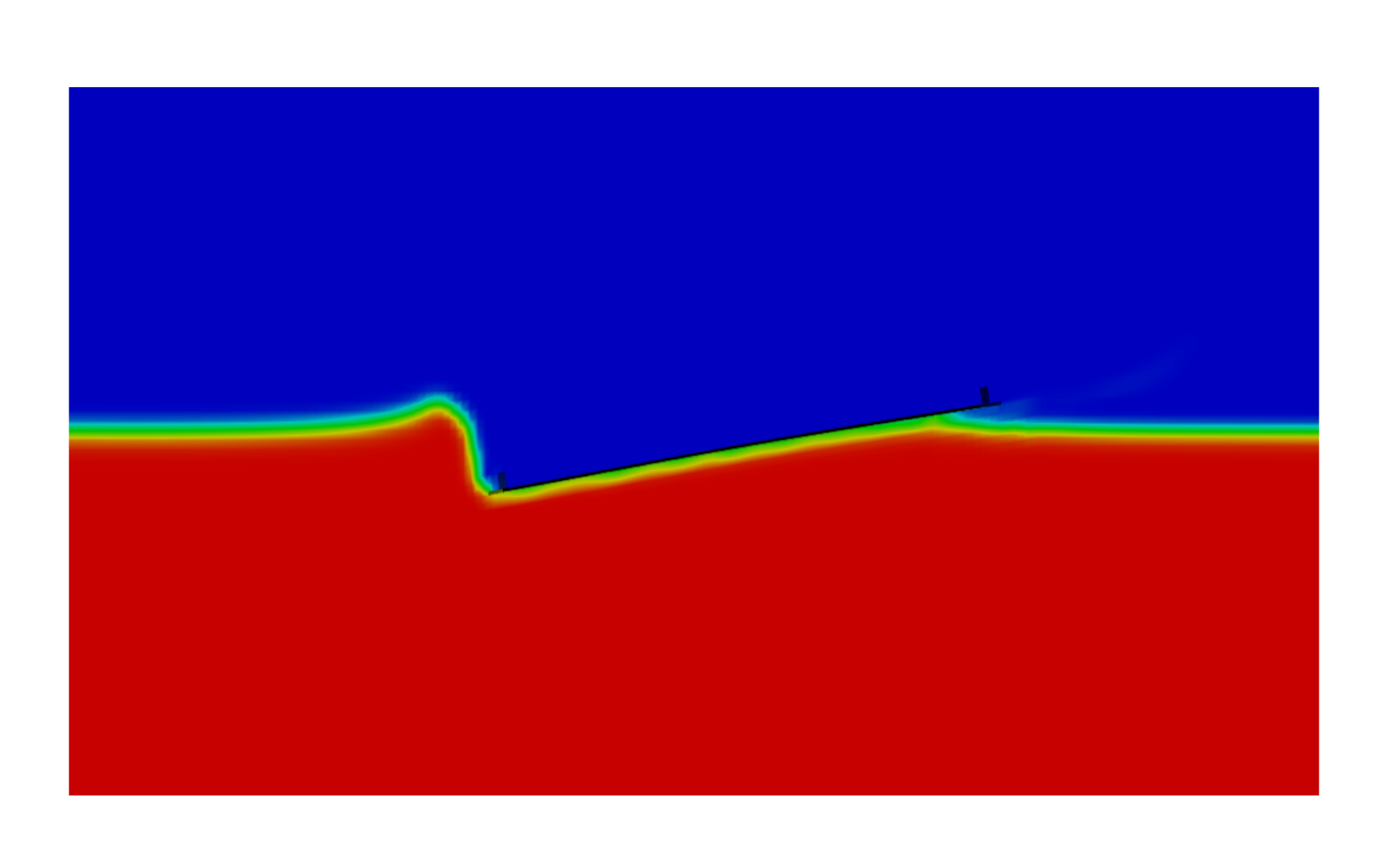}}}
\subfloat[dz/Z = 0.8]{\scalebox{0.25}{\includegraphics[width=1\textwidth]{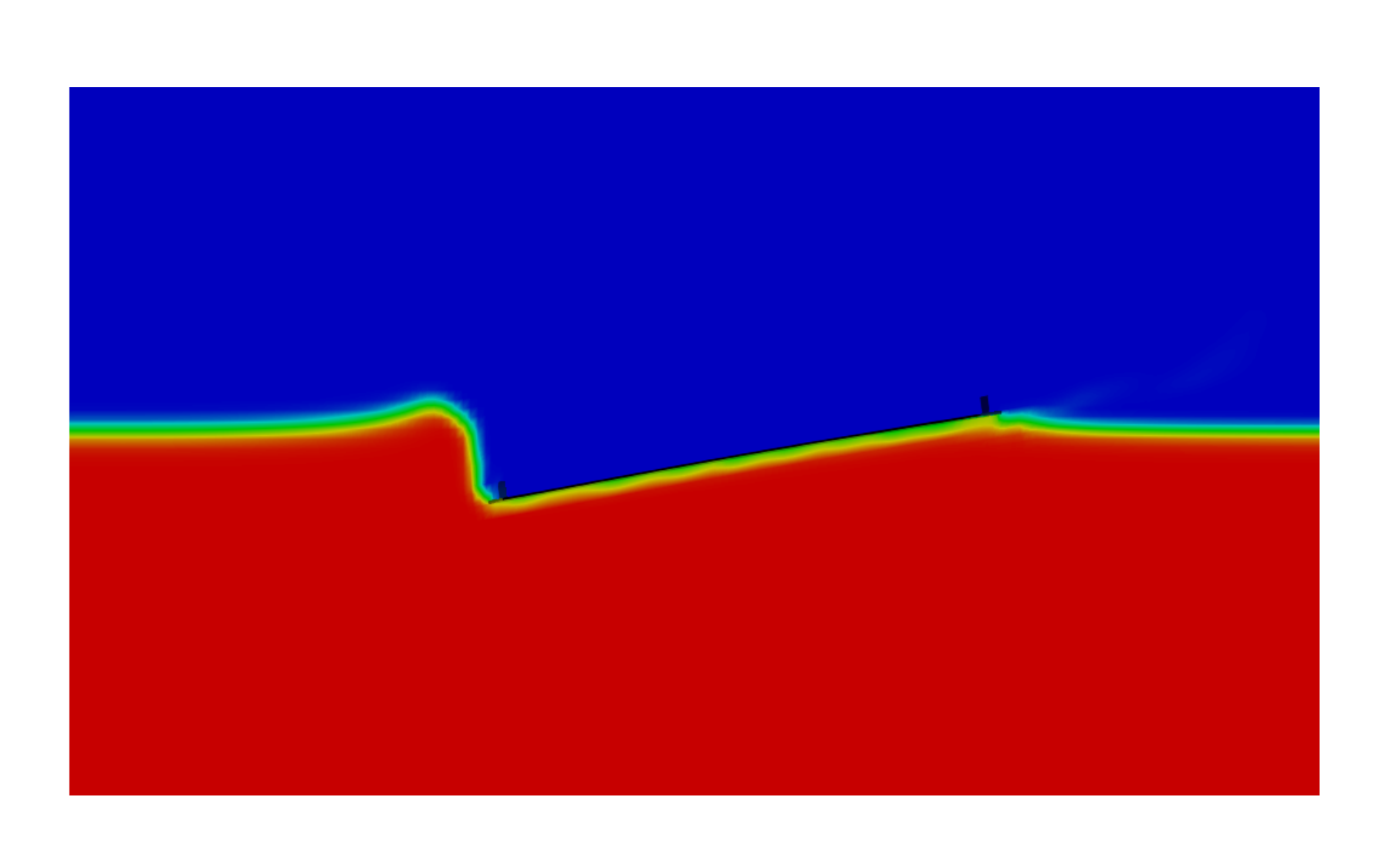}}}
\subfloat[dz/Z = 0.9]{\scalebox{0.25}{\includegraphics[width=1\textwidth]{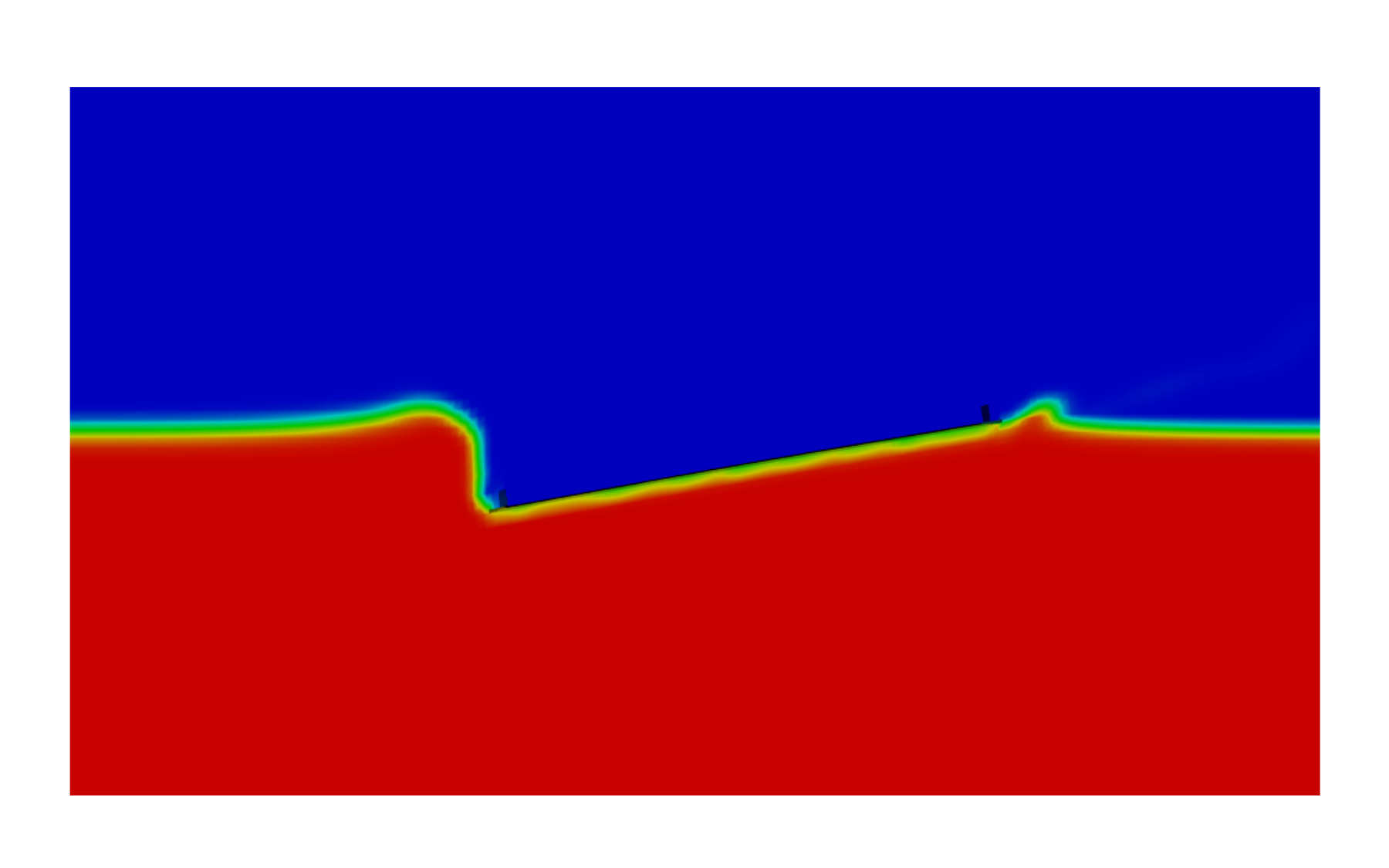}}}
\subfloat[dz/Z = 1]{\scalebox{0.25}{\includegraphics[width=1\textwidth]{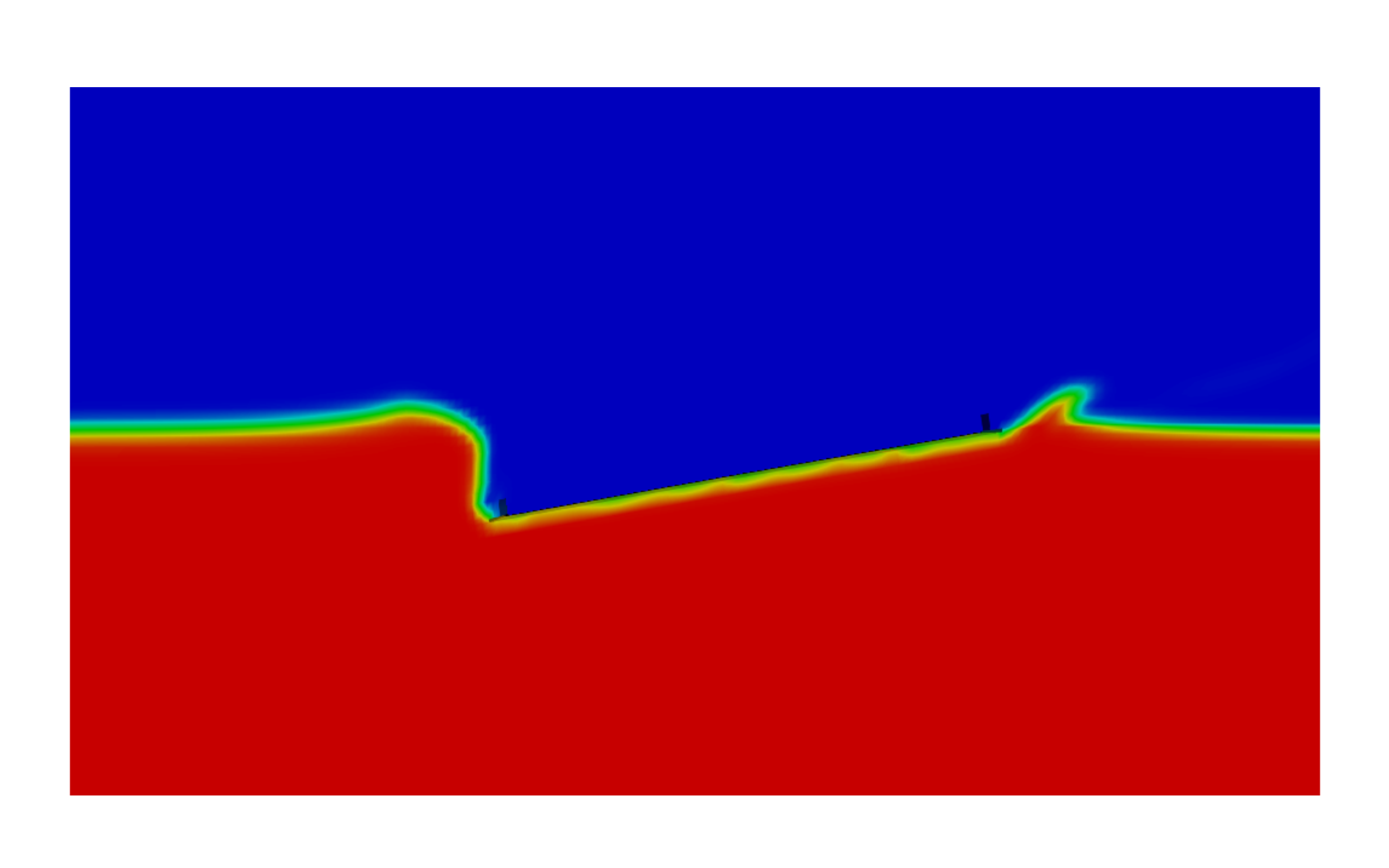}}} \\
\subfloat[dz/Z = 0.7]{\scalebox{0.25}{\includegraphics[width=1\textwidth]{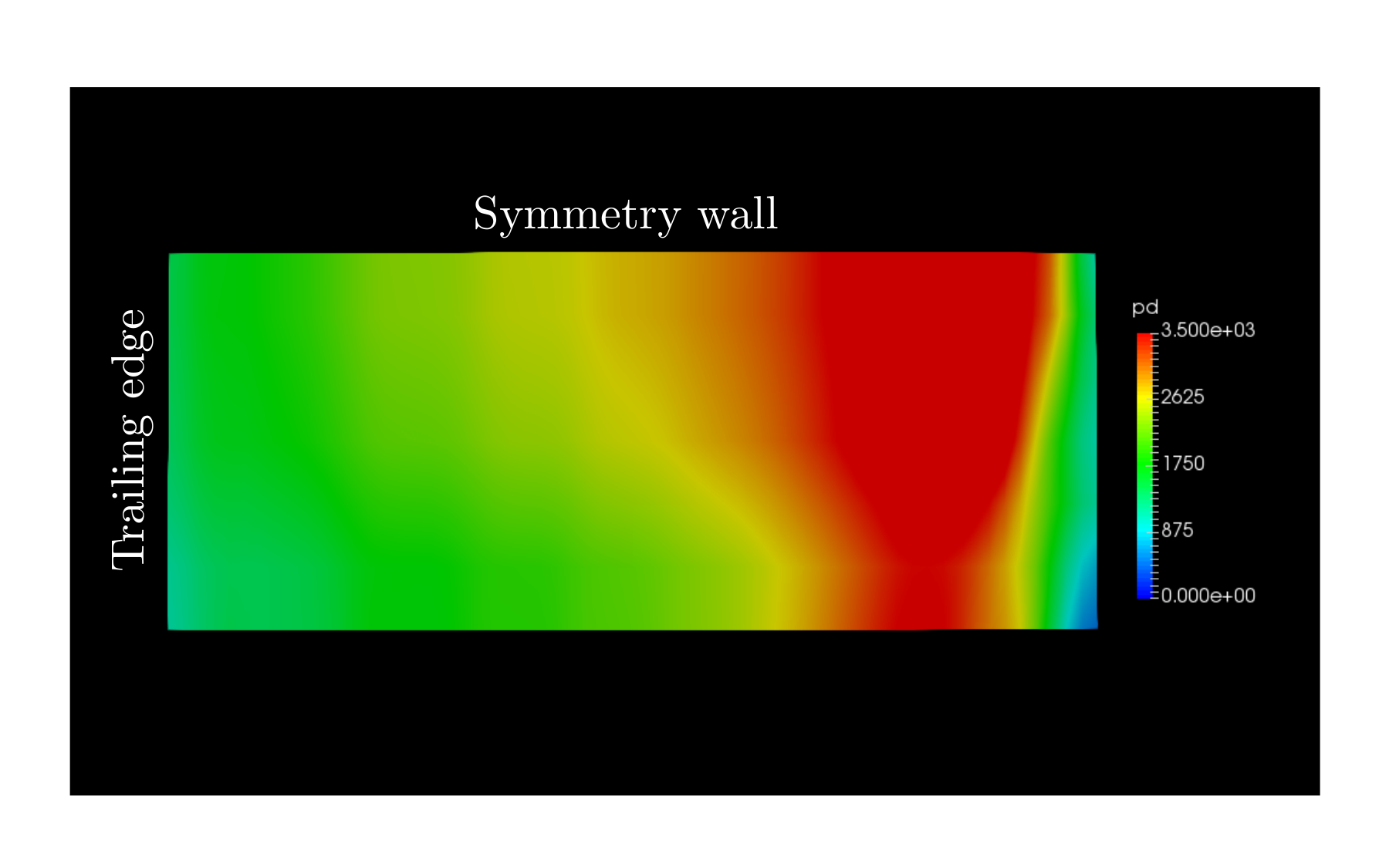}}}
\subfloat[dz/Z = 0.8]{\scalebox{0.25}{\includegraphics[width=1\textwidth]{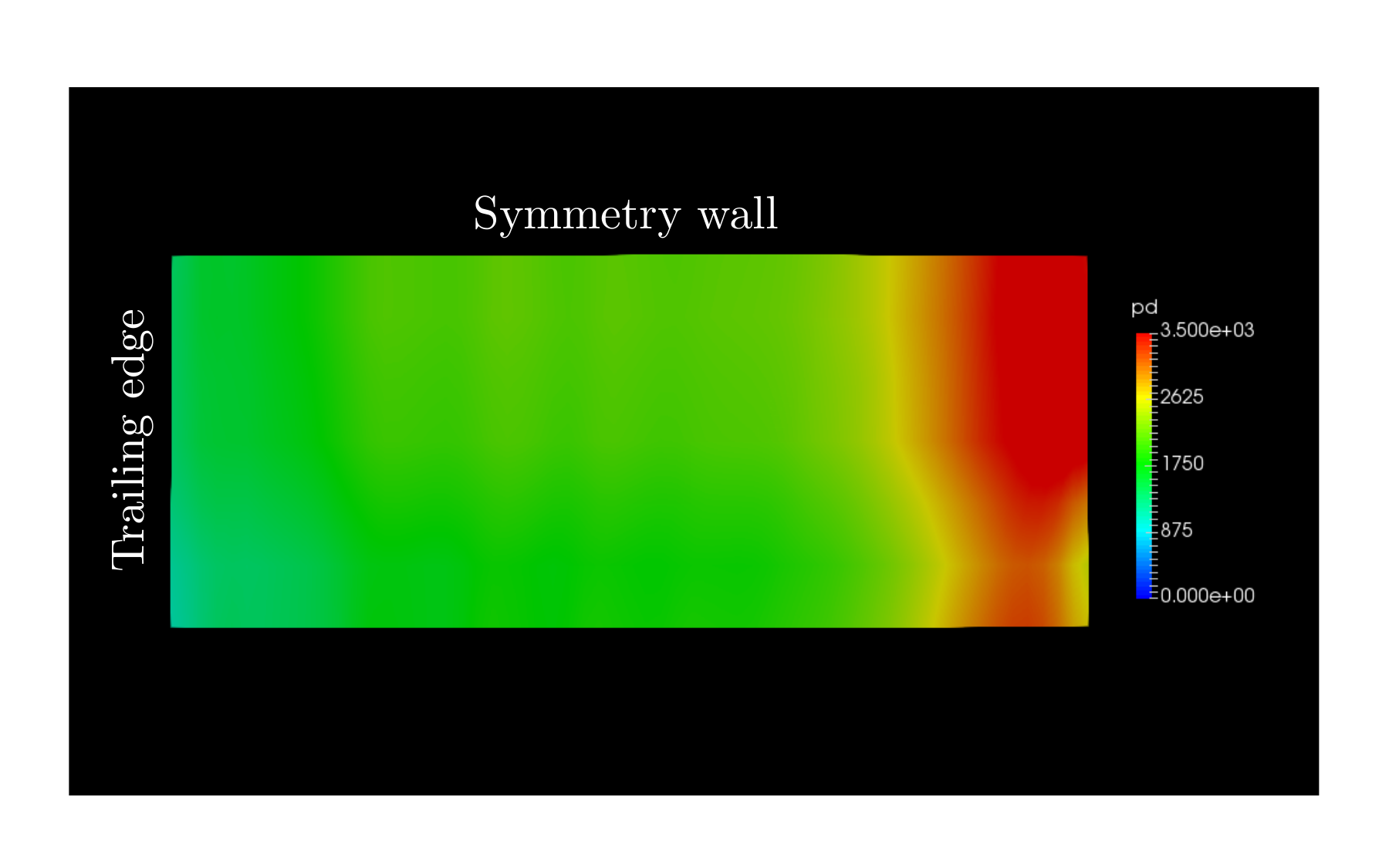}}}
\subfloat[dz/Z = 0.9]{\scalebox{0.25}{\includegraphics[width=1\textwidth]{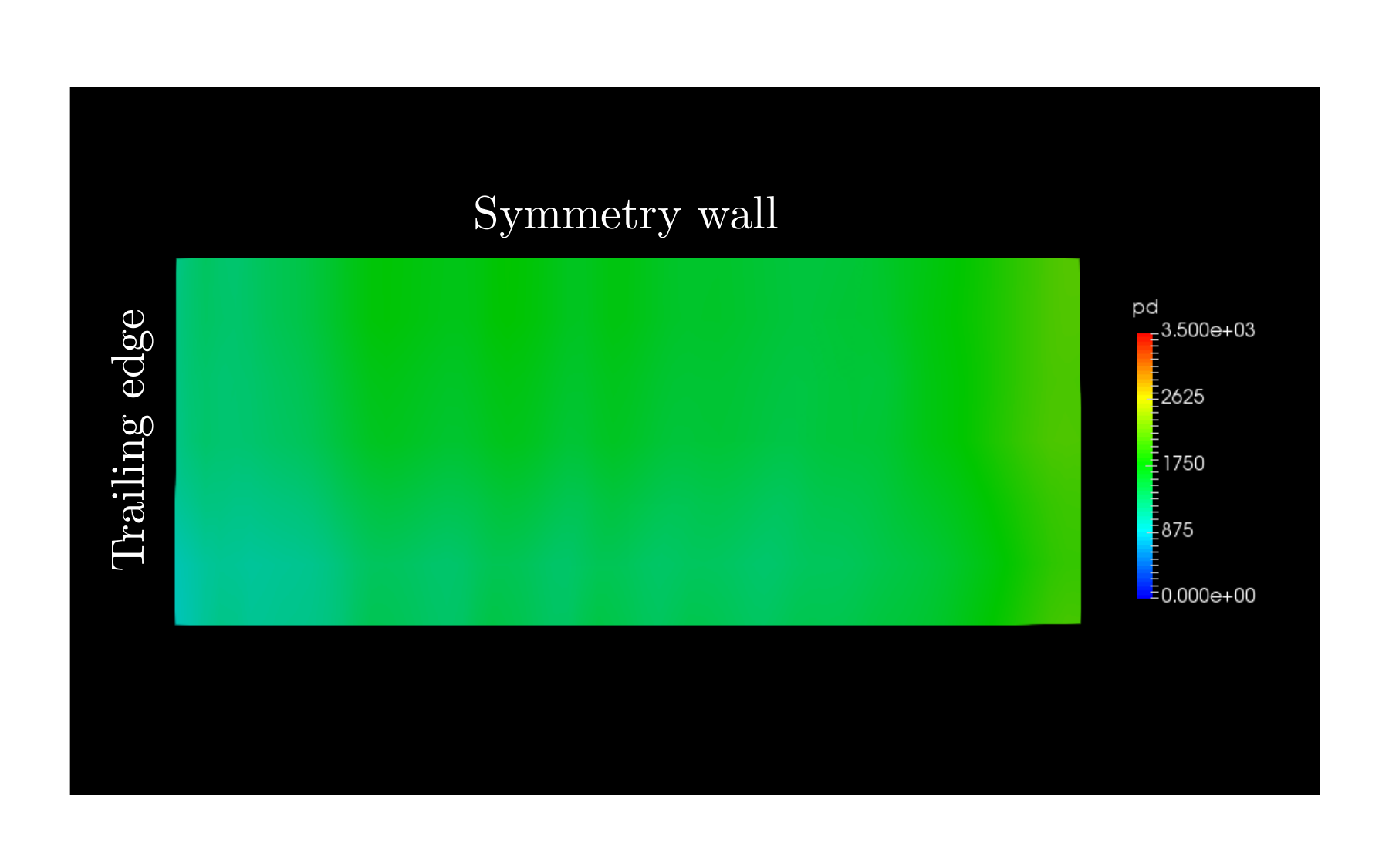}}}
\subfloat[dz/Z = 1]{\scalebox{0.25}{\includegraphics[width=1\textwidth]{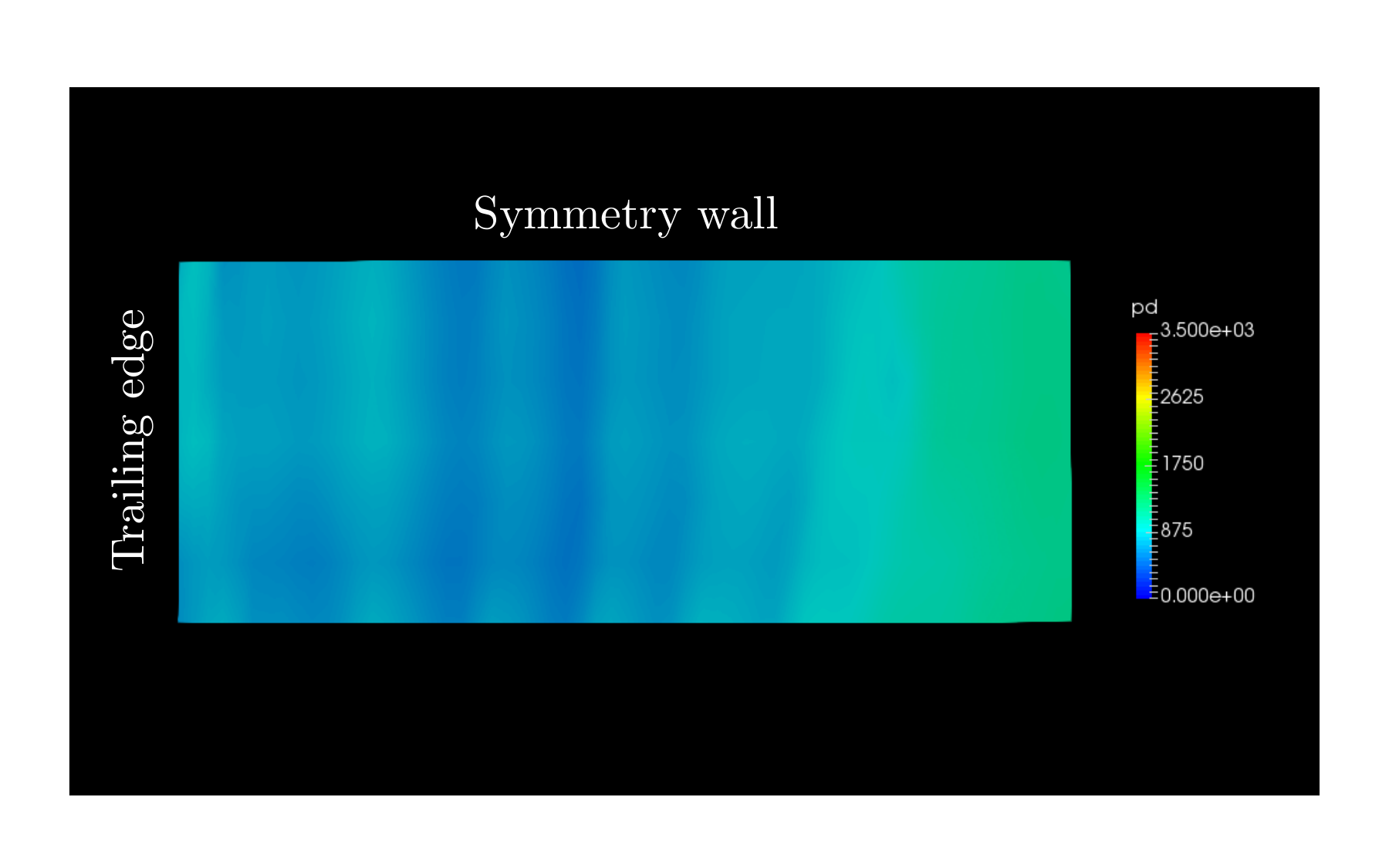}}} \\
\caption{Snapshots of the water free surface evolution and the corresponding pressure distribution over the plate from $dz/Z = \left[0.7, 1.0\right]$ of the slamming test with $W_n = 0.875$~m/s.} \label{W0p889}
\end{figure}

\begin{figure}[h!]
\centering
\captionsetup[subfigure]{justification=centering}
\subfloat[dz/Z = 0.7]{\scalebox{0.25}{\includegraphics[width=1\textwidth]{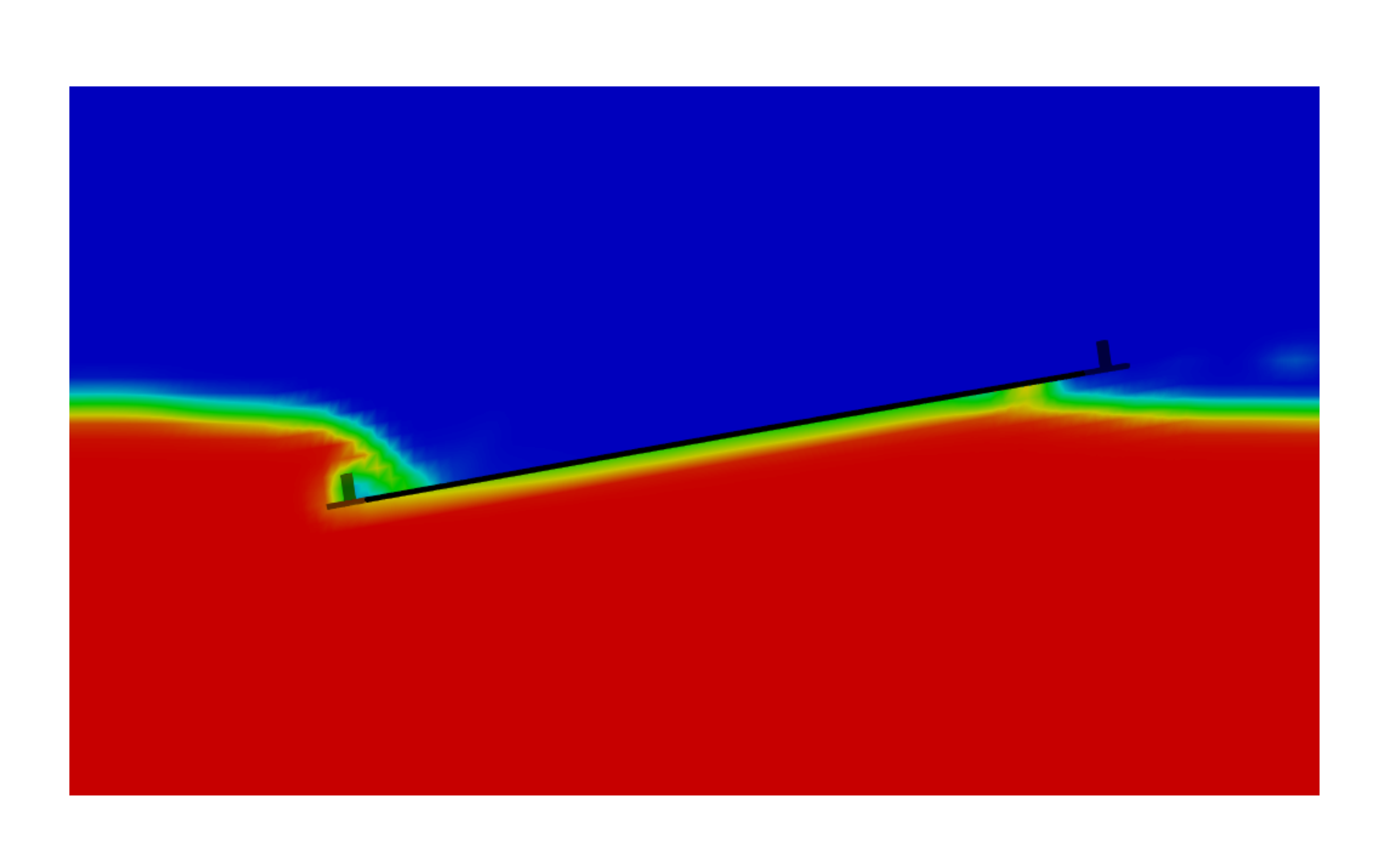}}}
\subfloat[dz/Z = 0.8]{\scalebox{0.25}{\includegraphics[width=1\textwidth]{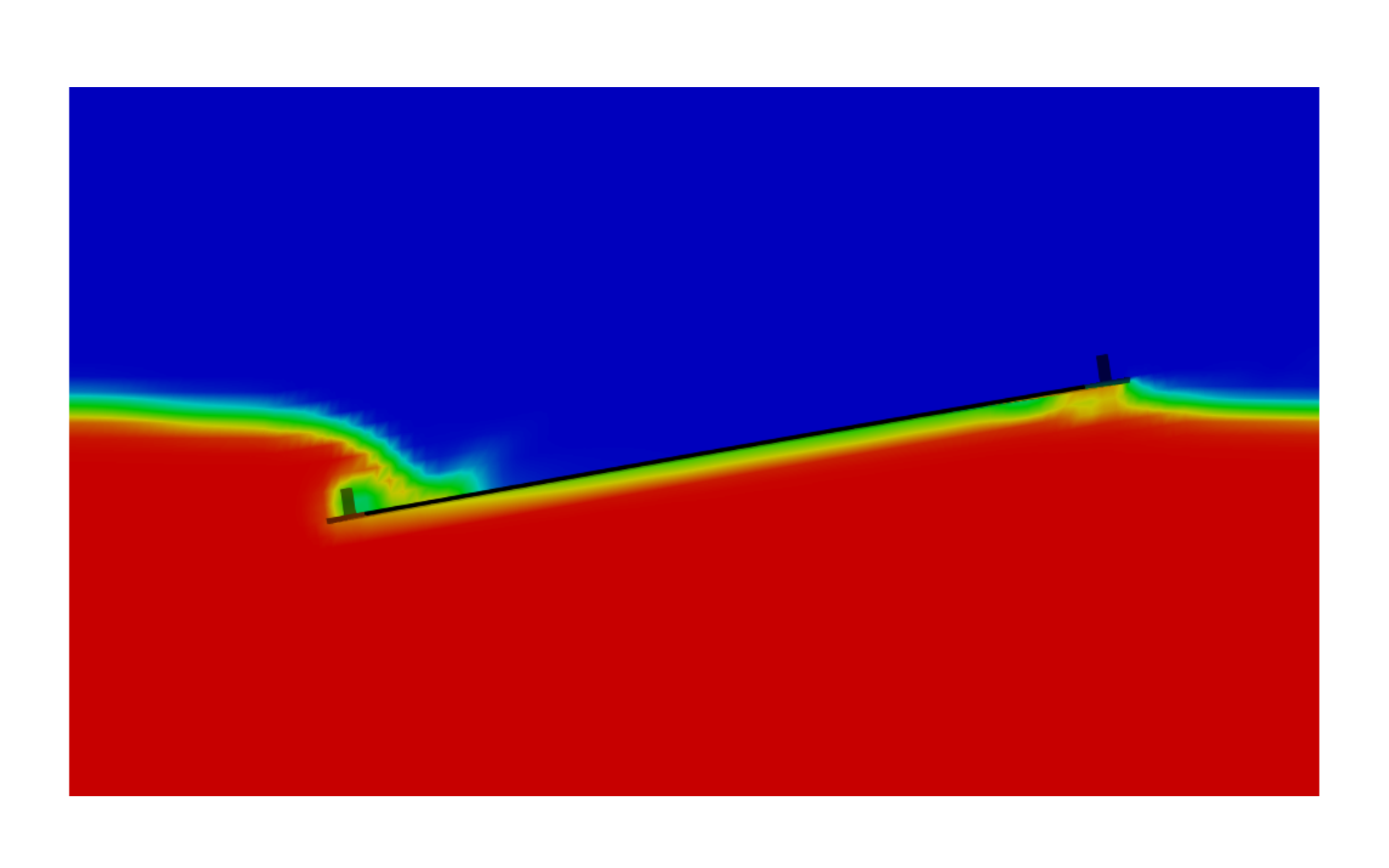}}}
\subfloat[dz/Z = 0.9]{\scalebox{0.25}{\includegraphics[width=1\textwidth]{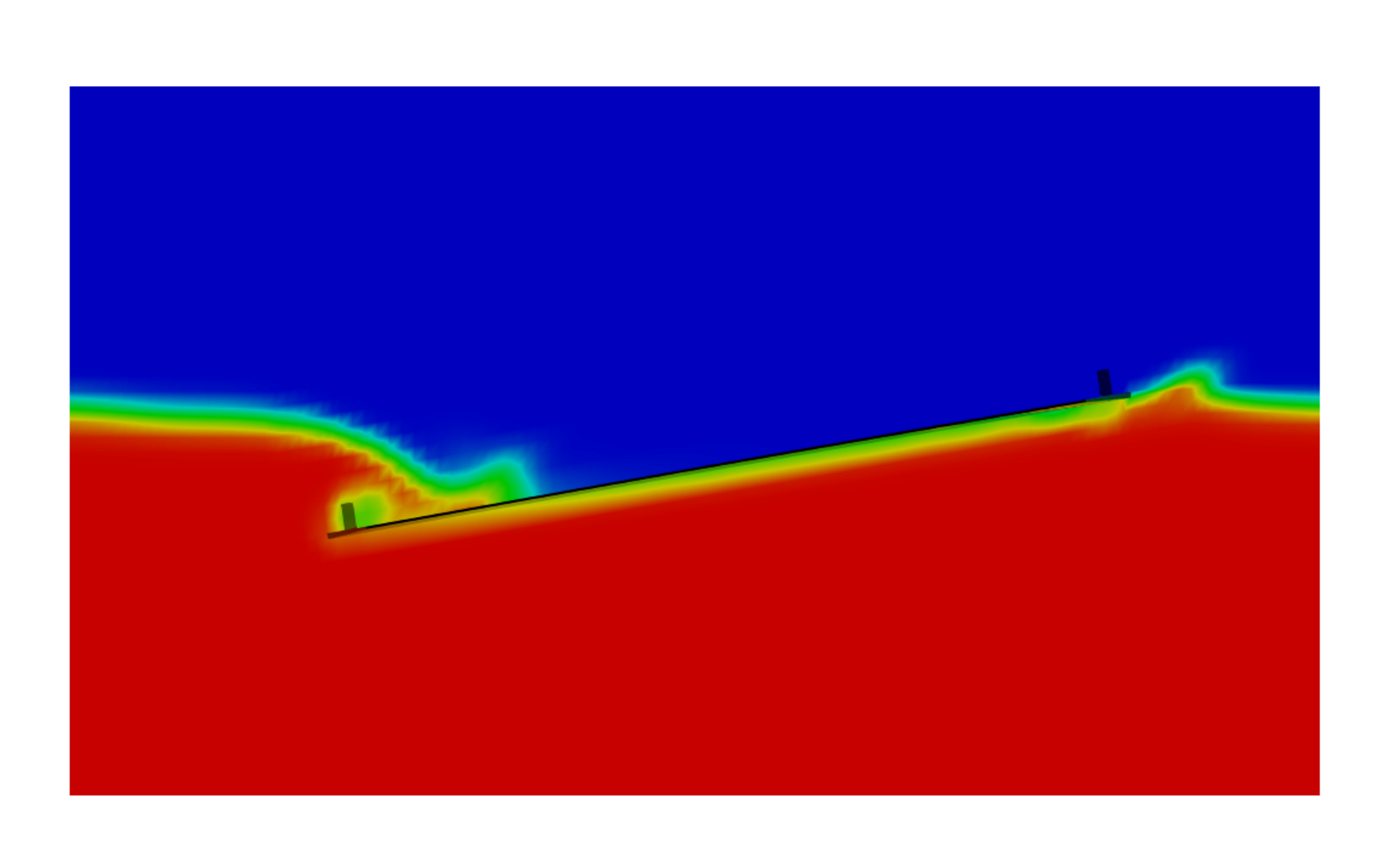}}}
\subfloat[dz/Z = 1]{\scalebox{0.25}{\includegraphics[width=1\textwidth]{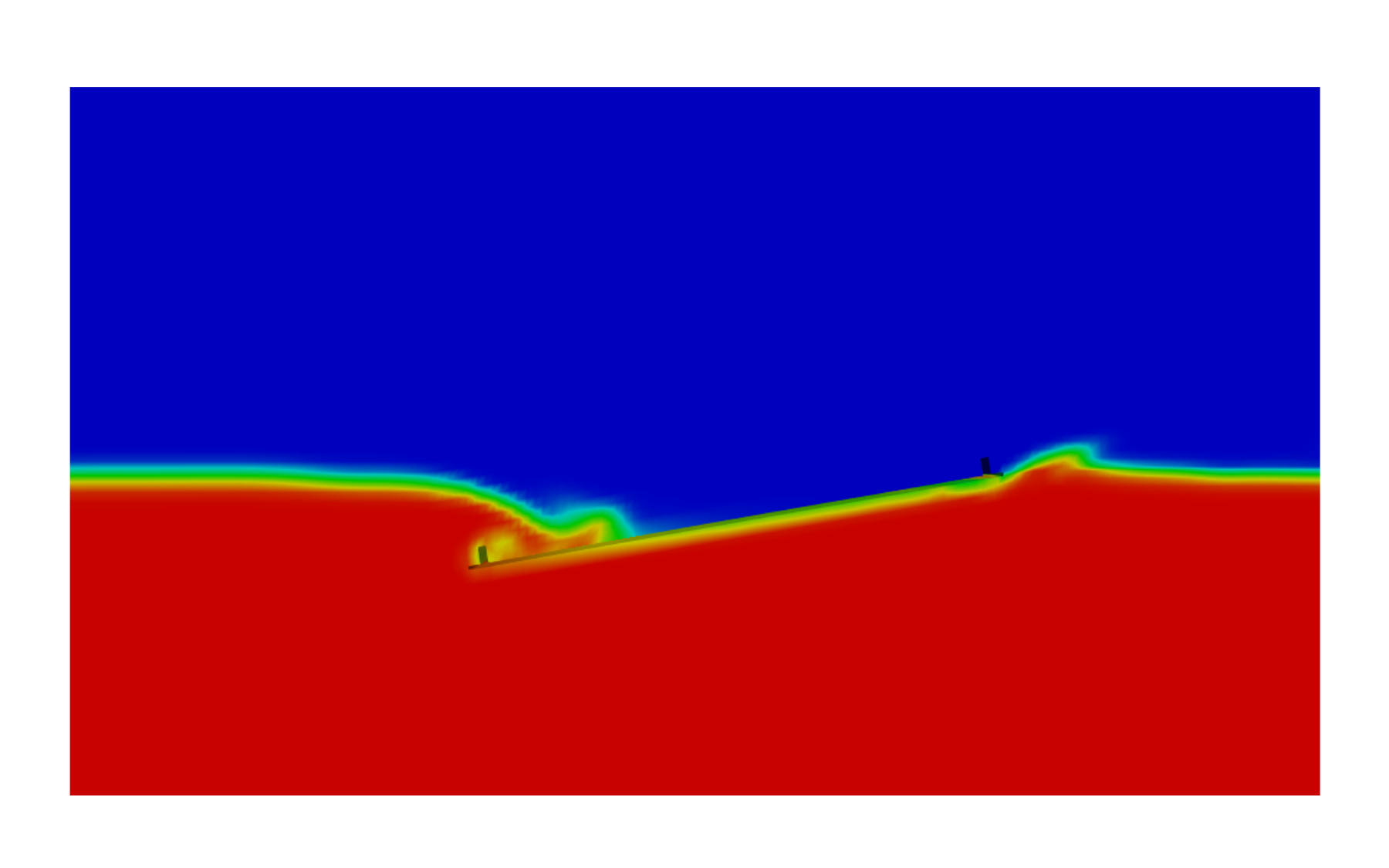}}} \\
\subfloat[dz/Z = 0.7]{\scalebox{0.25}{\includegraphics[width=1\textwidth]{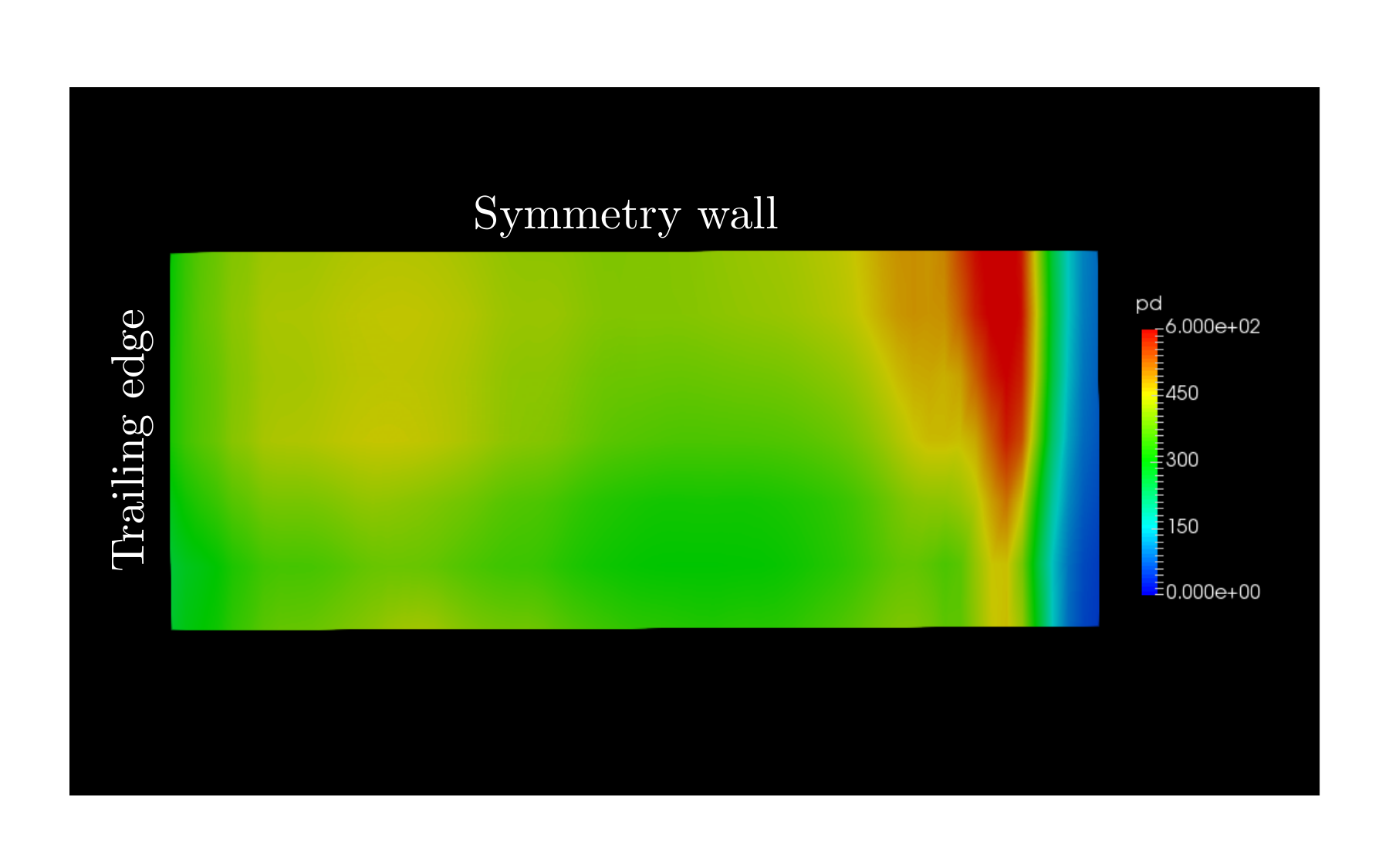}}}
\subfloat[dz/Z = 0.8]{\scalebox{0.25}{\includegraphics[width=1\textwidth]{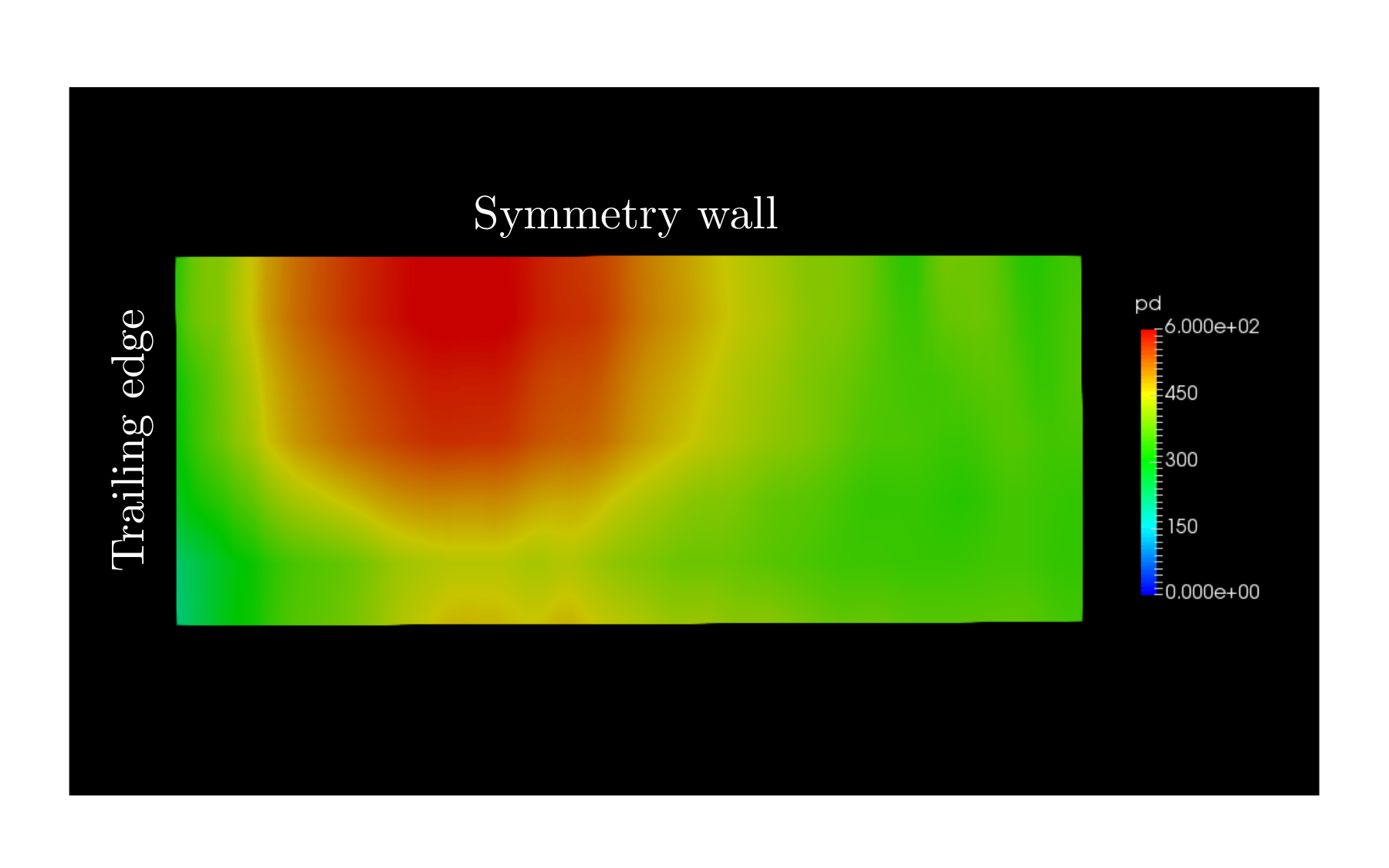}}}
\subfloat[dz/Z = 0.9]{\scalebox{0.25}{\includegraphics[width=1\textwidth]{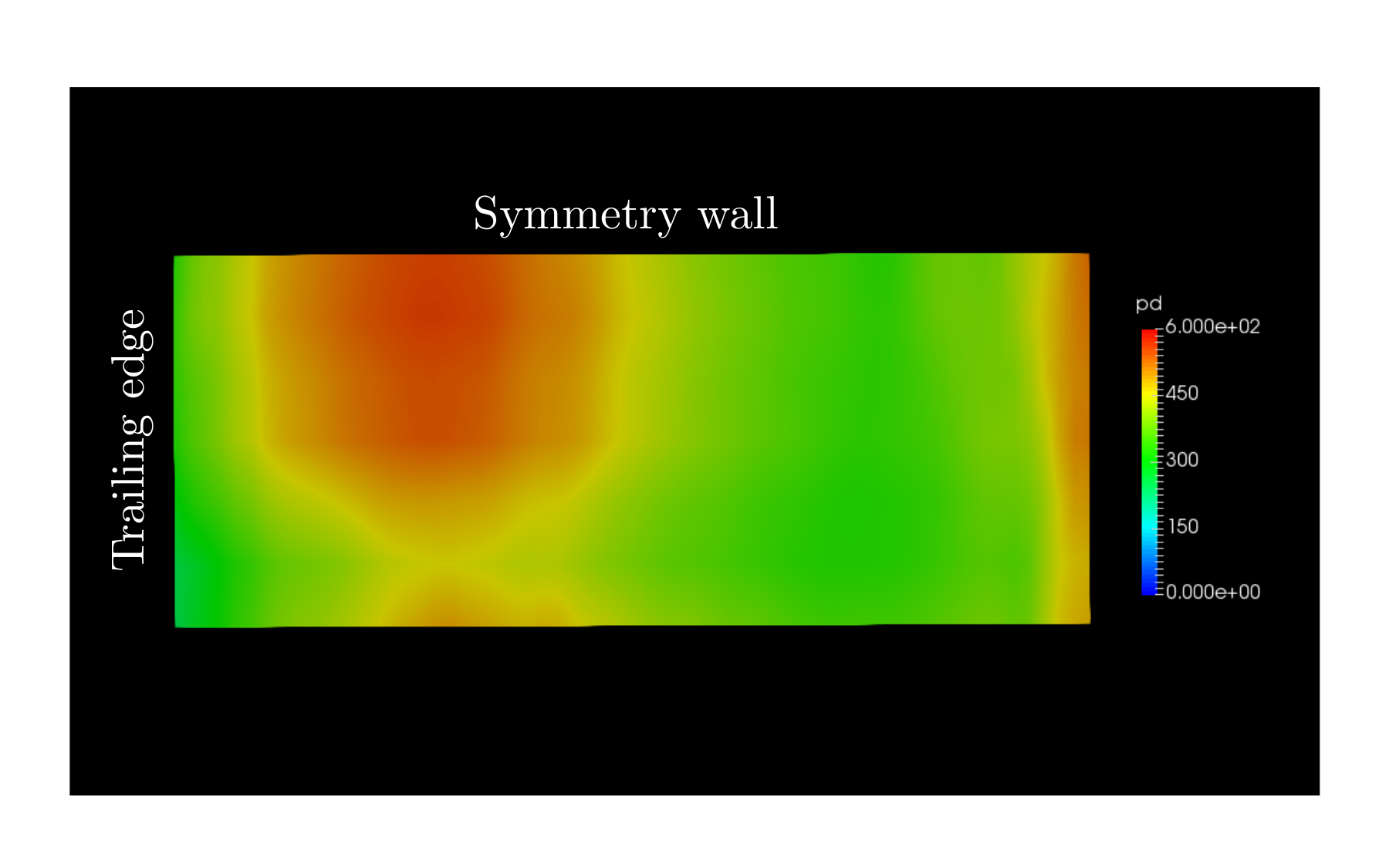}}}
\subfloat[dz/Z = 1]{\scalebox{0.25}{\includegraphics[width=1\textwidth]{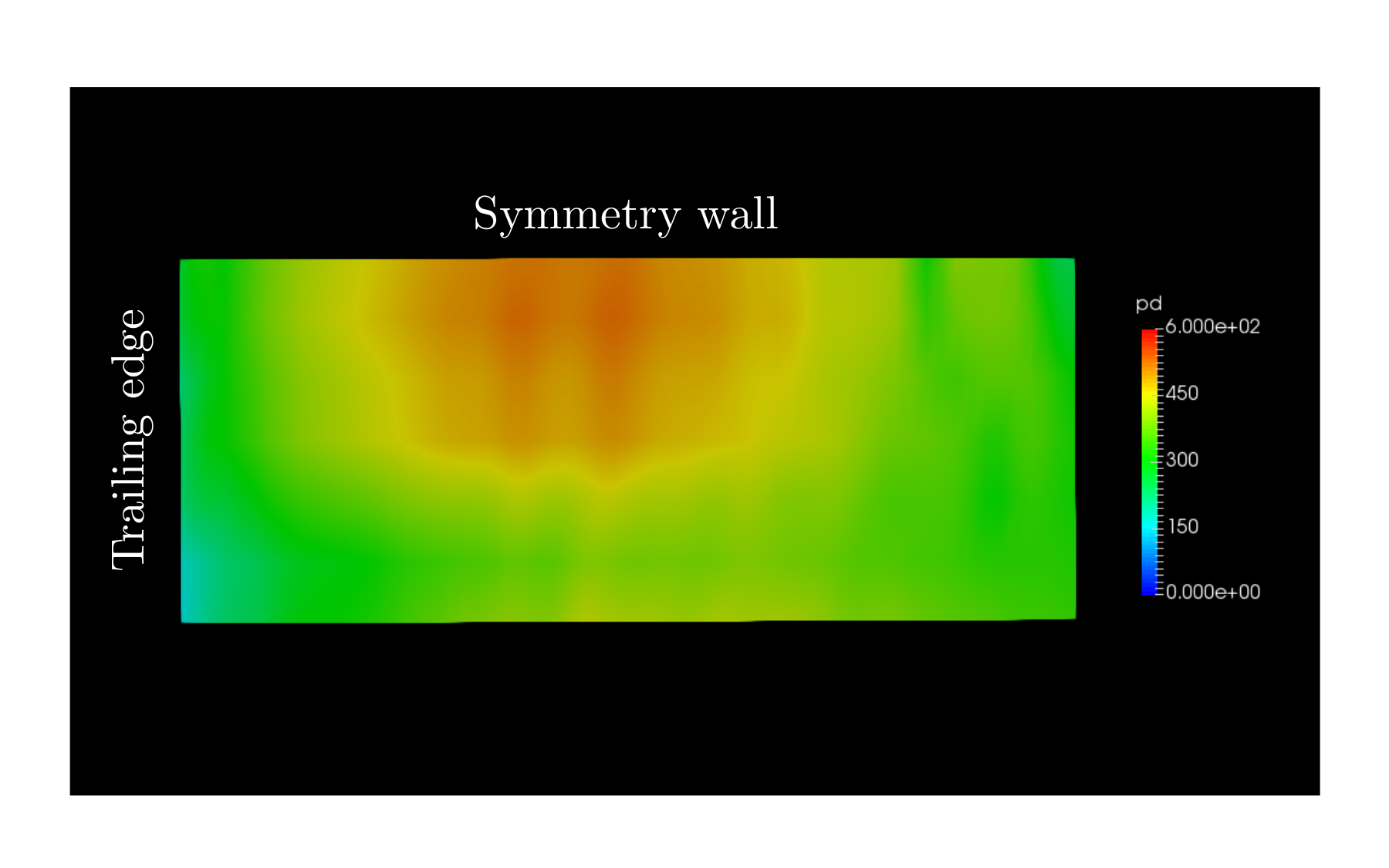}}} \\
\caption{Snapshots of the water free surface evolution and the corresponding pressure distribution over the plate from $dz/Z = \left[0.7, 1.0\right]$ of the slamming test with $W_n = 0.292$~m/s.} \label{W0p296}
\end{figure}

Figure \ref{W0p889} and \ref{W0p296} provide snapshots of the slamming events for $W_n = $ 0.875~m/s and 0.292~m/s within the interval $dz/Z=[0.7, 1]$, as well as the corresponding pressure distributions on the plate; to better understand the discrepancy in force behaviors in Figure \ref{UW0}. As the plate is submerged into the water, it experiences a highly concentrated pressure at the spray root (\textit{i.e.,} the intersection point between the water free surface and the plate); the concentrated pressure is represented by red within the pressure distribution contour in Figure \ref{W0p889}e and f. As shown in Figure \ref{W0p889}f, this highly concentrated pressure leaves the plate at $dz/Z=0.8$, as the spray root passes the leading edge of the plate. Once the high pressure leaves the plate, the fluid pressure on the plate is substantially reduced; thus leading to force unloading in the case $W_n = 0.875$~m/s. 

Contrasting to the $W_n = 0.875$~m/s case, the $W_n = 0.292$~m/s case displays a different fluid behavior, as shown in Figure \ref{W0p296}. Since the downward velocity in $W_n = 0.292$~m/s is much lower, the impact resulting from the plate entering the water free surface does not generate enough of the momentum for the water to splash away from the trailing edge. Rather, the water from trailing edge splashes over onto the plate as shown in Figure \ref{W0p296}a to d. The water splash over induces a second pressure source onto the plate; this time counteracting the slamming pressure. As a result, force unloading is offset by the second high pressure impact in the $W_n=0.292~m/s$ case. 
 
\begin{figure}[h!]
\centering
\captionsetup[subfigure]{justification=centering}
\subfloat[]{\scalebox{0.35}{\includegraphics[width=1\textwidth]{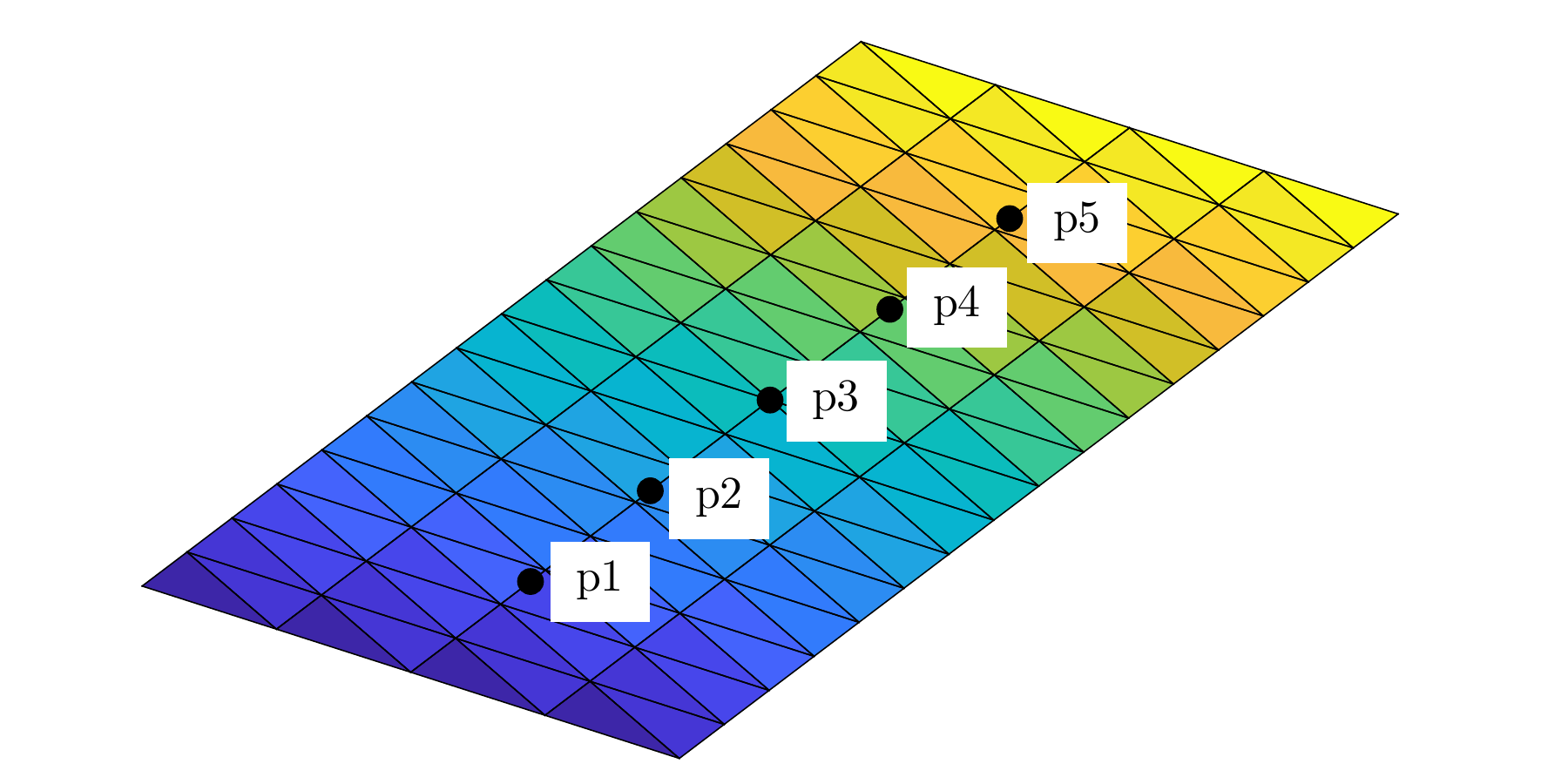}}}
\subfloat[$W_n$ = 0.292~m/s]{\scalebox{0.3}{\includegraphics[width=1\textwidth]{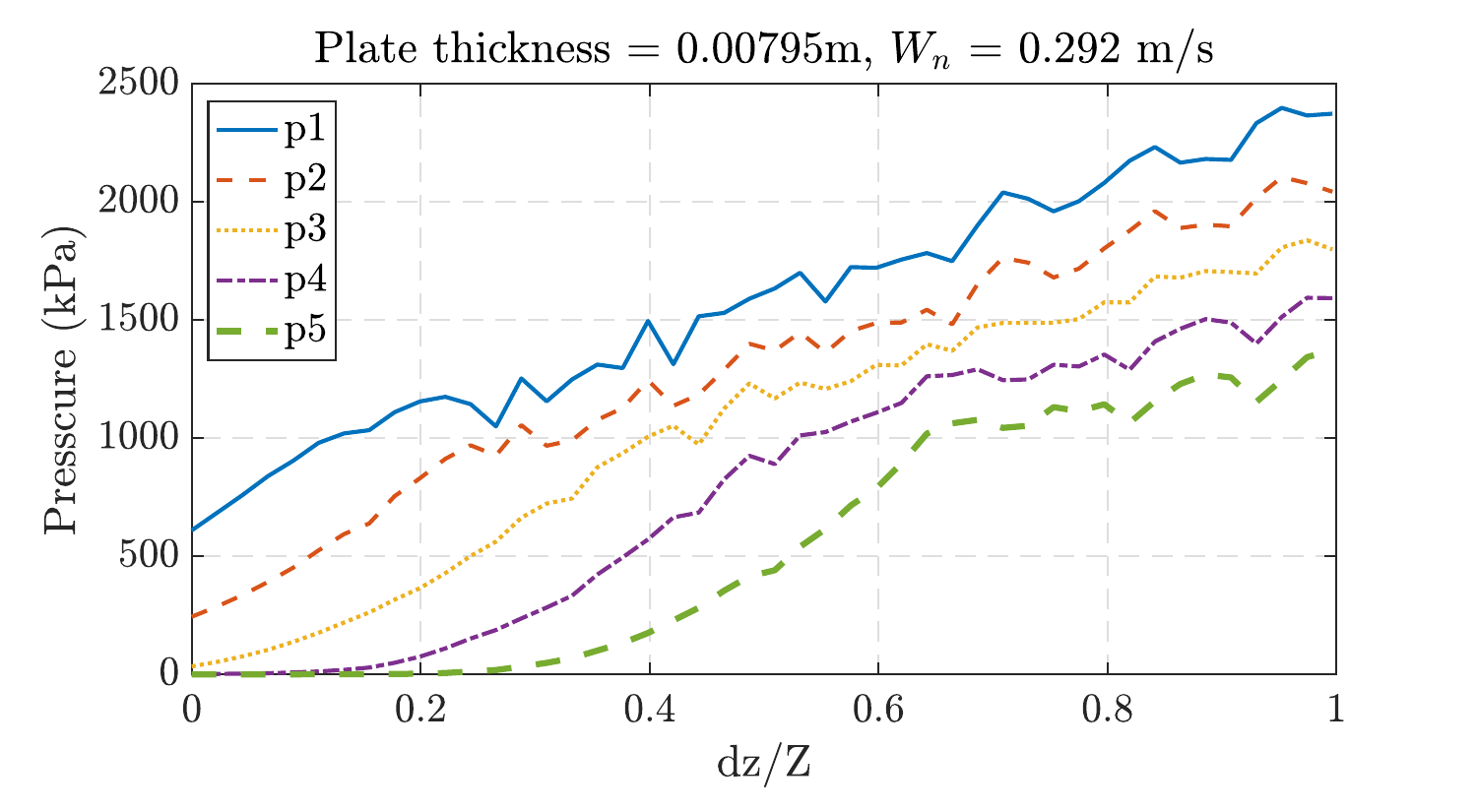}}}
\subfloat[$W_n$ = 0.875~m/s]{\scalebox{0.3}{\includegraphics[width=1\textwidth]{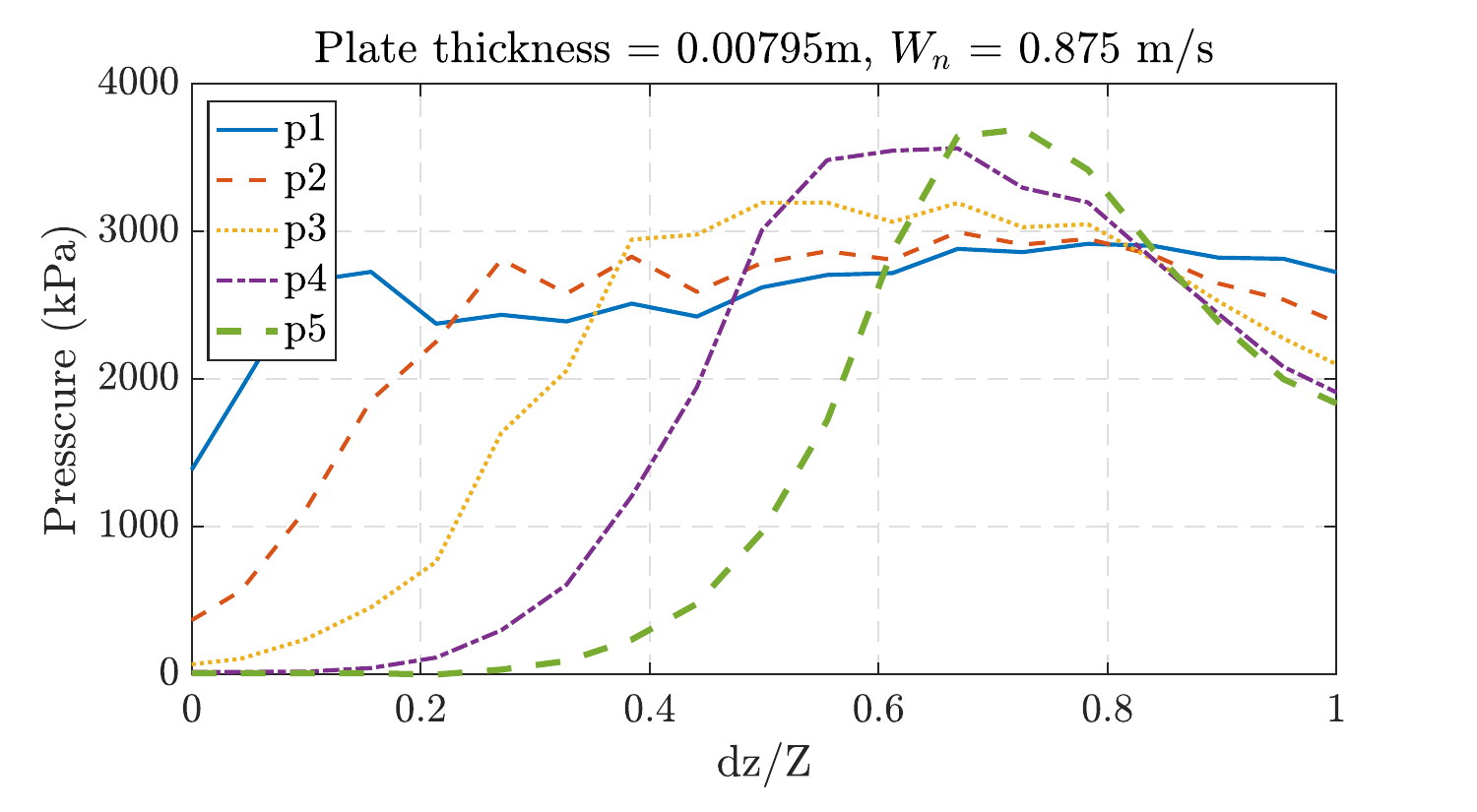}}}\\
\caption{The pressure histories of slamming impacts with $W_n = 0.292$~m/s and $W_n = 0.875$~m/s. Figure (a) displays the locations at which the pressures are measured in both cases.} \label{pressure}
\end{figure}

The pressure histories of the $W_n = 0.875~m/s$ and $W_n = 0.292~m/s$ cases on the 0.00795~m plate are presented in Figure \ref{pressure}. The pressure histories are extracted at five different locations along the plate as shown in Figure \ref{pressure}a (the black dots signal the locations of the pressure probes). In Figure \ref{pressure}b, we see that, due to the water splashing from the trailing edge, the pressure in all five locations of the plate is monotonically increasing. This leads to the monotonically increasing force behavior in Figure \ref{UW0} in $W_n = 0.292$~m/s. While in Figure \ref{pressure}c, pressure unloading is observed where the pressure drops after the spray root passes the locations of the pressure probes. This reflects the bilinear impact force behavior shown in Figure \ref{UW0} in $W_n = 0.875$~m/s. 

In this section, we have presented three sets of validation tests. The first testing set examines the influence of $U/W$ ratio on structural behaviors in cases with $W_n = 1.386$~m/s; the second testing set investigates the effects of $F_r$ on cases with $U/W = 8.33$; and the third testing set studies the structural behaviors of the plate without influences of forward velocity, $U$. In all these validation tests, our proposed FSI solver successfully captures the global force and plate deformation features in these cases. Furthermore, the FSI solver offers satisfactory agreements with experimental displacement data, especially for the 0.00795~m and 0.0127~m plates (the relative errors in the present work for forces are comparable to \cite{Pellegrini2020} and our displacement errors are substantially lower). Validation results of all slamming tests listed in Figure \ref{test_cases} can be found in \ref{appendix:a}.

\section{Additional FSI investigations into oblique slamming impact} \label{applications}

In this section, we use our validated FSI solver to carry out a series of complementary FSI slamming analyses, in an effort to elucidate some missing physical insight from the flexible plate oblique impact experiments. Specifically, we apply our validated FSI simulation method to the study of global trends that arise within the engineering context of oblique slamming of flexible plates.

\subsection{The effect of horizontal velocity}
\begin{figure}[h!]
\centering
\captionsetup[subfigure]{justification=centering}
\subfloat[]{\scalebox{0.33}{\includegraphics[width=1\textwidth]{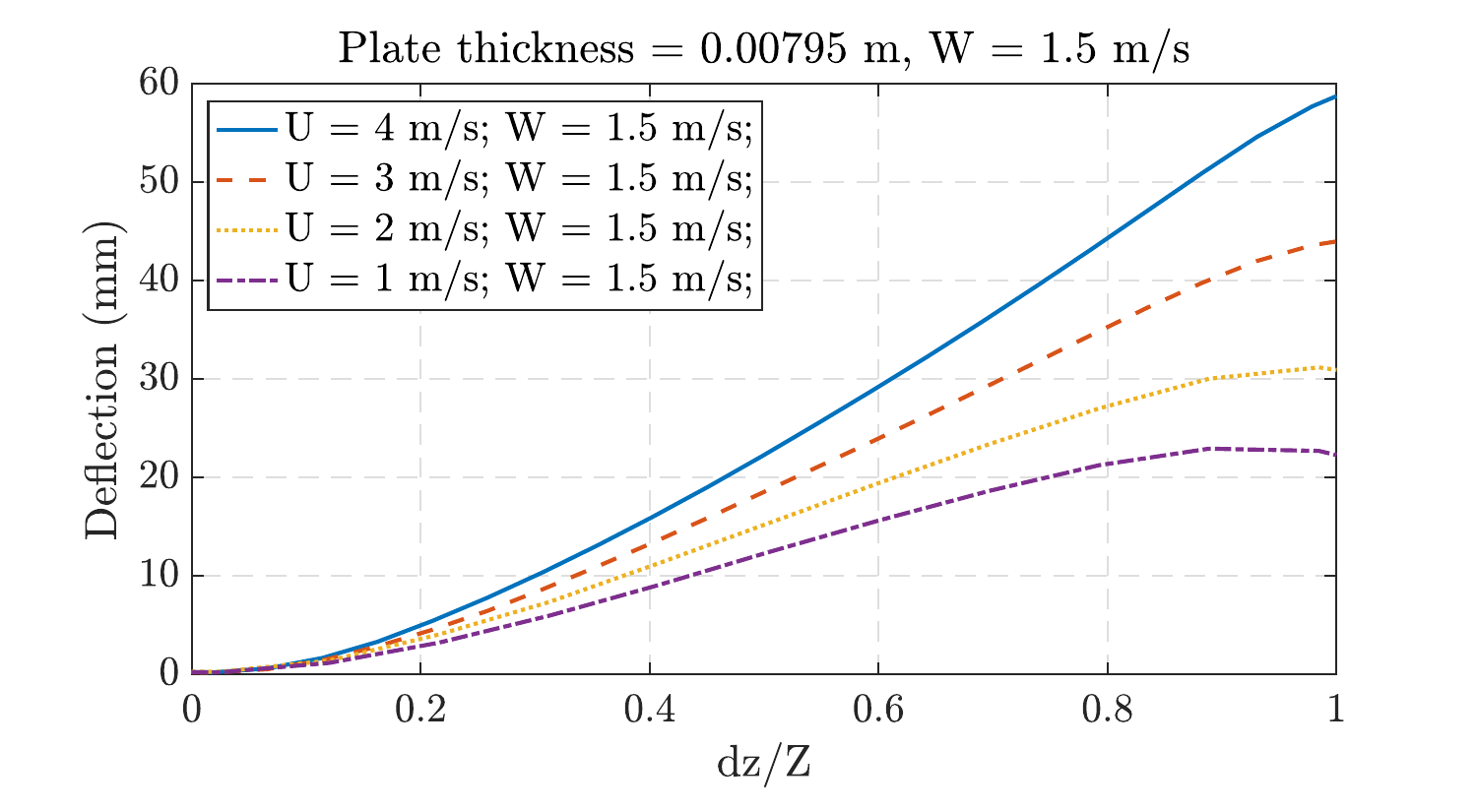}}}
\subfloat[]{\scalebox{0.33}{\includegraphics[width=1\textwidth]{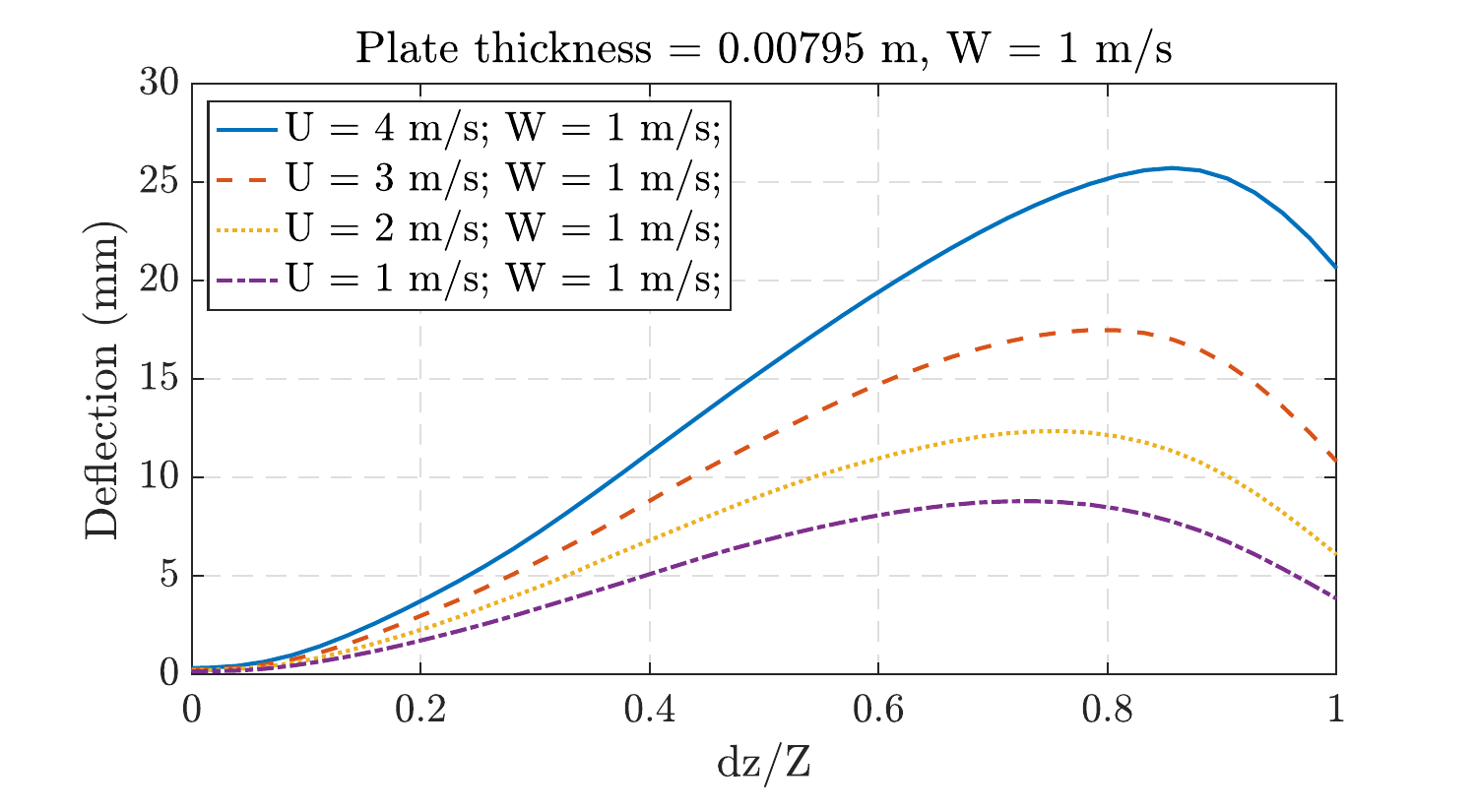}}}
\subfloat[]{\scalebox{0.33}{\includegraphics[width=1\textwidth]{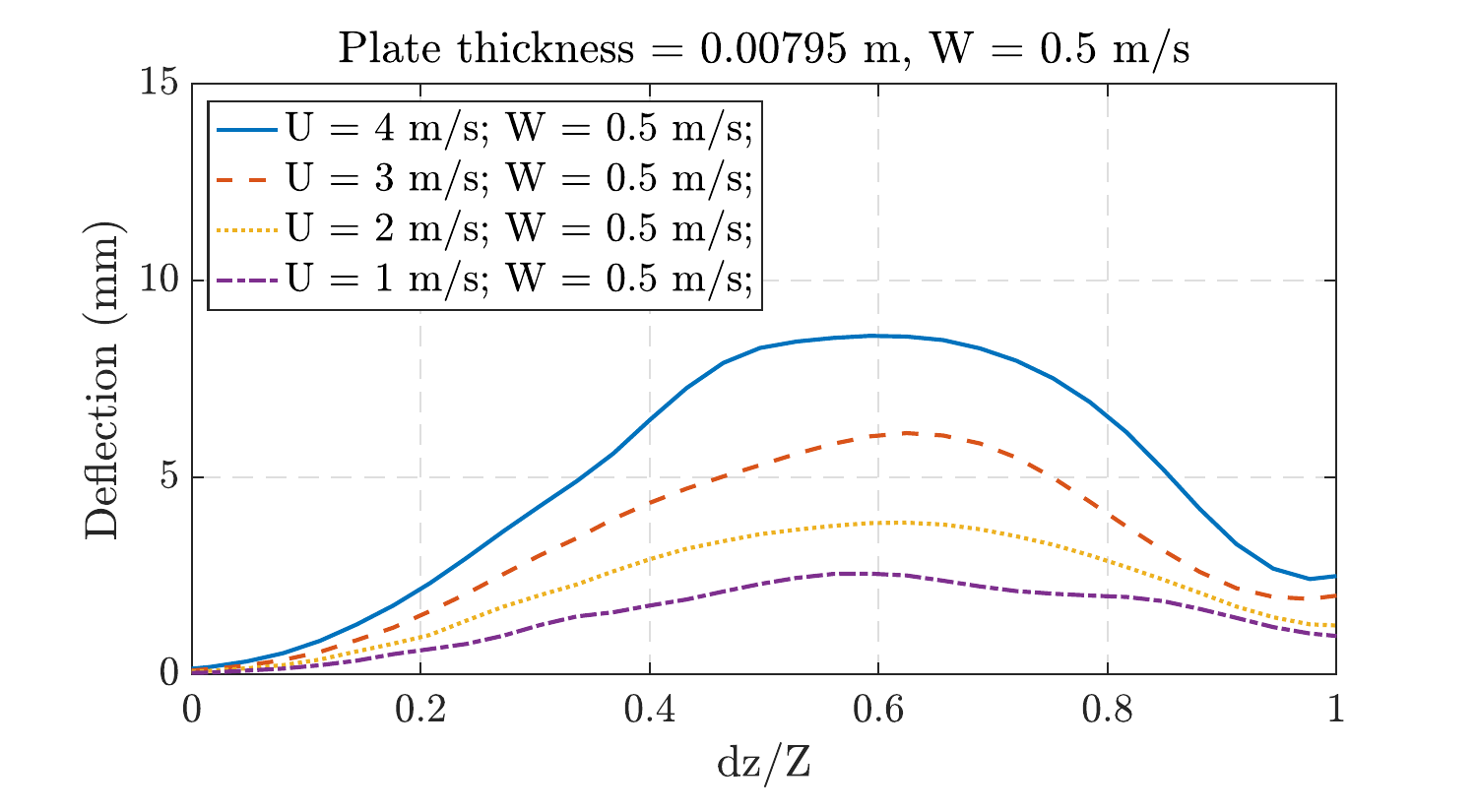}}}\\
\subfloat[]{\scalebox{0.33}{\includegraphics[width=1\textwidth]{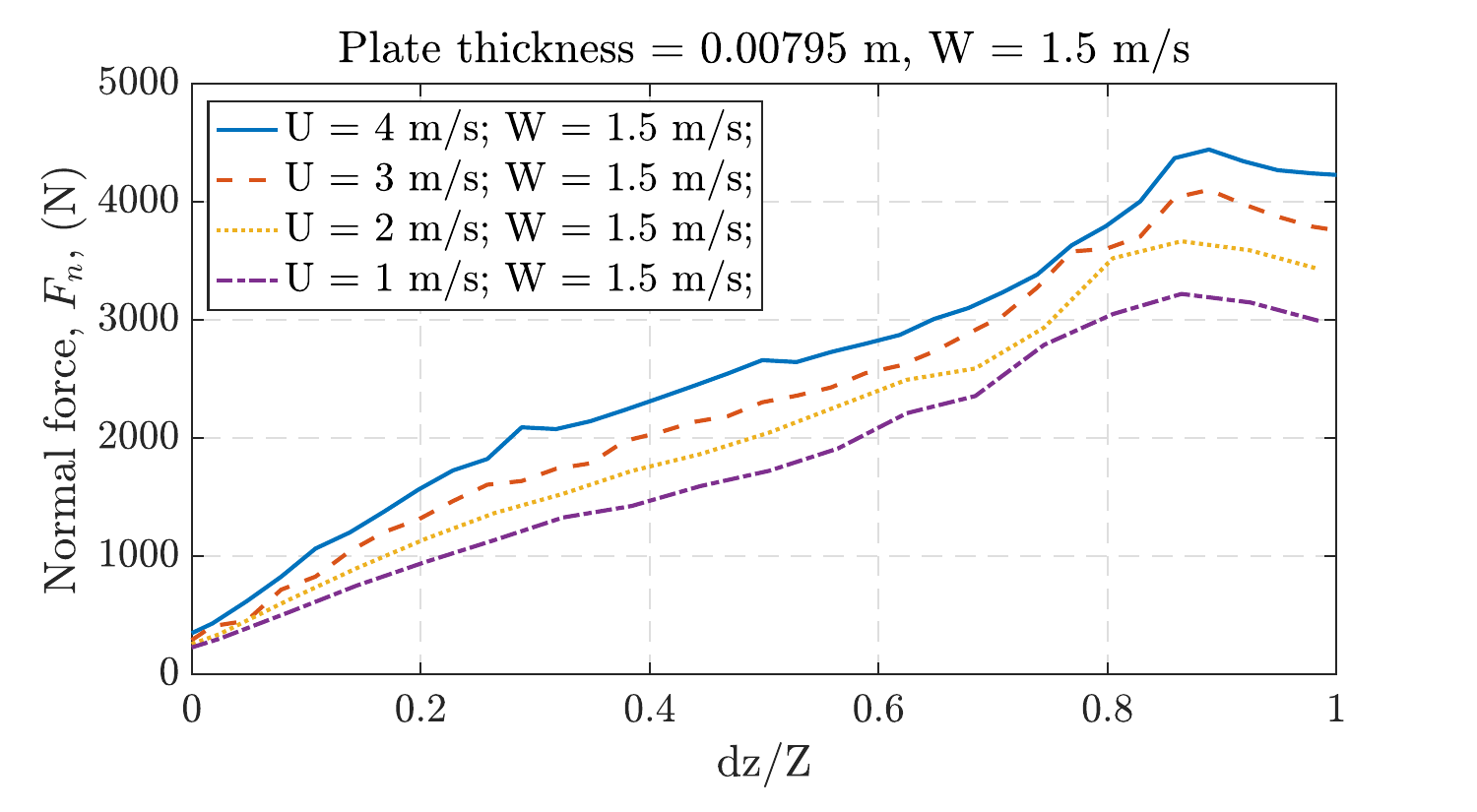}}}
\subfloat[]{\scalebox{0.33}{\includegraphics[width=1\textwidth]{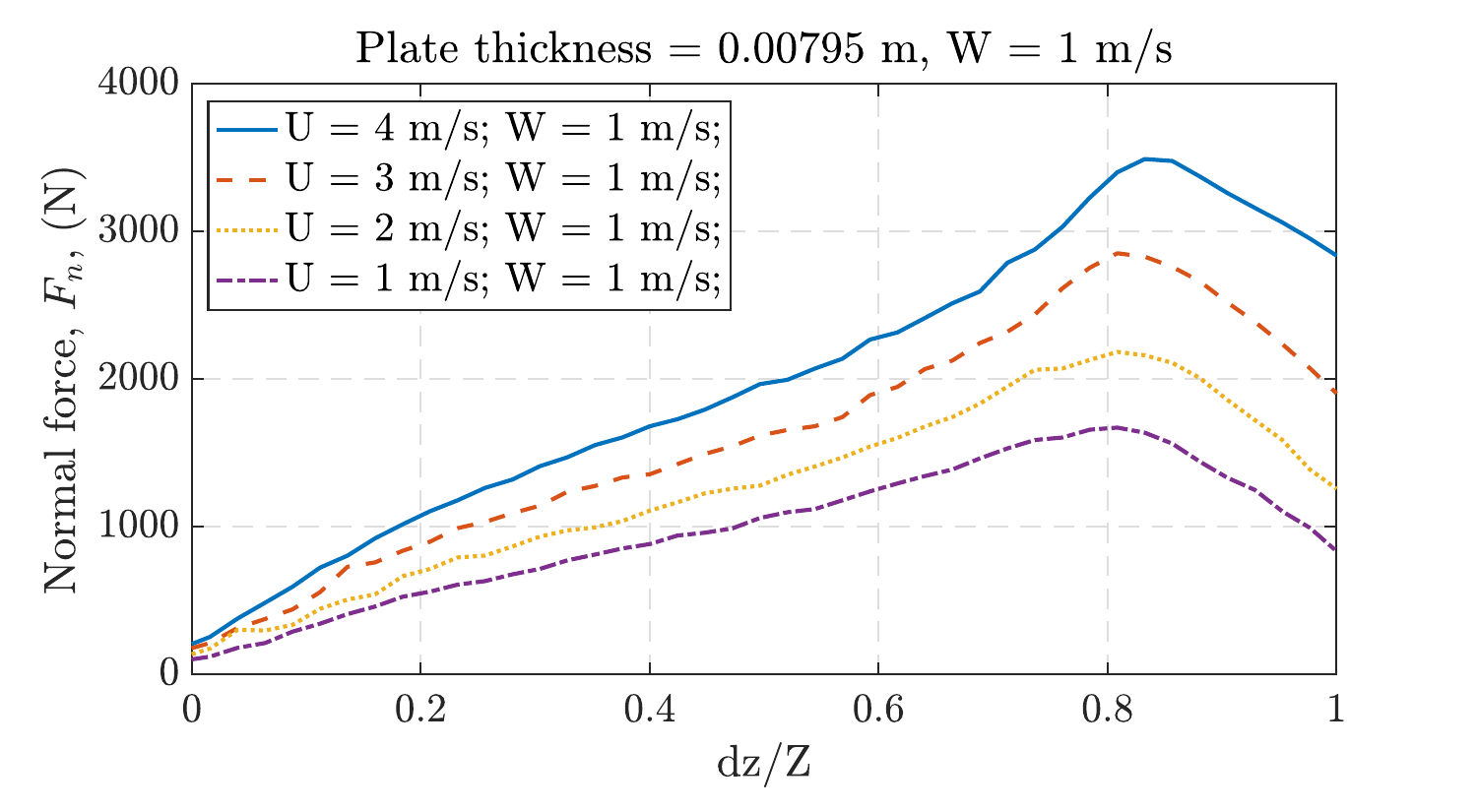}}}
\subfloat[]{\scalebox{0.33}{\includegraphics[width=1\textwidth]{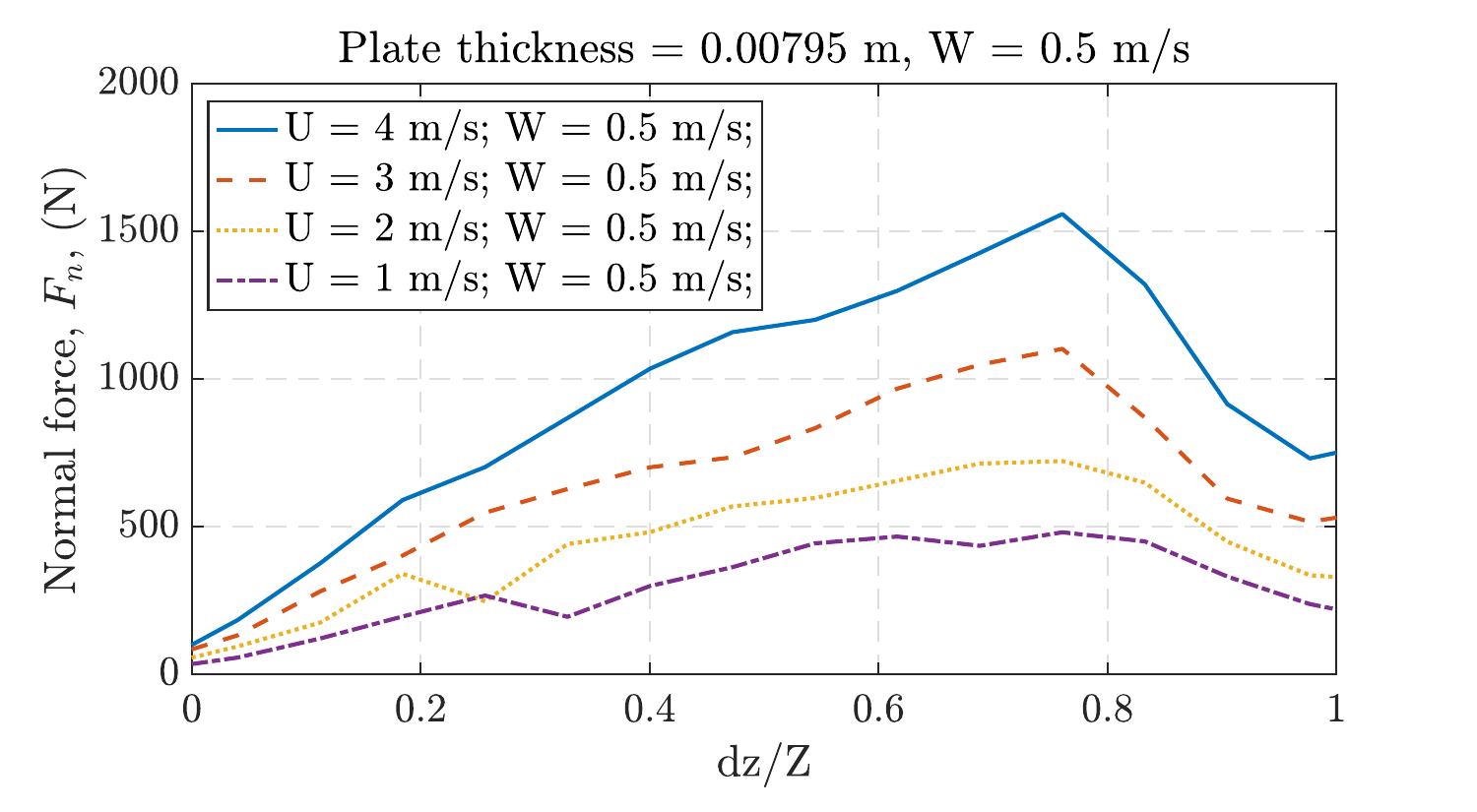}}}
\caption{Plate deflection and normal impact force histories of the 0.00795~m plate subjected to slamming impacts with $W = 1.5, 1, \text{and } 0.5$~m/s.} \label{Wfixed}
\end{figure}

In the previous section, we presented and discussed the structural behaviors of the plates under slamming cases with fixed $W_n$, $U/W$, and $U$. In this section, we discuss FSI results with fixed $W$, to offer a more comprehensive picture of how velocities influence the normal impact force and plate deformation. The FSI analyses in this section focus on the 0.00795~m plate thickness case. We perform a total of 12 FSI slamming simulations; the slamming tests are divided into three groups with $W = 1.5$~m/s, $W = 1$~m/s, and $W = 0.5$~m/s, respectively. The $U$ and $W$ of the slamming tests are listed in Table \ref{app_test1}. The plate deflection and impact force histories are shown in Figure \ref{Wfixed}. As expected, by keeping $W$ fixed, higher $U$ results in higher $W_n$, and therefore leads to higher plate deflection and impact force. In addition, we observe in Figure \ref{Wfixed}a-c that the plate reaches its peak deformation faster (indicated by smaller $dz/Z$ value) and with lower $W$. Similarly, the impact force also reaches its peak sooner, with lower $W$. However, since the plate deformation is affected by the pressure distribution on the plate, rather than the normal impact force (the sum of pressure on the plate), the time of peak plate deformation, and peak force, do not necessarily coincide. 

\begin{table}[h!]
\centering
\caption{Forward and downward velocities of slamming simulations tests with fixed $W$.} \label{app_test1} 
\resizebox{\textwidth}{!}{\begin{tabular}{l|cccc|cccc|cccc}
 \hline  
 & \multicolumn{4}{c}{Group 1} &  \multicolumn{4}{c}{Group 2} &  \multicolumn{4}{c}{Group 3} \\
 \hline   
$W_n$ (m/s)  &  2.172 &  1.998     & 1.824  & 1.650 & 1.679 & 1.506 & 1.332 & 1.158 &  1.187 & 1.013& 0.840 & 0.666    \\
  U/W        &   2.67    & 2      & 1.33 & 0.67 &  4 & 3& 2& 1&  8 & 6& 4 & 2\\
  U (m/s)   &   4         &  3     & 2& 1 &   4     &  3 & 2& 1 & 4     &  3 & 2& 1   \\ 
  W (m/s)  &   1.5      &1.5    & 1.5  & 1.5 & 1.0 & 1.0 & 1.0 & 1.0& 0.5 & 0.5 & 0.5 & 0.5    \\ 
  $F_r$    & 0.668 & 0.614 & 0.561 & 0.506 & 0.516 & 0.463& 0.409 & 0.356&  0.365&  0.311 & 0.258 & 0.205 \\
  \hline
\end{tabular}}
\end{table}

 \begin{figure}[h!]
\centering
\captionsetup[subfigure]{justification=centering}
\subfloat[]{\scalebox{0.5}{\includegraphics[width=1\textwidth]{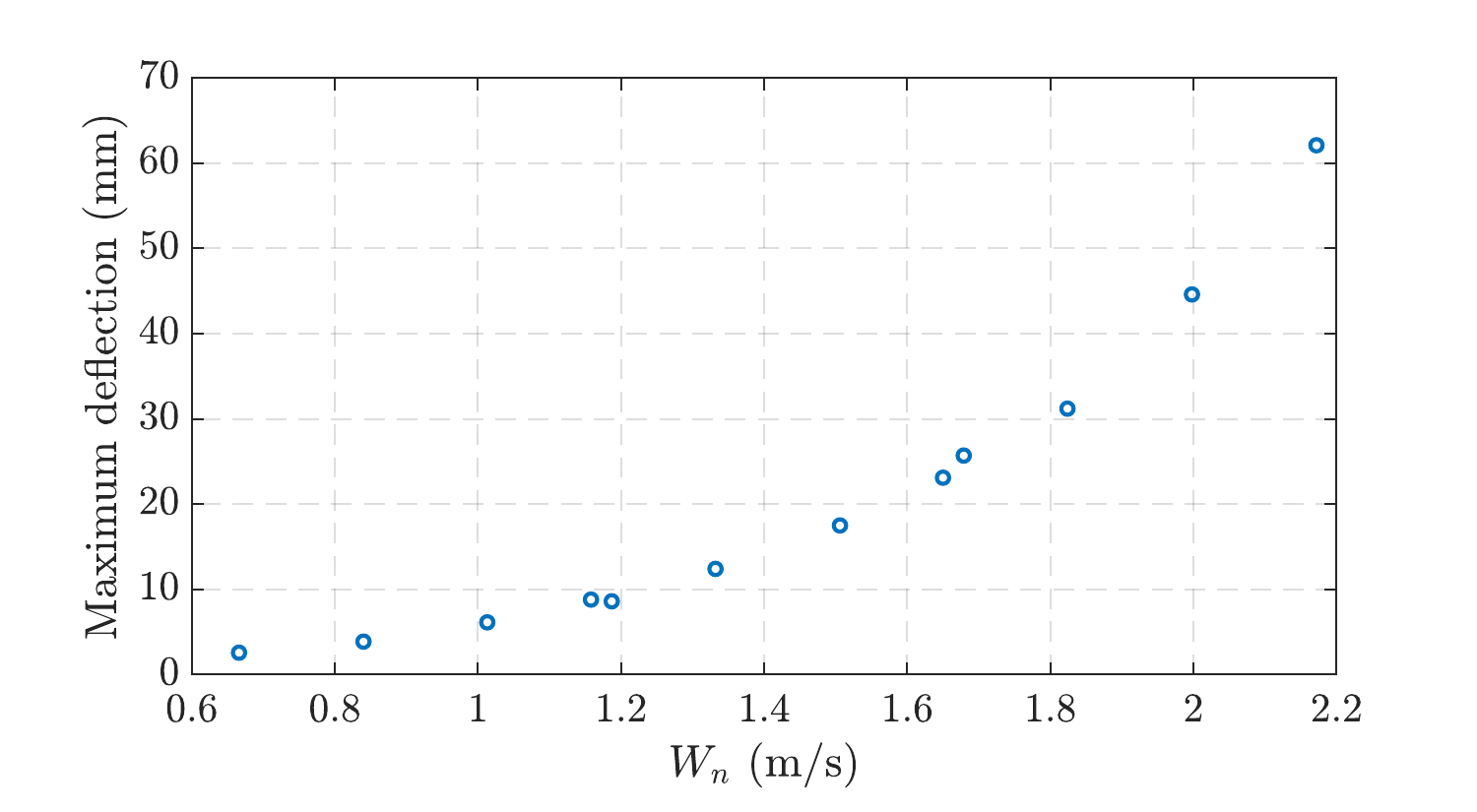}}}
\subfloat[]{\scalebox{0.5}{\includegraphics[width=1\textwidth]{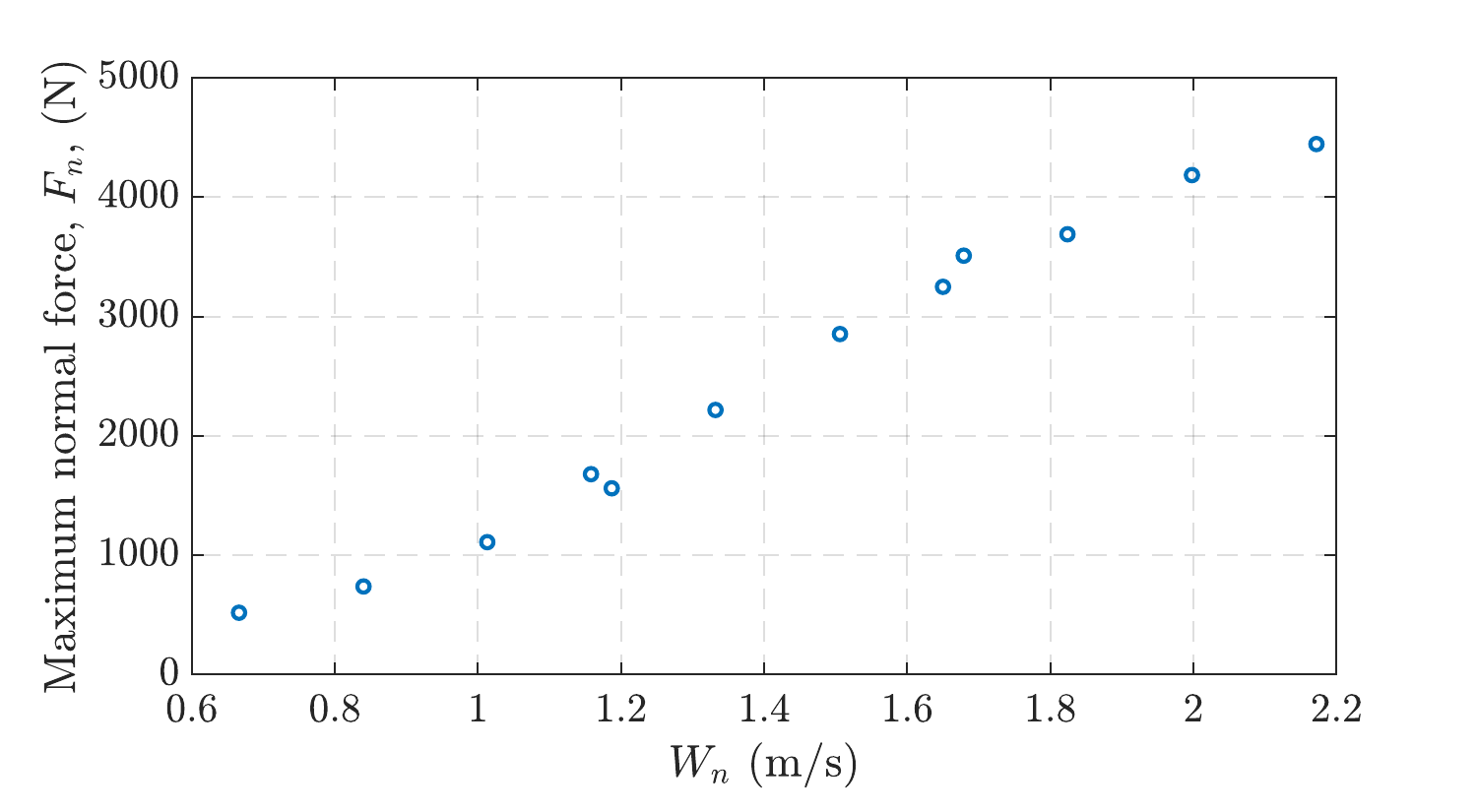}}}
\caption{Peak displacement and impact force as a function of $W_n$.} \label{peak}
\end{figure} 

In naval application, peak deflections and impact forces are of most interest, as these engineering responses directly connect to the adequacy in design. The peak deflections and normal impact forces, w.r.t $W_n$, from the 12 FSI slamming simulations are presented in Figure \ref{peak}, to better understand the relations between peak deflection/normal impact force and $W_n$. In Figure \ref{peak}, we observe positive relation between maximum plate deflections/normal impact forces and $W_n$; the maximum plate deflection increases in an approximately quadratic manner, with increasing $W_n$, while the maximum impact force appears to increase linearly with increasing $W_n$. 

\subsection{The effect of U/W ratios and plate flexural rigidity}
\begin{figure}[h!]
\centering
\captionsetup[subfigure]{justification=centering}
\subfloat[]{\scalebox{0.33}{\includegraphics[width=1\textwidth]{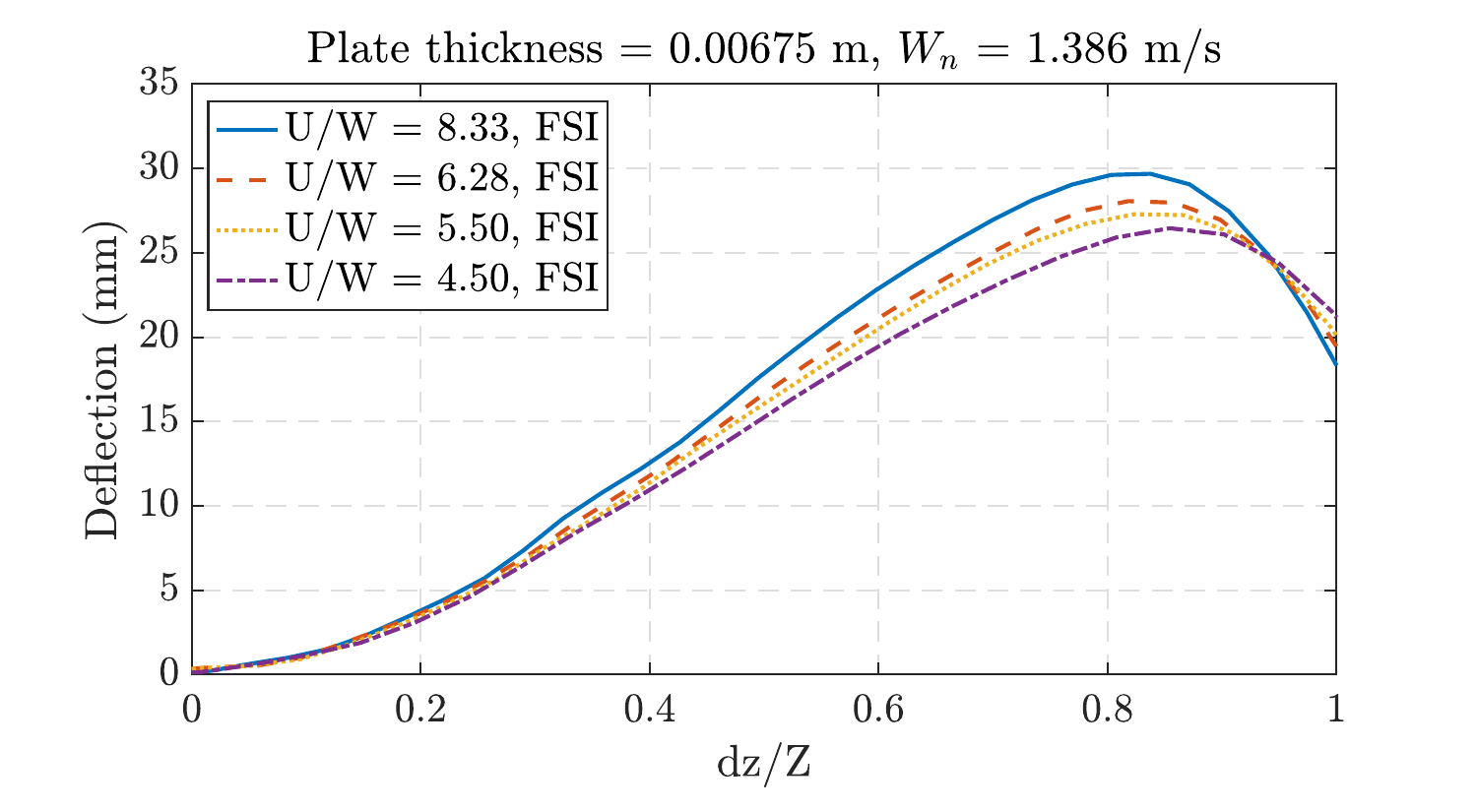}}}
\subfloat[]{\scalebox{0.33}{\includegraphics[width=1\textwidth]{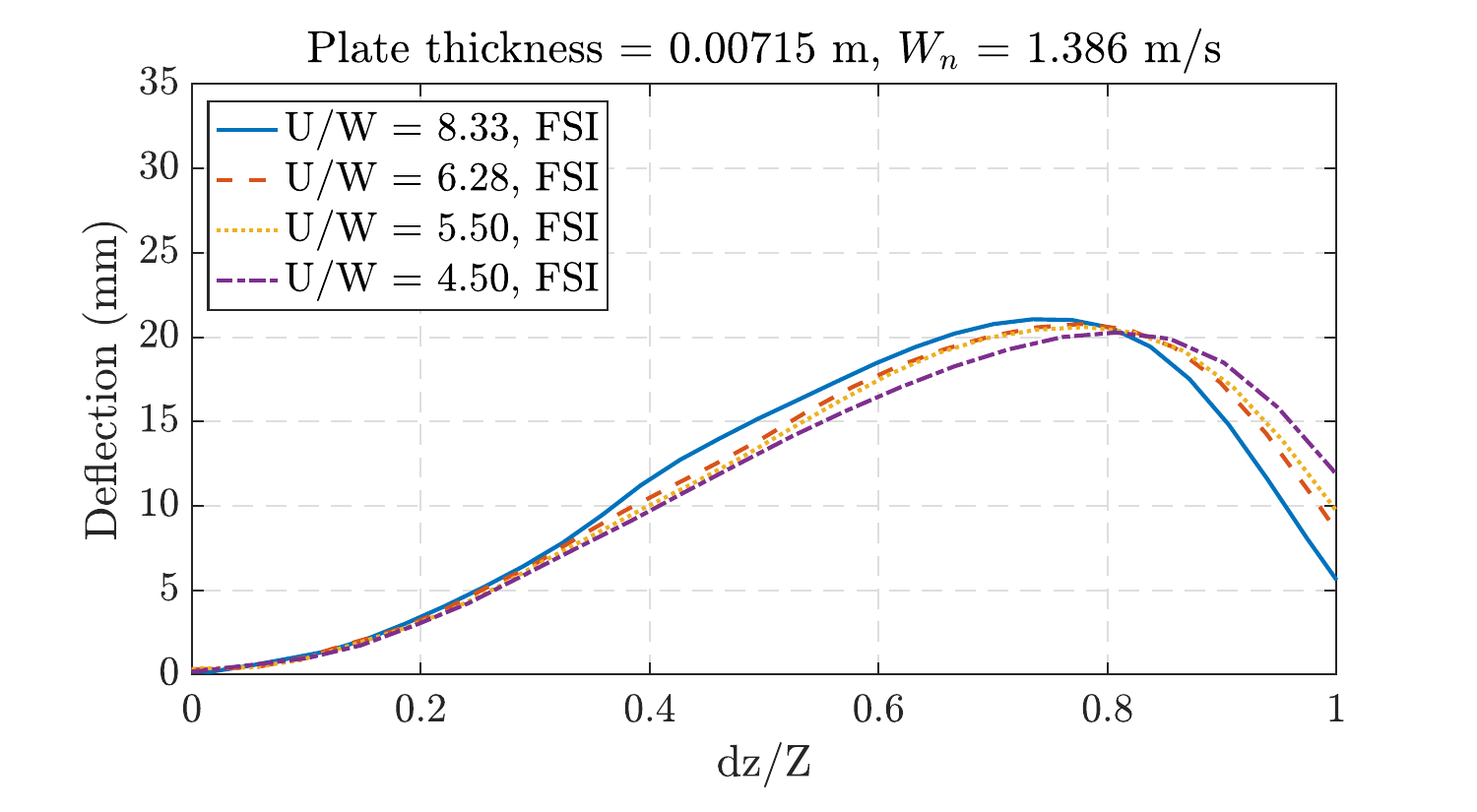}}}
\subfloat[]{\scalebox{0.33}{\includegraphics[width=1\textwidth]{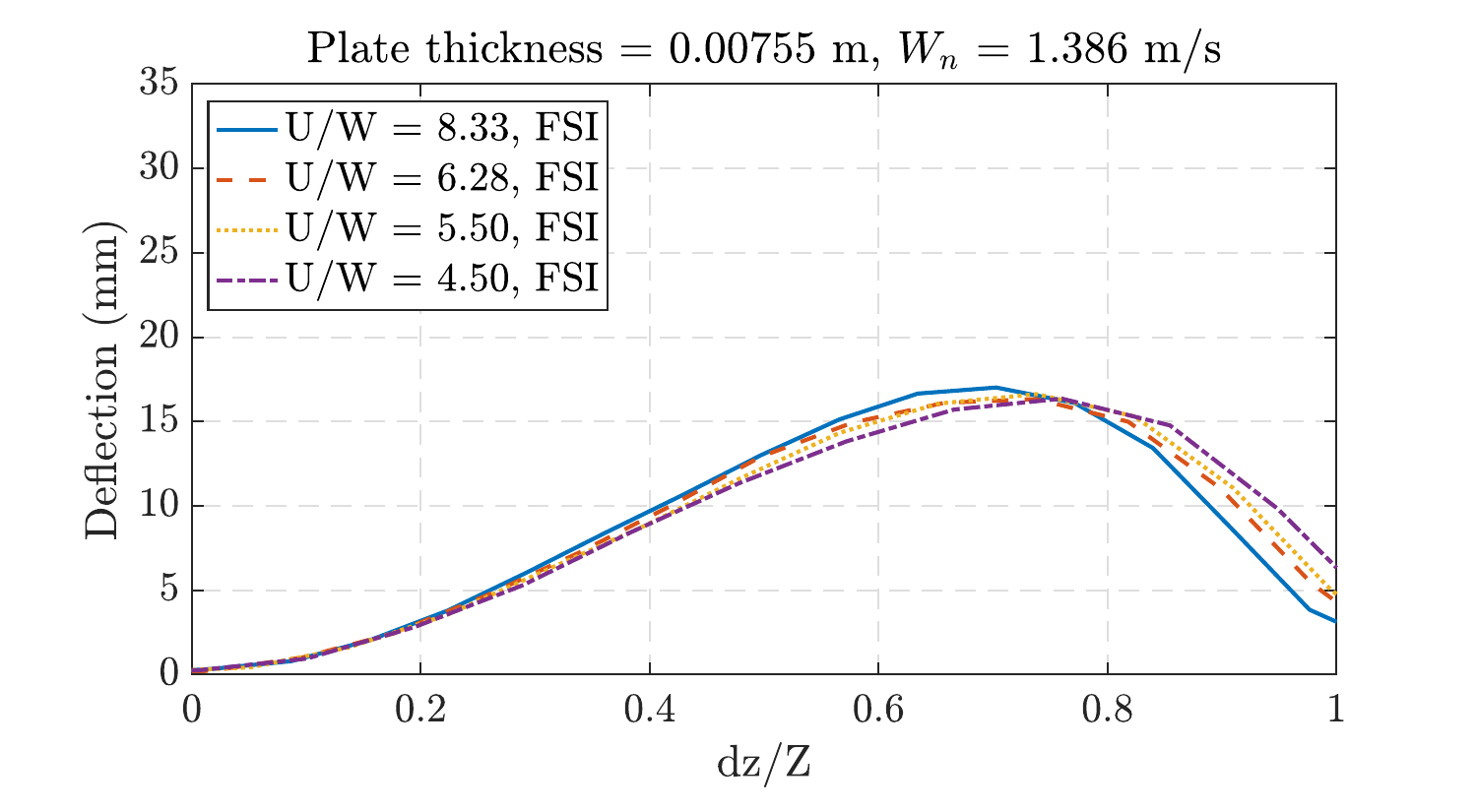}}}\\ 
\subfloat[]{\scalebox{0.33}{\includegraphics[width=1\textwidth]{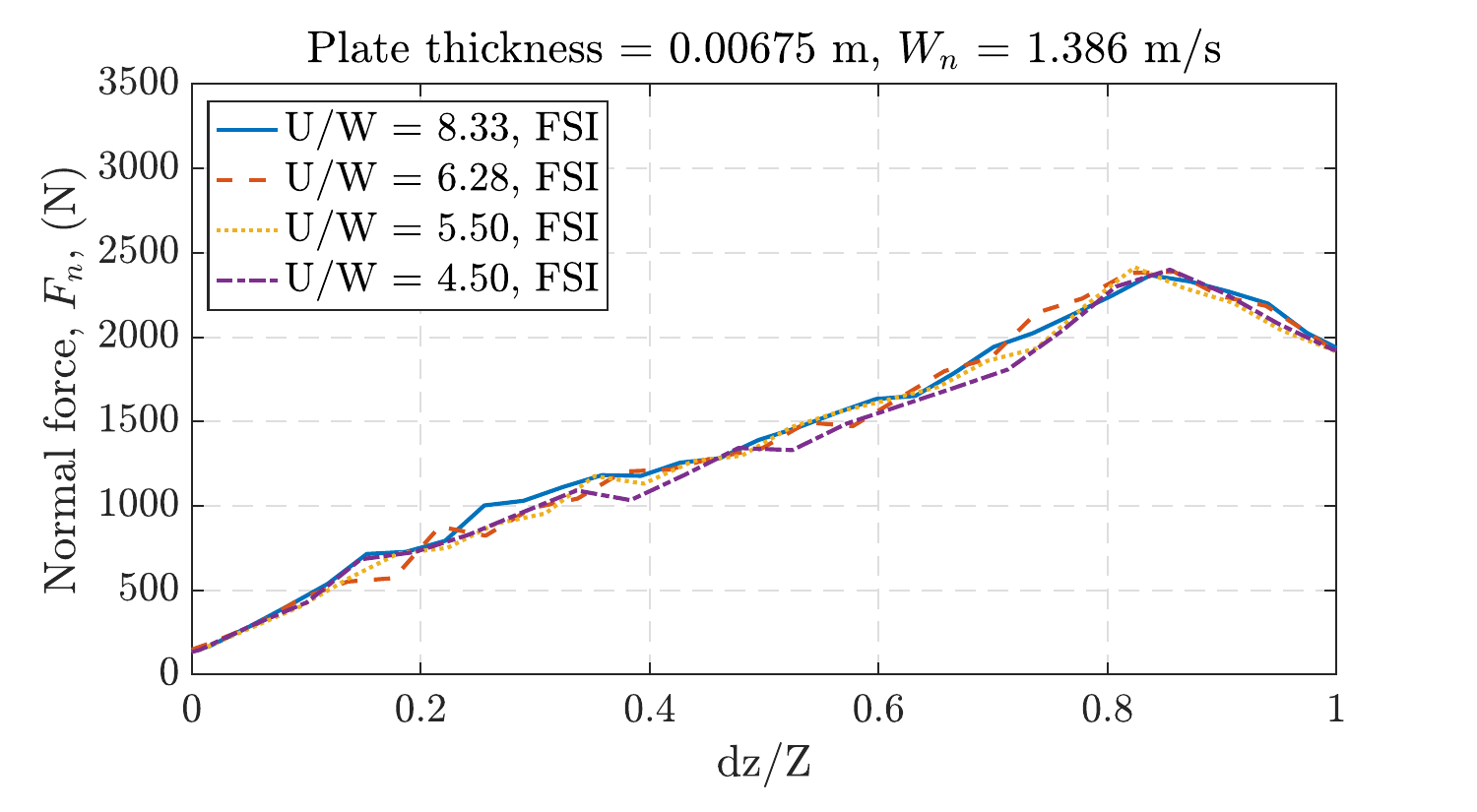}}}
\subfloat[]{\scalebox{0.33}{\includegraphics[width=1\textwidth]{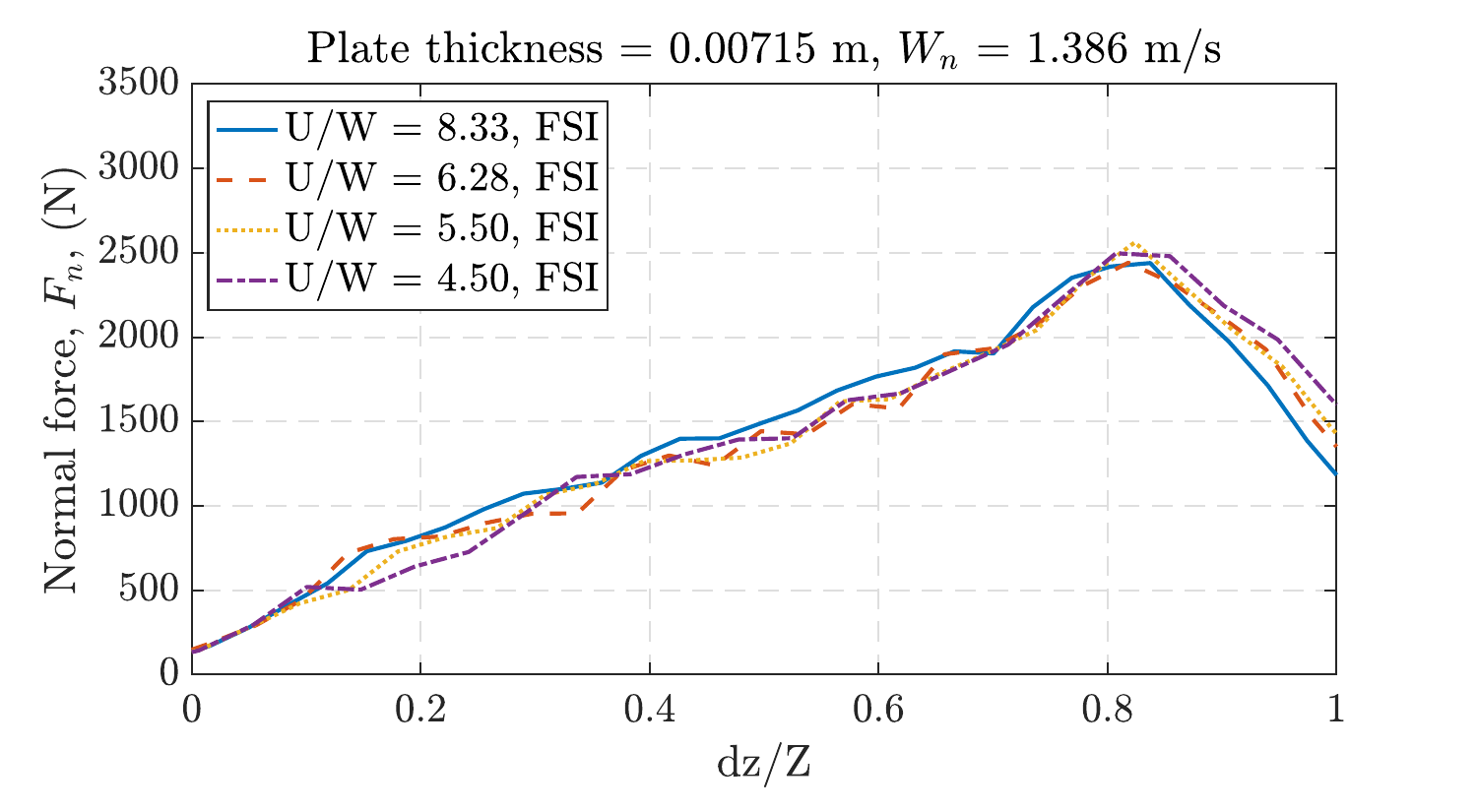}}}
\subfloat[]{\scalebox{0.33}{\includegraphics[width=1\textwidth]{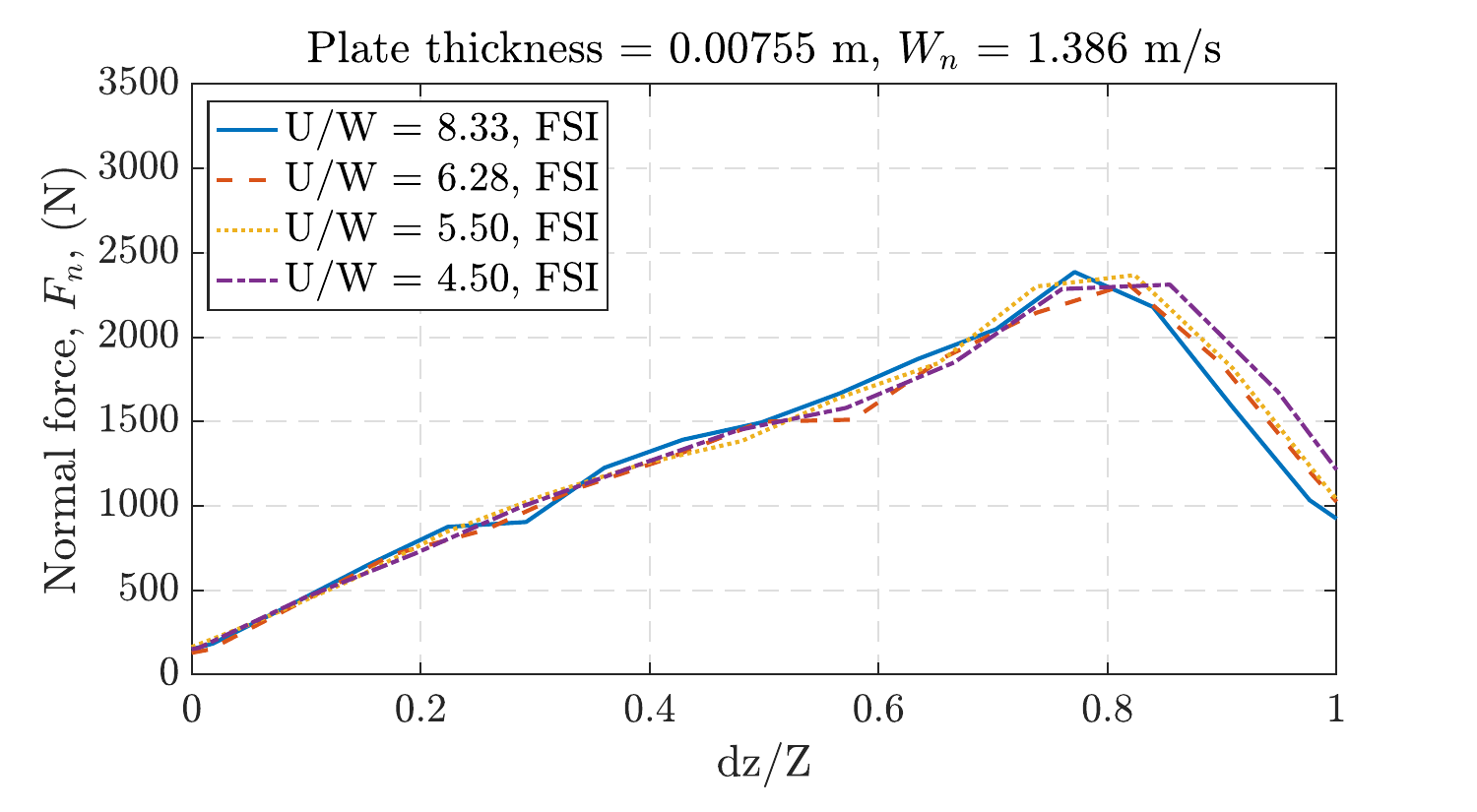}}} 
\caption{Plate deflection and normal impact force histories of flexible plates with thickness 0.00675~m, 0.00715~m, and 0.00755~m subjected to slamming impacts with $W_n = 1.386$~m/s.} \label{app_Wn1p386}
\end{figure}

Motivated by the validation results from the first set of FSI validation tests, the influence of U/W ratios on plates in relation to their rigidities is now examined. We prescribe slamming conditions $W_n = 1.386$~m/s and $U/W = 8.33, 6.28, 5.50,$ and $4.50$ on aluminum plates with thickness 0.00675~m, 0.00715~m, and 0.00755~m, respectively. Although the cases of 0.00795~m and 0.0127~m thick plates appear insensitive to $U/W$ ratio, we observed that higher $U/W$ ratio yields higher deformations on the 0.00635~m thick plate. The plate deflection and normal impact force of the 0.00675~m, 0.00715~m, and 0.00755~m thick plates are presented in Figure \ref{app_Wn1p386}. In the figures, we observed that the deformation discrepancy between $U/W$ increases with decreasing plate thickness. While the deviation between impact force seems insignificant in each of the plates, it appears that the thicker plate precipitates a steeper force unloading. 

\begin{figure}[h!]
\centering
\includegraphics[width=1\textwidth]{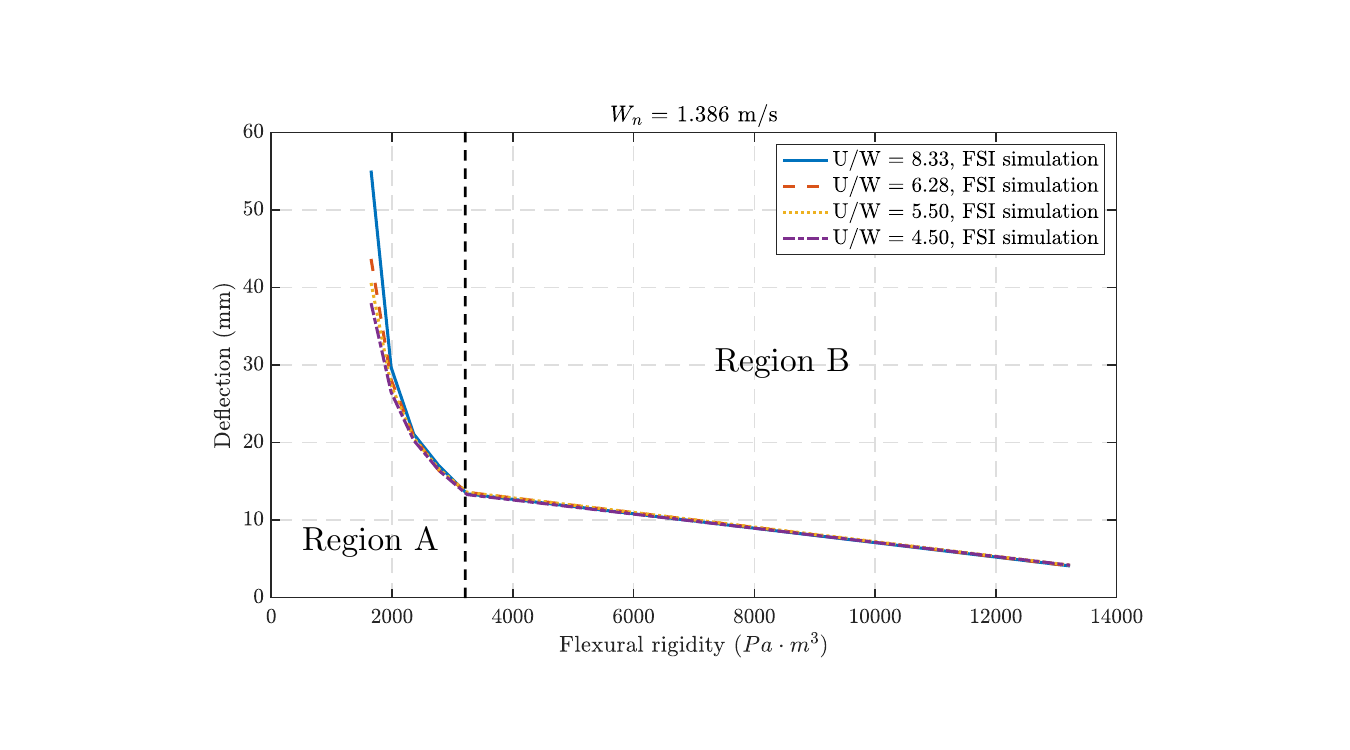}
\caption{Comparison of plate deflection among various $U/W$ as a function of flexural rigidity.} \label{rigidity_disp}
\end{figure}

The plate deformation as a function of flexural rigidity for all six plates (three from FSI validation test and three from current modeling) is presented in Figure \ref{rigidity_disp}. The flexural rigidity of each plate is computed following Equation \ref{flexural}, and used in the plots of Figure \ref{rigidity_disp}, where an inverse relationship between plate deflection and flexural rigidity is observed, as expected. The relationship can be divided into two regions, demarcated by a flexural rigidity of 3200 $Pa\cdot m^3$. Within Region A (\textit{i.e.,} flexural rigidity $\leq$ 3200 $Pa\cdot m^3$), plate deformation appears to decrease quadratically with increasing flexural rigidity. On the other hand, within Region B (\textit{i.e.,} flexural rigidity $>$ 3200 $Pa\cdot m^3$), plate deformation appears to decrease approximately linearly with increasing flexural rigidity. Moreover, the influence of $U/W$ is negligible in Region B, while visible differences in deflection are observed in Region A, between various $U/W$ (the influence of $U/W$ is more prominent with smaller flexural rigidity). Note that the threshold 3200 $Pa\cdot m^3$ is purely phenomenological, based on the flexible plate geometry in the present work. More extensive studies are needed to identify a general threshold for differentiating the linear and nonlinear region in the deformation vs flexural rigidity relation.  

To summarize this section, a series of complementary FSI slamming simulations are performed to better understand the effects of downward impact velocity, $W$, and plate flexural rigidity on the structural behaviors of the flexible plates (\textit{i.e.,} we are using our experimentally validated simulations to study global trends in oblique slamming behavior). Through the analyses, we have learned 1) both the plate's peak displacement and peak impact force are positively related to $W$, with peak displacement increasing quadratically and peak impact force increases linearly with increasing $W$; and 2) the plate's peak displacement is negatively related to flexural rigidity. The plate deflection decreases quadratically with increasing flexural rigidity up to 3200 $Pa\cdot m^3$, afterwards, the plate deflection reduces nearly linearly as a function of flexural rigidity. 

\section{Engineering theory for flexible plates under oblique slamming impacts}\label{theory}

In this section, we leverage our insights from Section \ref{applications}, obtained using our validated simulation approach, to pursue useful design-equation-type mathematical models. Previous works on added mass approximation have been reported in \cite{Faltinsenbook, scolan2003, korobkin2017}. However, the reported works are limited to rigid, symmetric bodies (\textit{e.g.,} circular cylinder, elliptic paraboloid, and horizontal plate). Additionally, many of the relevant theories in the literature are purely theoretical, without the benefit of experimental observation. In the current work, we aim to use our experimental validated software to offer unprecedented mechanistic insight as we develop a simple engineering theory for flexible plate oblique slamming. We begin from a mathematical model based on Newton's second law, to characterize the normal impact force resulting from a flexible plate entering a quiescent water free surface at constant downward velocity. As we move the plate into the water, the water surrounding the plate is disturbed by the motion of the plate; this creates an unsteady flow. The unsteady flow generates an additional inertial force, known as the \textit{added mass force}, that contributes to the plate deformation. To approximate the total normal impact force acting on the plate, we begin from Newton's second law, written as 

\begin{equation} \label{newton}
F_n = \frac{d}{dt}\left[(m_p+m_a)W_n^t\right],
\end{equation}
where $F_n$ is the normal impact force, $m_p$ is the mass of the plate, $m_a$ is the added mass from the unsteady flow, and $W_n^t$ is the impact velocity orthogonal to plate surface at time $t$. Surface tension, buoyancy, and the steady state drag force are assumed negligible in the present work. 

In the case of constant impact velocity, the added mass is varying throughout the slamming impact event, while the mass of the plate is fixed, Equation \ref{newton} becomes  

\begin{equation} \label{newton2}
F_n = \frac{dm_a}{dt}W_n^t.
\end{equation}

Equation \ref{newton2}, suggests the normal impact force of flexible plate slamming at a constant impact velocity is solely dependent on the change of added mass in time, as well as the impact velocity. However, to the Authors' knowledge, a general analytical solution for determining the exact added mass has yet to be established -- indeed the added mass will take a suitable form, depending on the geometry of the structure at play, along with other factors such as the engineering properties of the structure, the viscosity of the fluid, \textit{etc}. That said, it is also very difficult to measure the so called added mass force systematically in experimental settings. In this work, we aim to ascertain the added mass influence by studying the pressure contours within the water phase that accompanies the slamming events within the context of the FSI validation testing set in Subsection \ref{validation_results} (\textit{i.e.,} the testing cases subjected to vertical impact only). From these ``empirical" results, obtained through pressure contour investigations within our simulations, we uncover the evolutions of the added mass: this allows us to generalize the shape of the added mass for our context. Thereafter, we derive a simple, yet accurate mathematical formulation, to approximate the normal impact forces in slamming under constant vertical velocity, that may be suitable for use in design. 

\subsection{Added mass surface modeling} \label{acquisition}
\begin{figure}[h!]
\centering
\includegraphics[width=1\textwidth]{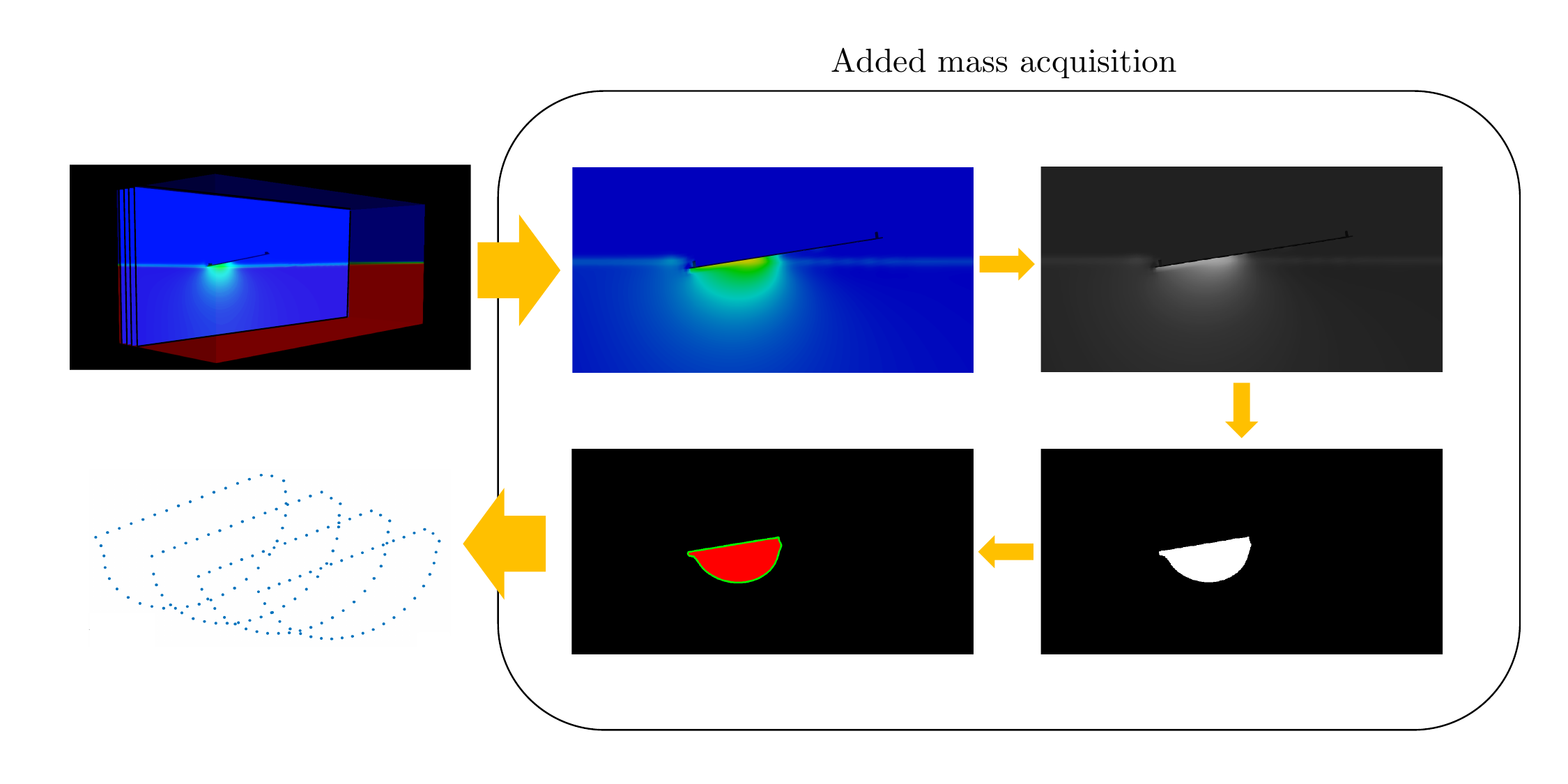}
\caption{The image processing procedure adopted in the present work to extract the point cloud which surrounds the added mass.} \label{added_mass_acq}
\end{figure}

Theoretically, the added mass shape can be uncovered either through the velocity or the dynamic pressure contours (since dynamic pressure is defined as $p_d = \frac{\rho_w(\mathbf{u}^f)^2}{2}$). However, during post-processing, the color gradient of the velocity contour falls within a very small range, which makes it nearly impossible to identify the added mass boundary beneath the plate from this measure. Consequently, dynamic pressure contours are preferred in the present work, for their wider range of color gradient, and therefore are used to identify the flow motion and the size of the added mass. The adopted image processing procedure supporting the added mass identification is illustrated in Figure \ref{added_mass_acq}. At each $dz/Z$, we obtain four slices of pressure contours along the longitudinal direction of the plate. The width of the plate is bounded by $y = [0, 0.406]$, and so the four pressure slices are made at $y = 0.05, 0.15, 0.25, \text{and }0.35$, respectively. The top left figure in Figure \ref{added_mass_acq} shows the computational model with four pressure contour slices. For each pressure contour, we first convert the pressure contour into gray scale (top right). Then, we apply image thresholding to isolate the region bounded by the cyan layer in the original pressure contour (bottom right). Lastly, we create a boundary to enclose the region of interest (bottom center). Bottom left in Figure \ref{added_mass_acq} shows a point cloud of the added mass boundaries extracted from the four pressure contours. The choice of the cyan color as an indicator of the added mass boundary is based on numerical experimentations -- the cyan color provides a more accurate added mass prediction. The described procedure is repeated for $dz/Z = 0, 0.125, 0.25, ..., 1$, for each slamming test considered. Note, the pressure contours are scaled such that the pressure values correspond to the same color gradient across all $dz/Z$ values.

After we obtain an added mass point cloud, we adopt a stable and robust quadratic surface fitting method, the nonlinear Koopman method \cite{li2004}, to recover the shape of the added mass. The nonlinear Koopman method described in \ref{appendix:b} is formulated specifically to fit our data to an ellipsoid, which suits our needs given the shape of the added mass point cloud observed within the totality of simulations undertaken in this work. 

\subsection{Added mass surface fitting results}
\begin{figure}[h!]
\centering
\captionsetup[subfigure]{justification=centering}
\subfloat[dz/Z=0.125]{\scalebox{0.25}{\includegraphics[width=1\textwidth]{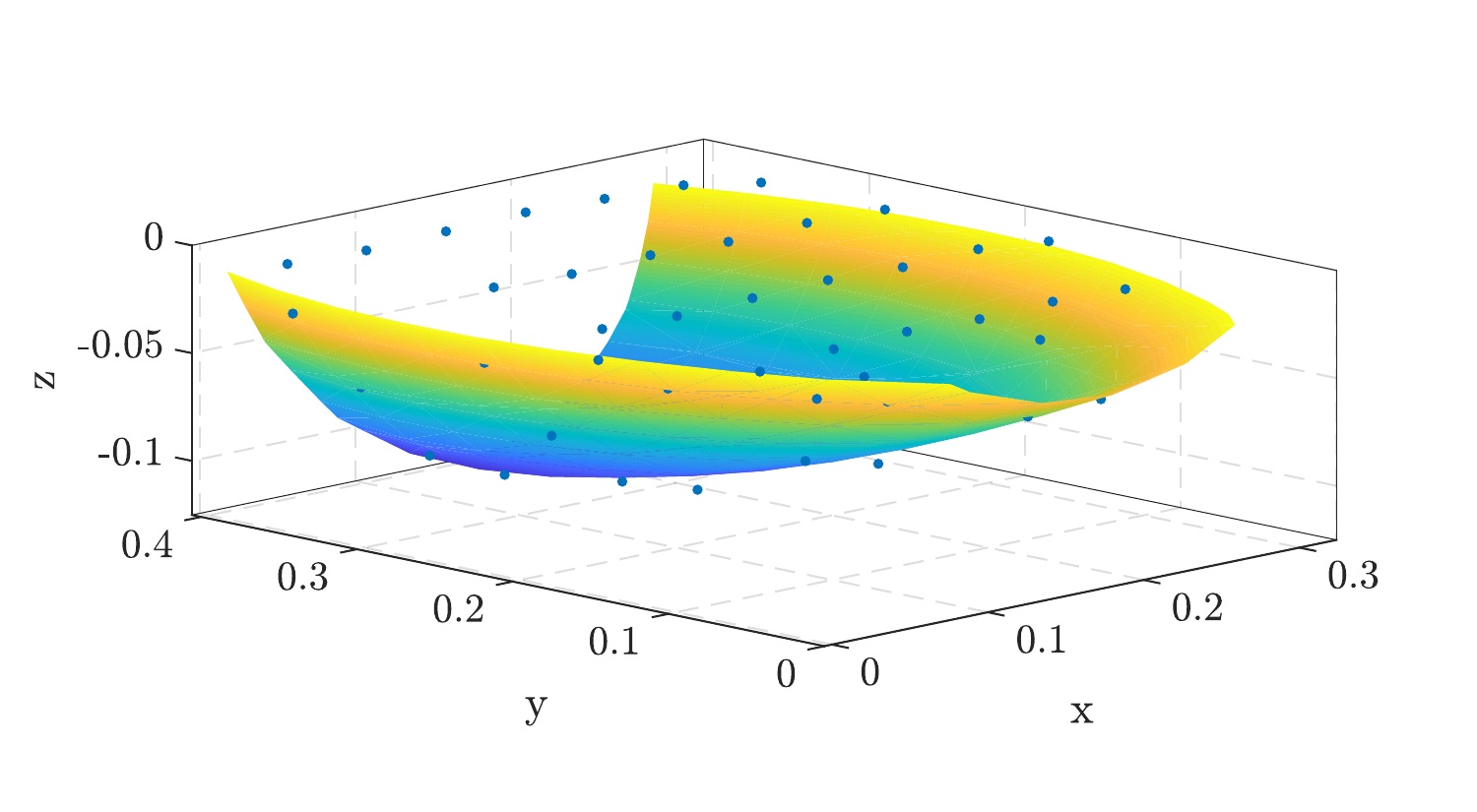}}}
\subfloat[dz/Z=0.25]{\scalebox{0.25}{\includegraphics[width=1\textwidth]{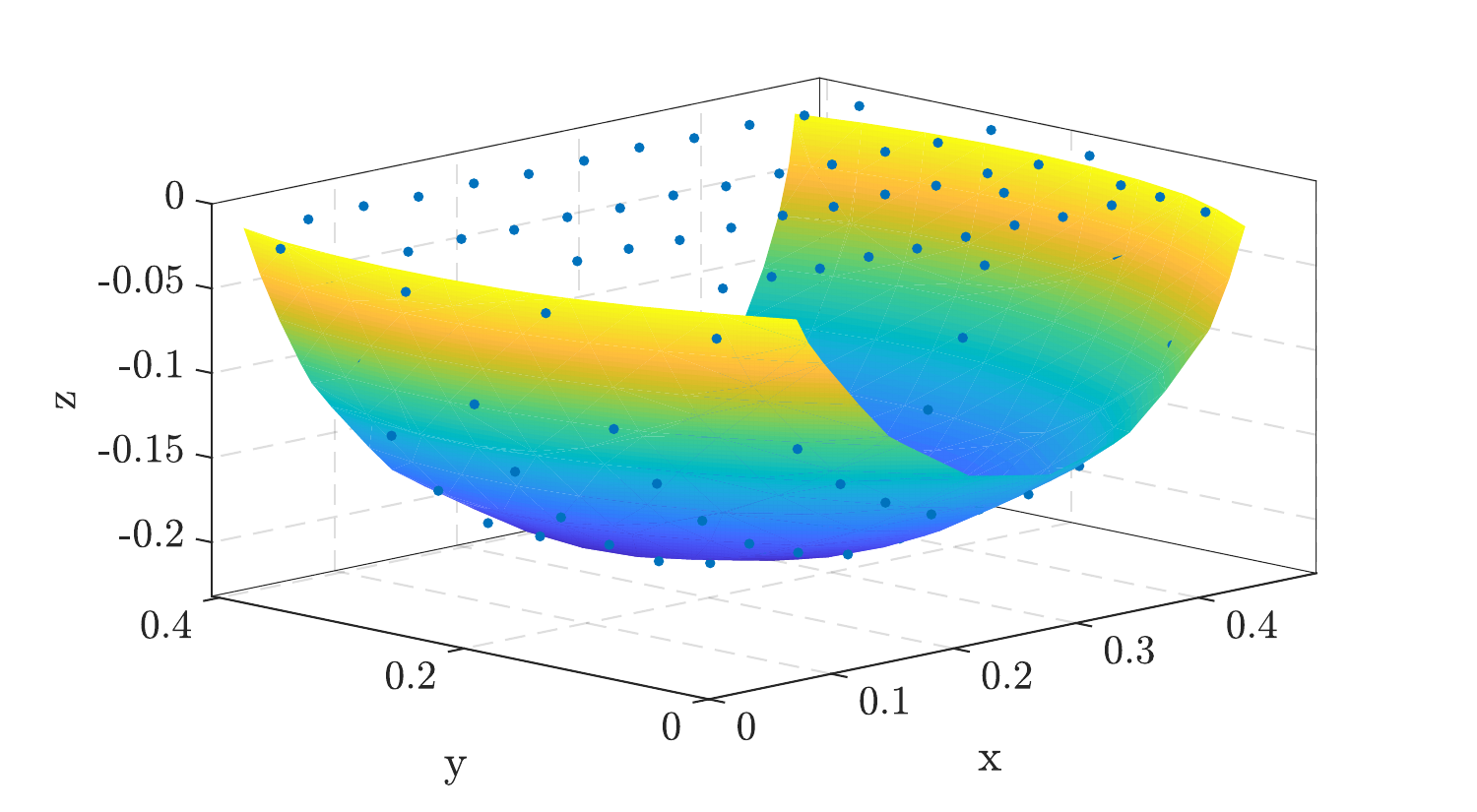}}}
\subfloat[dz/Z=0.375]{\scalebox{0.25}{\includegraphics[width=1\textwidth]{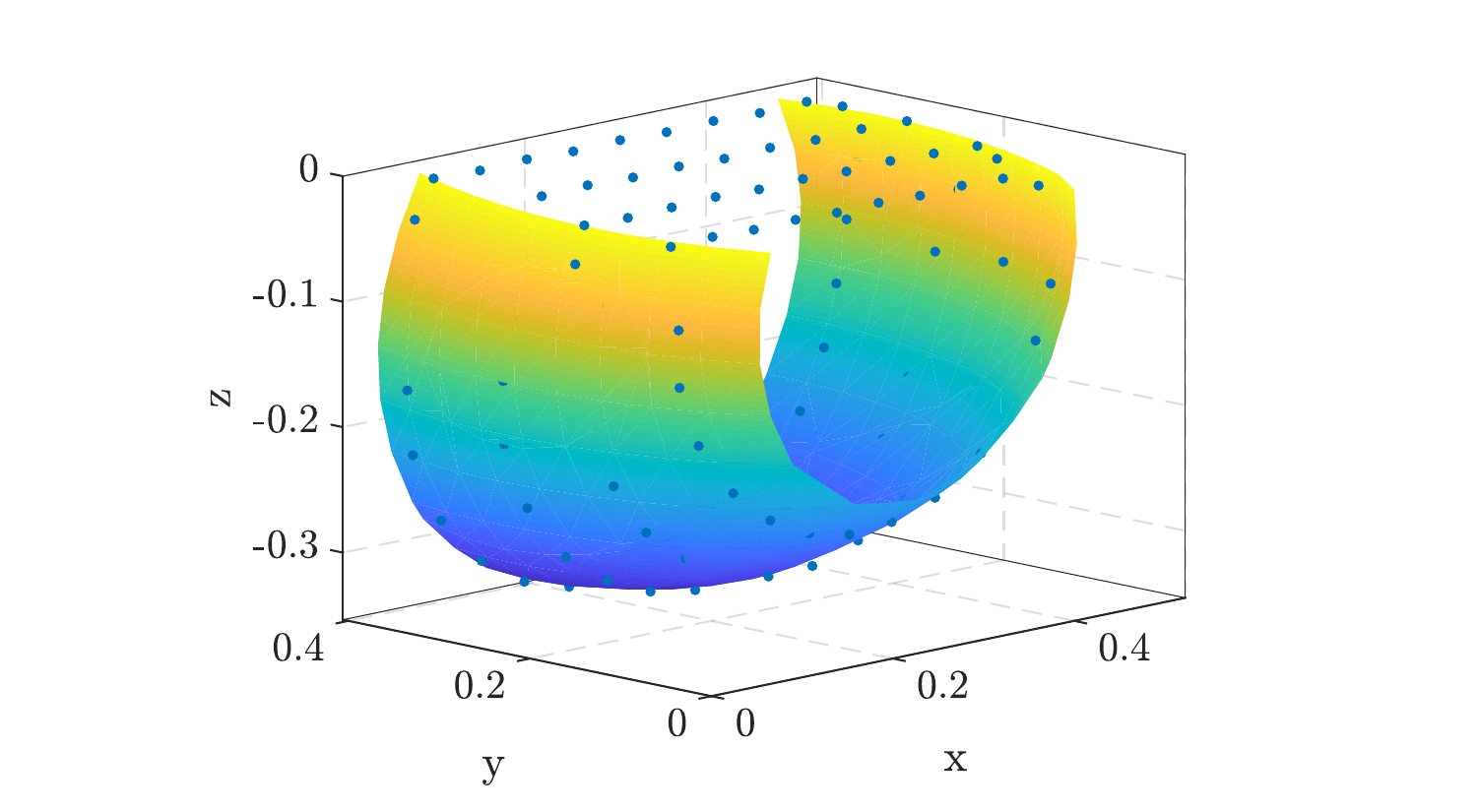}}}
\subfloat[dz/Z=0.5]{\scalebox{0.25}{\includegraphics[width=1\textwidth]{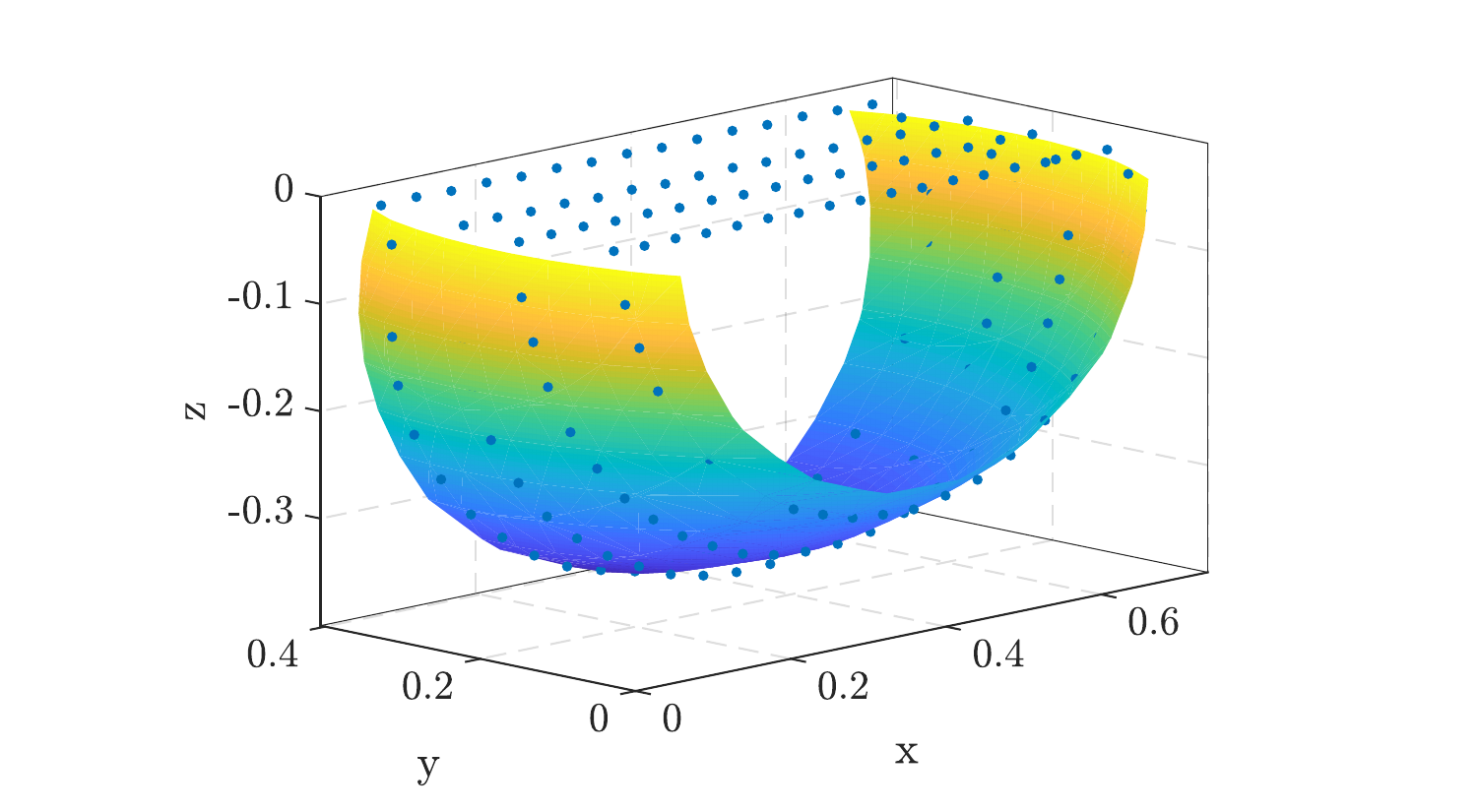}}}\\
\subfloat[dz/Z=0.625]{\scalebox{0.25}{\includegraphics[width=1\textwidth]{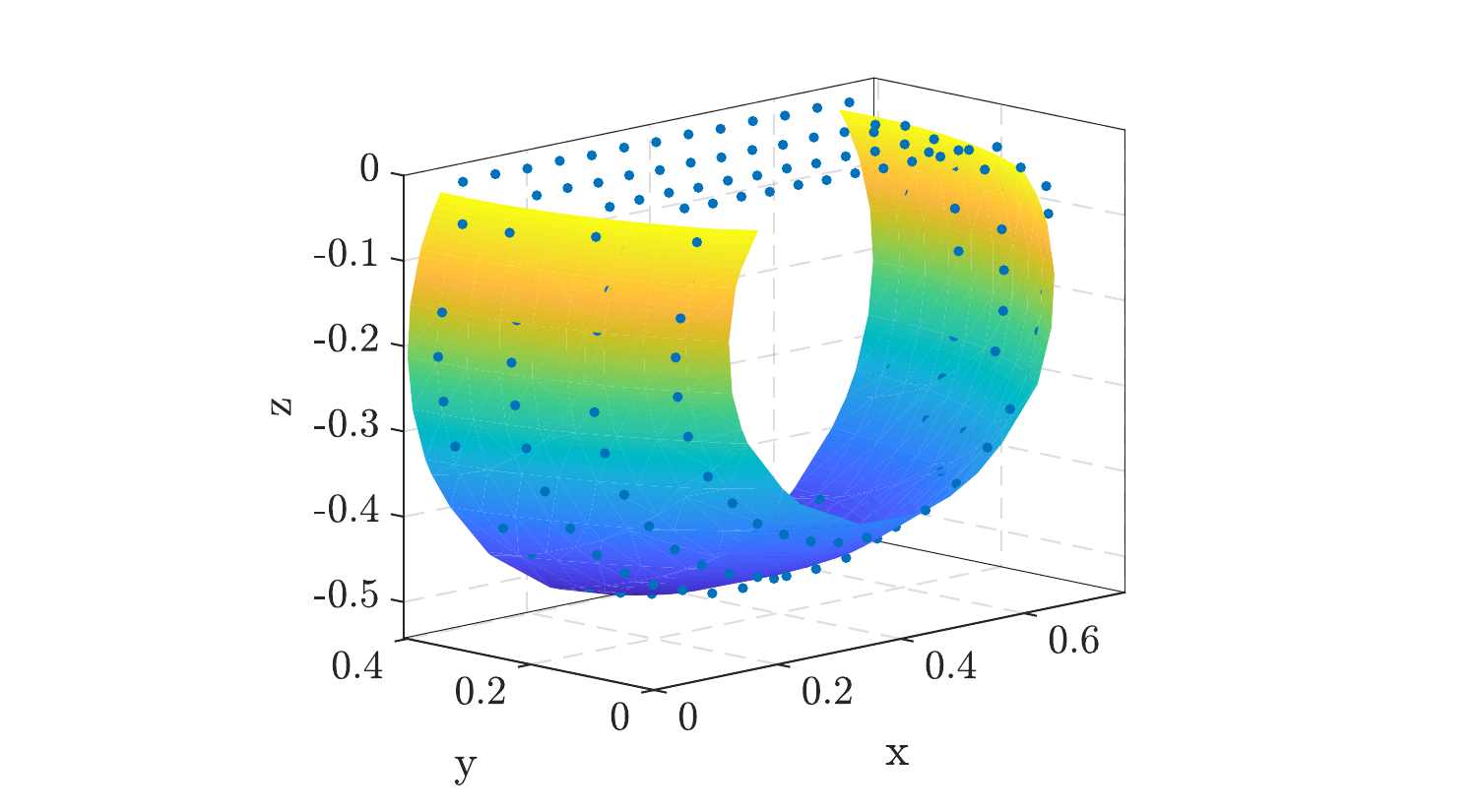}}}
\subfloat[dz/Z=0.75]{\scalebox{0.25}{\includegraphics[width=1\textwidth]{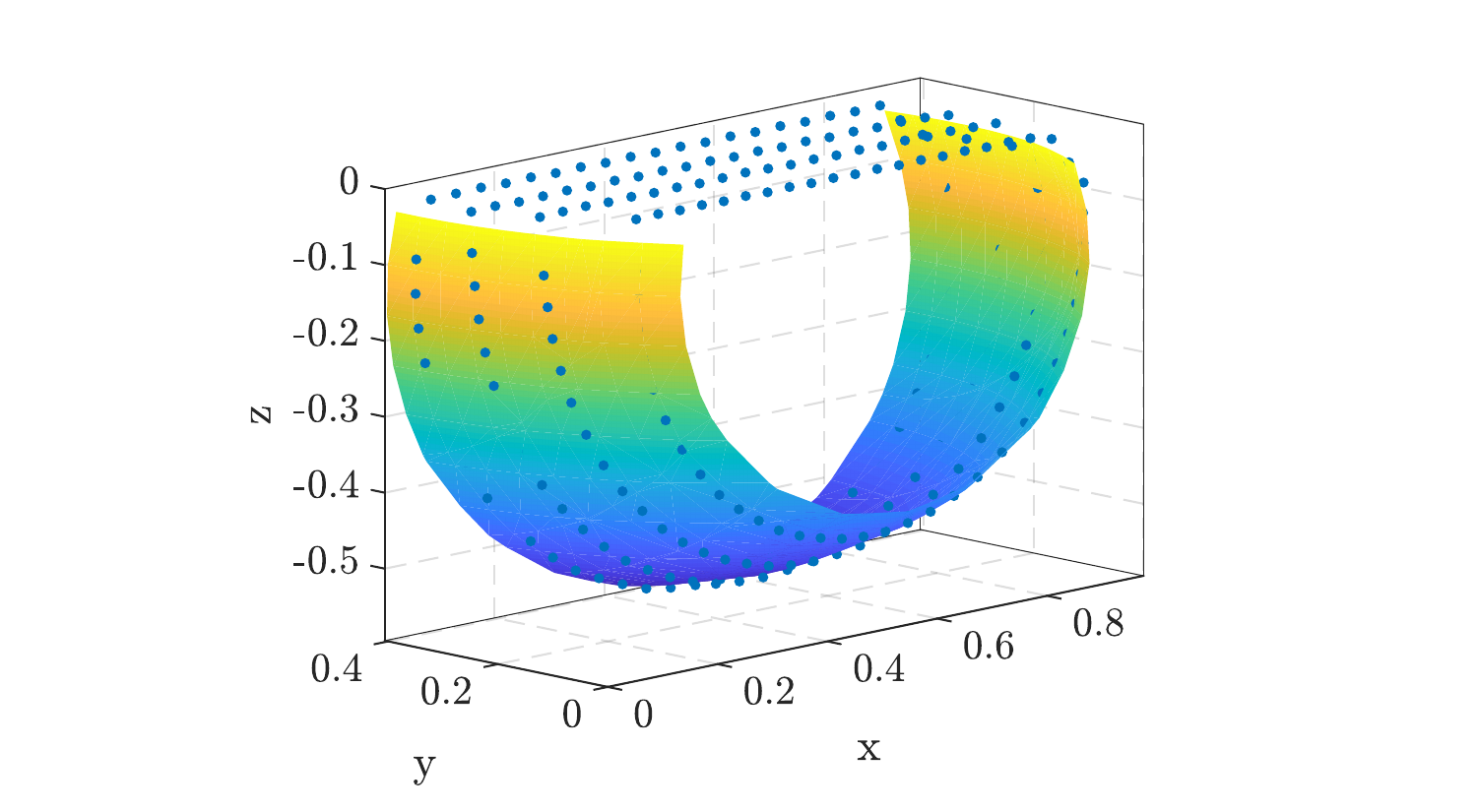}}}
\subfloat[dz/Z=0.875]{\scalebox{0.25}{\includegraphics[width=1\textwidth]{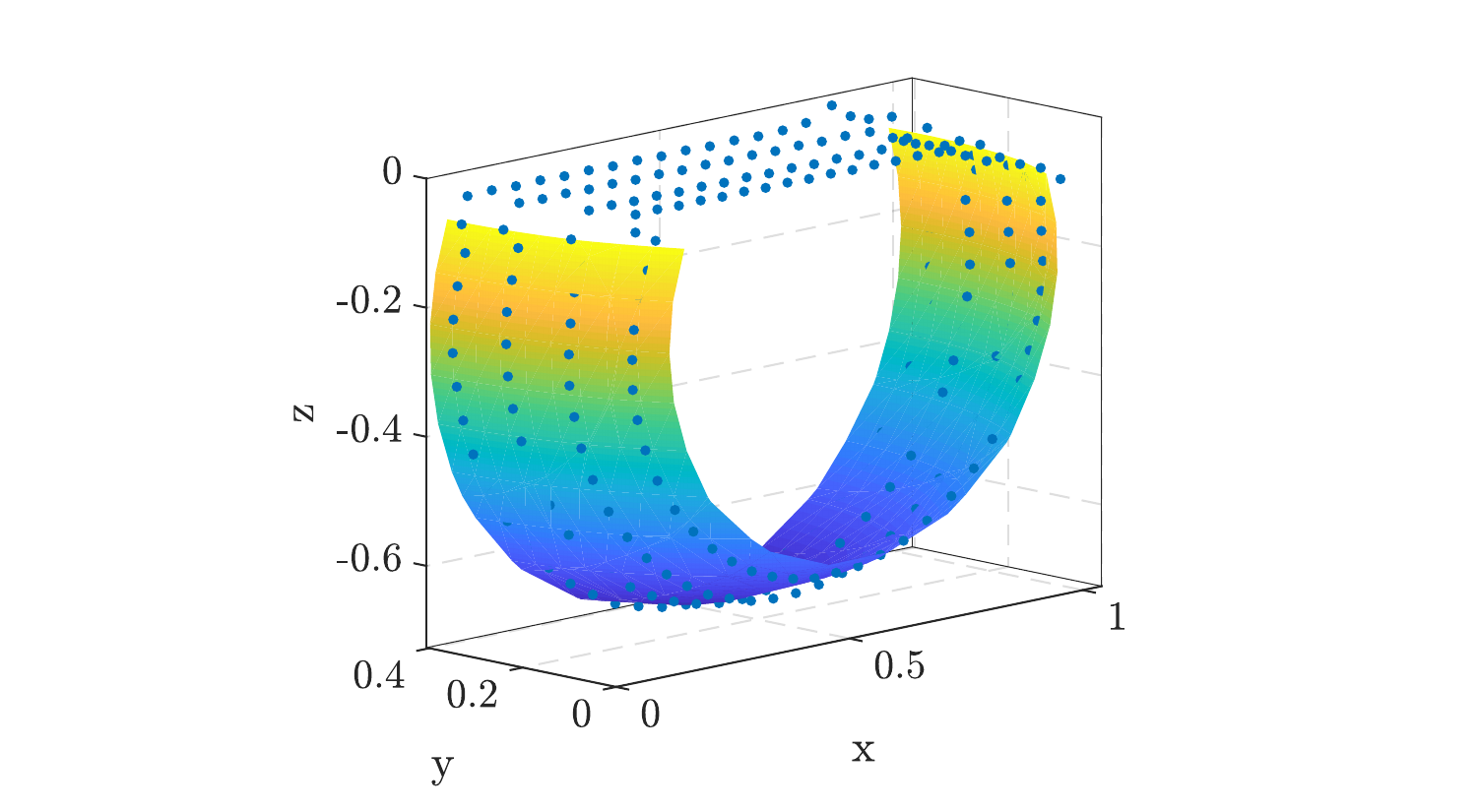}}}
\subfloat[dz/Z=1.0]{\scalebox{0.25}{\includegraphics[width=1\textwidth]{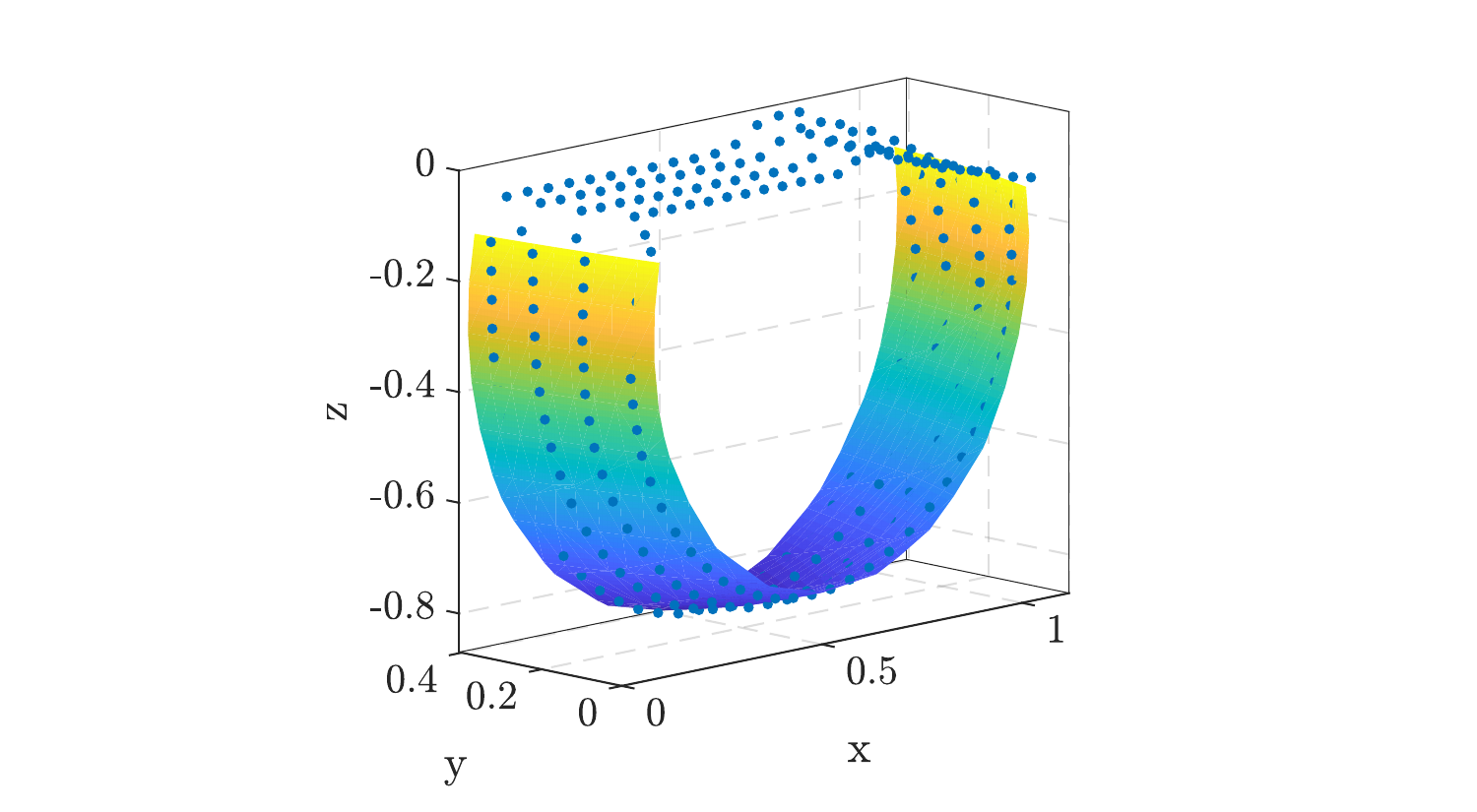}}}
\caption{Added mass surface approximation via ellipsoid surface fitting method (see \ref{appendix:b}) for the slamming cases with $W_n = 0.292$ m/s.} \label{ellipsoid_approx}
\end{figure}

\begin{figure}[h!]
\centering
\captionsetup[subfigure]{justification=centering}
\subfloat[dz/Z=0.125]{\scalebox{0.25}{\includegraphics[width=1\textwidth]{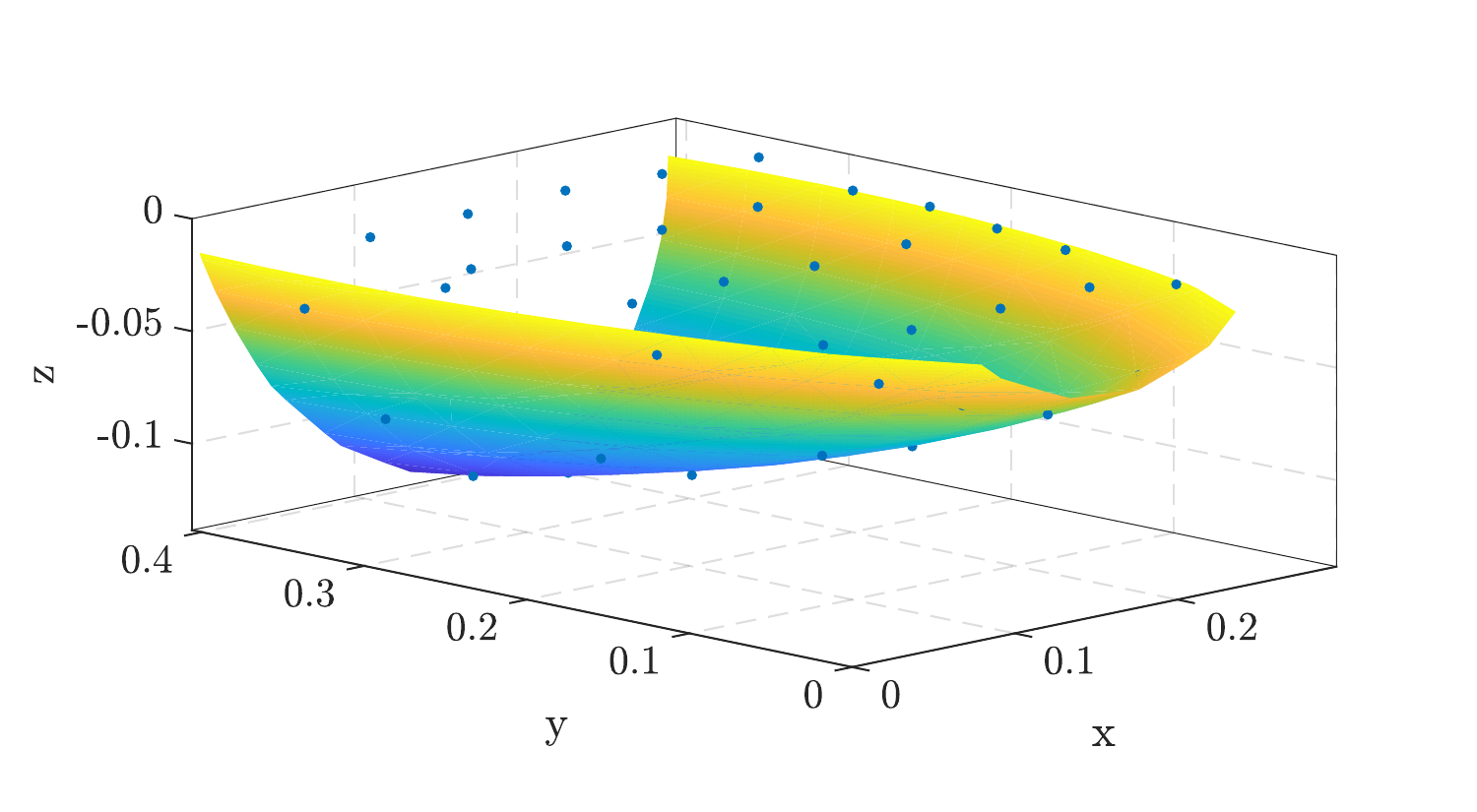}}}
\subfloat[dz/Z=0.25]{\scalebox{0.25}{\includegraphics[width=1\textwidth]{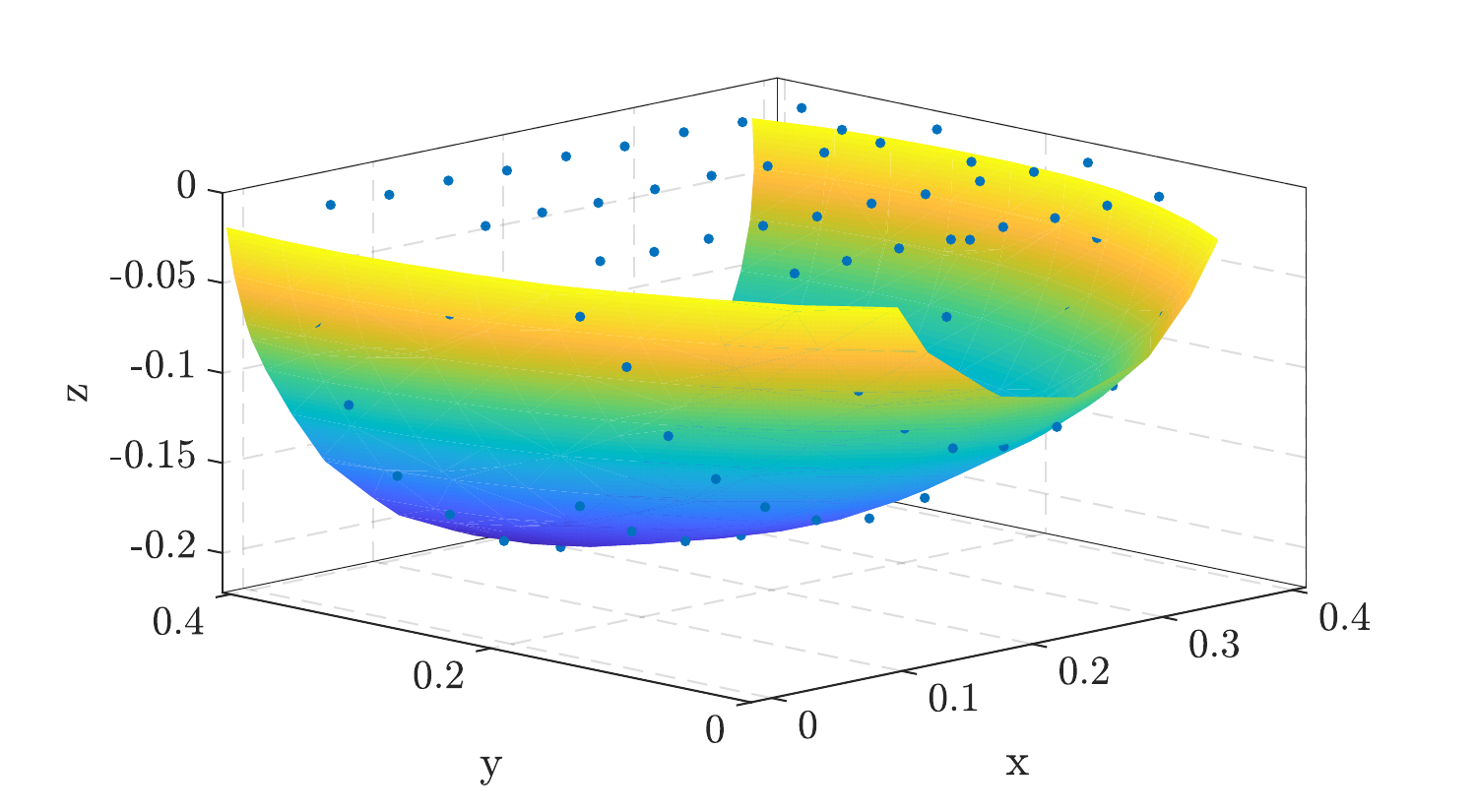}}}
\subfloat[dz/Z=0.375]{\scalebox{0.25}{\includegraphics[width=1\textwidth]{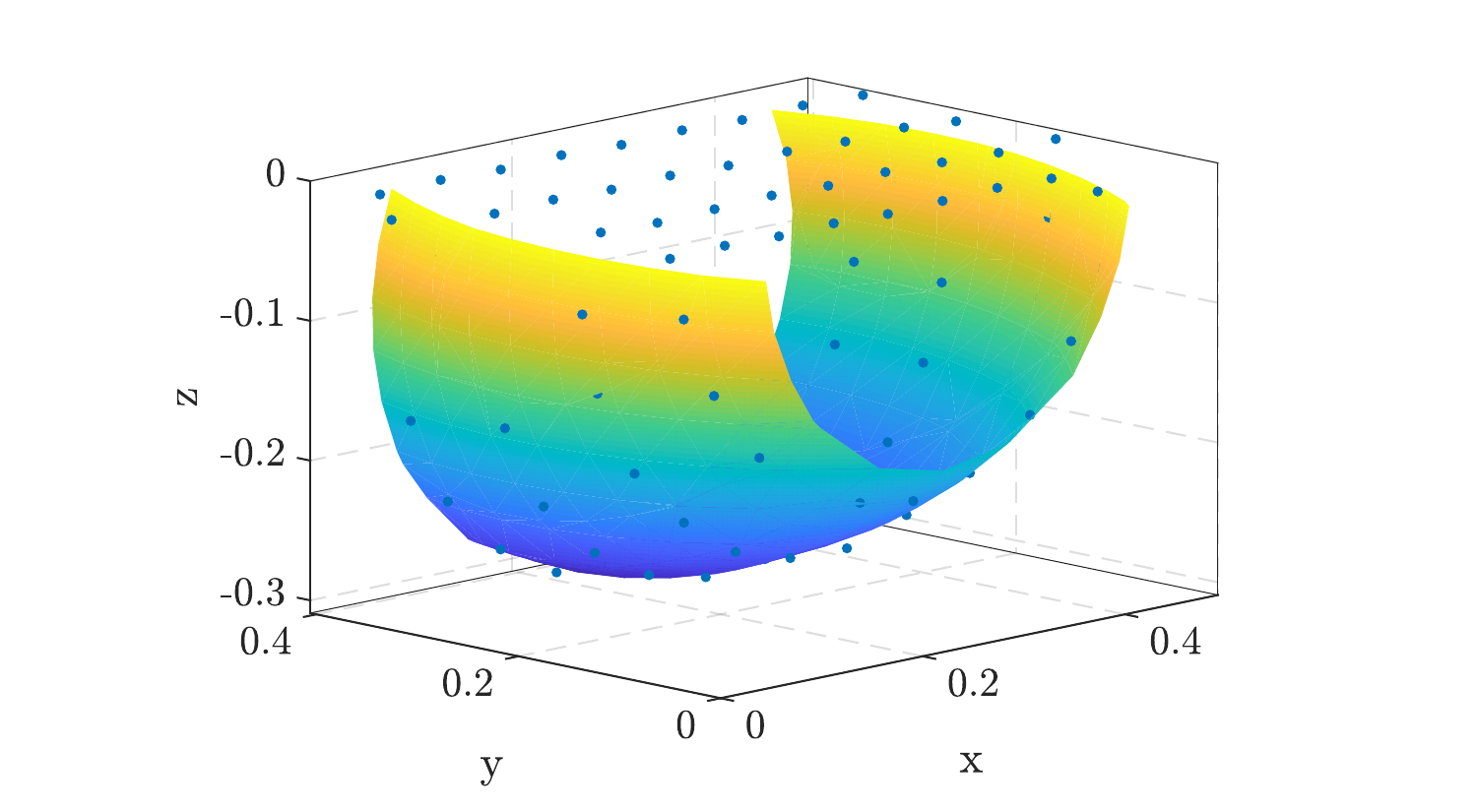}}}
\subfloat[dz/Z=0.5]{\scalebox{0.25}{\includegraphics[width=1\textwidth]{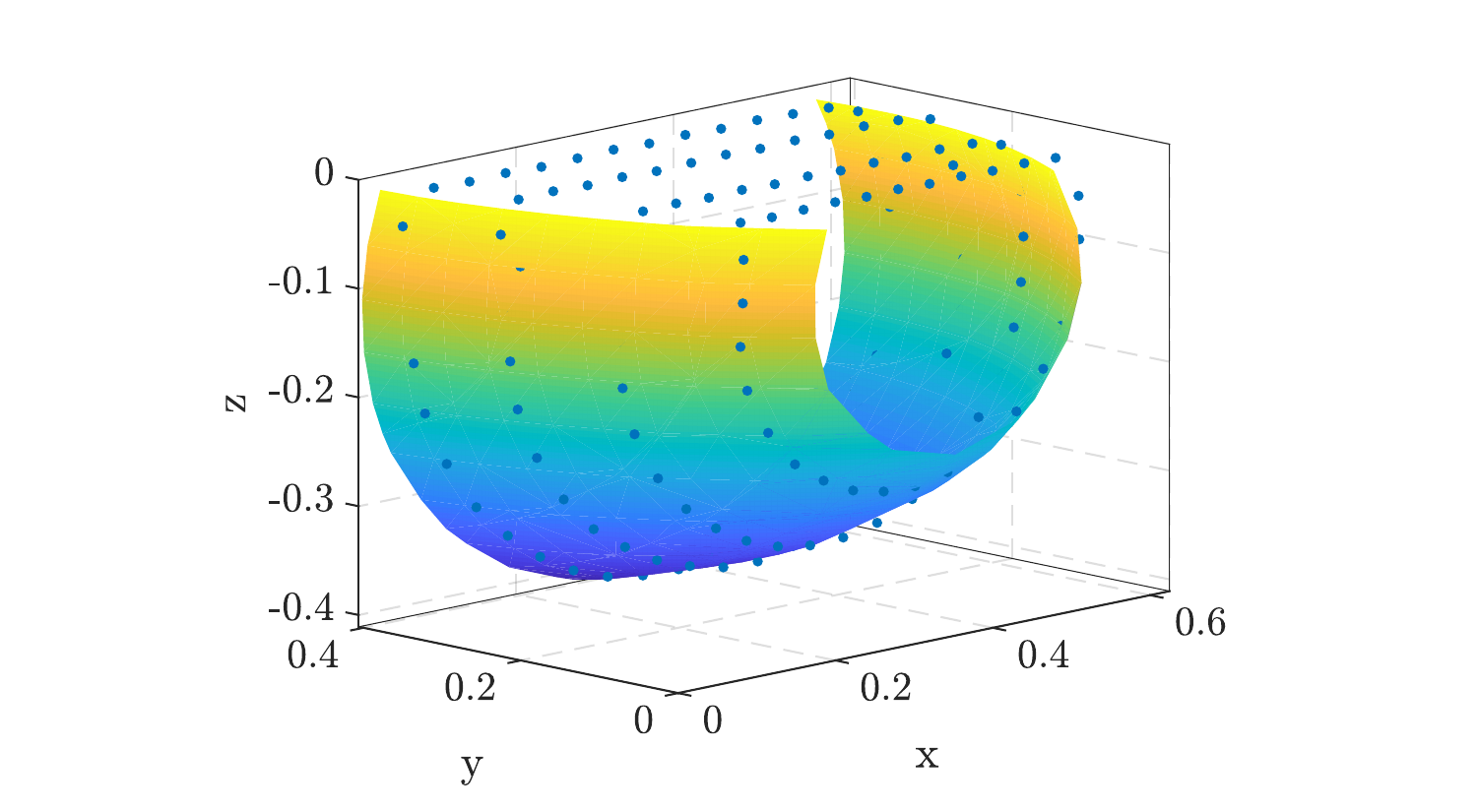}}}\\
\subfloat[dz/Z=0.625]{\scalebox{0.25}{\includegraphics[width=1\textwidth]{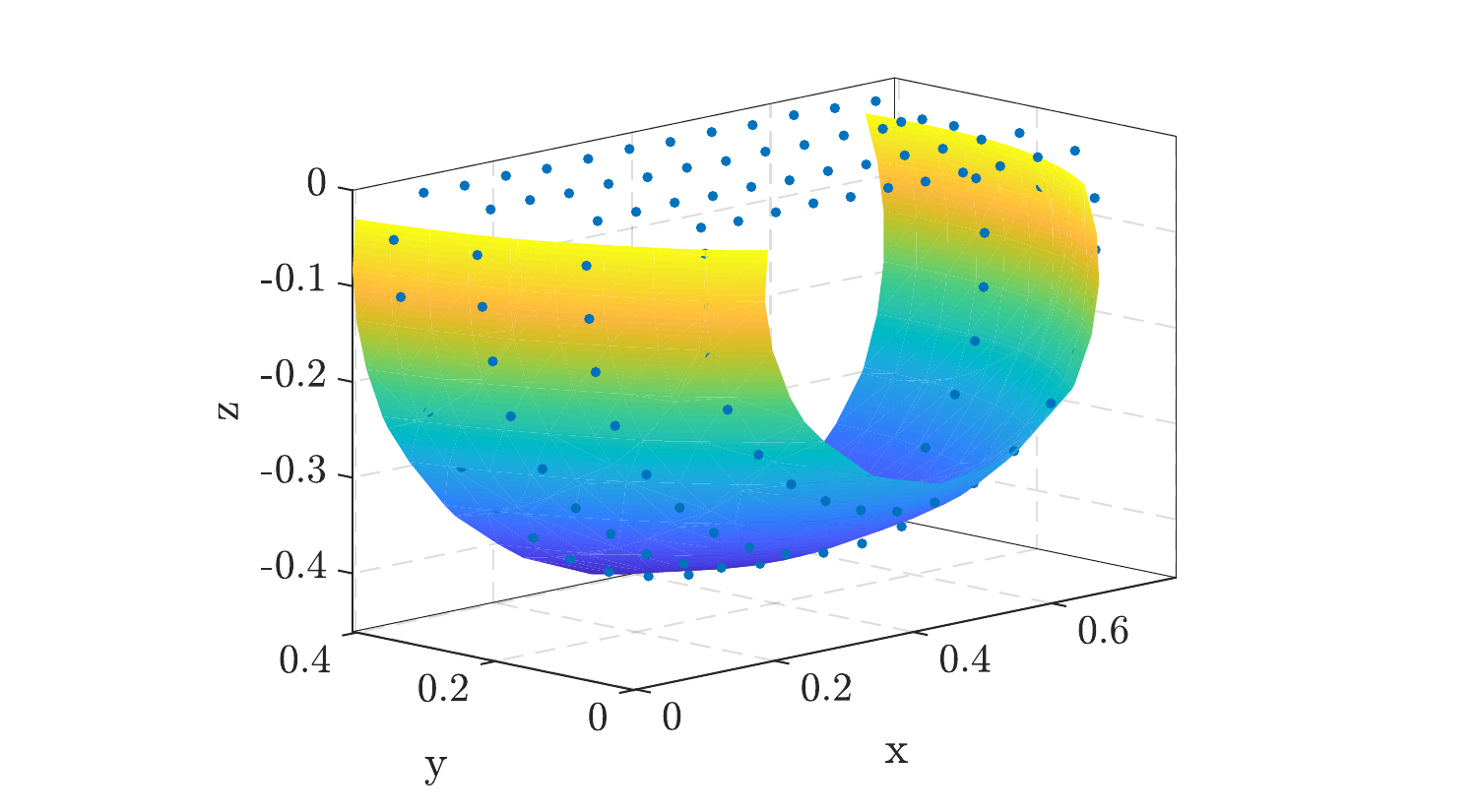}}}
\subfloat[dz/Z=0.75]{\scalebox{0.25}{\includegraphics[width=1\textwidth]{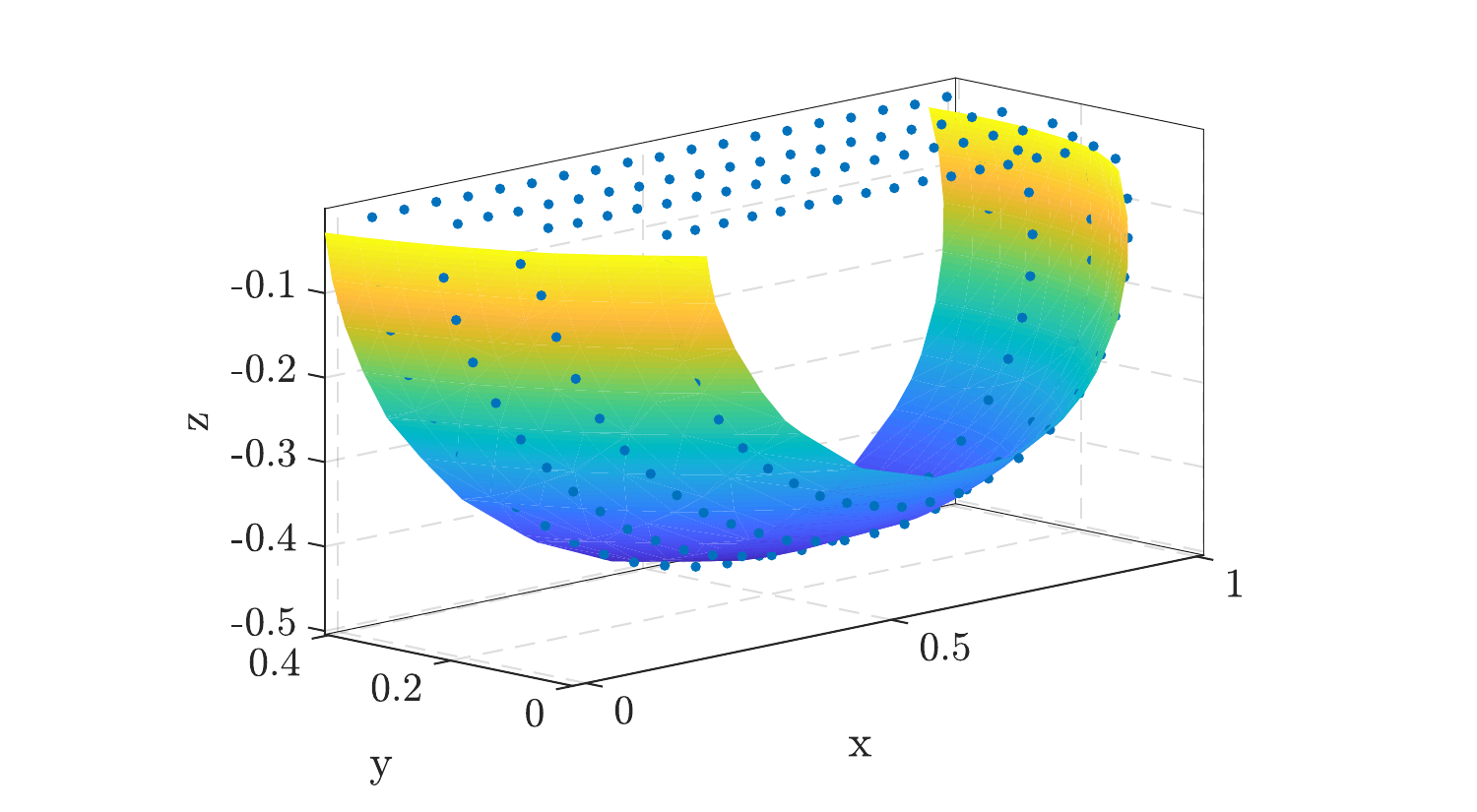}}}
\subfloat[dz/Z=0.875]{\scalebox{0.25}{\includegraphics[width=1\textwidth]{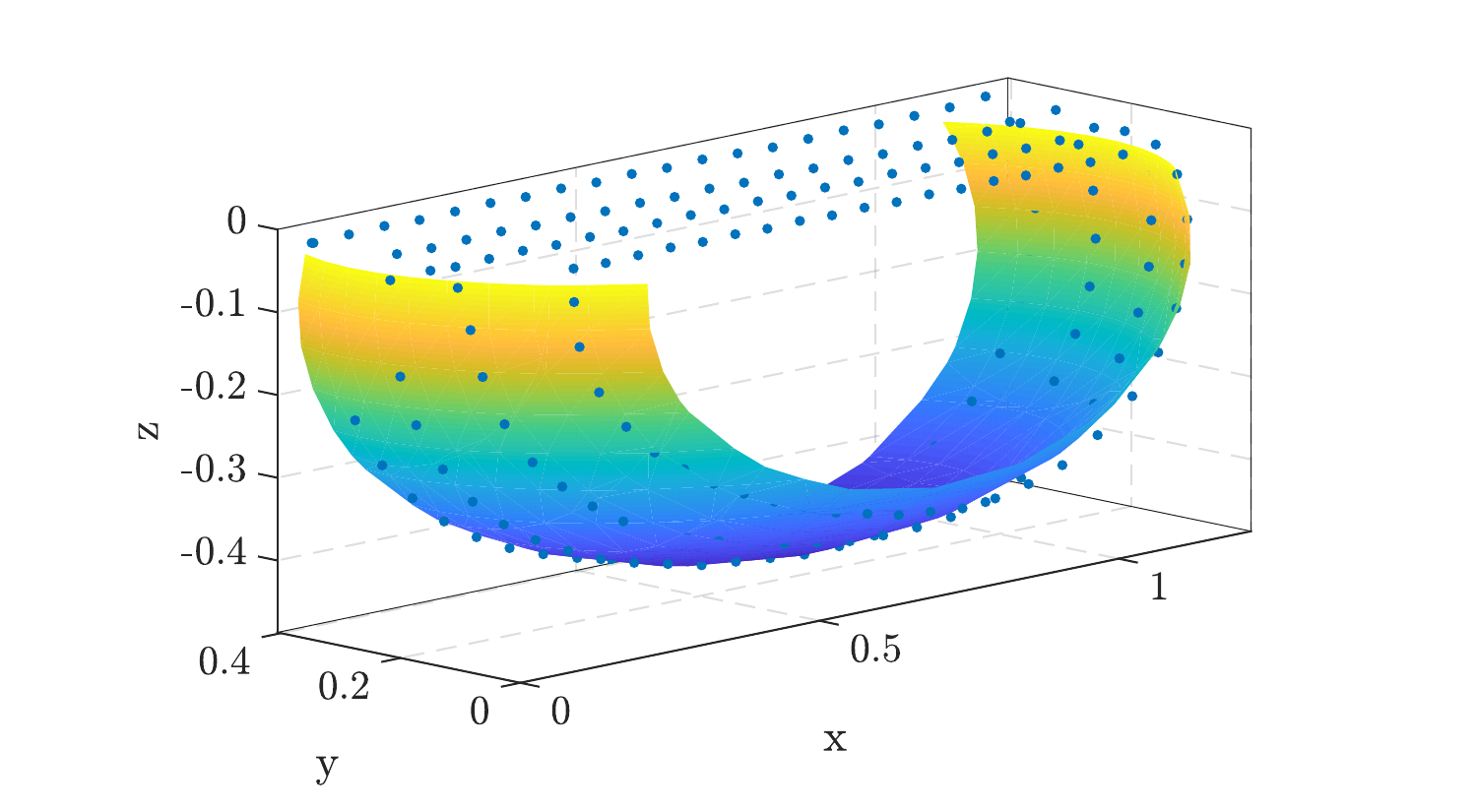}}}
\subfloat[dz/Z=1.0]{\scalebox{0.25}{\includegraphics[width=1\textwidth]{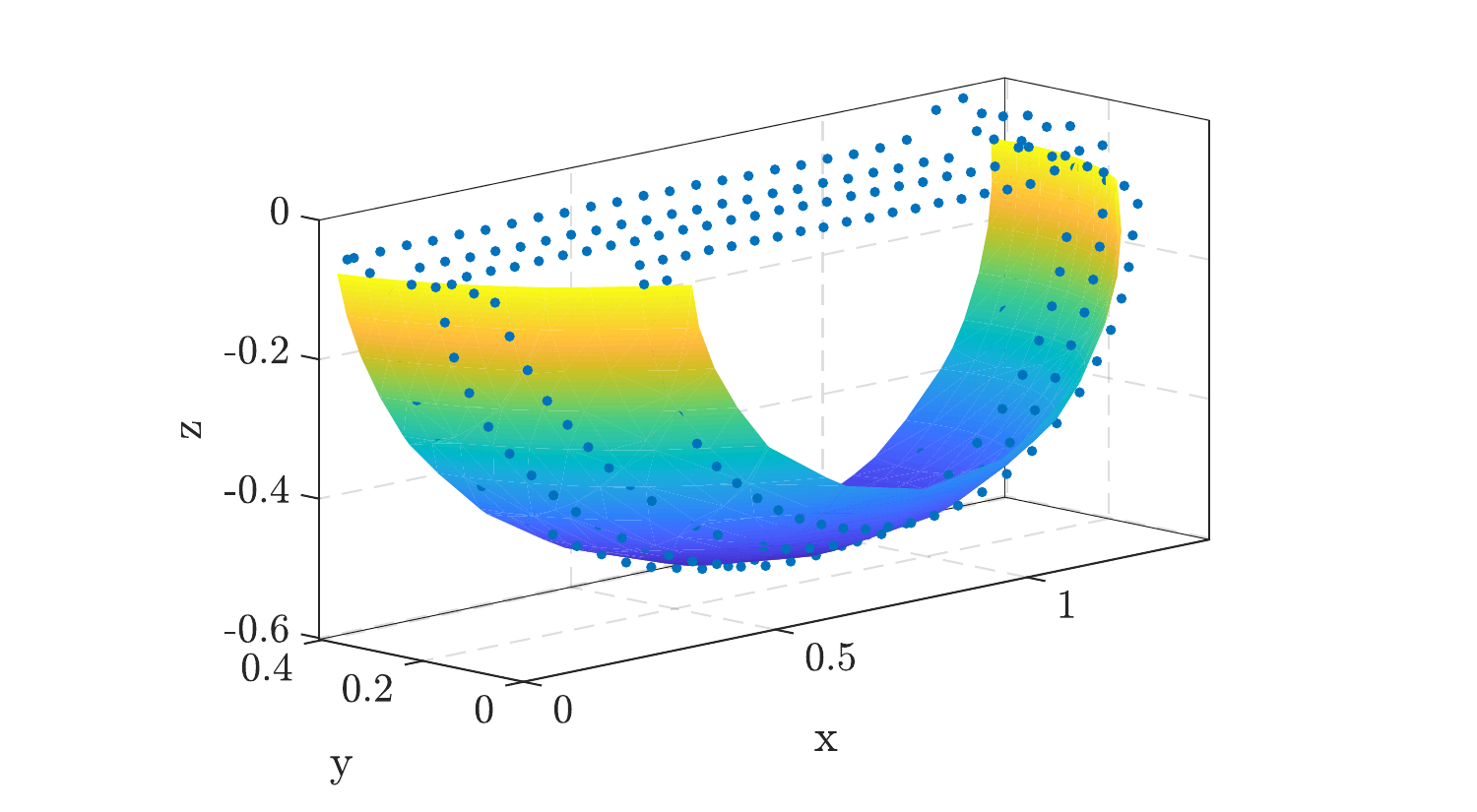}}}
\caption{Added mass surface approximation via ellipsoid surface fitting method (see \ref{appendix:b}) for the slamming cases with $W_n = 0.438$ m/s.} \label{ellipsoid_approx2}
\end{figure}

\begin{figure}[h!]
\centering
\captionsetup[subfigure]{justification=centering}
\subfloat[dz/Z=0.125]{\scalebox{0.25}{\includegraphics[width=1\textwidth]{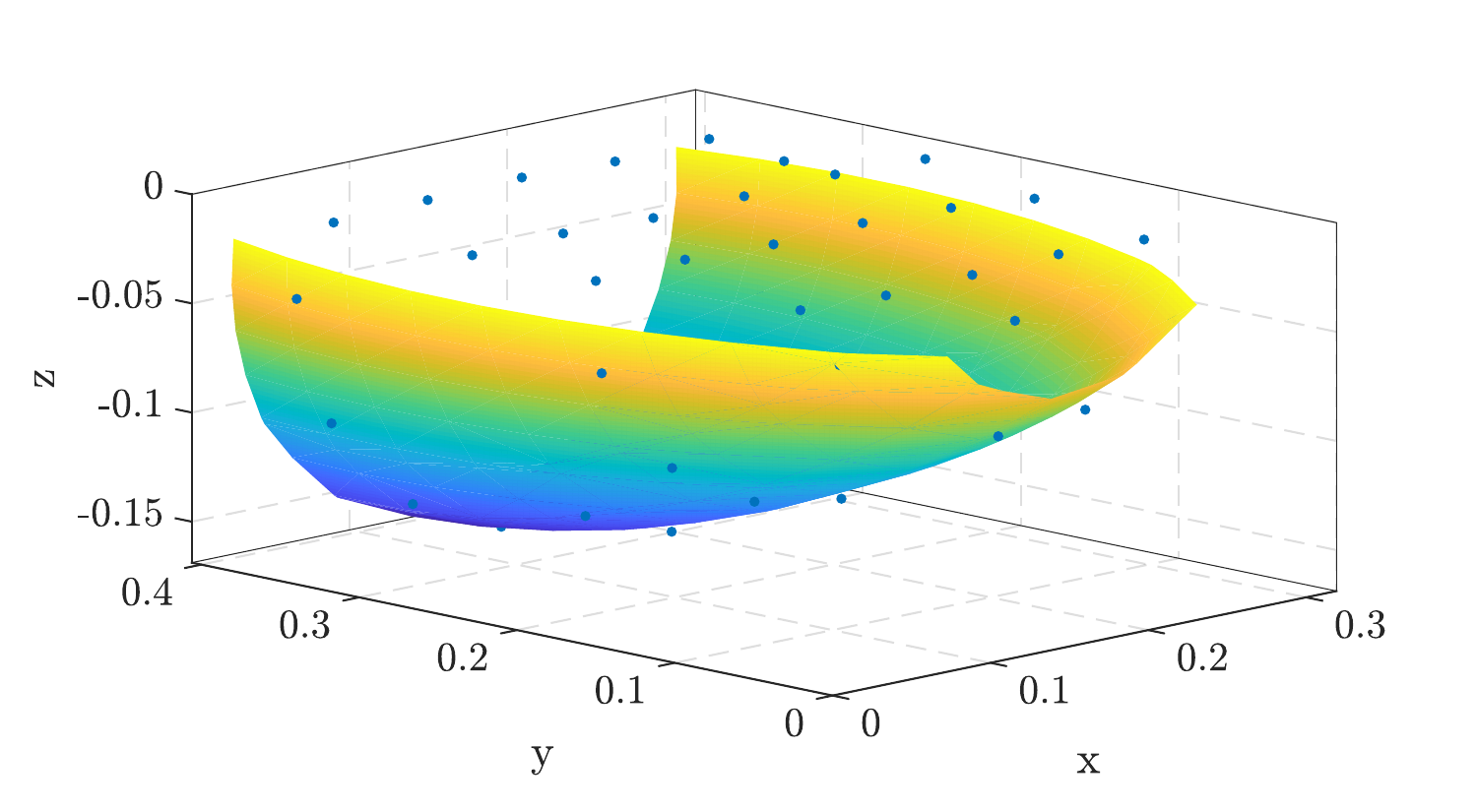}}}
\subfloat[dz/Z=0.25]{\scalebox{0.25}{\includegraphics[width=1\textwidth]{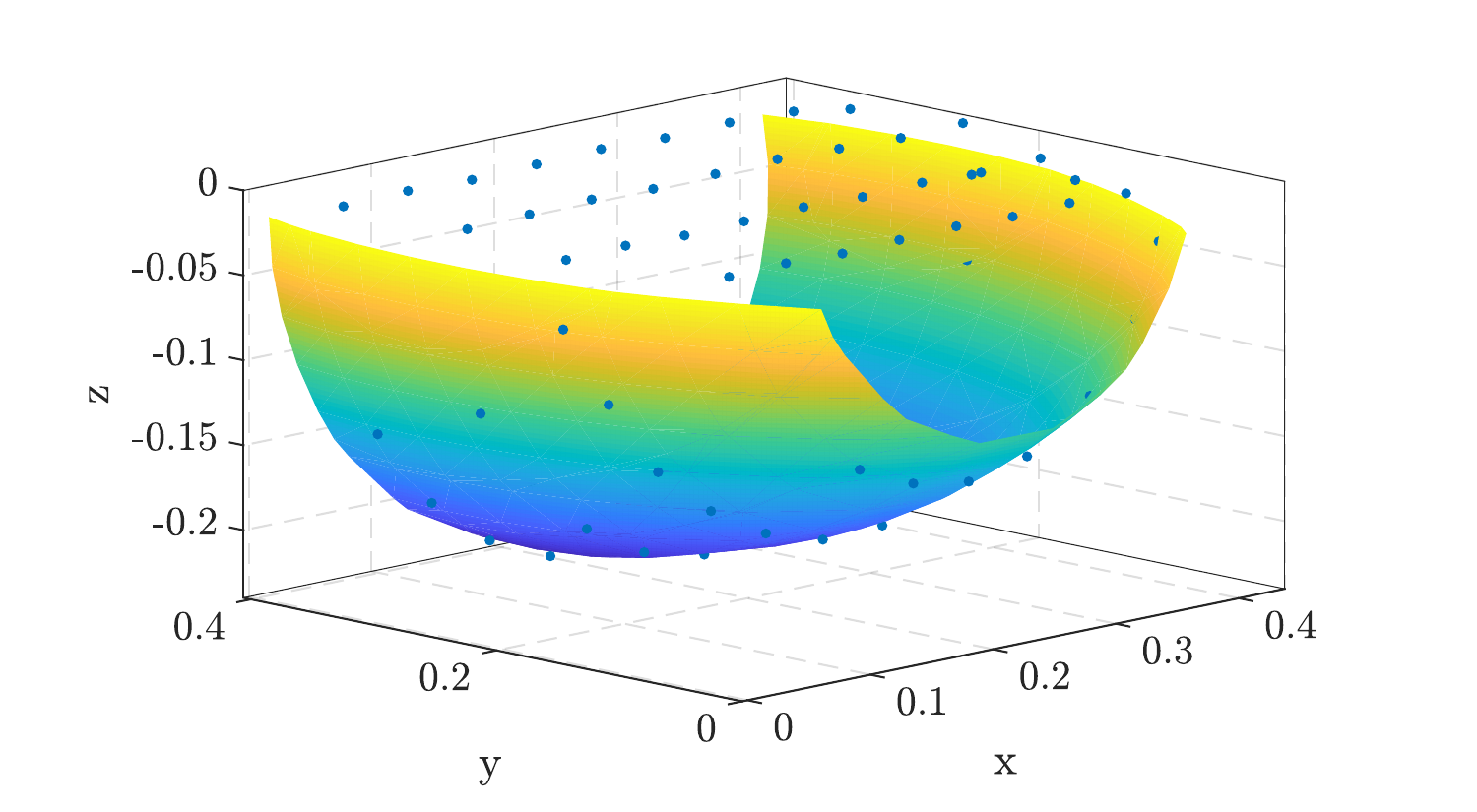}}}
\subfloat[dz/Z=0.375]{\scalebox{0.25}{\includegraphics[width=1\textwidth]{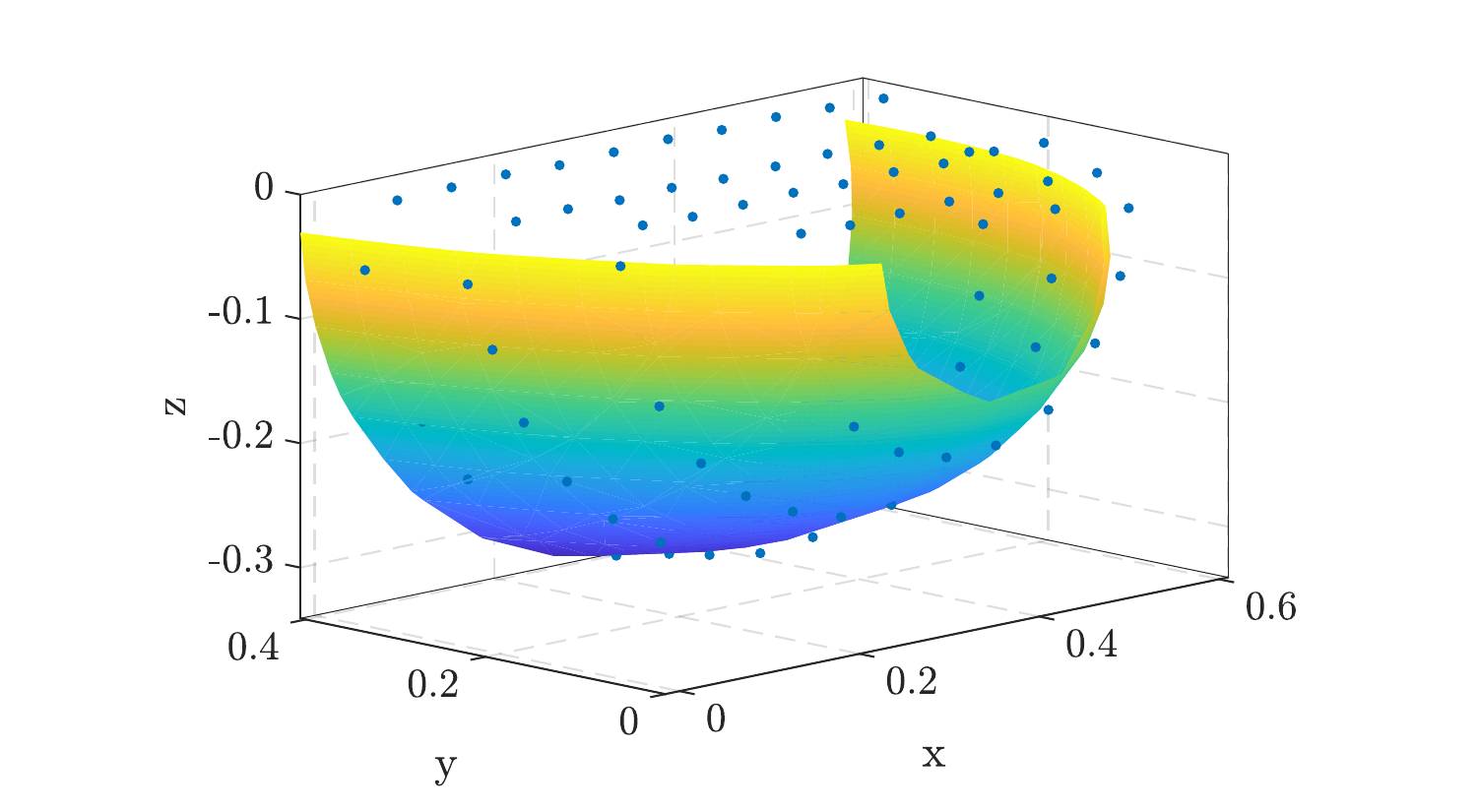}}}
\subfloat[dz/Z=0.5]{\scalebox{0.25}{\includegraphics[width=1\textwidth]{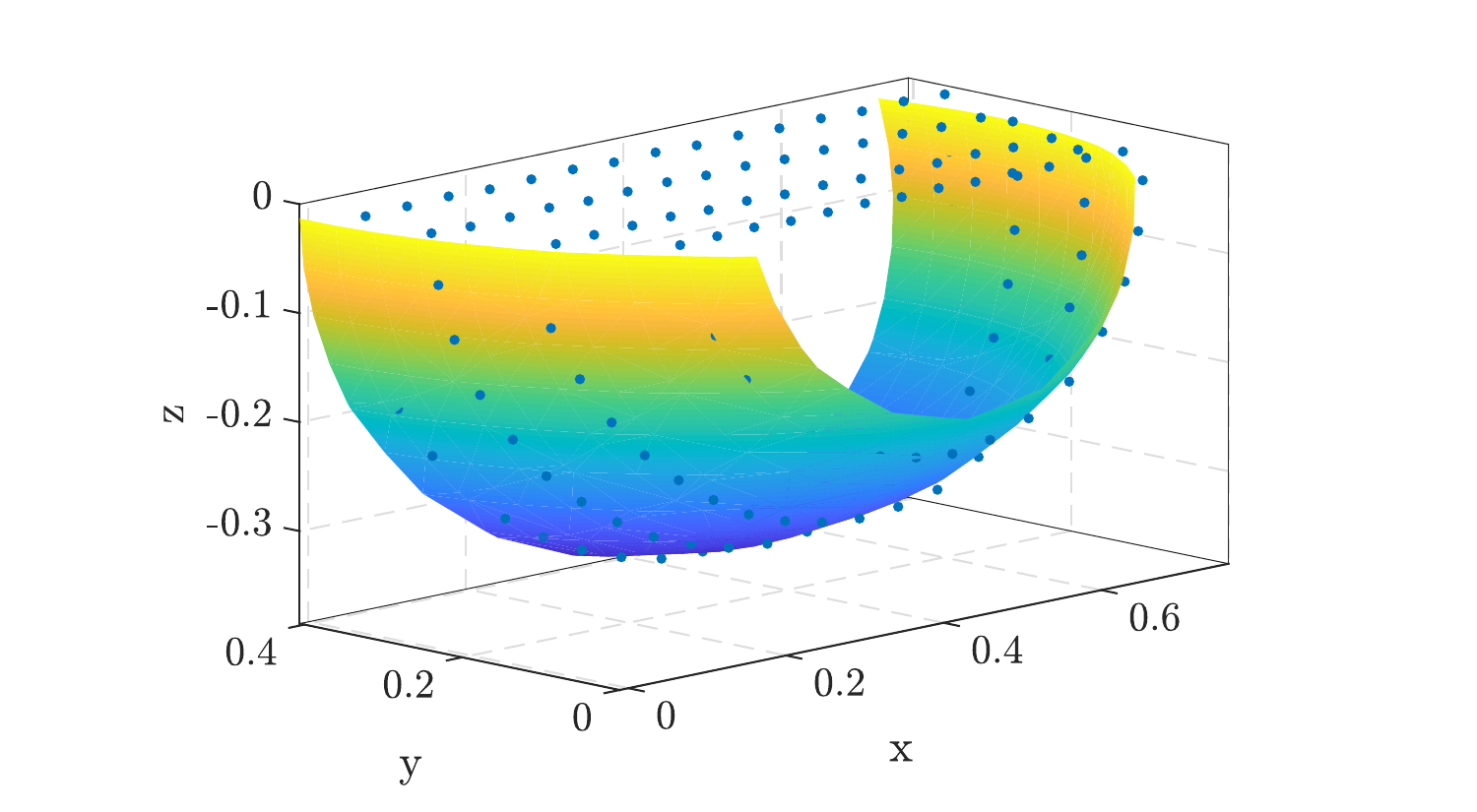}}}\\
\subfloat[dz/Z=0.625]{\scalebox{0.25}{\includegraphics[width=1\textwidth]{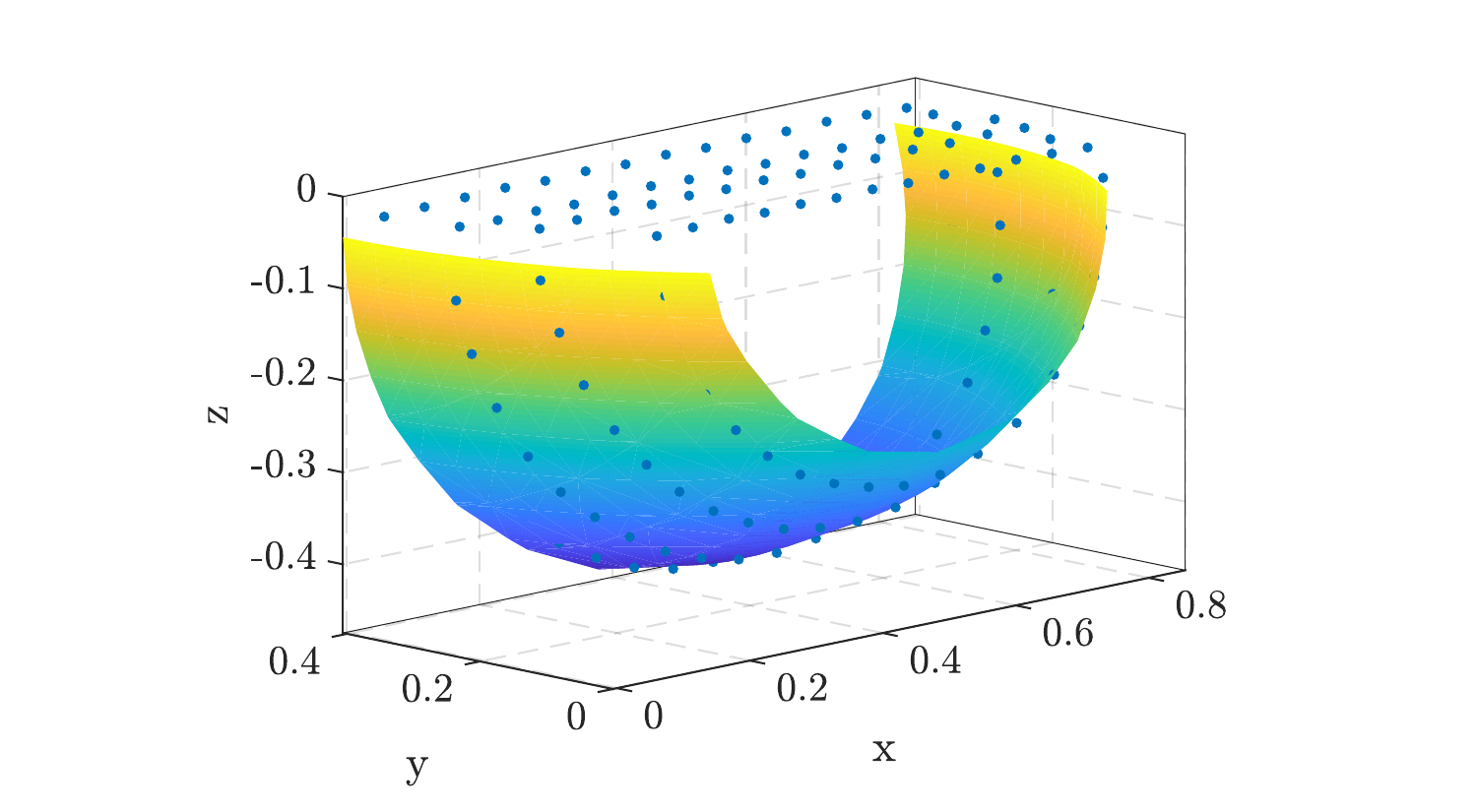}}}
\subfloat[dz/Z=0.75]{\scalebox{0.25}{\includegraphics[width=1\textwidth]{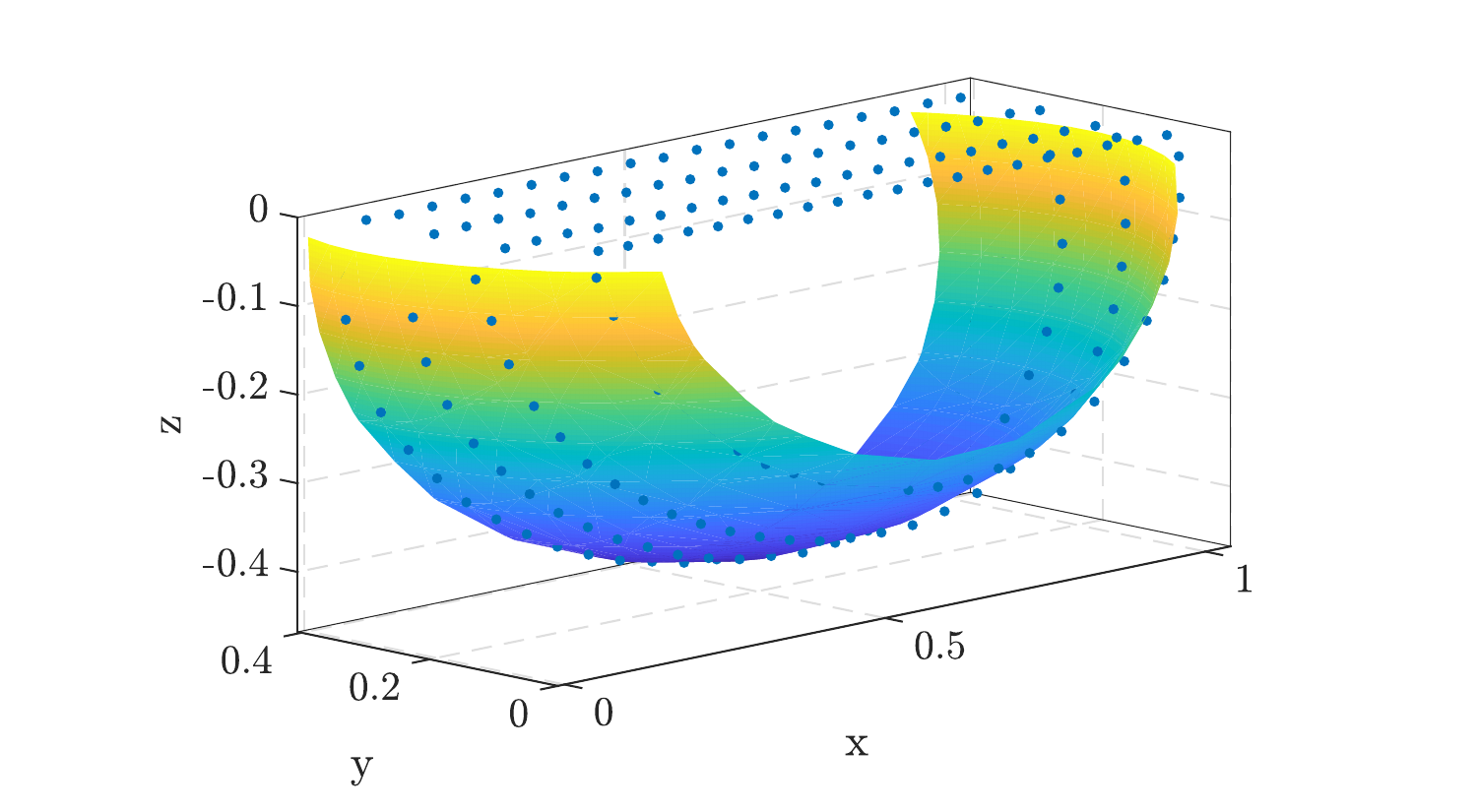}}}
\subfloat[dz/Z=0.875]{\scalebox{0.25}{\includegraphics[width=1\textwidth]{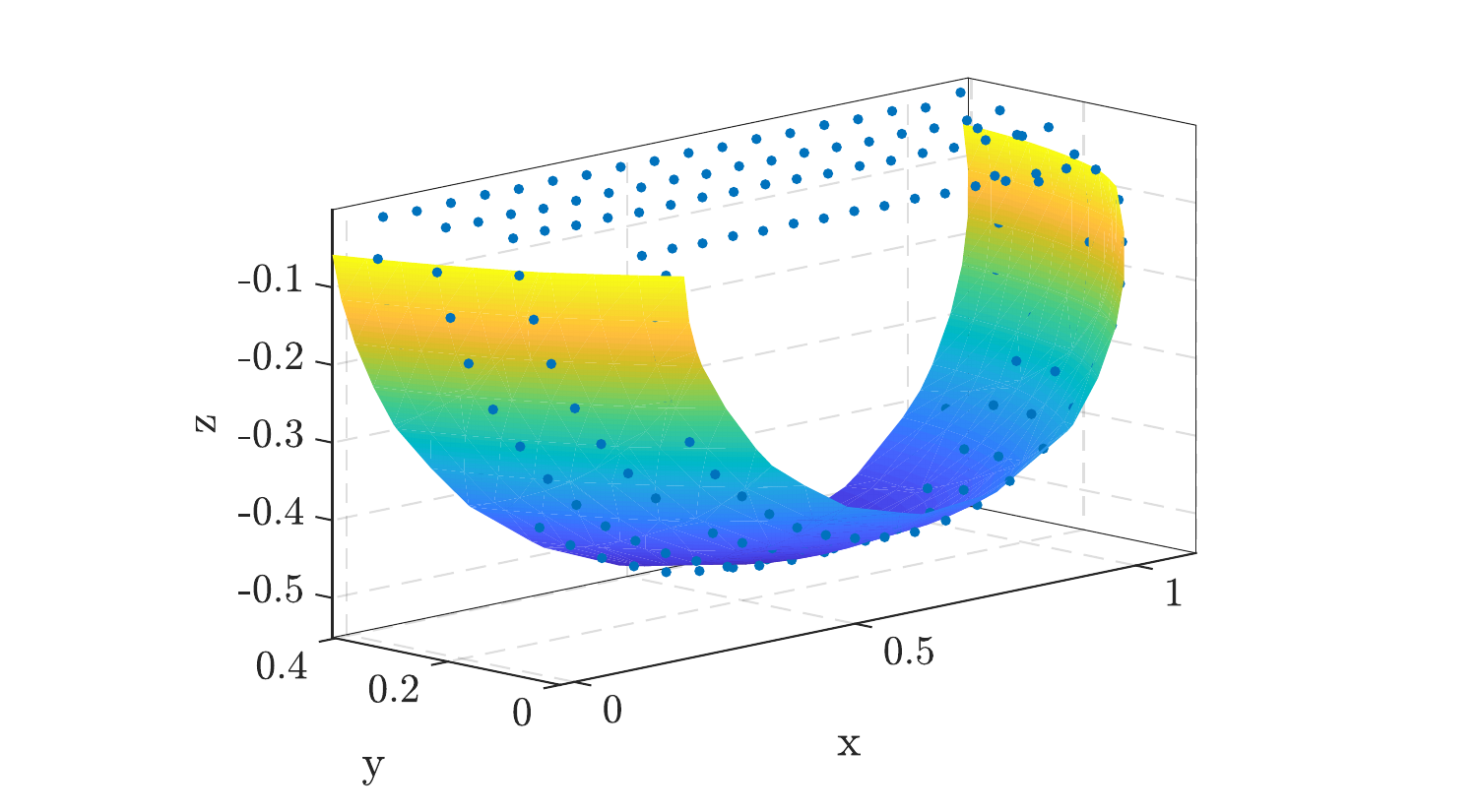}}}
\subfloat[dz/Z=1.0]{\scalebox{0.25}{\includegraphics[width=1\textwidth]{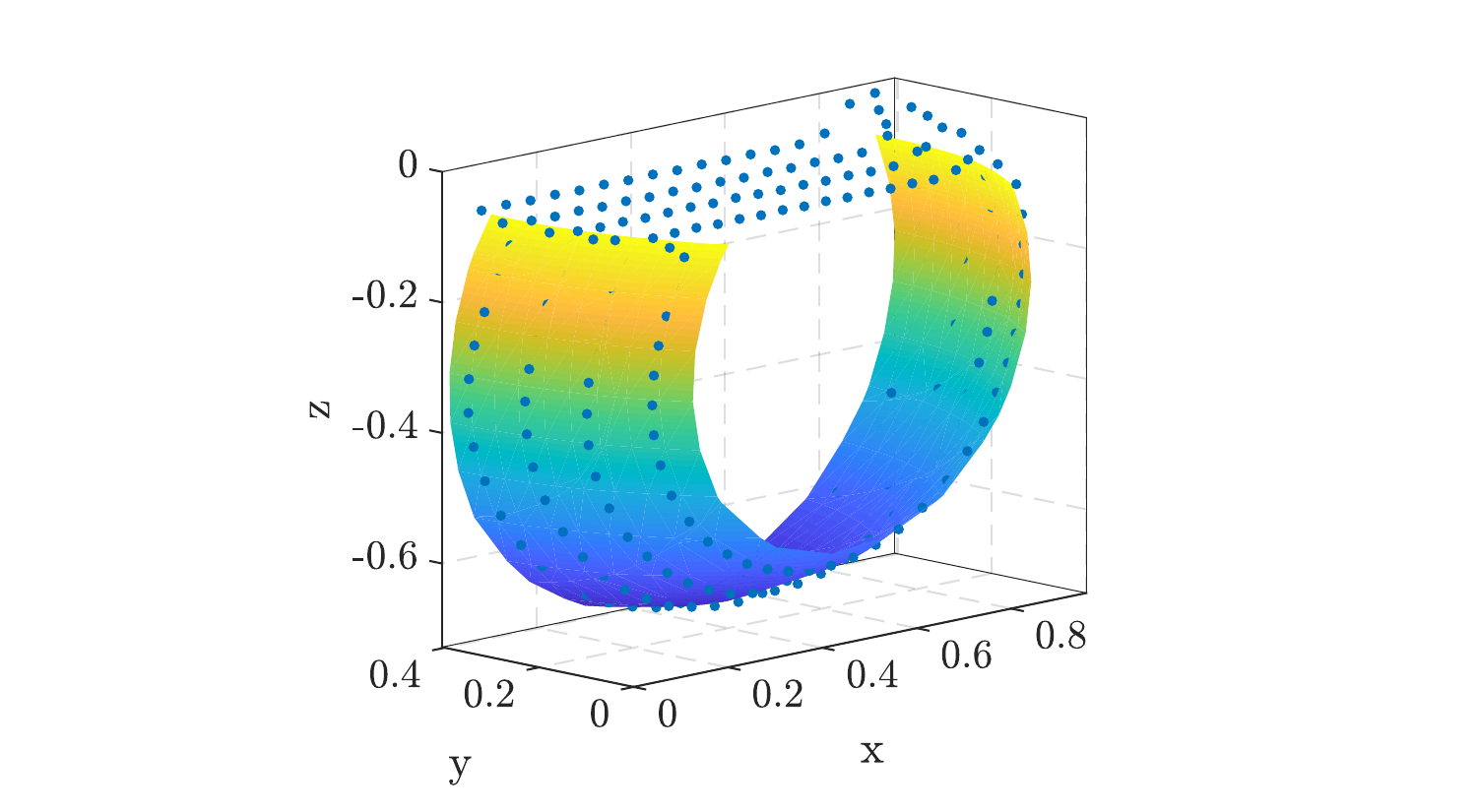}}}
\caption{Added mass surface approximation via ellipsoid surface fitting method (see \ref{appendix:b}) for the slamming cases with $W_n = 0.584$ m/s.} \label{ellipsoid_approx3}
\end{figure}

We now present the added masses obtained for the test cases in the last FSI validation set (slamming tests with impact velocities $W_n = 0.292, 0.438, 0.584$, and $0.875$~m/s) using the added mass acquisition procedure detailed in Section \ref{acquisition} and \ref{appendix:b}. We focus on recovering the normal impact force of the moderably deformable plate with plate thickness 0.00795 m. The approximated added mass boundaries for each of the slamming cases are presented in Figure \ref{ellipsoid_approx} to \ref{ellipsoid_approx3}. As shown, added mass among the four slamming cases, at each $dz/Z$, share similar shape. In all four cases, the added mass started off with the shape of a quarter of a long and narrow ellipsoid. As we submerge the plate in the water, both the major and minor radii of the ellipsoid are proportionally increasing in a way such that the semi-major and semi-minor radii of the ellipsoid are roughly in a $1:1$ ratio. 

 \begin{figure}[h!]
\centering
\captionsetup[subfigure]{justification=centering}
\subfloat[dz/Z=0.125]{\scalebox{0.25}{\includegraphics[width=1\textwidth]{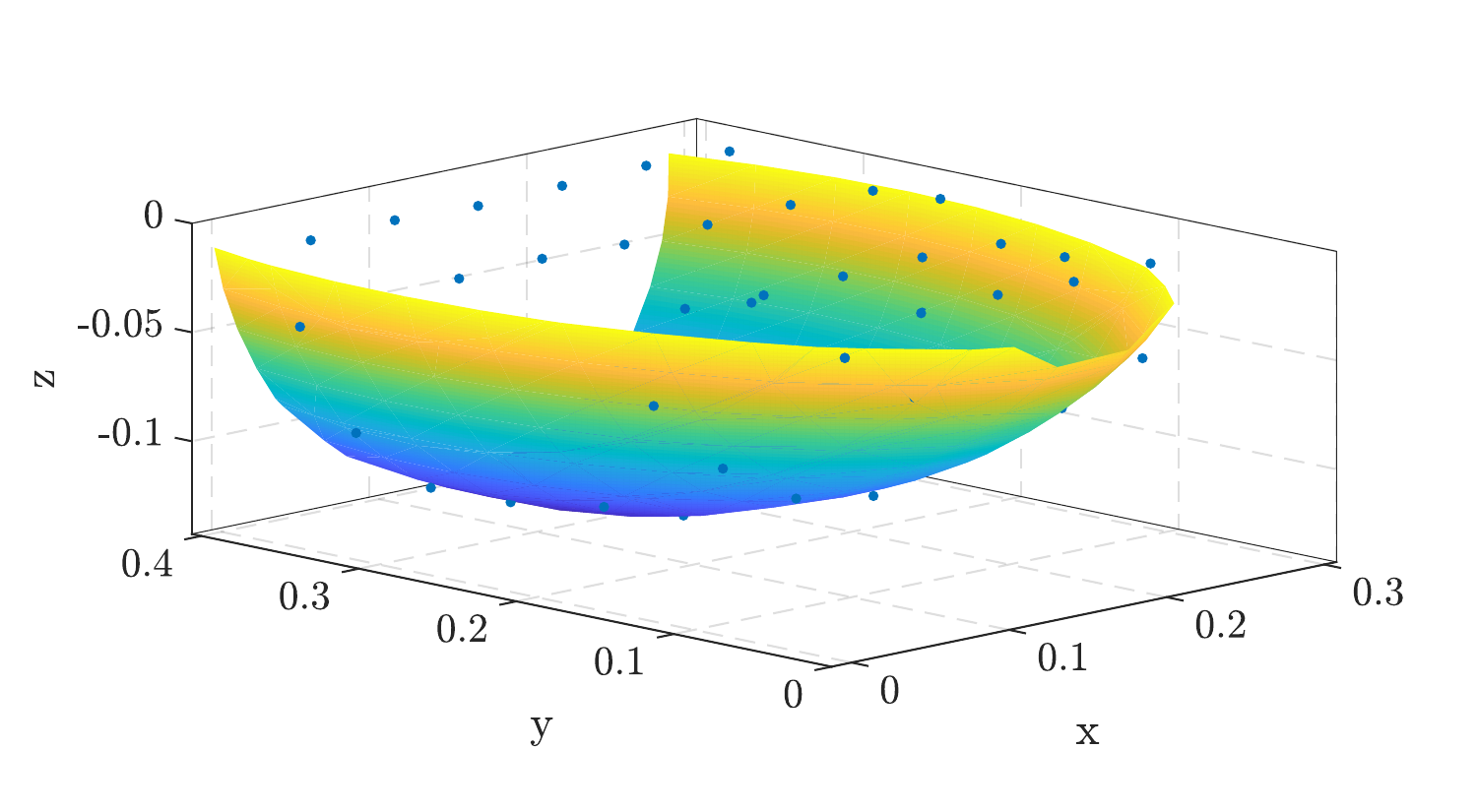}}}
\subfloat[dz/Z=0.25]{\scalebox{0.25}{\includegraphics[width=1\textwidth]{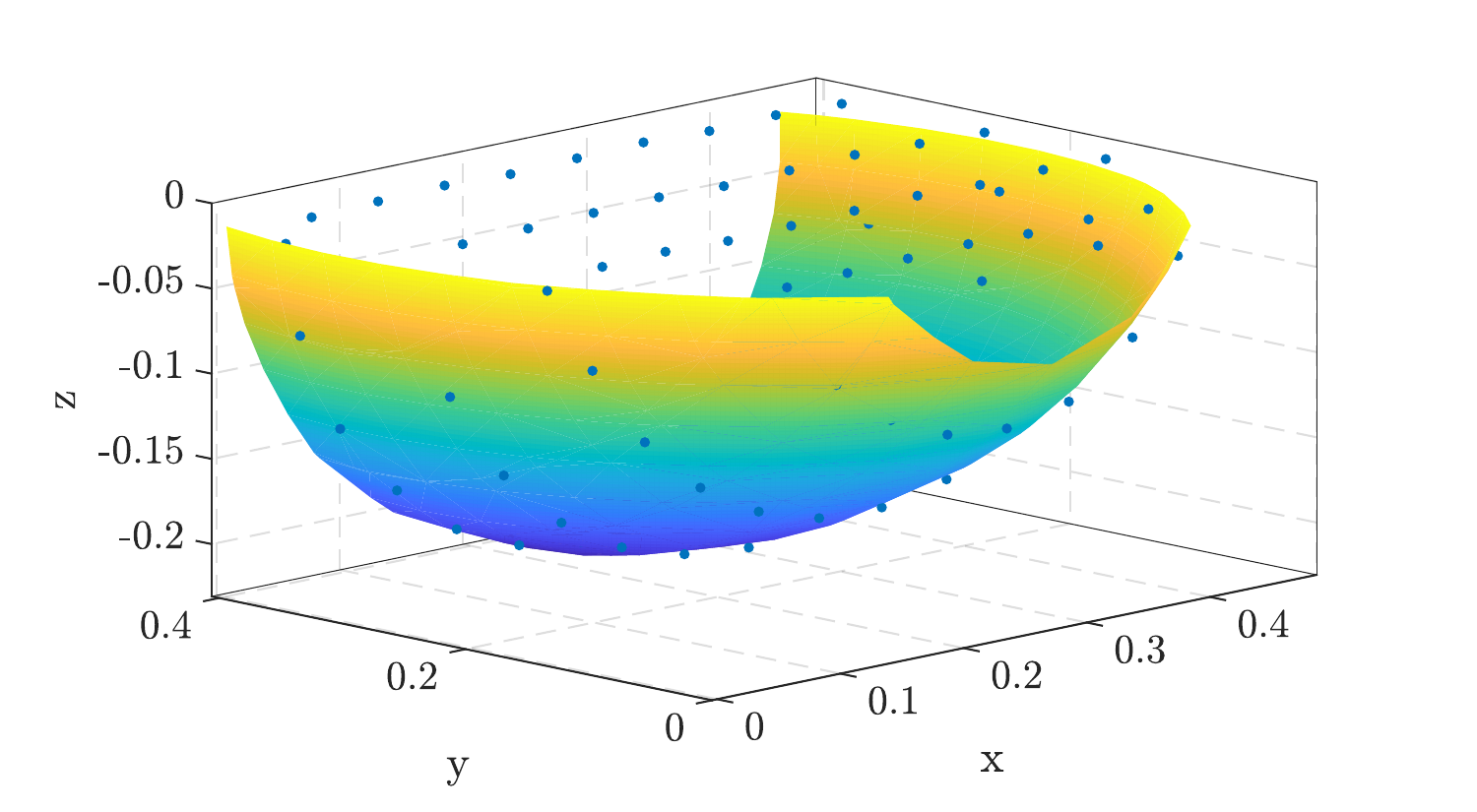}}}
\subfloat[dz/Z=0.375]{\scalebox{0.25}{\includegraphics[width=1\textwidth]{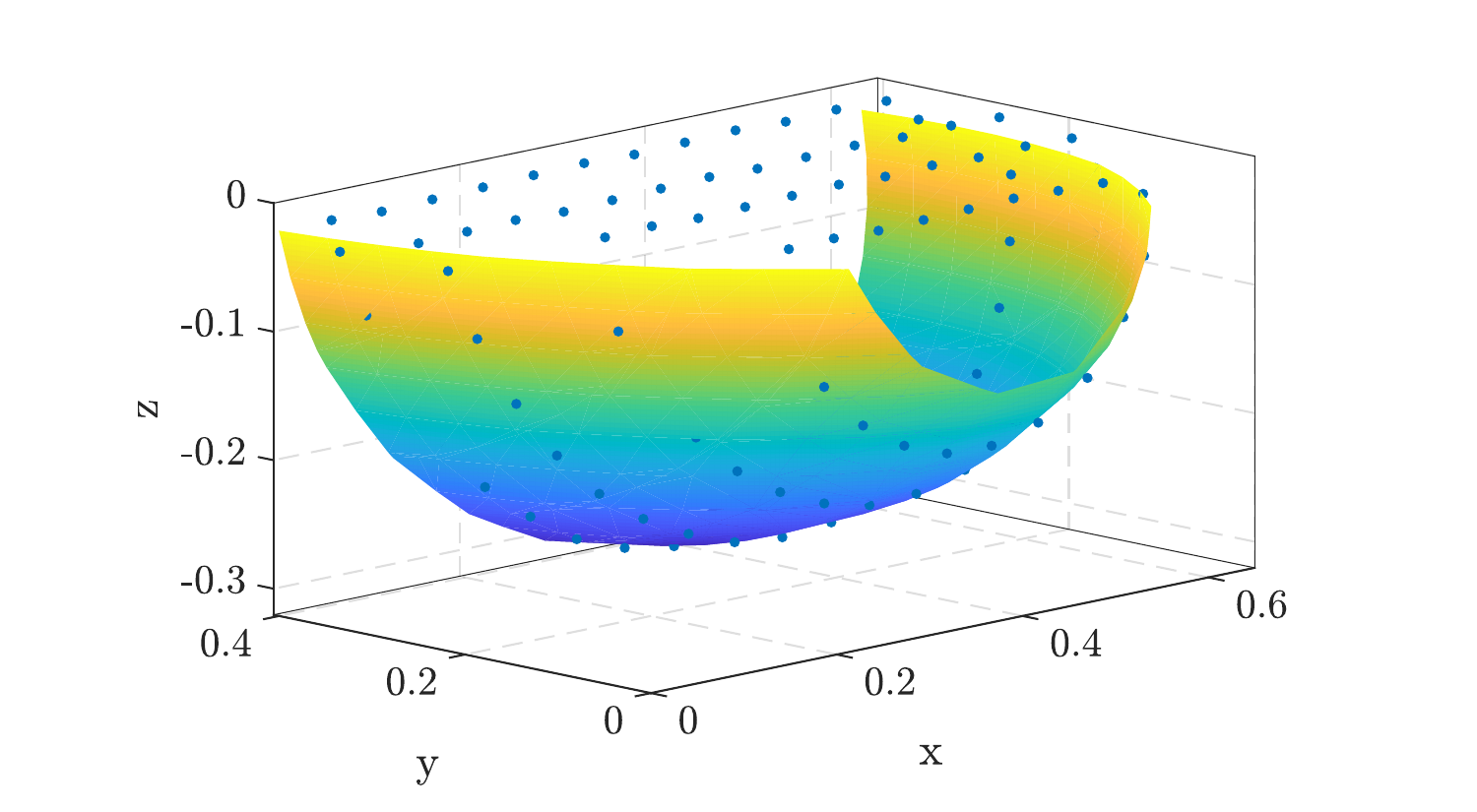}}}
\subfloat[dz/Z=0.5]{\scalebox{0.25}{\includegraphics[width=1\textwidth]{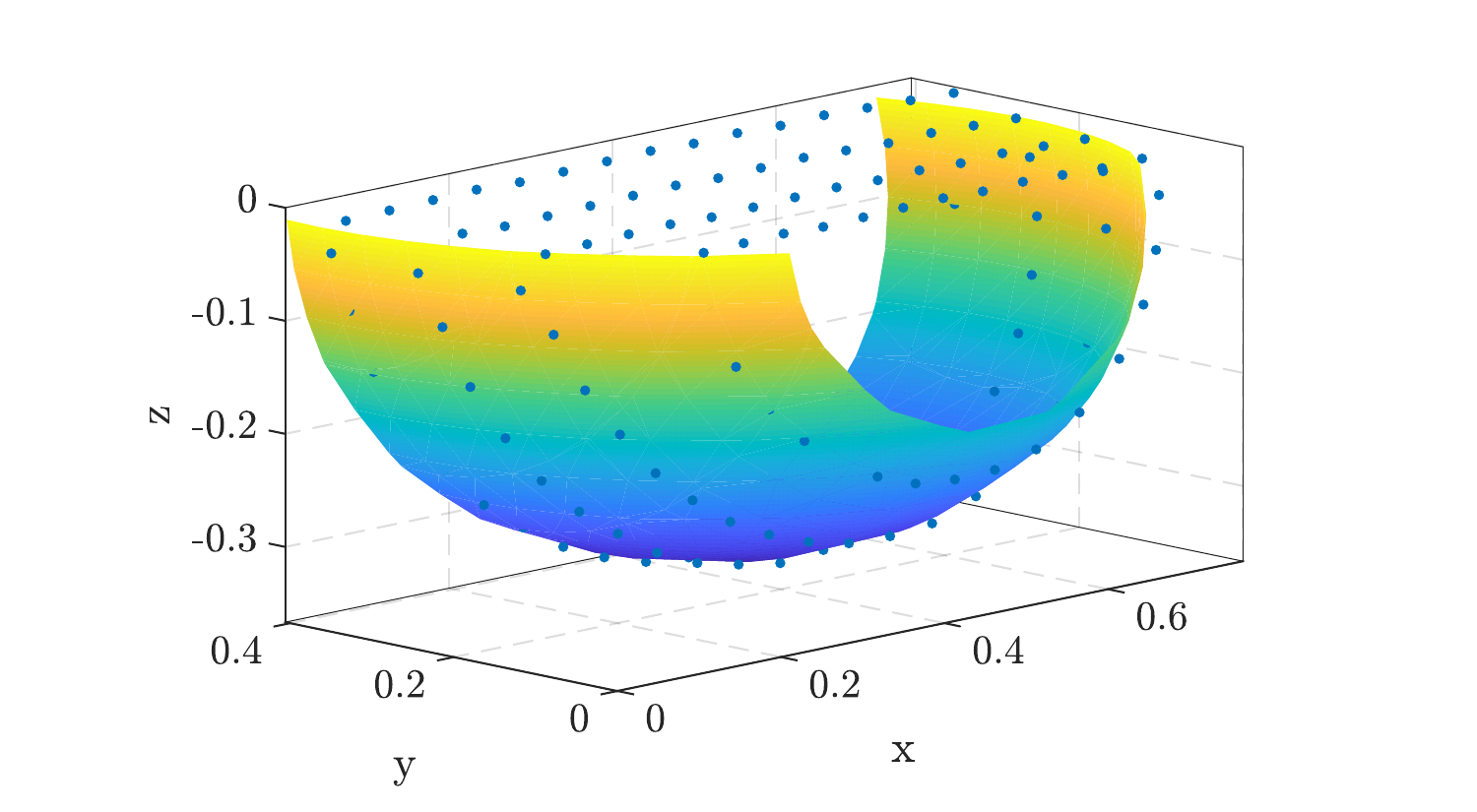}}}\\
\subfloat[dz/Z=0.625]{\scalebox{0.25}{\includegraphics[width=1\textwidth]{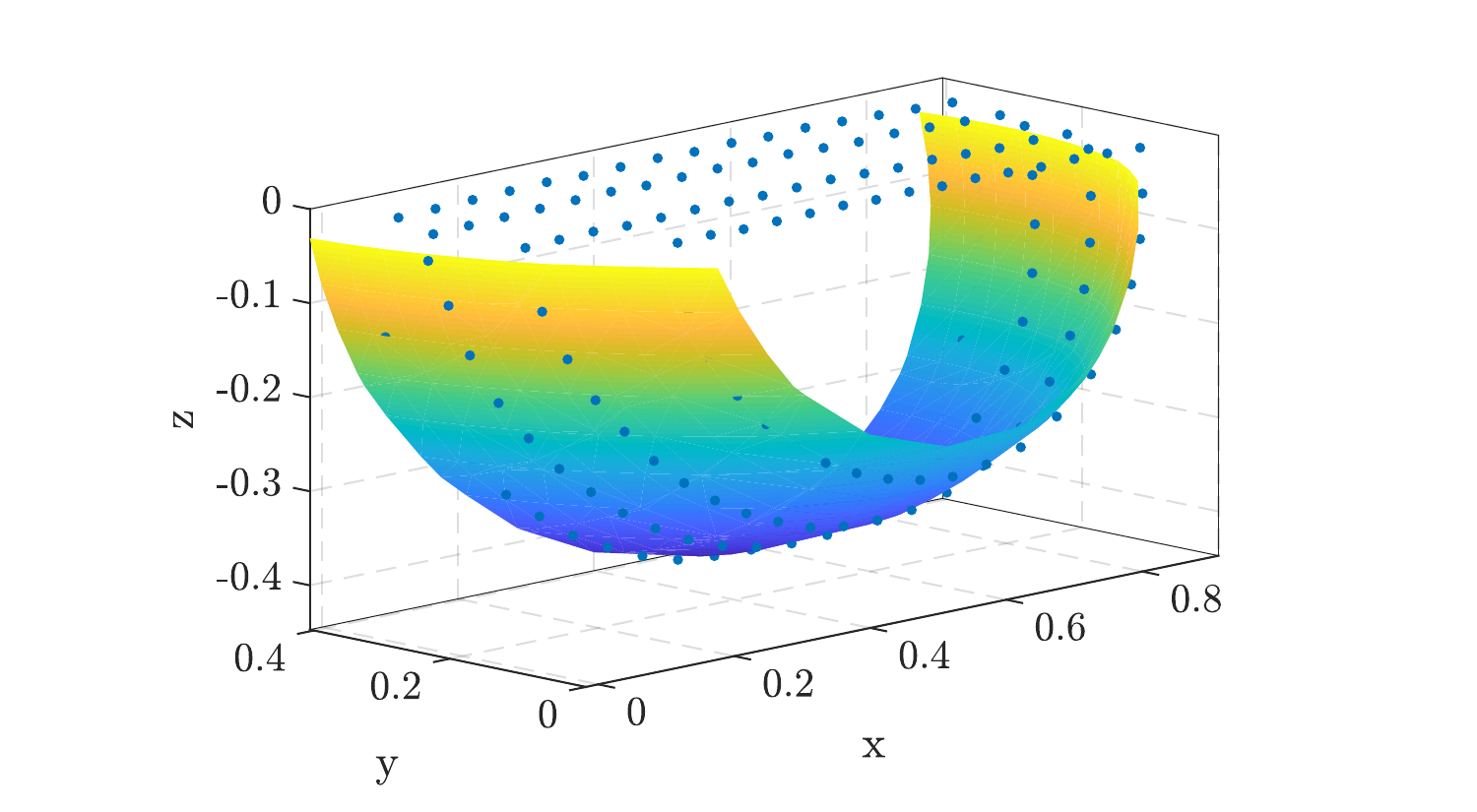}}}
\subfloat[dz/Z=0.75]{\scalebox{0.25}{\includegraphics[width=1\textwidth]{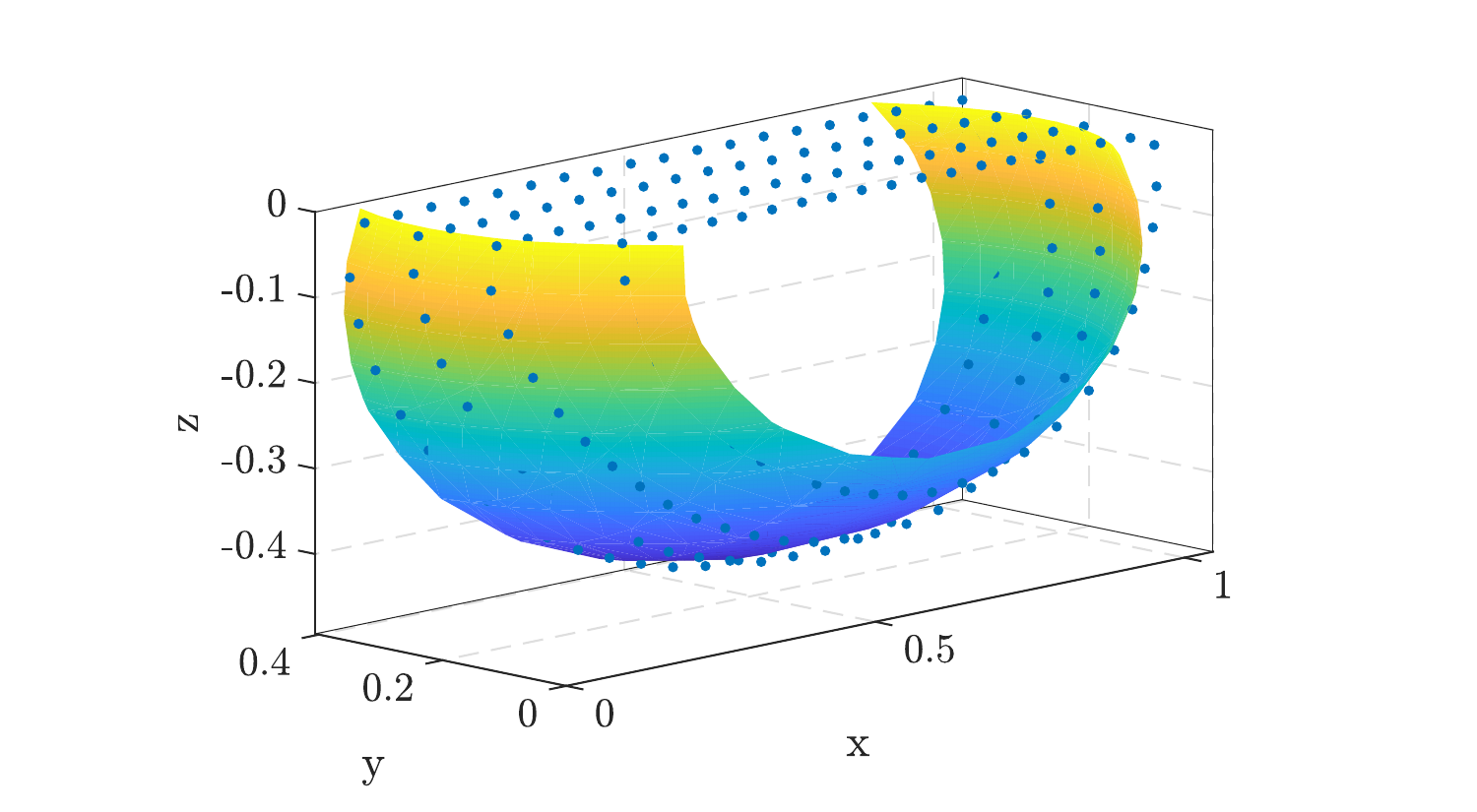}}}
\subfloat[dz/Z=0.875]{\scalebox{0.25}{\includegraphics[width=1\textwidth]{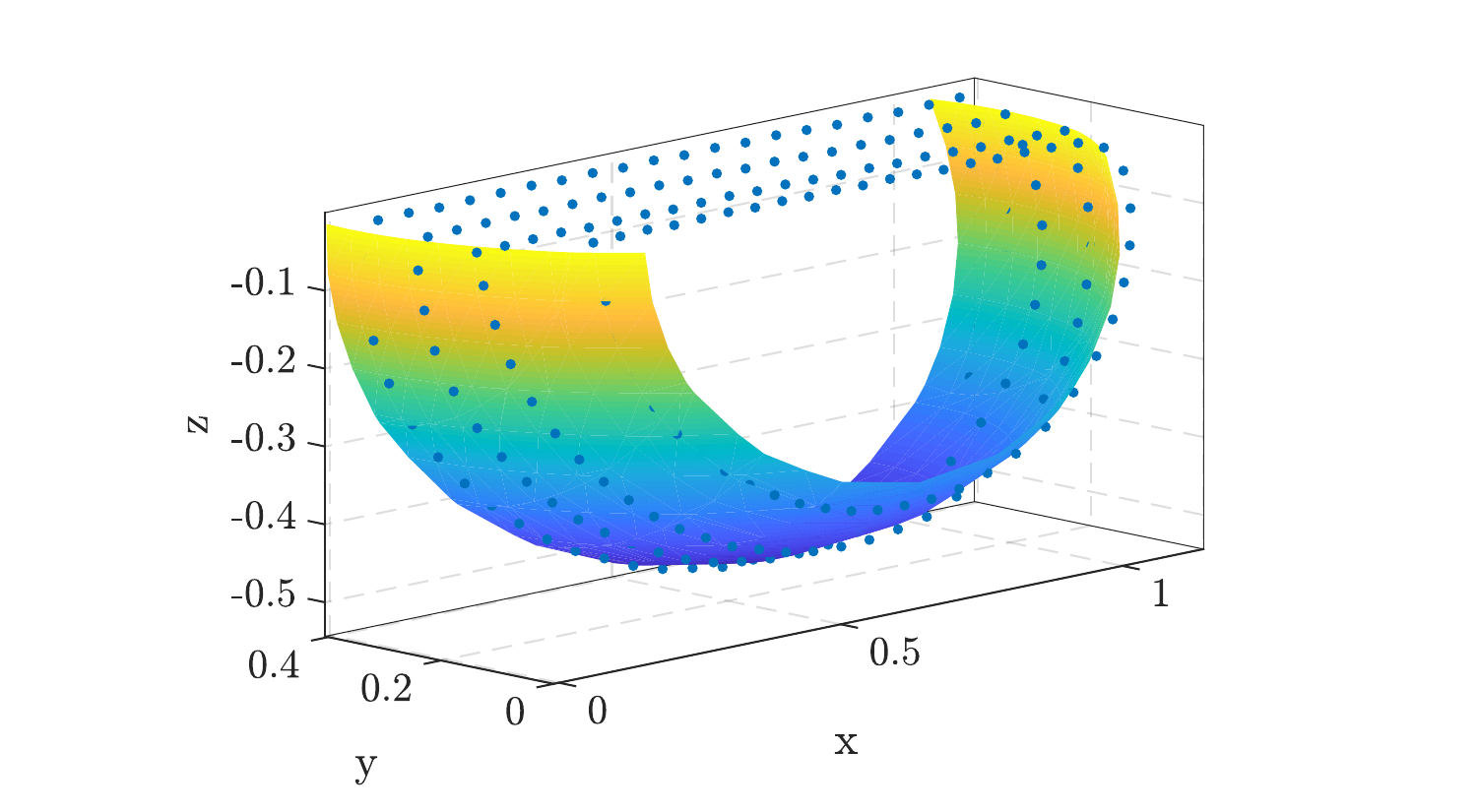}}}
\subfloat[dz/Z=1.0]{\scalebox{0.25}{\includegraphics[width=1\textwidth]{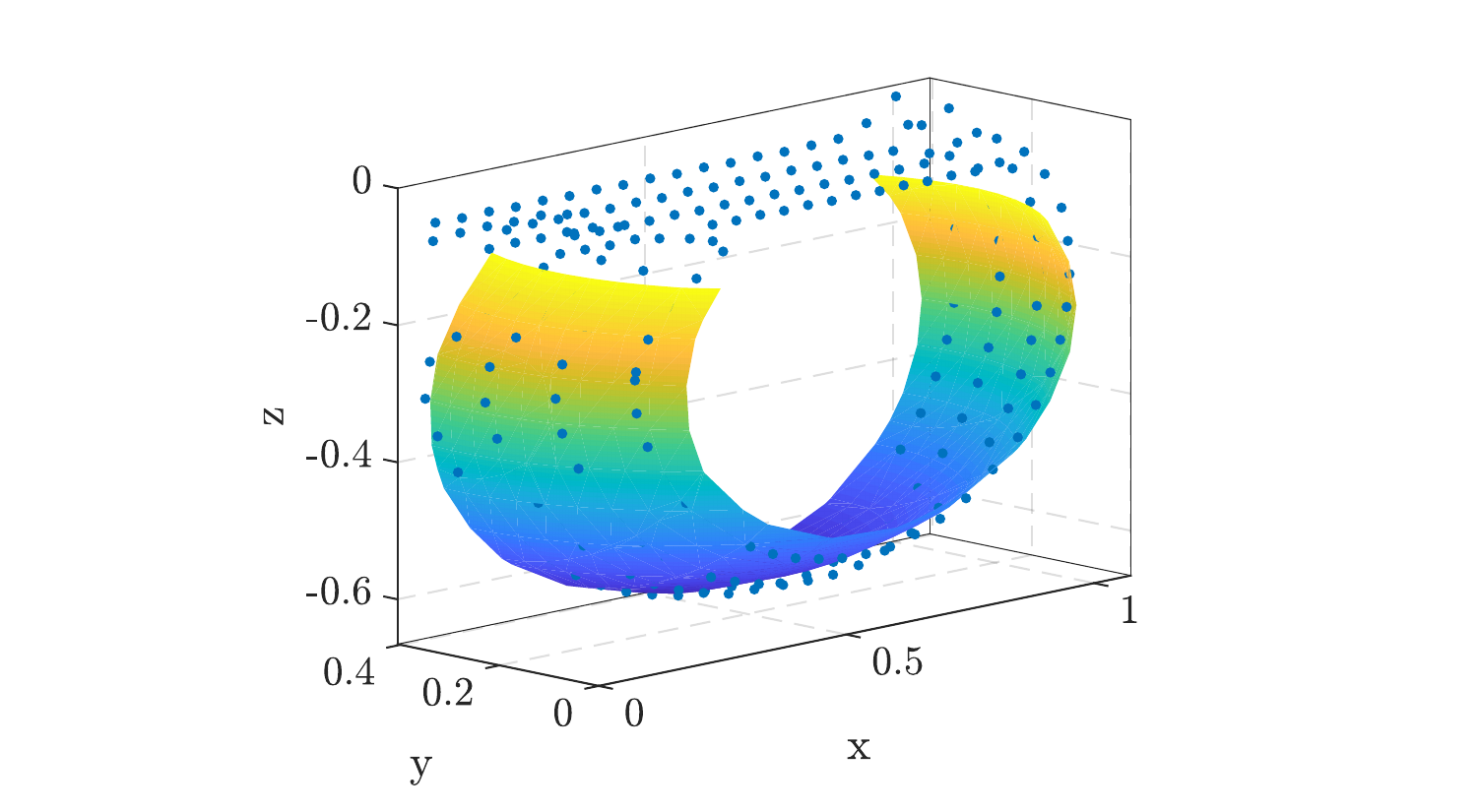}}}
\caption{Added mass surface approximation via ellipsoid surface fitting method (see \ref{appendix:b}) for the slamming cases with $W_n = 0.875$ m/s.} \label{ellipsoid_approx4}
\end{figure}  

\begin{figure}[h!]
\centering
\includegraphics[width=0.7\textwidth]{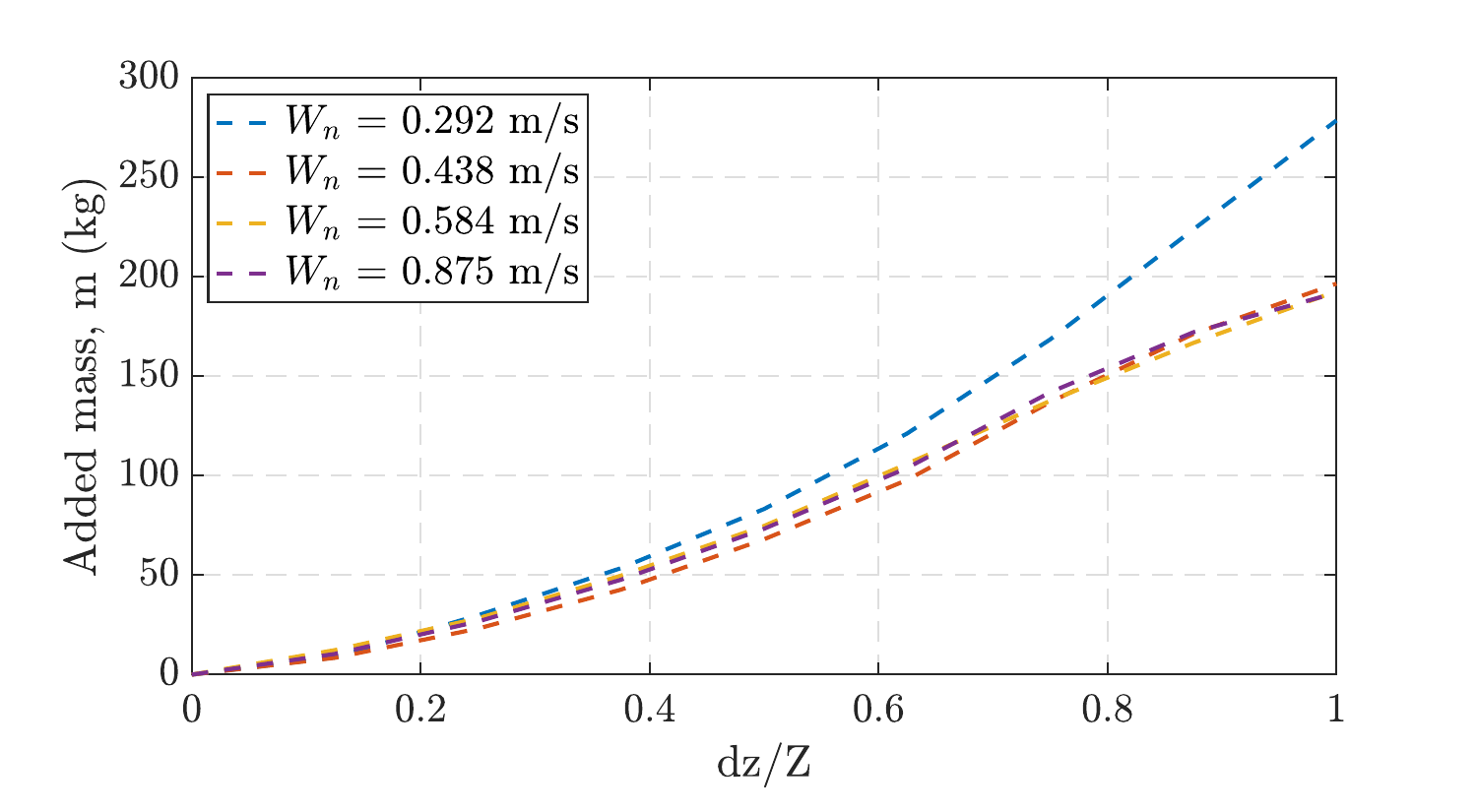}
\caption{Added mass obtained from pressure contour inspections for slamming cases with $W_n$ = 0.292, 0.438, 0.584, and 0.875~m/s.} \label{added_mass}
\end{figure}

\begin{figure}[h!]
\centering
\captionsetup[subfigure]{justification=centering}
\subfloat[$W_n$ = 0.292 m/s]{\scalebox{0.5}{\includegraphics[width=1\textwidth]{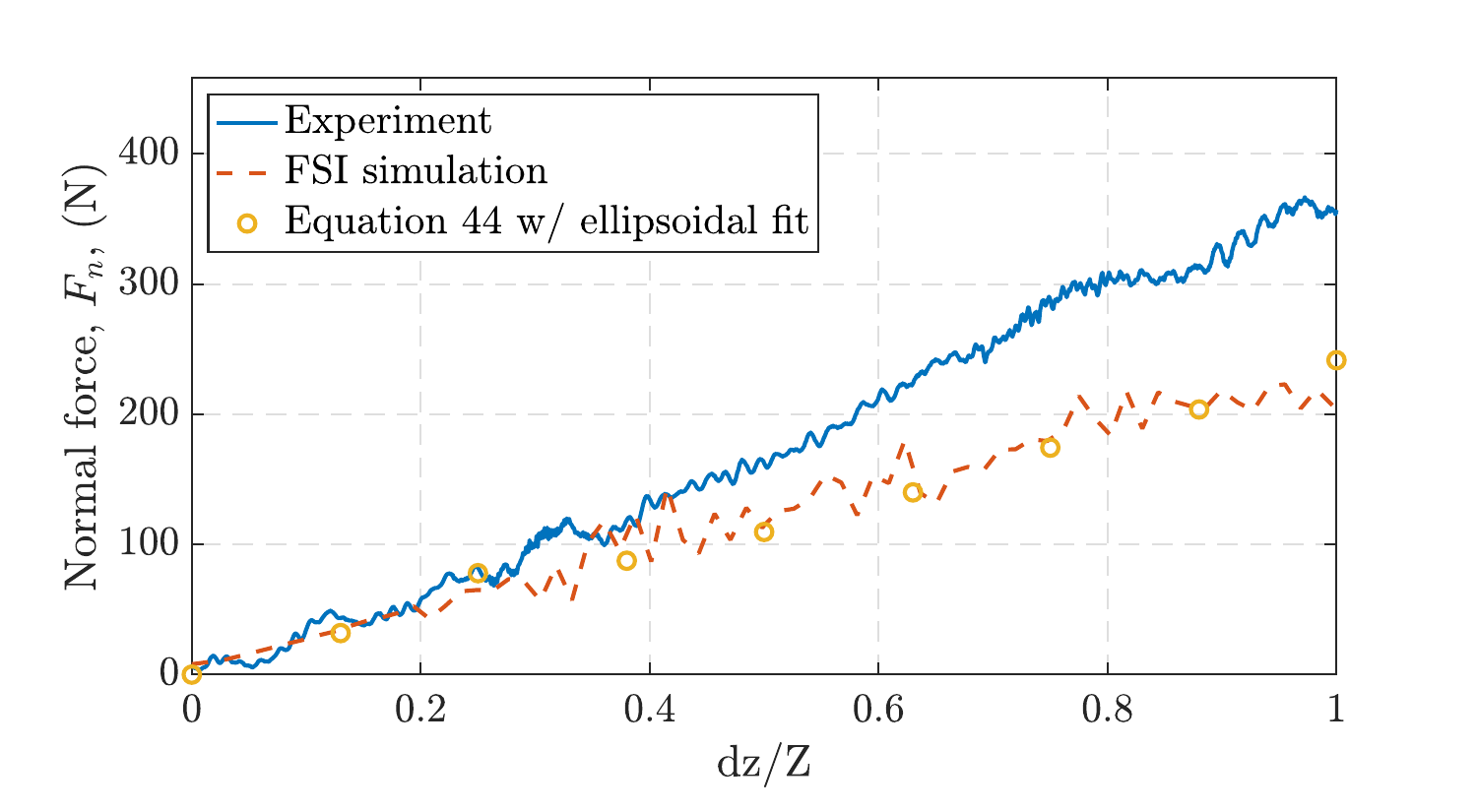}}}
\subfloat[$W_n$ = 0.438 m/s]{\scalebox{0.5}{\includegraphics[width=1\textwidth]{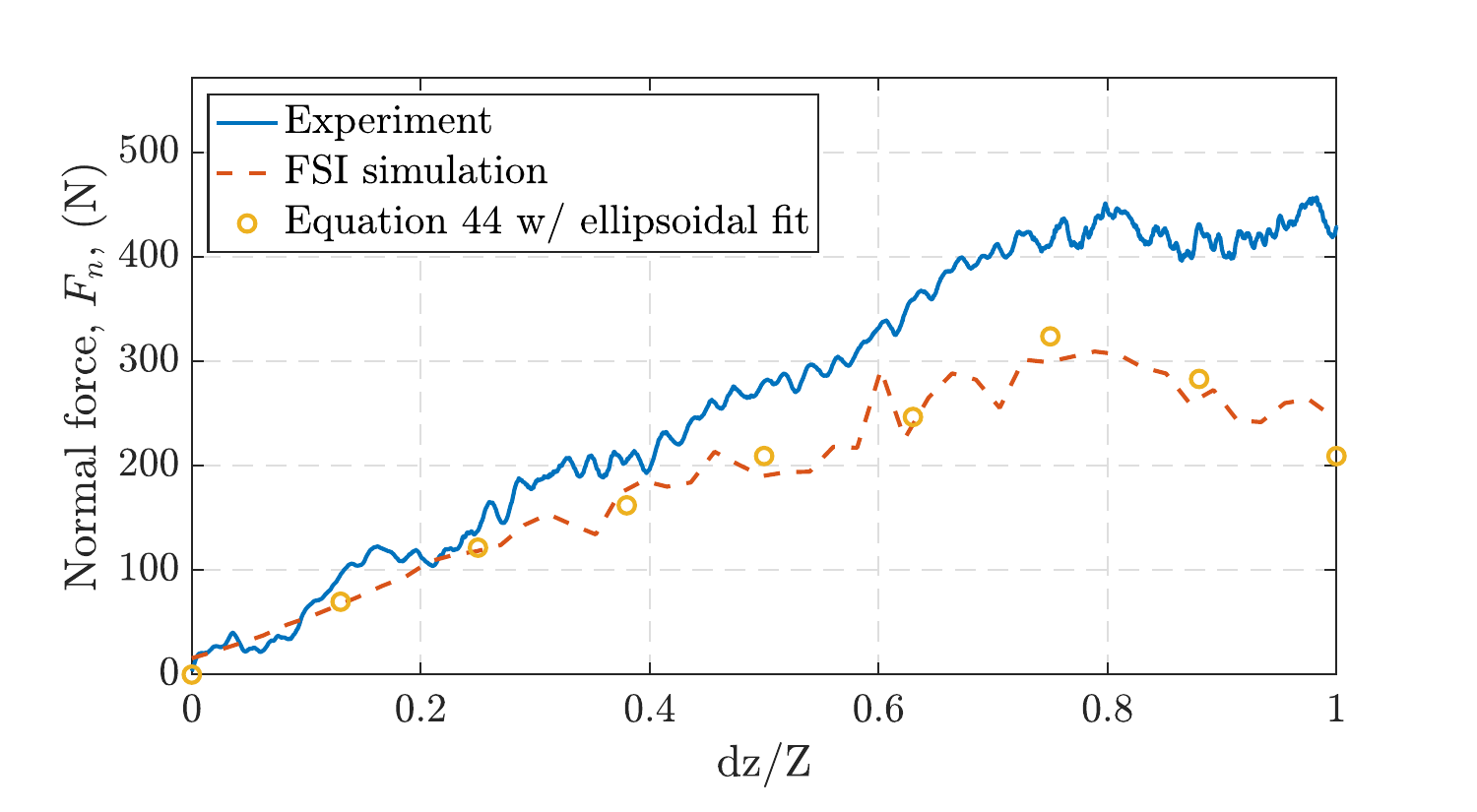}}}\\
\subfloat[$W_n$ = 0.584 m/s]{\scalebox{0.5}{\includegraphics[width=1\textwidth]{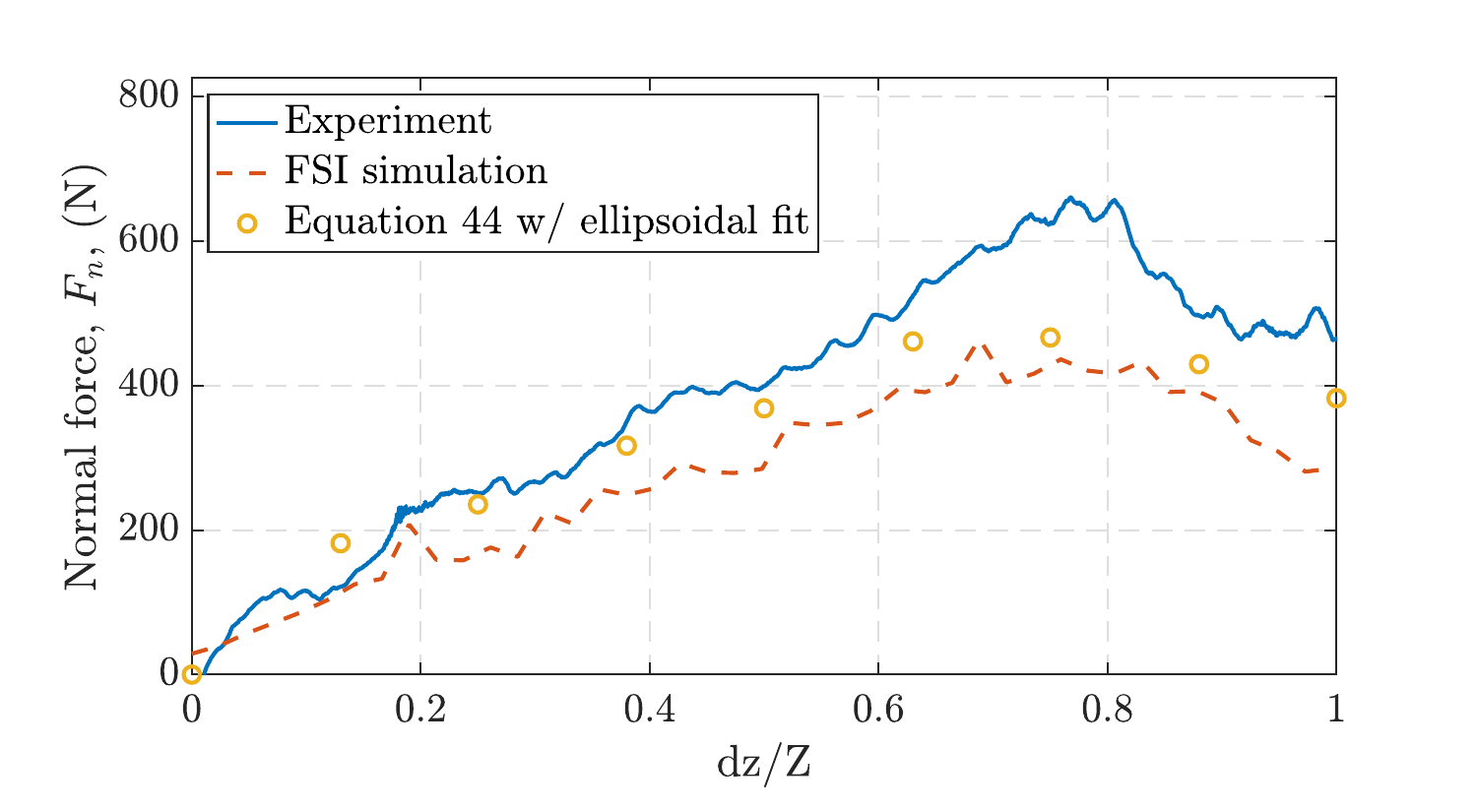}}}
\subfloat[$W_n$ = 0.875 m/s]{\scalebox{0.5}{\includegraphics[width=1\textwidth]{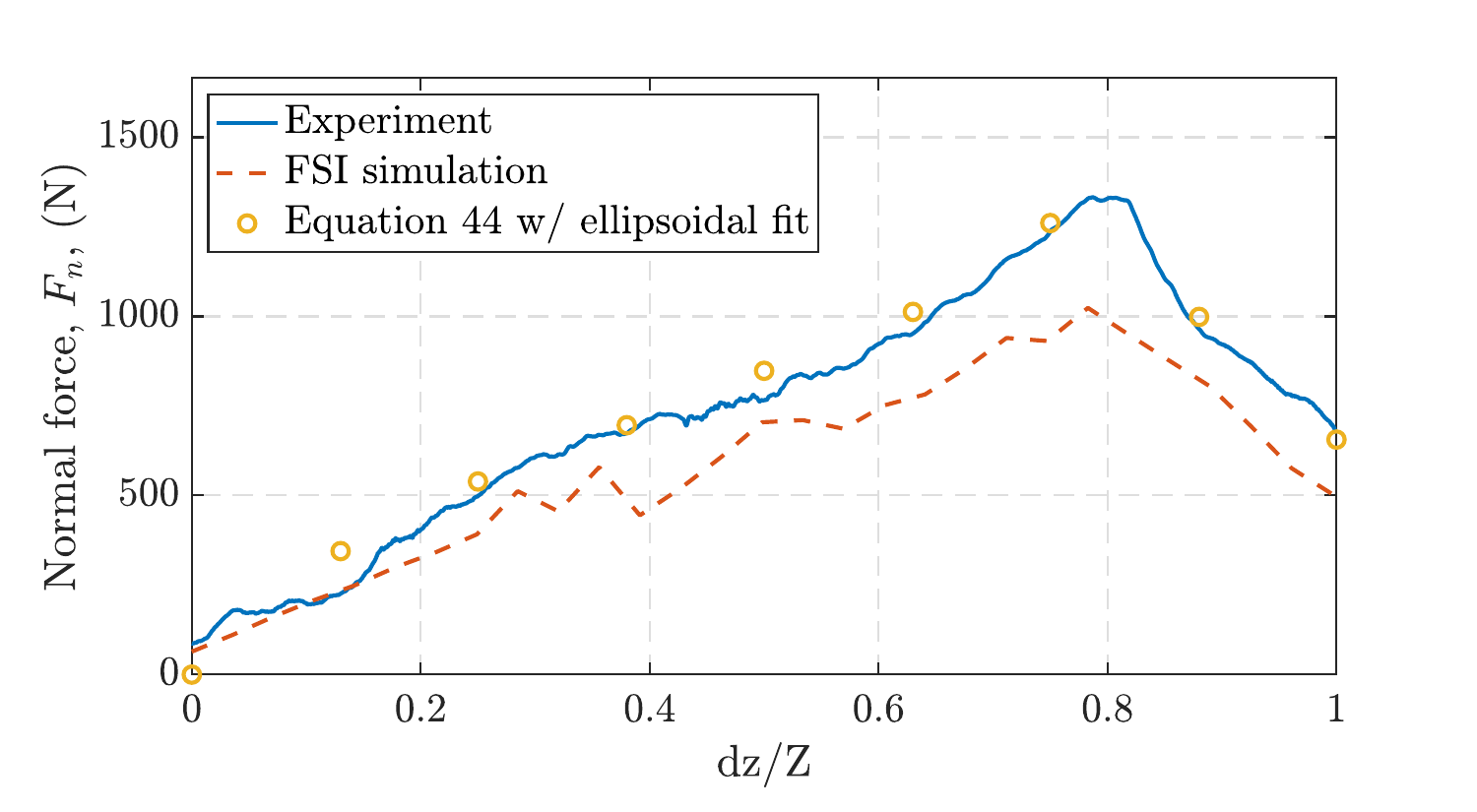}}}
\caption{Normal impact force comparison for slamming cases with $W_n$ = 0.292, 0.438, 0.584, and 0.875~m/s.} \label{theory_val}
\end{figure}

The added mass can be obtained by multiplying the volume bounded by the approximated added mass surface, by the water density, shown in Figure \ref{added_mass}. All four cases have similar added mass, except slamming case $W_n = 0.292$~m/s. The $W_n = 0.292$~m/s exhibits higher added mass due to water splashing over from the trailing edge as discussed in Section \ref{validation_results}. We adopt the added mass obtained using these ellipsoidal fits to approximate the normal impact force, $F_n$, using Equation \ref{newton2}. The empirical results are compared against experimental and FSI simulation results, presented in Figure \ref{theory_val}. Using the ellipsoidal fits for the added mass with Equation \ref{newton2}, we observe good agreement between validation experiments, high fidelity simulations, and Equation \ref{newton2} (the latter using ellipsoidal added mass fits).
 
\newpage
\subsection{Proposed added mass generalization}
\begin{figure}[h!]
\centering
\includegraphics[width=0.7\textwidth]{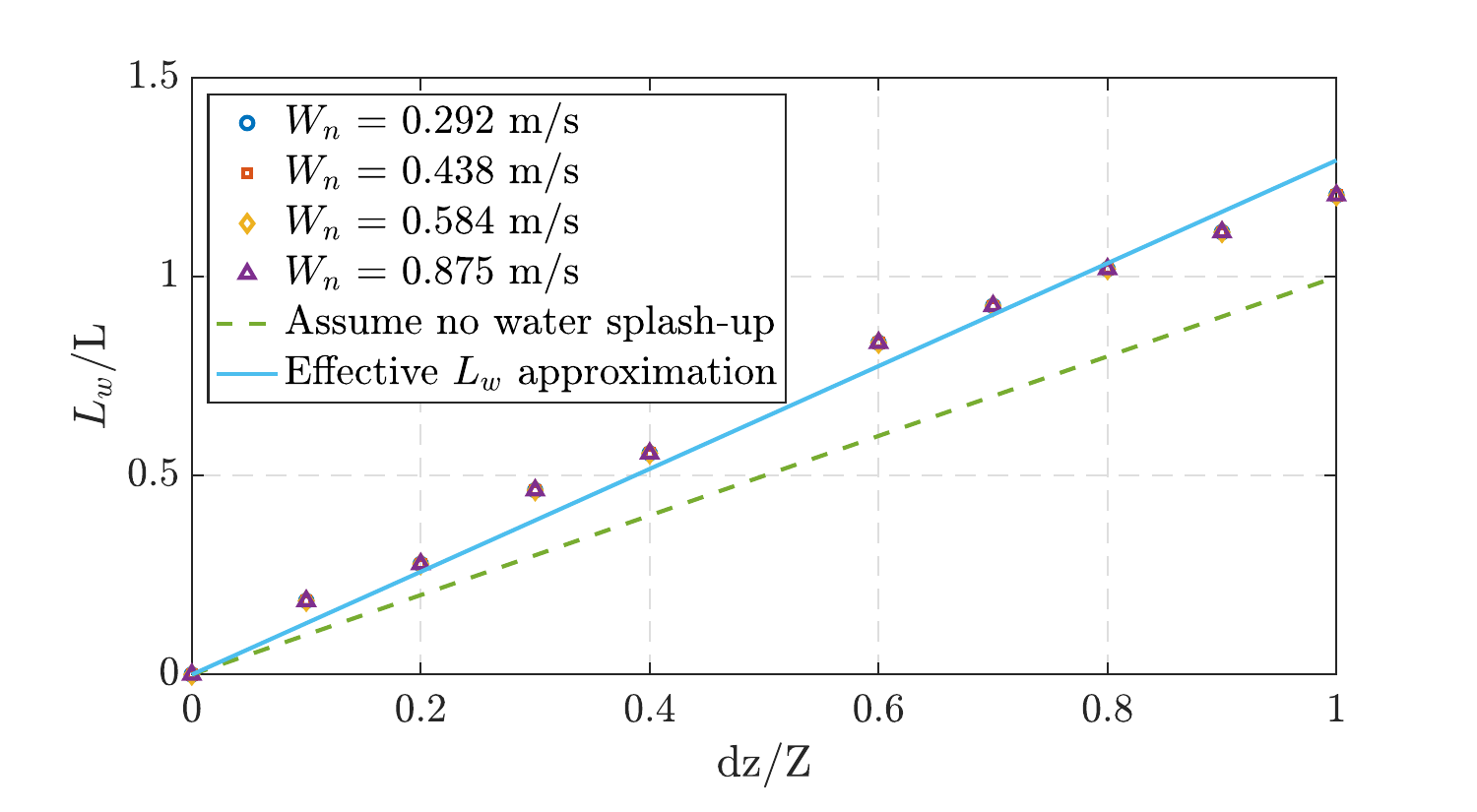}
\caption{Effective wetted length approximation from measured data. $L_w/L$ represents the length of the water spray sheet relative to the length of the plate.} \label{wetted_length}
\end{figure}

In the sequel, we aim to develop a generalized mathematical formulation to characterize the impact force arises from slamming. The force behavior up to force unloading is of interest, as this corresponds to the maximal loading for design. Referring back to the summary of the slamming analyses for the case of $W_n = 0.875$~m/s in Figure \ref{W0p889}, we observe that the force unloading is directly related to the location of the spray root. Specifically, the spray root generates a highly concentrated pressure that positively relates to the slamming impact force. As this spray root leaves the plate, the high pressure is released, and as a result, the impact force is reduced. Given this information, the lengths of the spray root (\textit{i.e.,} the effective wetted length of the plate) are measured within our simulations, to determine the instant when the spray root departs the plate. In Figure \ref{wetted_length}, $L_w$ denotes the wetted length, and $L$ denotes the plate's overall length. The effective wetted length is approximated by linear fitting to the measured data. The data show that the spray root exits the plate (when $L_w/L$ equals to one) at $1.293dz/Z$. In other words, the effective $L_w$ is approximately 1.293 times longer than the nominal $L_w$. The wetted length coefficient is likely to be higher in cases with smaller angle of attack, $\alpha$, as higher pressure is generated between the plate and the water free surface. 

\begin{figure}[h!]
\centering
\includegraphics[width=0.8\textwidth]{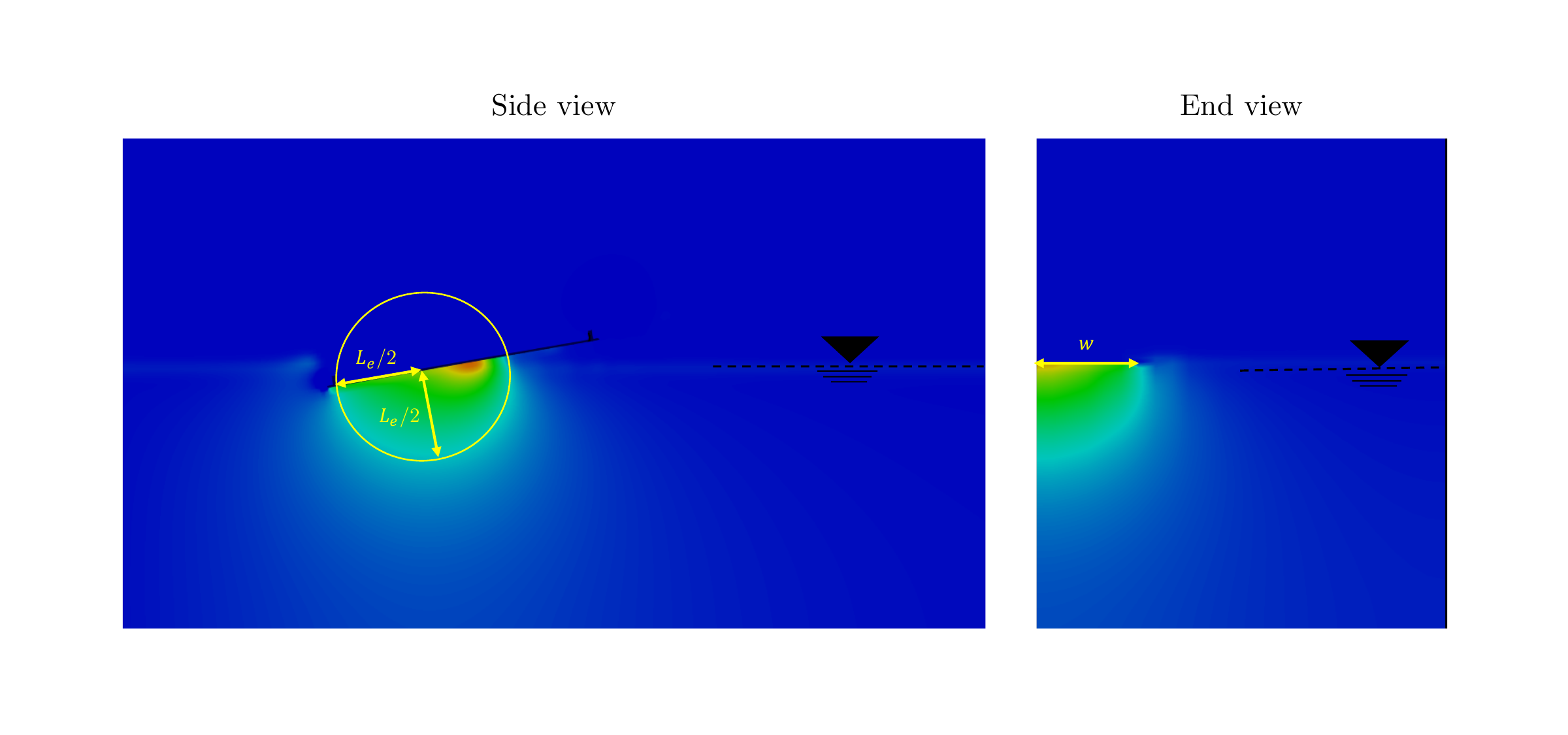}
\caption{The side view and end view of the pressure dynamic contour at $dz/Z = 0.5$ of a representative slamming case.} \label{ellipsoid_gen}
\end{figure}

Here, we propose to approximate the added mass volume as a quarter ellipsoid with dimension $L_e/2 \times L_e/2 \times w$, where $L_e$ is the effective $L_w$, and $w$ is the width of the plate, as shown in Figure \ref{ellipsoid_gen}. The added mass of the proposed quarter ellipsoid is formulated as 

\begin{equation} \label{mass}
m_a = \frac{\rho_w\pi w}{3}(\frac{L_e}{2})^2,
\end{equation}
where $L_e = 1.293L_n$ and $L_n=\frac{Wt}{sin(\alpha)}$. Substitute in $L_e$, Equation \ref{mass} is re-formulated as 

\begin{equation} \label{mass2}
m_a = \frac{\rho_w\pi w}{3}(\frac{1.293Wt}{2sin(\alpha)})^2.
\end{equation}

Integrating Equation \ref{mass2} w.r.t time, we obtain an expression for the change of added mass in time as,

\begin{equation} \label{dmdt}
\frac{dm_a}{dt} = \frac{1.293^2\rho_w\pi w}{6}\frac{W^2}{sin^2(\alpha)}t.
\end{equation}
Substituting Equation \ref{dmdt} into Equation \ref{newton2}, the generalized form of the normal impact force is arrived at:

\begin{equation} \label{Fn}
F_n = \frac{1.293^2\rho_w\pi w}{6}\frac{W^3cos(\alpha)}{sin^2(\alpha)}t \,\,\,\,\,\text{for}\,\,\,\,\,dz/Z \leq \frac{1}{1.293}.
\end{equation}
This mathematical form is a compact expression suitable for design. It leverages deep phenomenological insights, gleaned from our validated FSI simulations, within the classical context of Newton's second law of motion, to furnish a useful engineering theory. 

\begin{figure}[h!]
\centering
\captionsetup[subfigure]{justification=centering}
\subfloat[$W_n$ = 0.292 m/s]{\scalebox{0.5}{\includegraphics[width=1\textwidth]{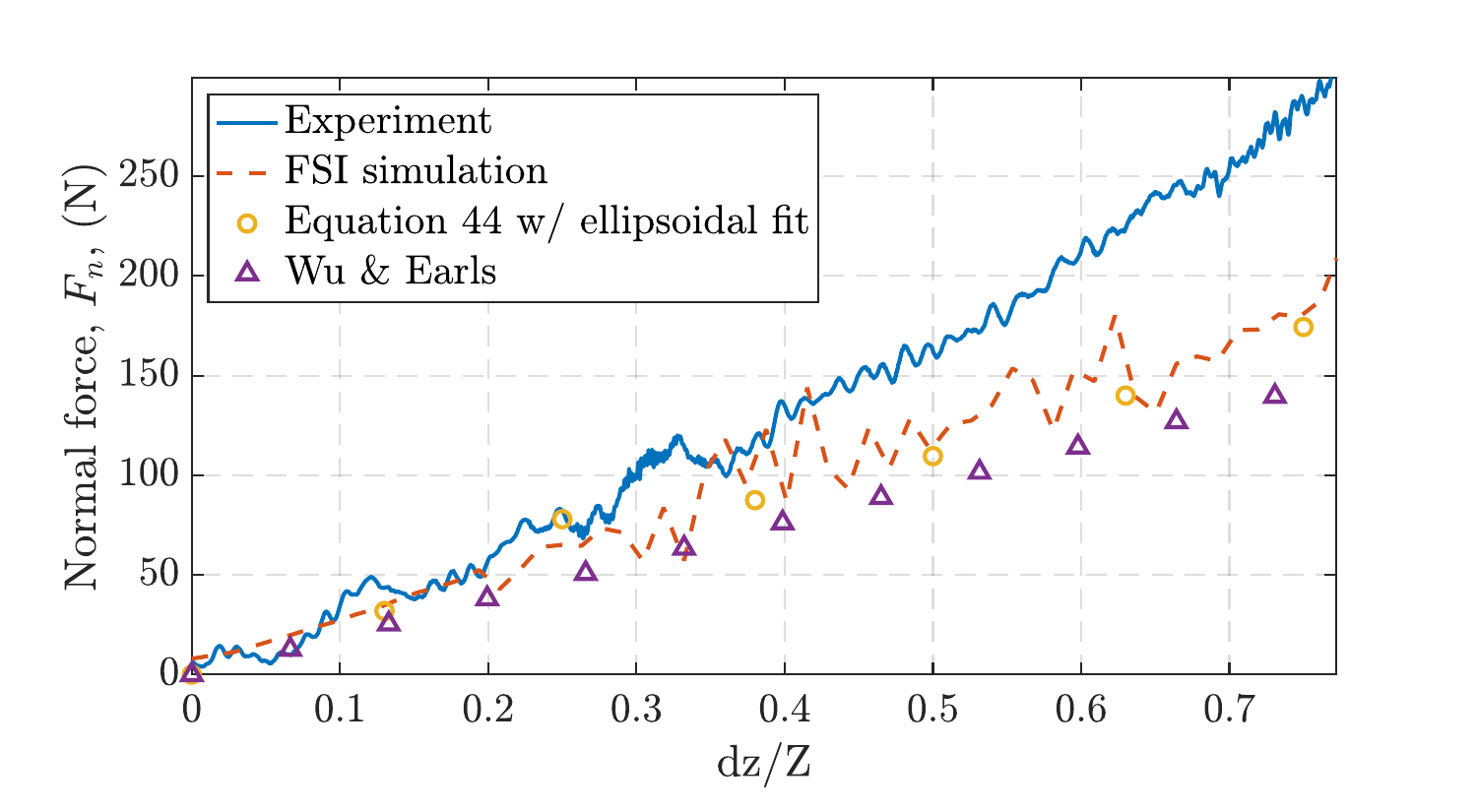}}}
\subfloat[$W_n$ = 0.438 m/s]{\scalebox{0.5}{\includegraphics[width=1\textwidth]{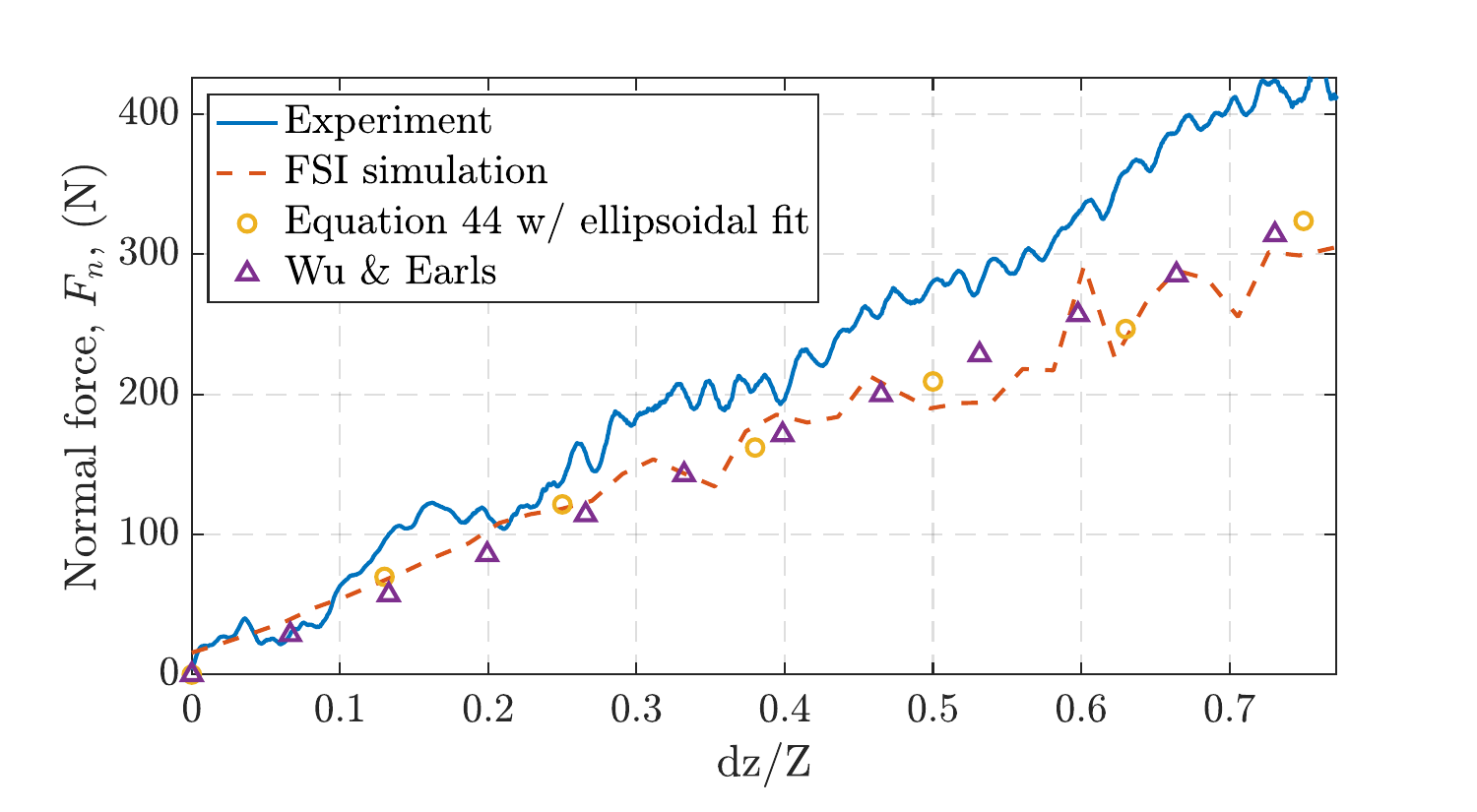}}}\\
\subfloat[$W_n$ = 0.584 m/s]{\scalebox{0.5}{\includegraphics[width=1\textwidth]{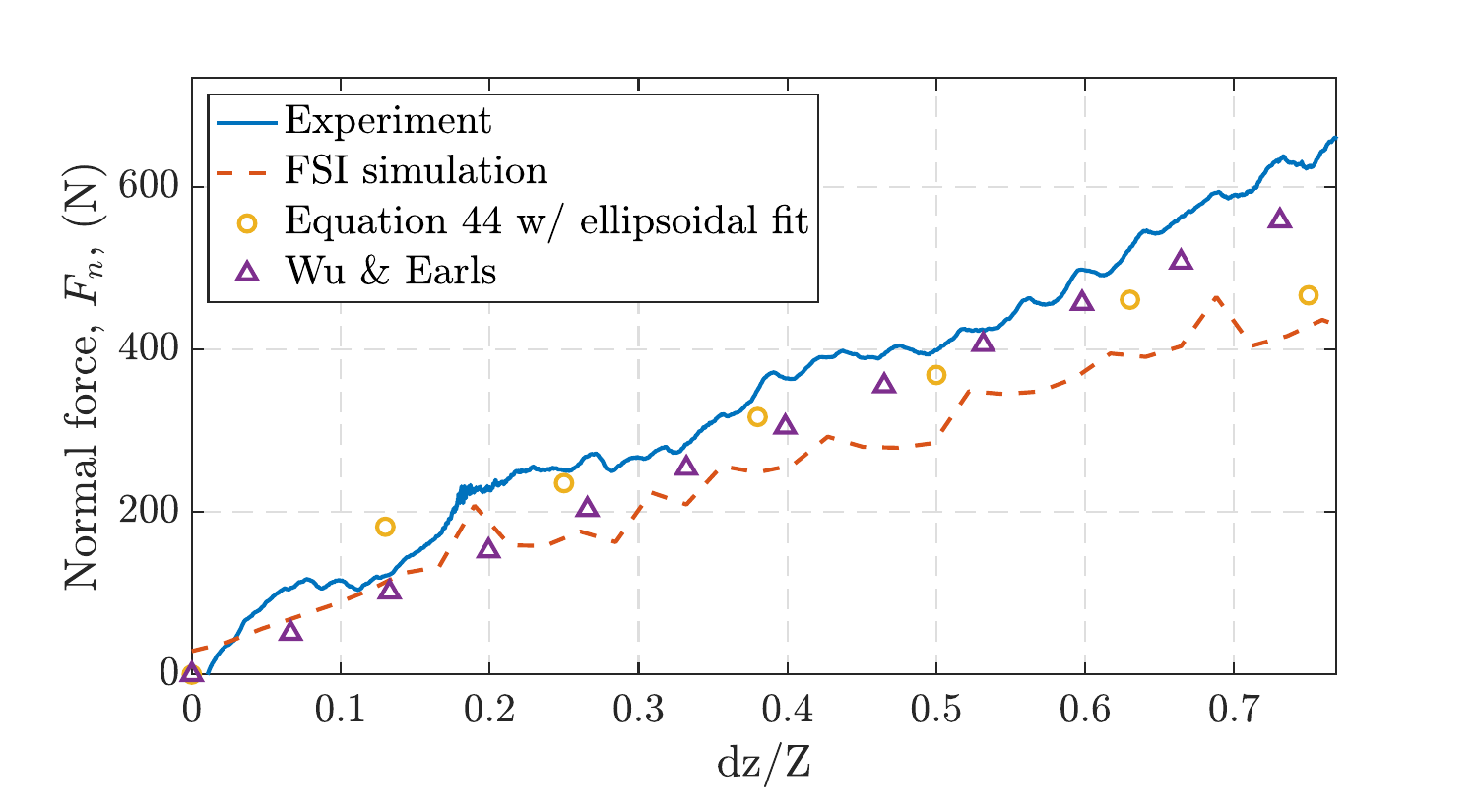}}}
\subfloat[$W_n$ = 0.875 m/s]{\scalebox{0.5}{\includegraphics[width=1\textwidth]{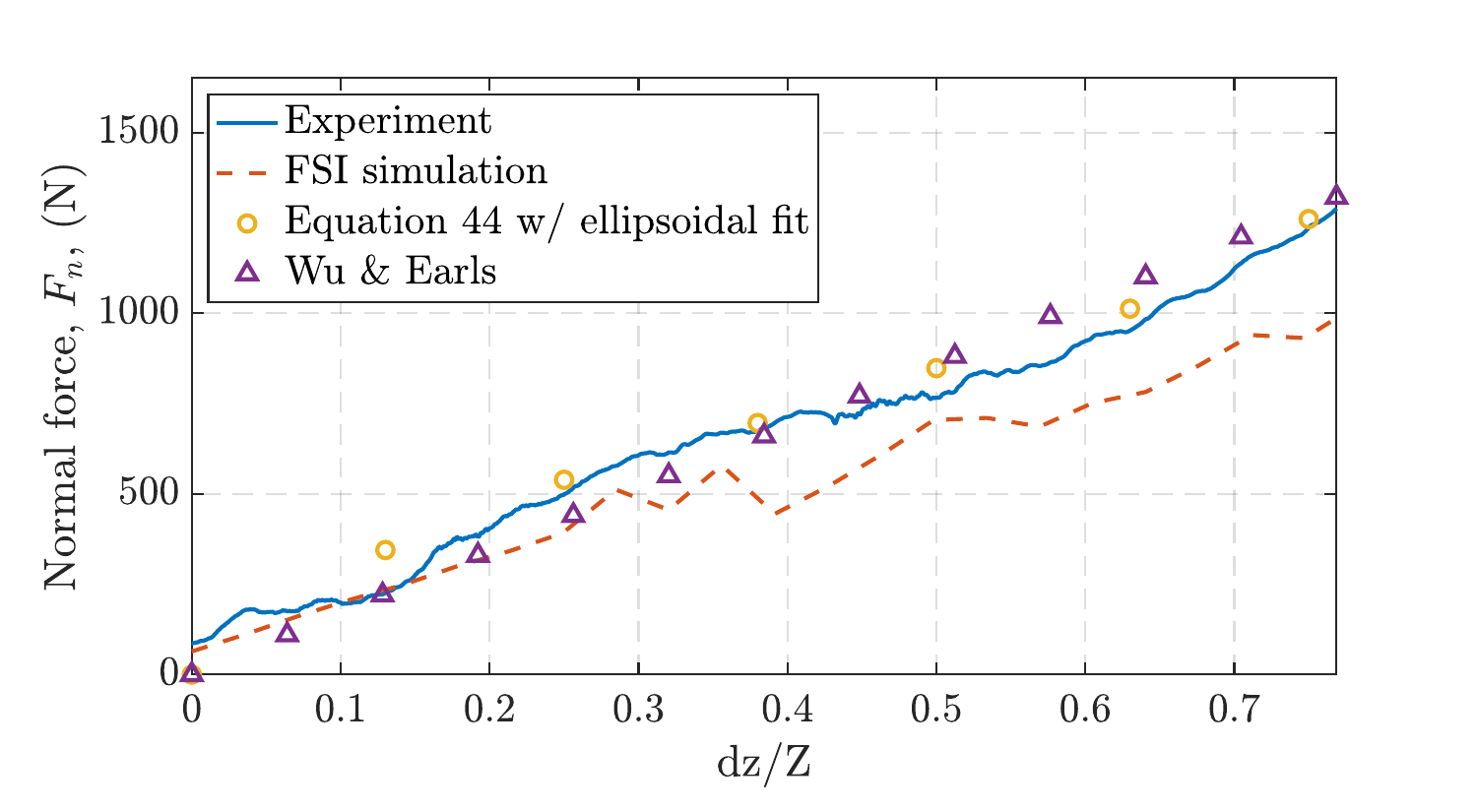}}}
\caption{Normal impact force comparison for slamming cases with $W_n$ = 0.292, 0.438, 0.584, and 0.875~m/s.} \label{theory_val2}
\end{figure}

We adopt the normal impact force formulation in Equation \ref{Fn} and approximate the normal impact forces for slamming cases with $W_n = 0.292, 0.438, 0.584,$ and $0.875$~m/s. The theoretical results are compared against FSI simulation and pressure contour inspection results. As shown in Figure \ref{theory_val2}, Equation \ref{Fn} exhibits excellent agreements in comparison to FSI simulations and pressure contour inspections, but with a greatly reduced effort in compression with full scale experiments, or even high fidelity FSI simulations (which consume up to 82 core hours in the representative case).

\section{Conclusion} \label{conclusion}
In this work, we have developed a novel engineering theory for the practically important slamming context of a flexible flat plate obliquely impacting the water free surface. To develop this theory, we leverage detailed phenomenological insights derived from experimentally validated FSI simulations. To confirm the veracity of the proposed implicit, partitioned FSI framework, we first validated our FSI solver against experimental data reported in \cite{Wang2020} (the FSI validation results have displayed excellent agreement with experiments). Subsequently, we applied our experimentally validated FSI framework to carry out a program of complementary FSI simulations in an effort to gain comprehensive engineering insight into the context of oblique slamming impact. Finally, we leveraged these insights to obtain a compact and simple mathematical model, derived from Newton's second law of motion, for flexible plates obliquely impacting the water free surface. Further slamming analyses are needed in order to finalize the parameters obtained in the current study. Nonetheless, the novel approach of deriving a closed form solution for slamming in the present work is potentially transformative and lays down a foundation towards a generalized slamming theory. This new engineering theory requires much less effort in comparison to experiments and FSI simulations, yet it displays very promising predictive power. Therefore, it presents considerable benefit and potential for use in the design of planning hulls. 
\section{Acknowledgement} \label{Acknowledgement}
This work is supported by the Office of Naval Research (ONR), under the grant N00014-19-1-2034. The authors thank Jonathan Stergiou, Ronald Miller, Hua Shan, and Dory Lummer from Naval Surface Warfare Center Carderock Division, who had provided guidance and suggestions in the development of the FSI software. The authors would also like to acknowledge An Wang and Professor James Duncan from the University of Maryland, for providing the experimental data reported within this paper.

\newpage
\appendix
\section{Validation Results}\label{appendix:a}
Here, we expand on the FSI validation results from all slamming test cases listed in Figure \ref{test_cases}.

\begin{figure}[h!]
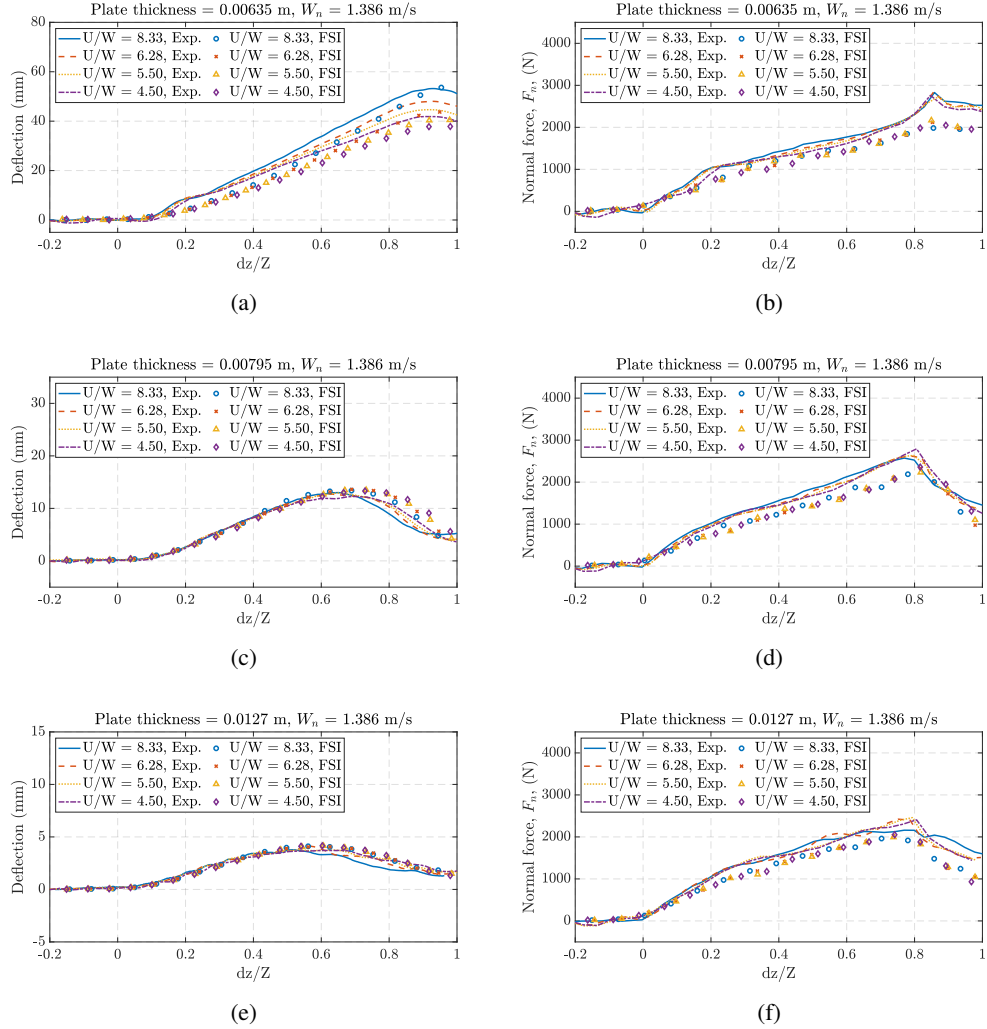

\centering
\captionsetup[subfigure]{justification=centering}
\subfloat[]{\scalebox{0.5}{\includegraphics[width=1\textwidth]{fig/th0p00635_Wn_1p386_disp.pdf}}}
\subfloat[]{\scalebox{0.5}{\includegraphics[width=1\textwidth]{fig/th0p00635_Wn_1p386_force.pdf}}}\\
\subfloat[]{\scalebox{0.5}{\includegraphics[width=1\textwidth]{fig/th0p00795_Wn_1p386_disp.pdf}}}
\subfloat[]{\scalebox{0.5}{\includegraphics[width=1\textwidth]{fig/th0p00795_Wn_1p386_force.pdf}}}\\
\subfloat[]{\scalebox{0.5}{\includegraphics[width=1\textwidth]{fig/th0p0127_Wn_1p386_disp.pdf}}}
\subfloat[]{\scalebox{0.5}{\includegraphics[width=1\textwidth]{fig/th0p0127_Wn_1p386_force.pdf}}}
\caption{Plate deflection and normal impact force histories of the highly flexible, moderately deformable, and nearly rigid plates subjected to slamming impacts with $W_n = 1.386$~m/s.}
\end{figure} 
\newpage
\begin{figure}[h!]
\centering
\captionsetup[subfigure]{justification=centering}
\subfloat[]{\scalebox{0.5}{\includegraphics[width=1\textwidth]{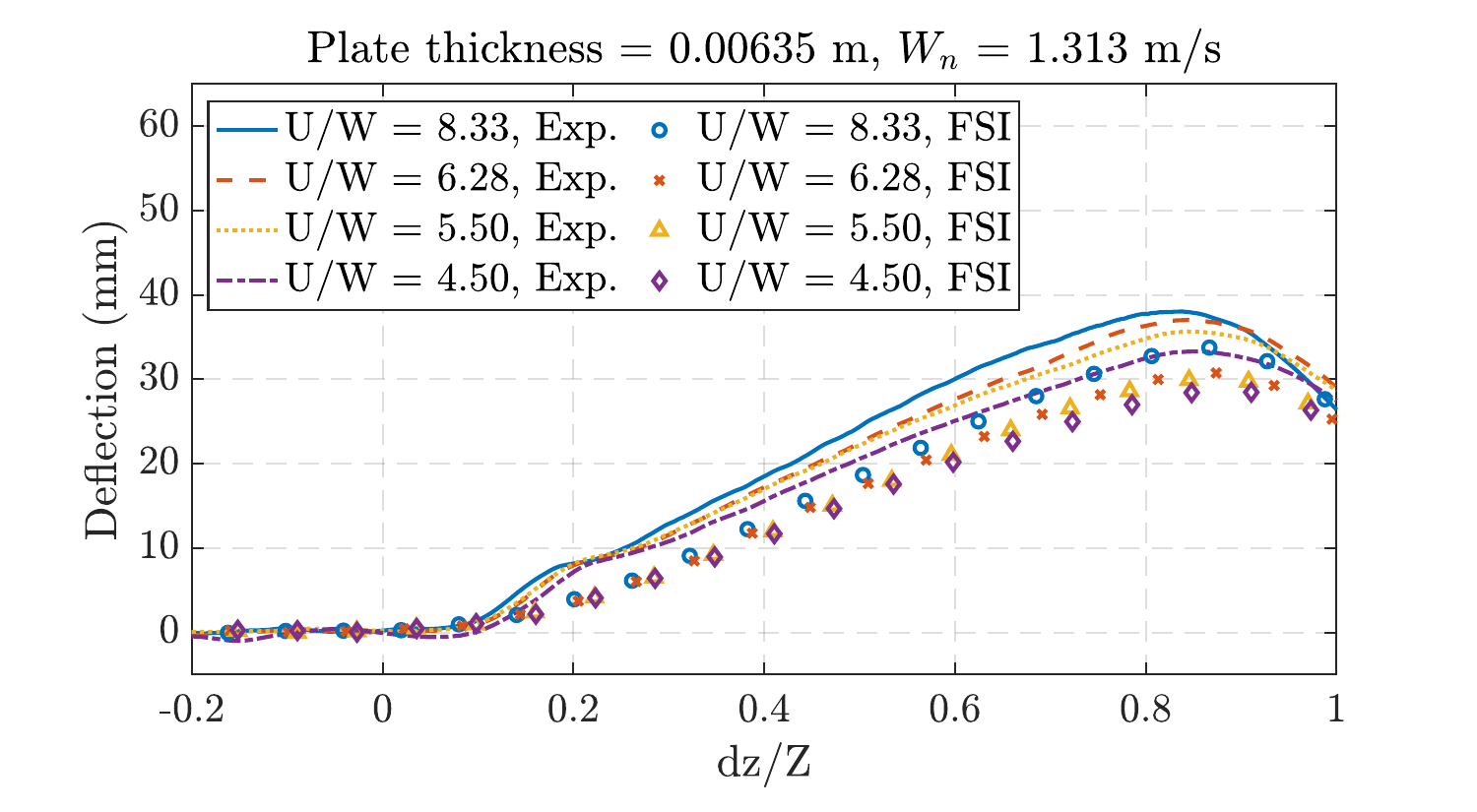}}}
\subfloat[]{\scalebox{0.5}{\includegraphics[width=1\textwidth]{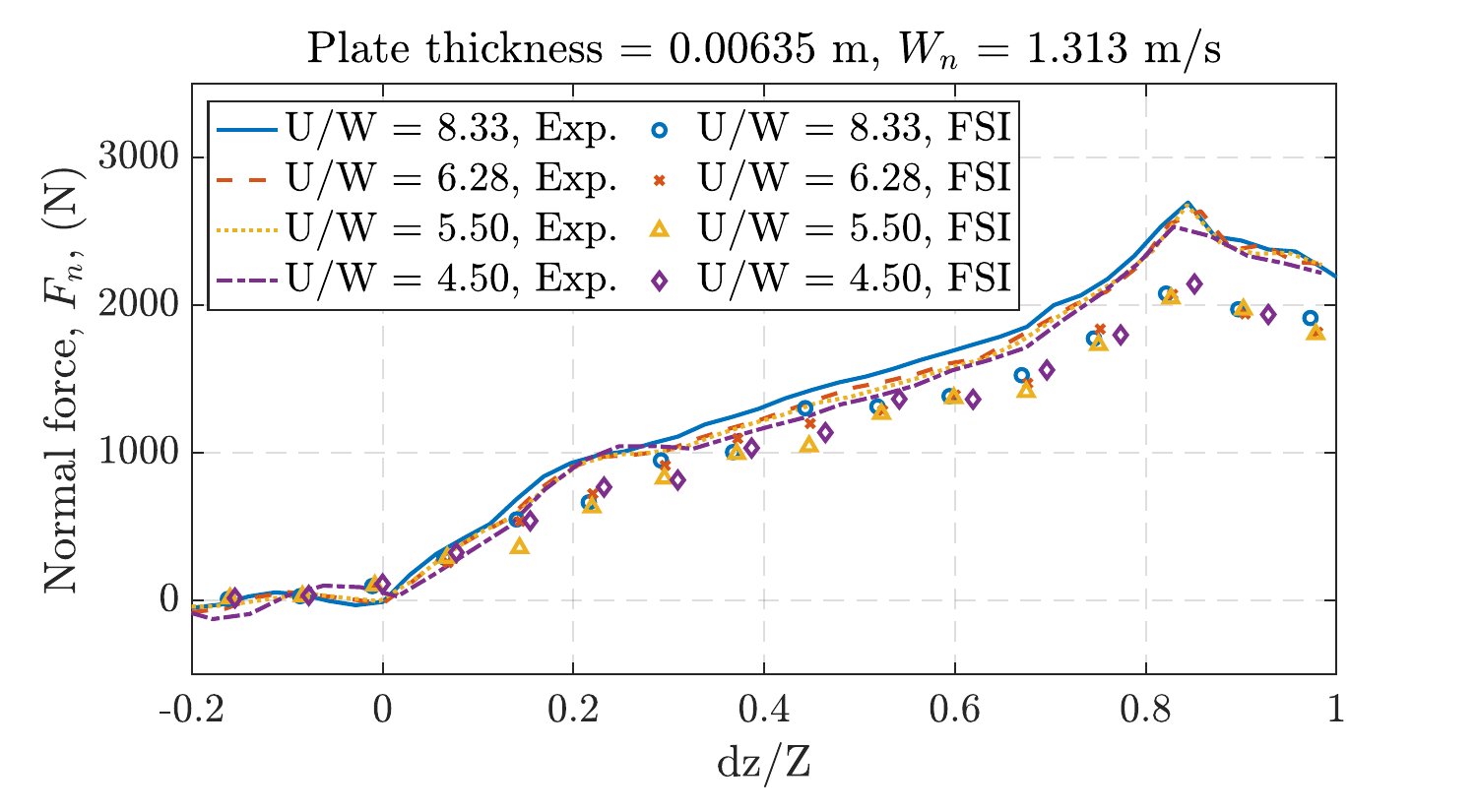}}}\\
\subfloat[]{\scalebox{0.5}{\includegraphics[width=1\textwidth]{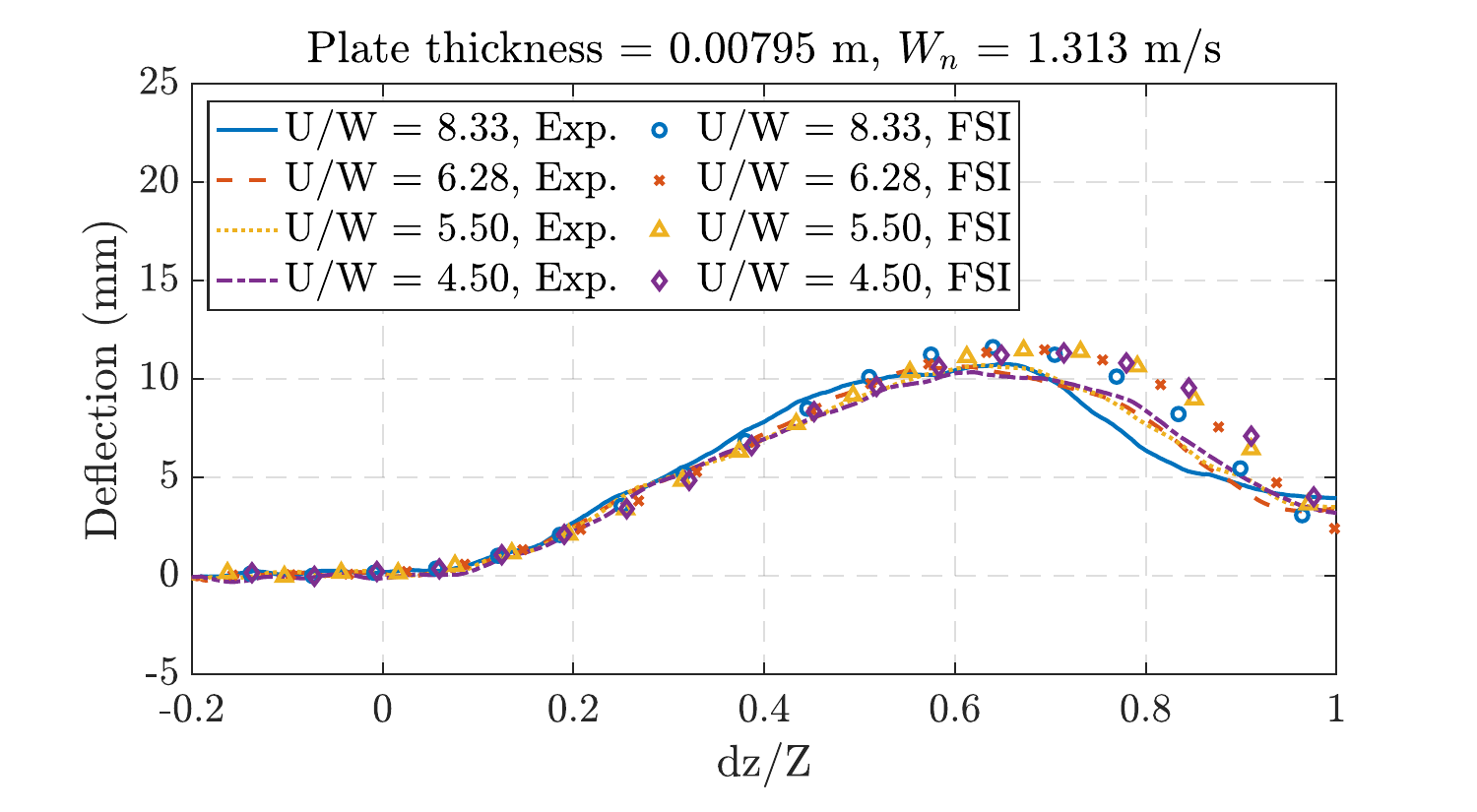}}}
\subfloat[]{\scalebox{0.5}{\includegraphics[width=1\textwidth]{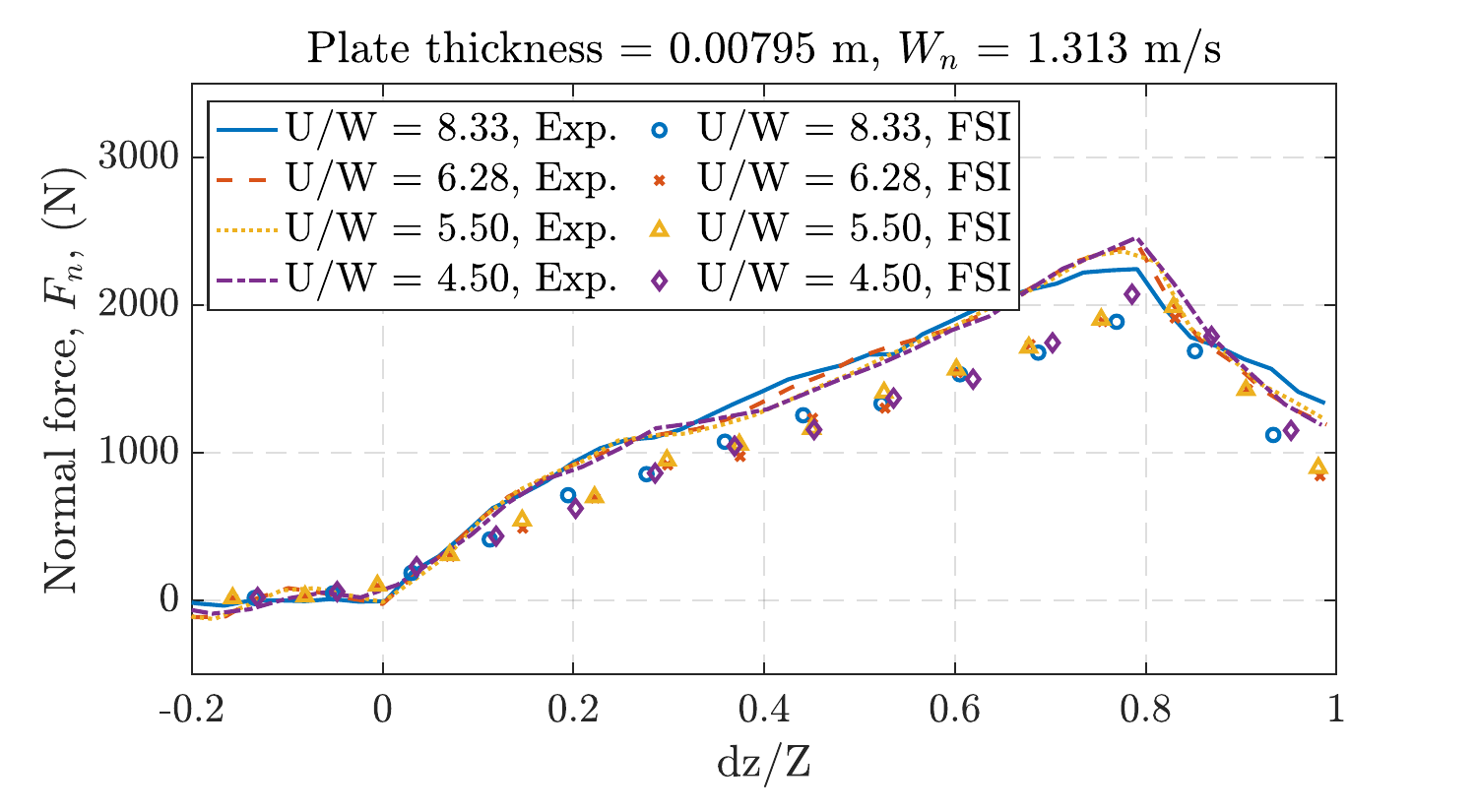}}}\\
\subfloat[]{\scalebox{0.5}{\includegraphics[width=1\textwidth]{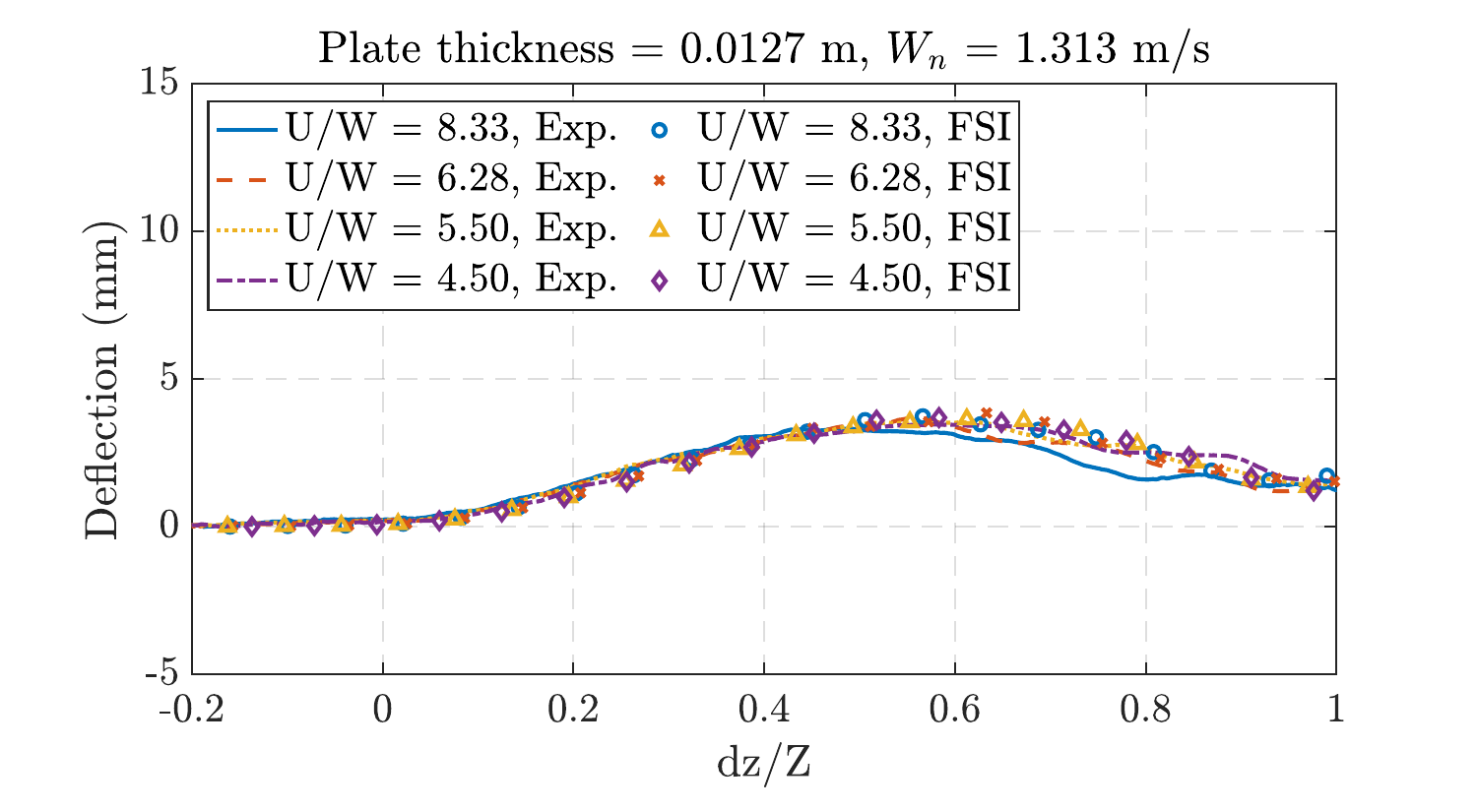}}}
\subfloat[]{\scalebox{0.5}{\includegraphics[width=1\textwidth]{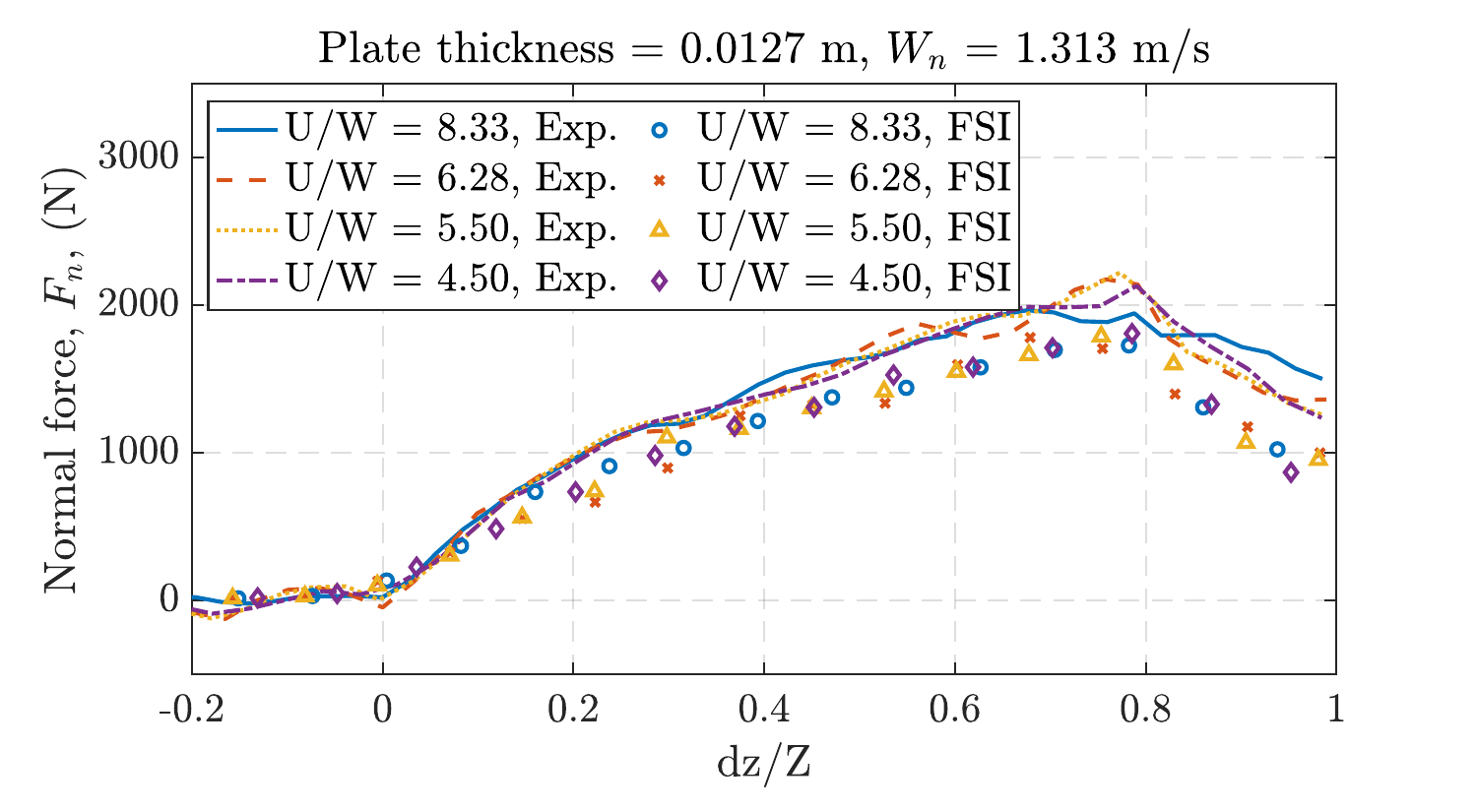}}}\\
\caption{Plate deflection and normal impact force histories of the highly flexible, moderately deformable, and nearly rigid plates subjected to slamming impacts with $W_n = 1.313$~m/s.}
\end{figure} 
\newpage
\begin{figure}[h!]
\centering
\captionsetup[subfigure]{justification=centering}
\subfloat[]{\scalebox{0.5}{\includegraphics[width=1\textwidth]{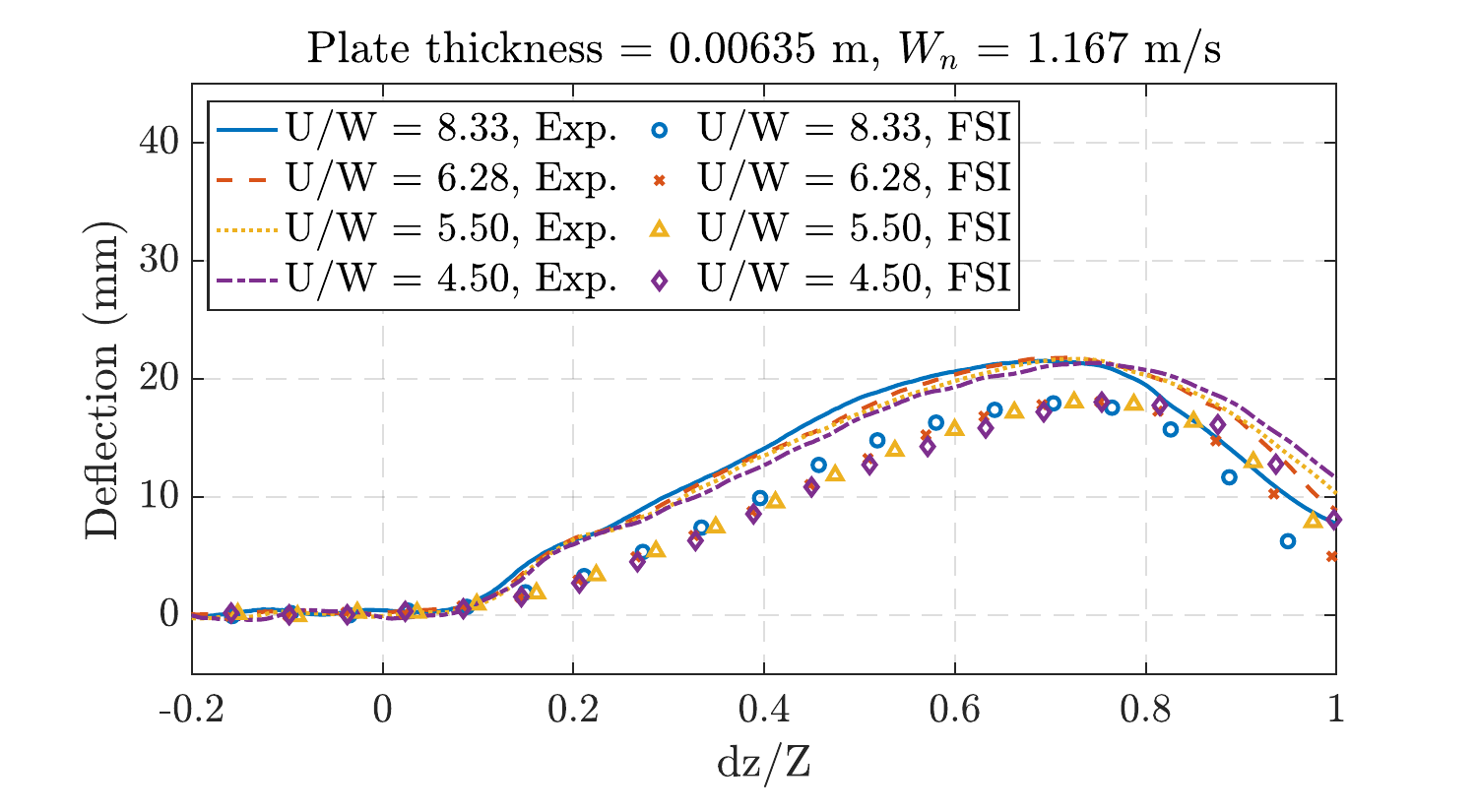}}}
\subfloat[]{\scalebox{0.5}{\includegraphics[width=1\textwidth]{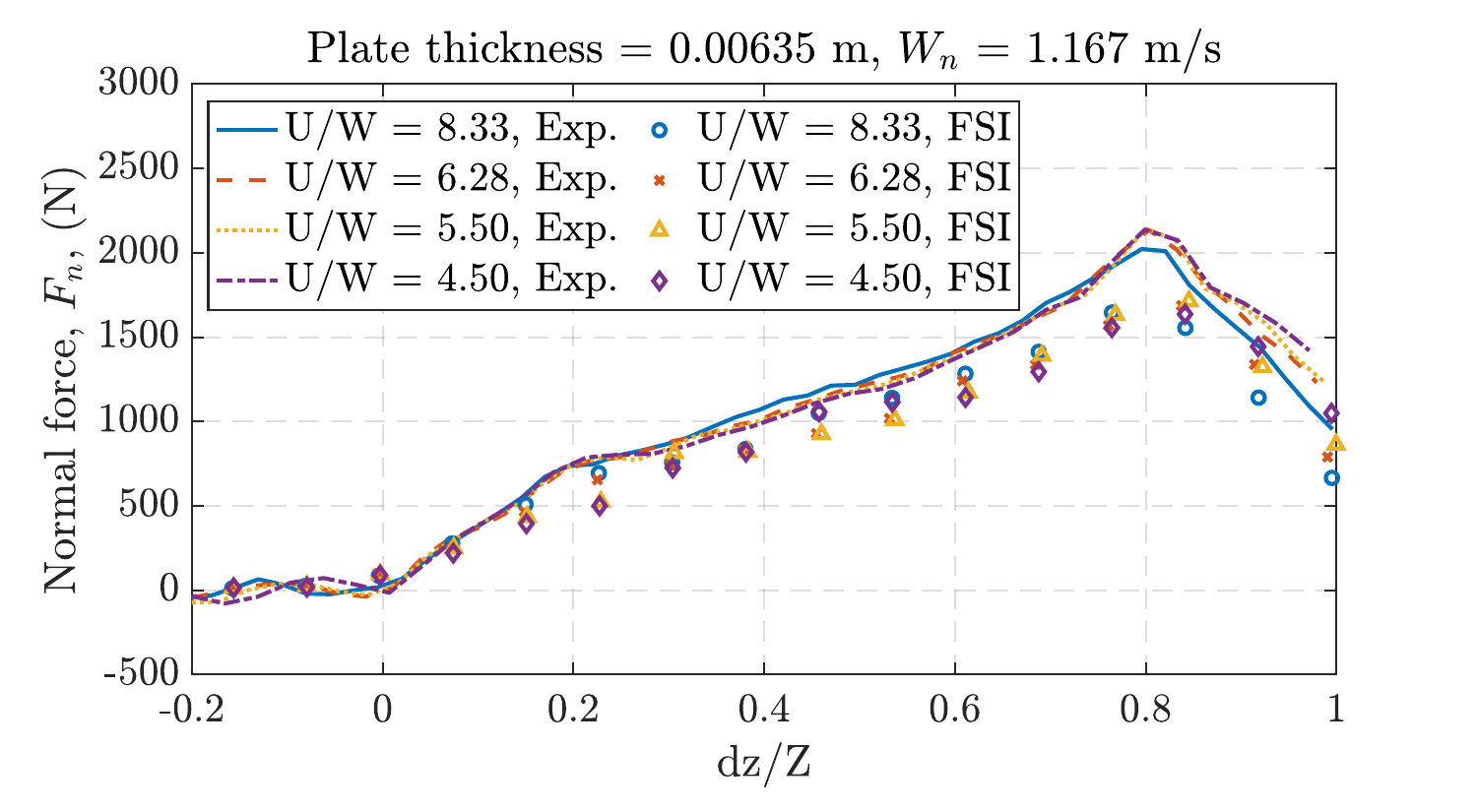}}}\\
\subfloat[]{\scalebox{0.5}{\includegraphics[width=1\textwidth]{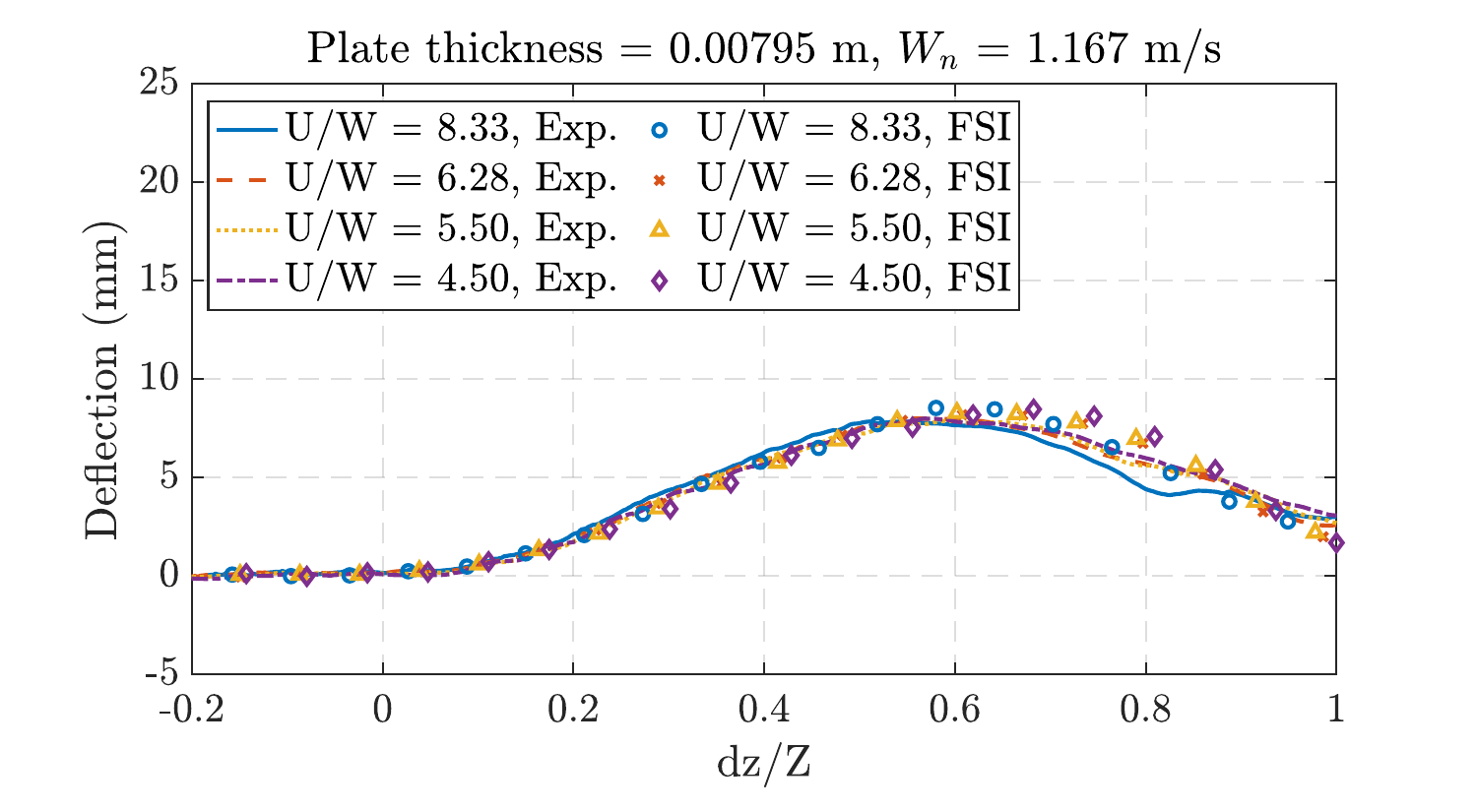}}}
\subfloat[]{\scalebox{0.5}{\includegraphics[width=1\textwidth]{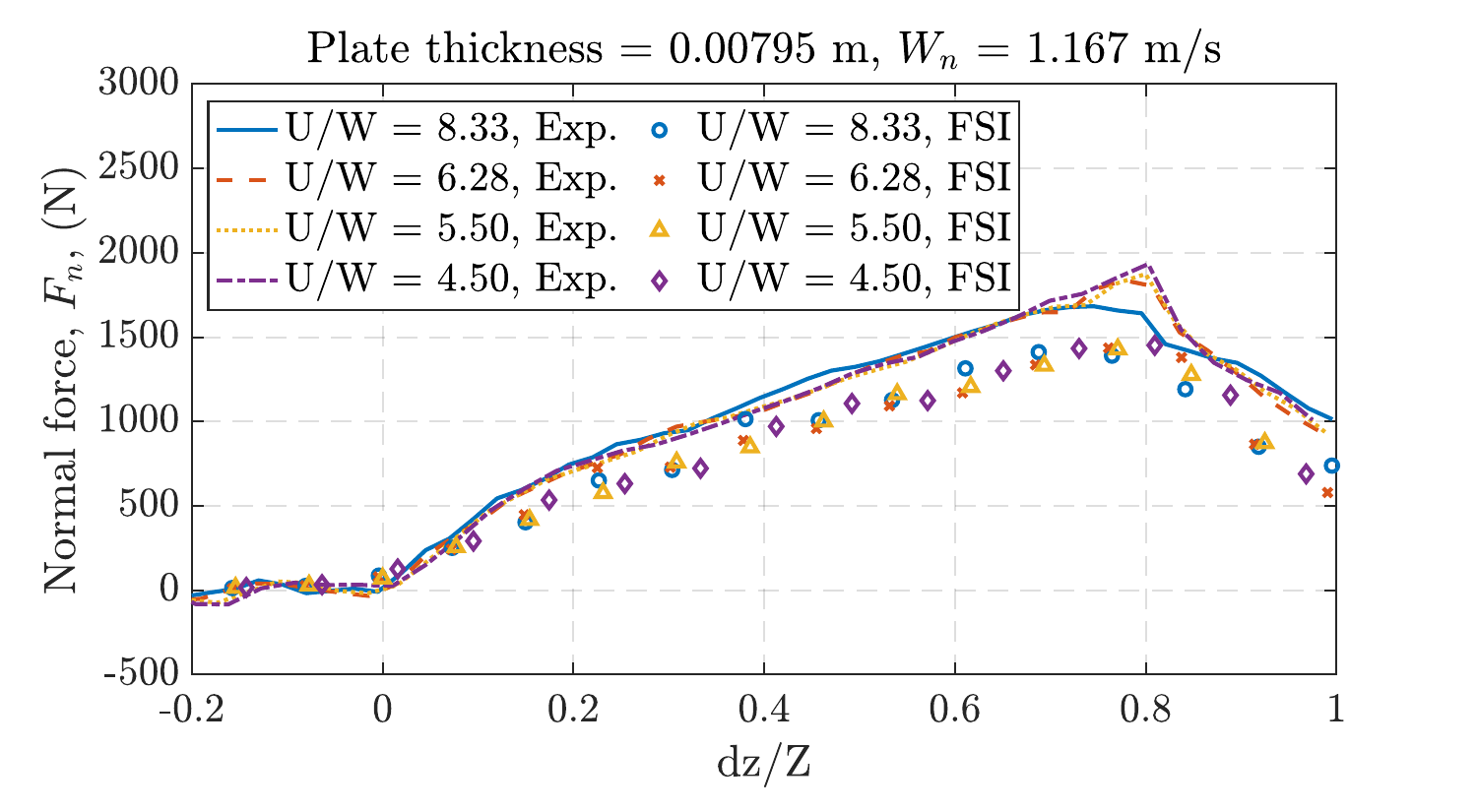}}}\\
\subfloat[]{\scalebox{0.5}{\includegraphics[width=1\textwidth]{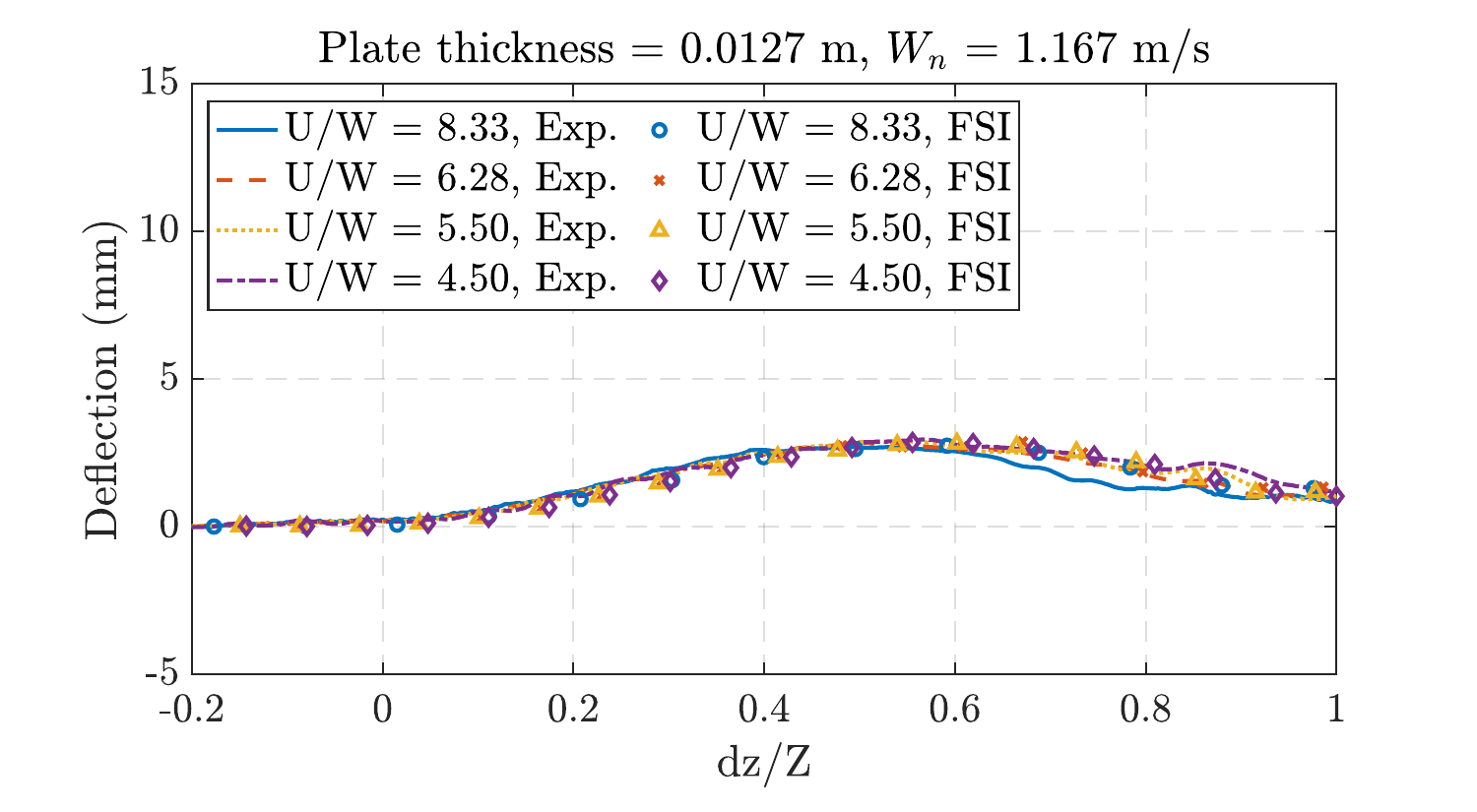}}} 
\subfloat[]{\scalebox{0.5}{\includegraphics[width=1\textwidth]{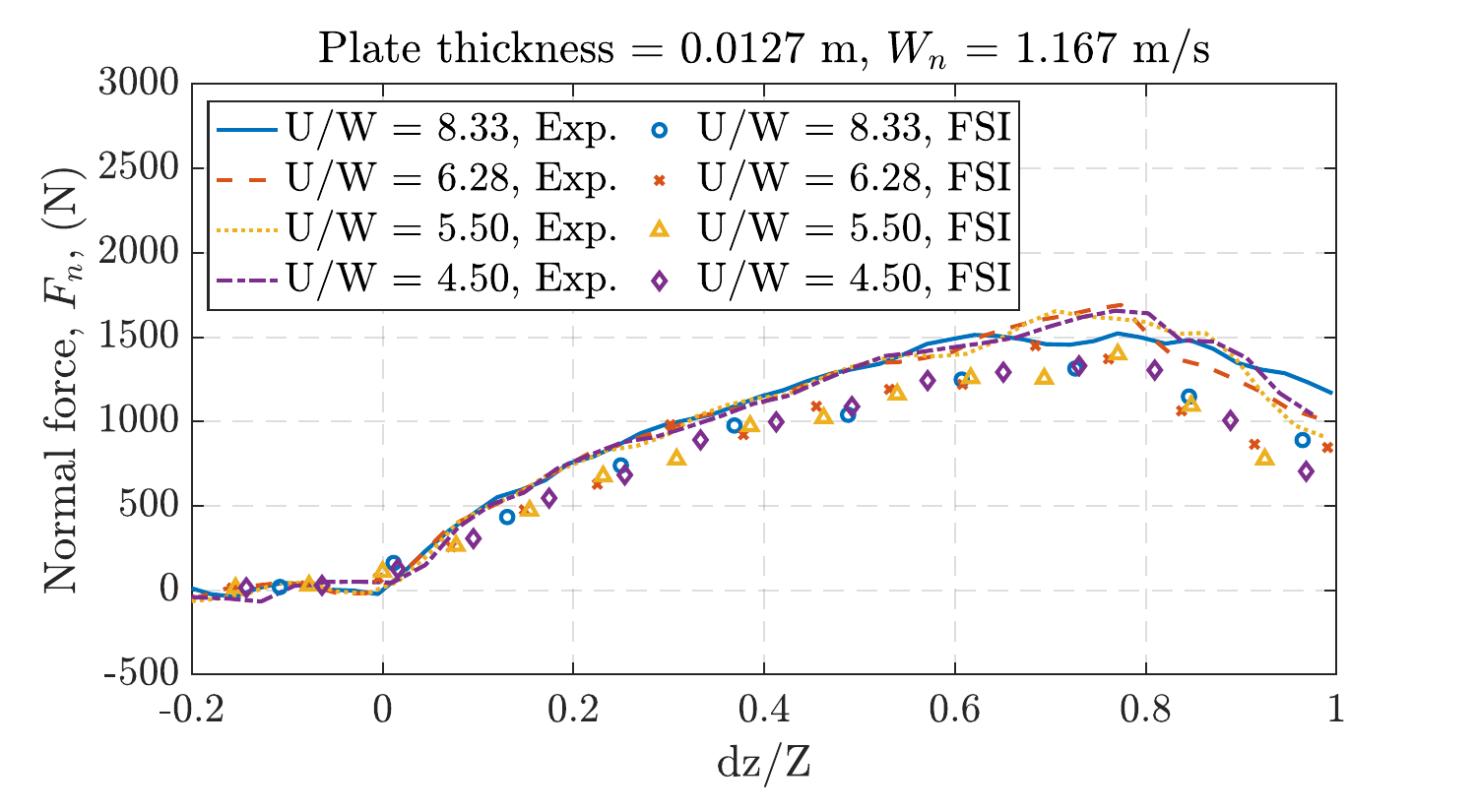}}}
\caption{Plate deflection and normal impact force histories of the highly flexible, moderately deformable, and nearly rigid plates subjected to slamming impacts with $W_n = 1.167$~m/s.}
\end{figure} 
\newpage
\begin{figure}[h!]
\centering
\captionsetup[subfigure]{justification=centering}
\subfloat[]{\scalebox{0.5}{\includegraphics[width=1\textwidth]{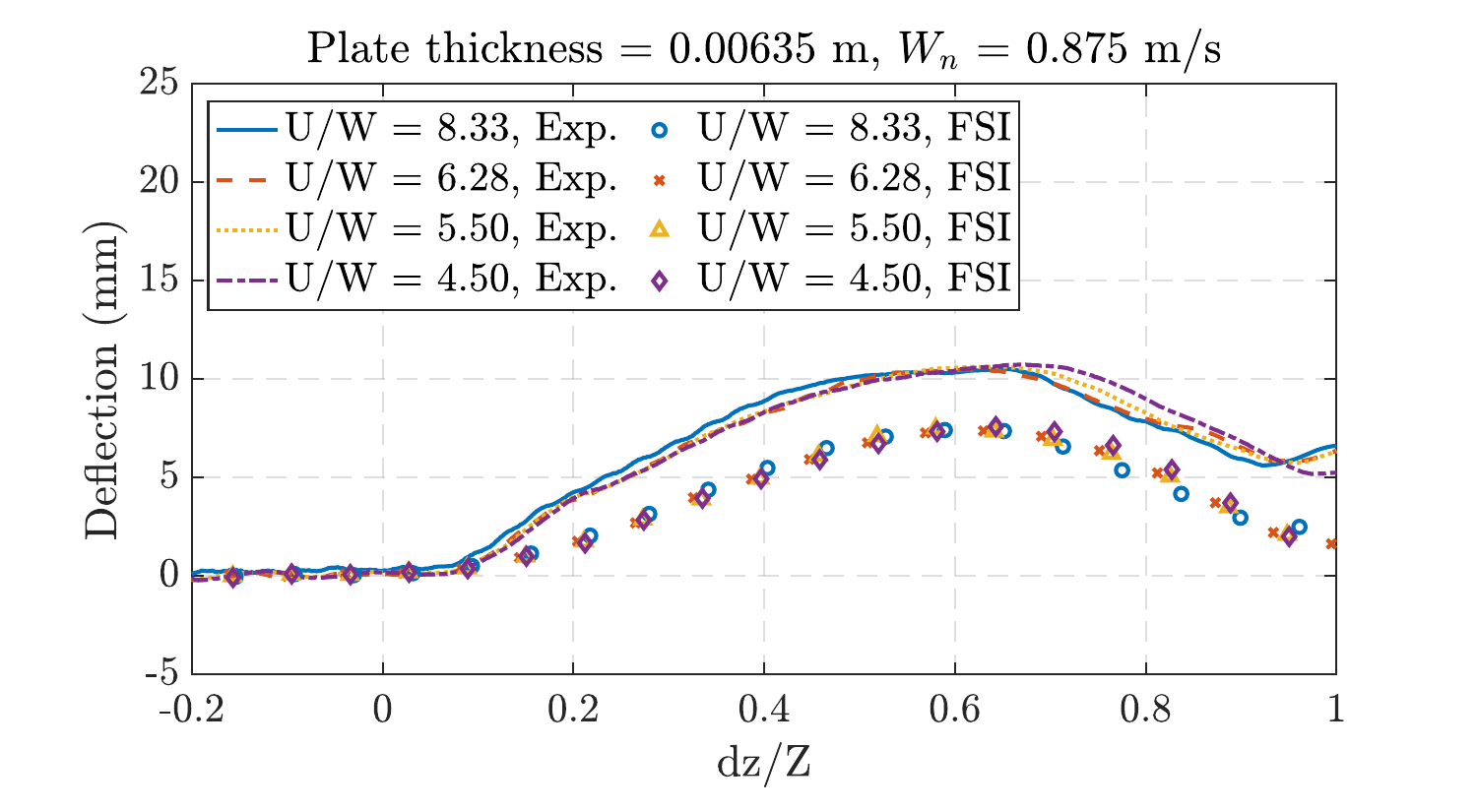}}}
\subfloat[]{\scalebox{0.5}{\includegraphics[width=1\textwidth]{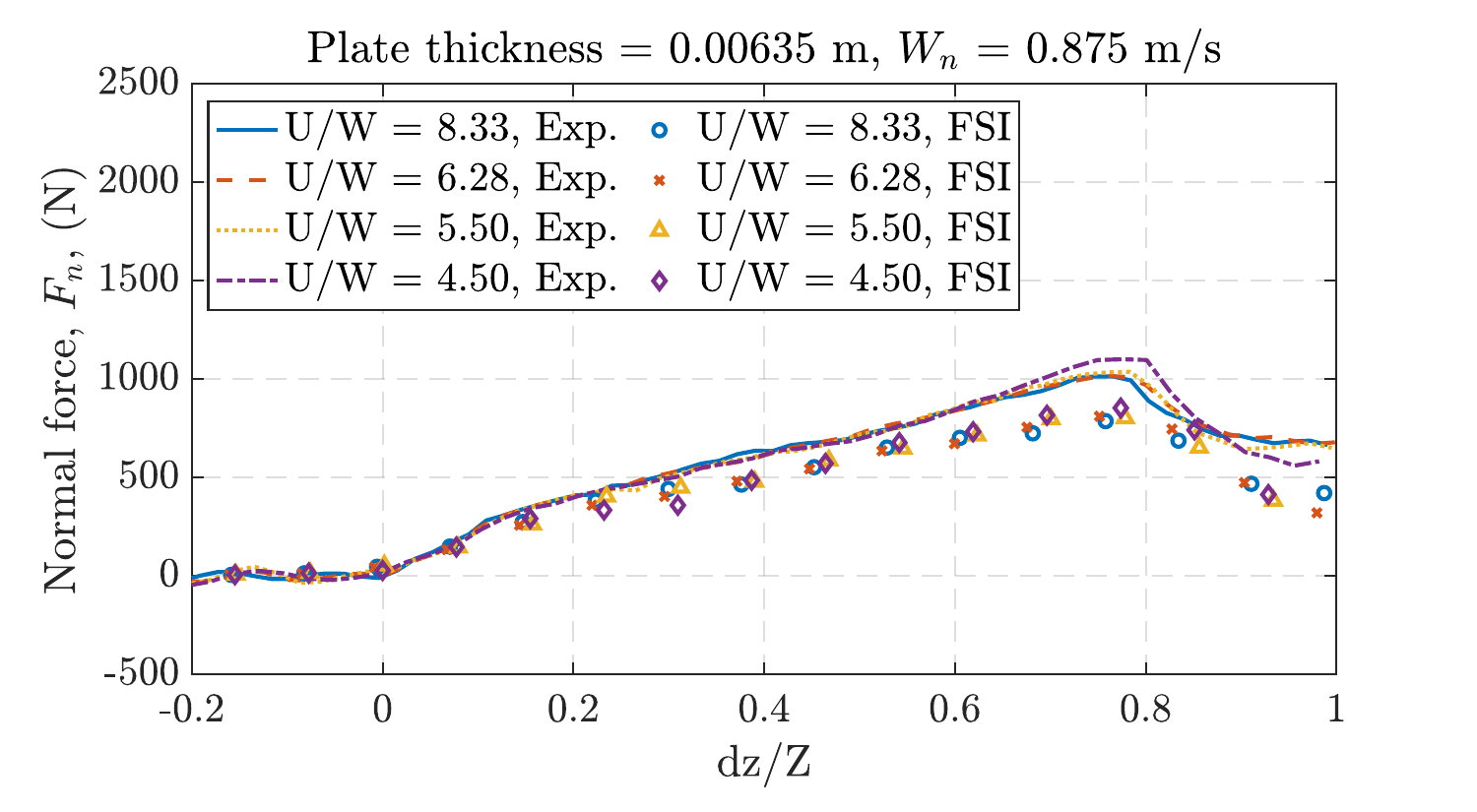}}}\\
\subfloat[]{\scalebox{0.5}{\includegraphics[width=1\textwidth]{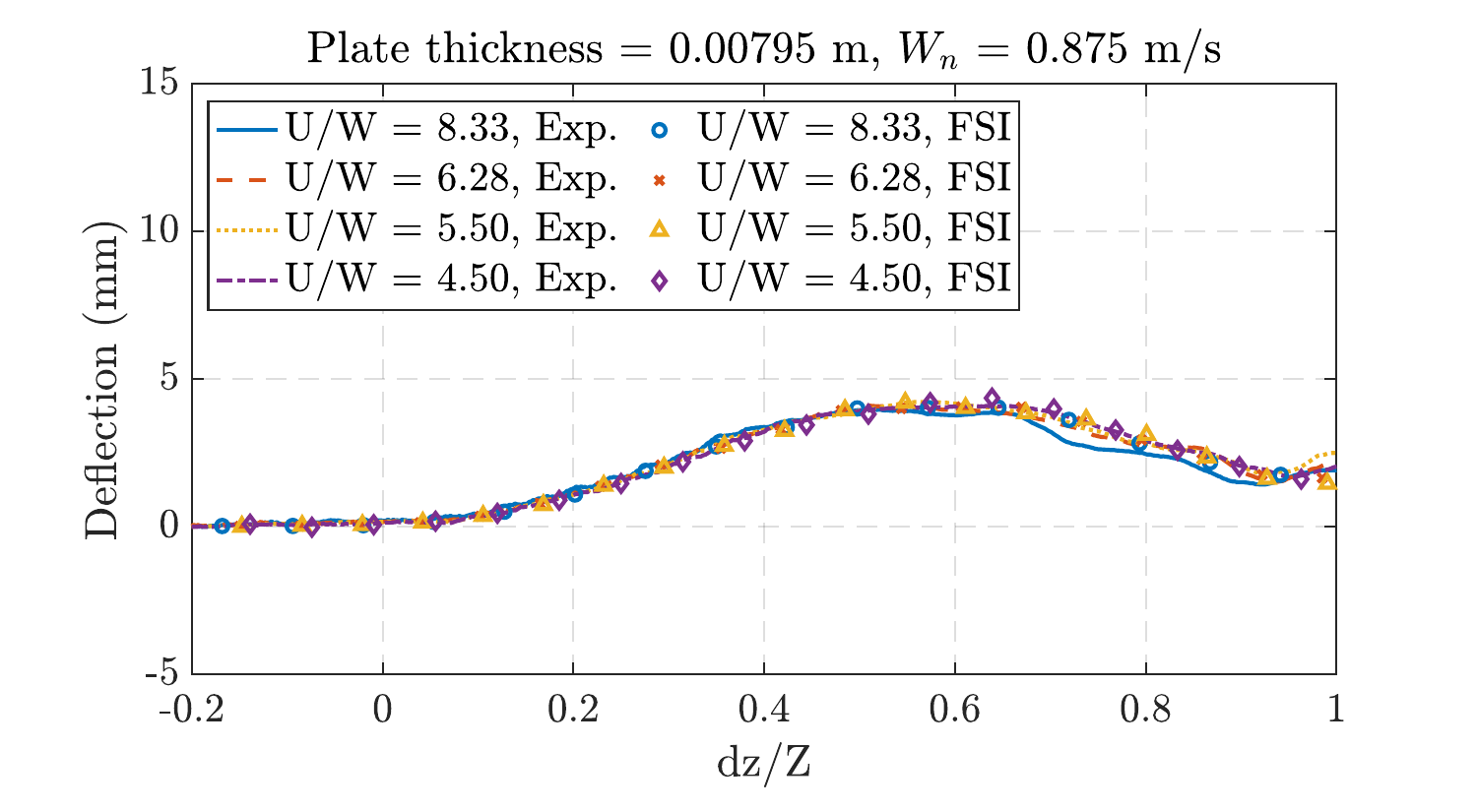}}}
\subfloat[]{\scalebox{0.5}{\includegraphics[width=1\textwidth]{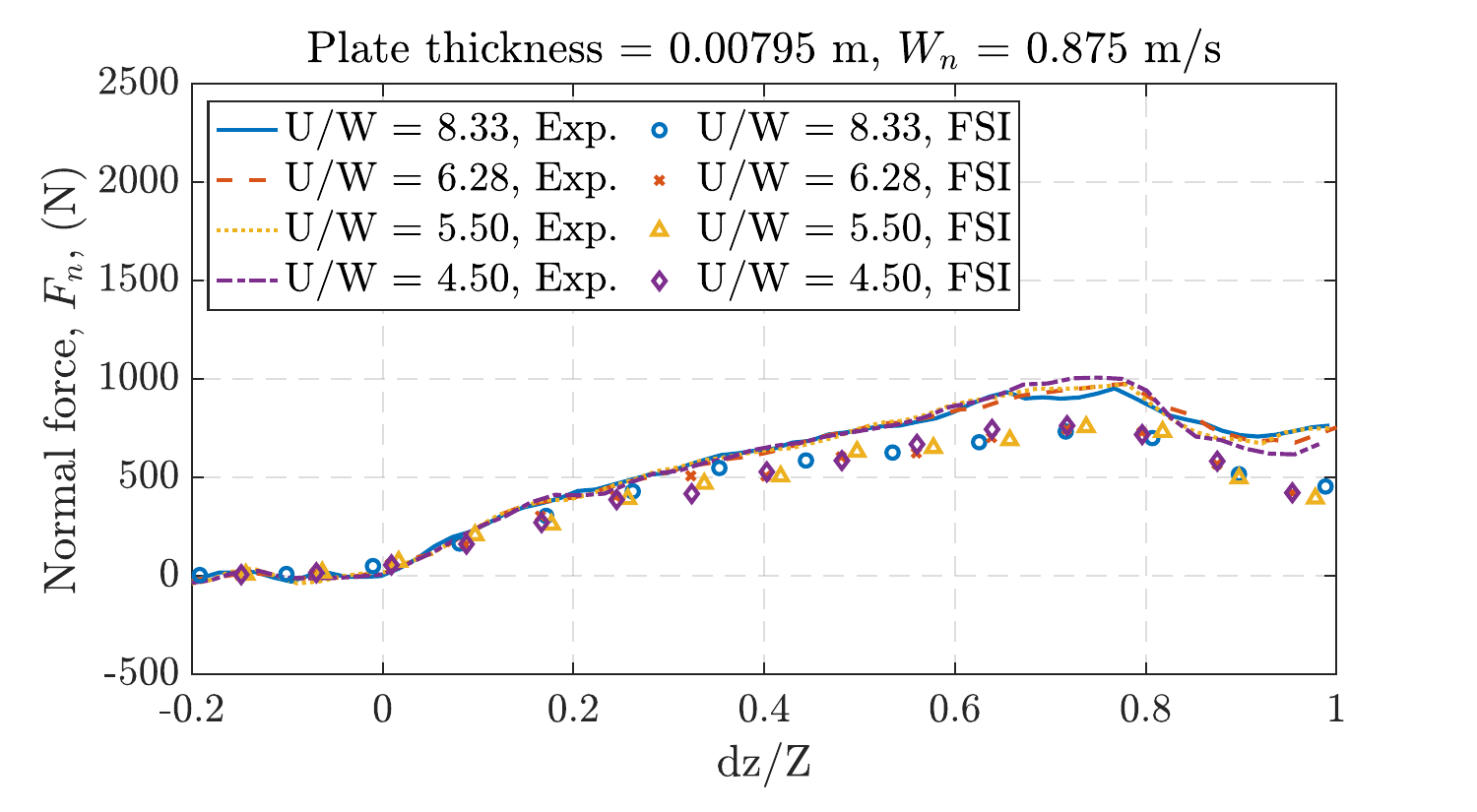}}}\\
\subfloat[]{\scalebox{0.5}{\includegraphics[width=1\textwidth]{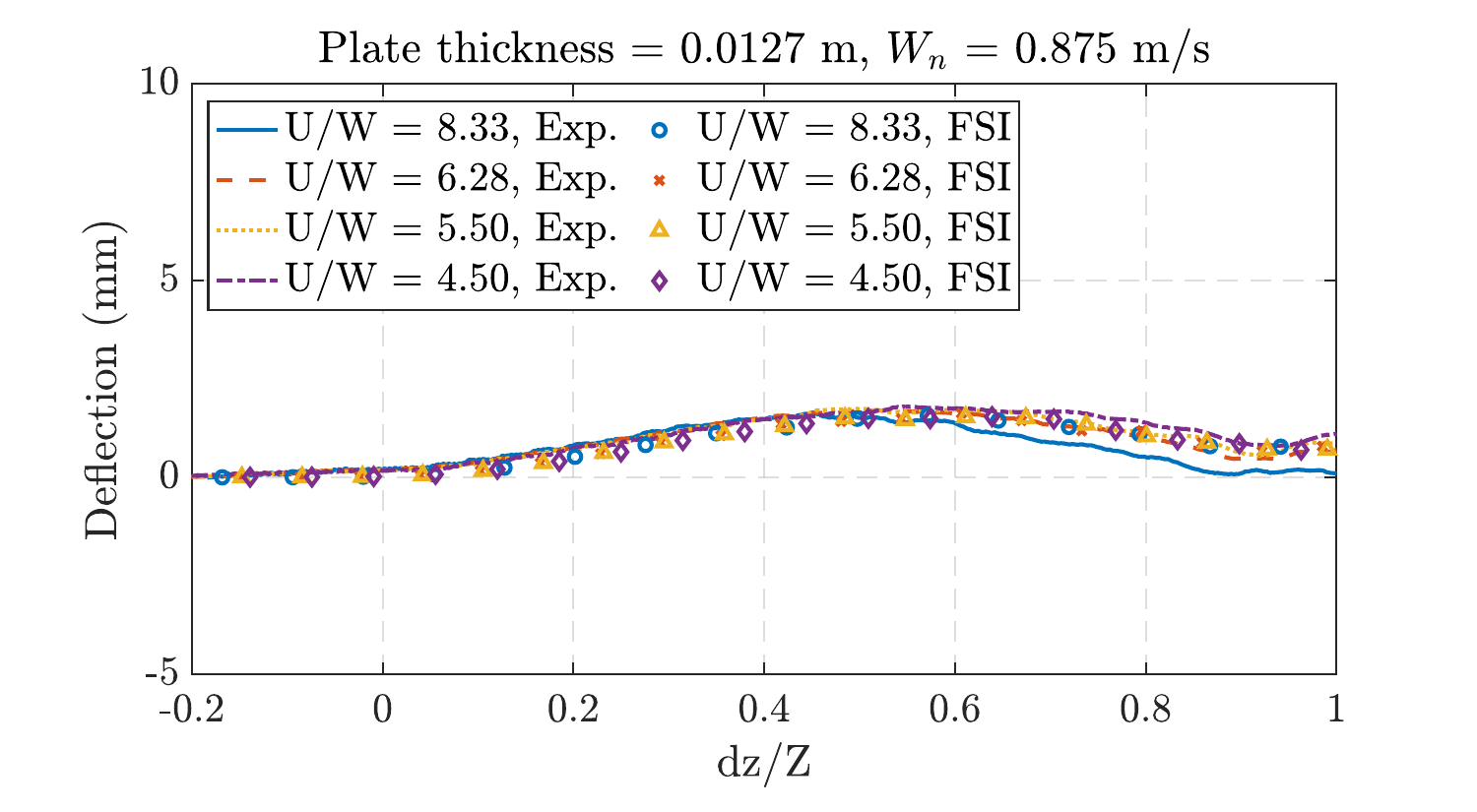}}} 
\subfloat[]{\scalebox{0.5}{\includegraphics[width=1\textwidth]{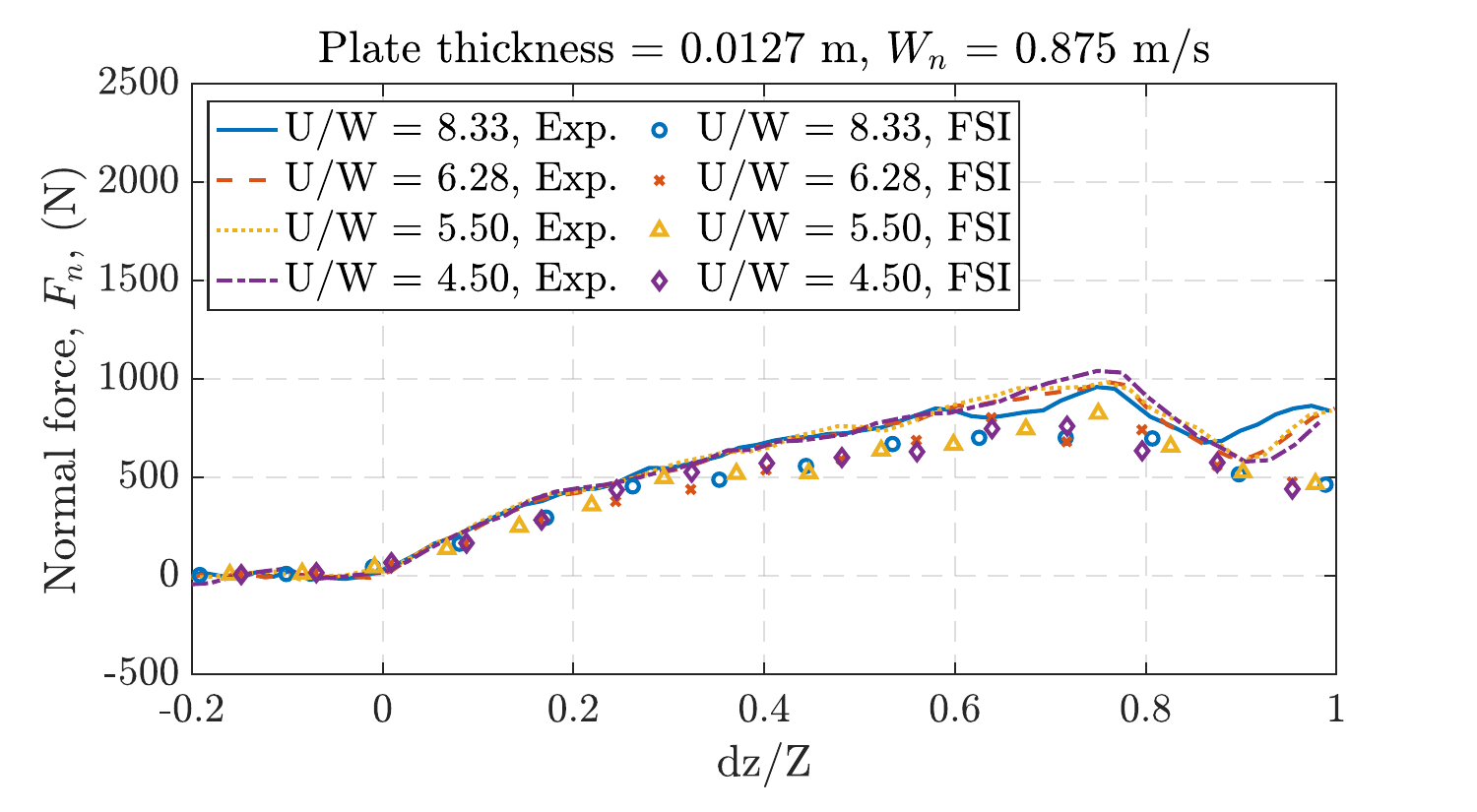}}}
\caption{Plate deflection and normal impact force histories of the highly flexible, moderately deformable, and nearly rigid plates subjected to slamming impacts with $W_n = 0.875$~m/s.}
\end{figure} 
\newpage
\begin{figure}[h!]
\centering
\captionsetup[subfigure]{justification=centering}
\subfloat[]{\scalebox{0.5}{\includegraphics[width=1\textwidth]{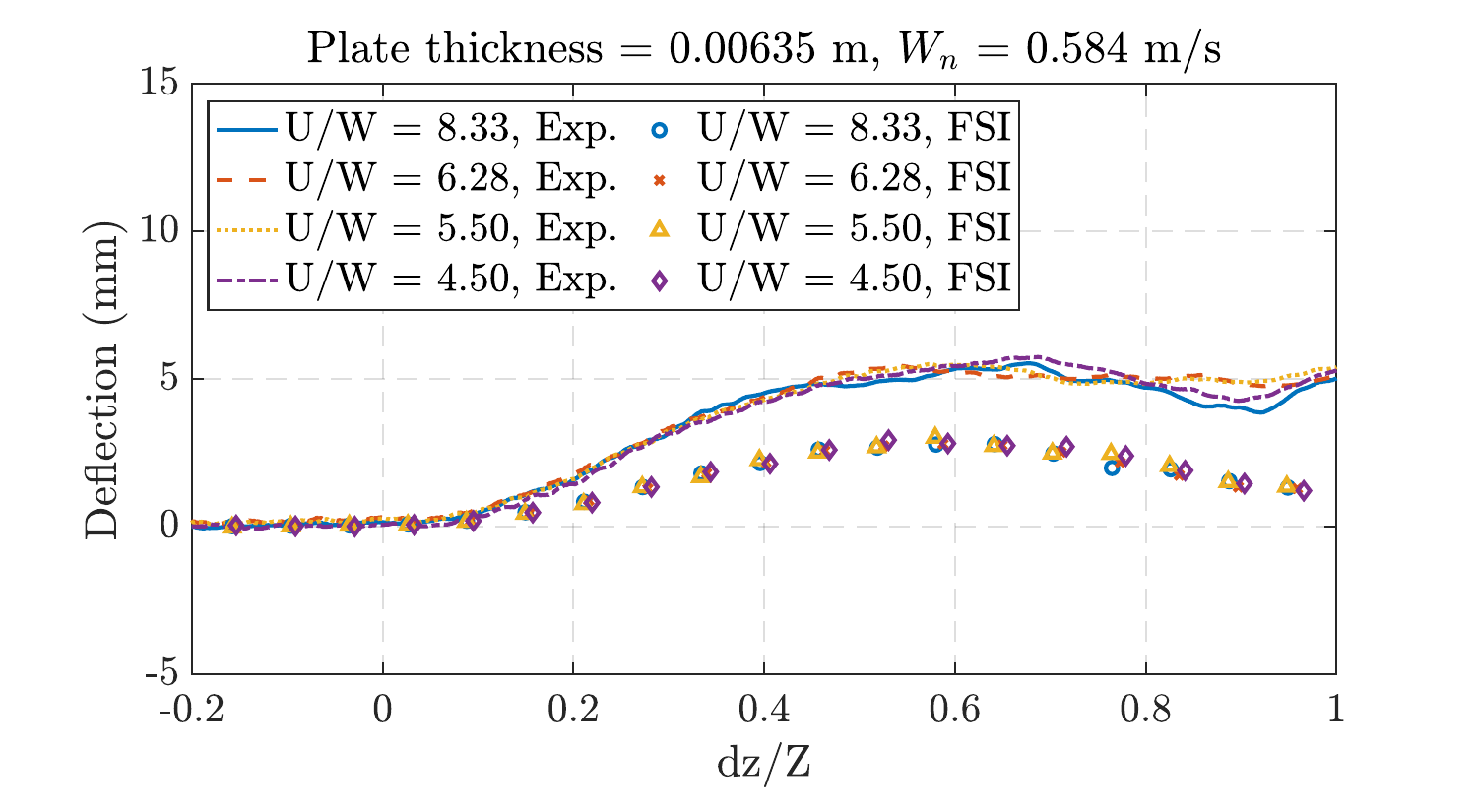}}}
\subfloat[]{\scalebox{0.5}{\includegraphics[width=1\textwidth]{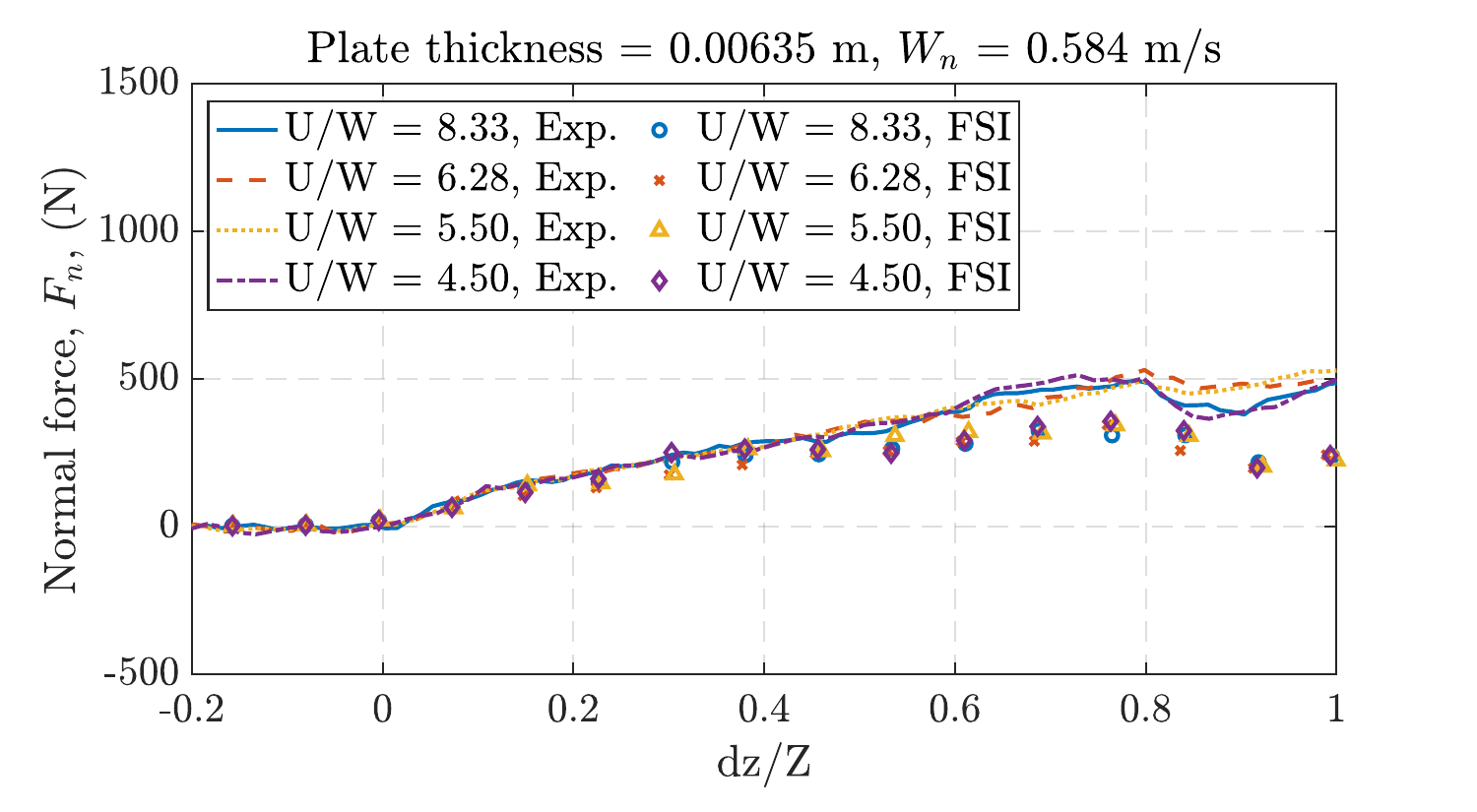}}}\\
\subfloat[]{\scalebox{0.5}{\includegraphics[width=1\textwidth]{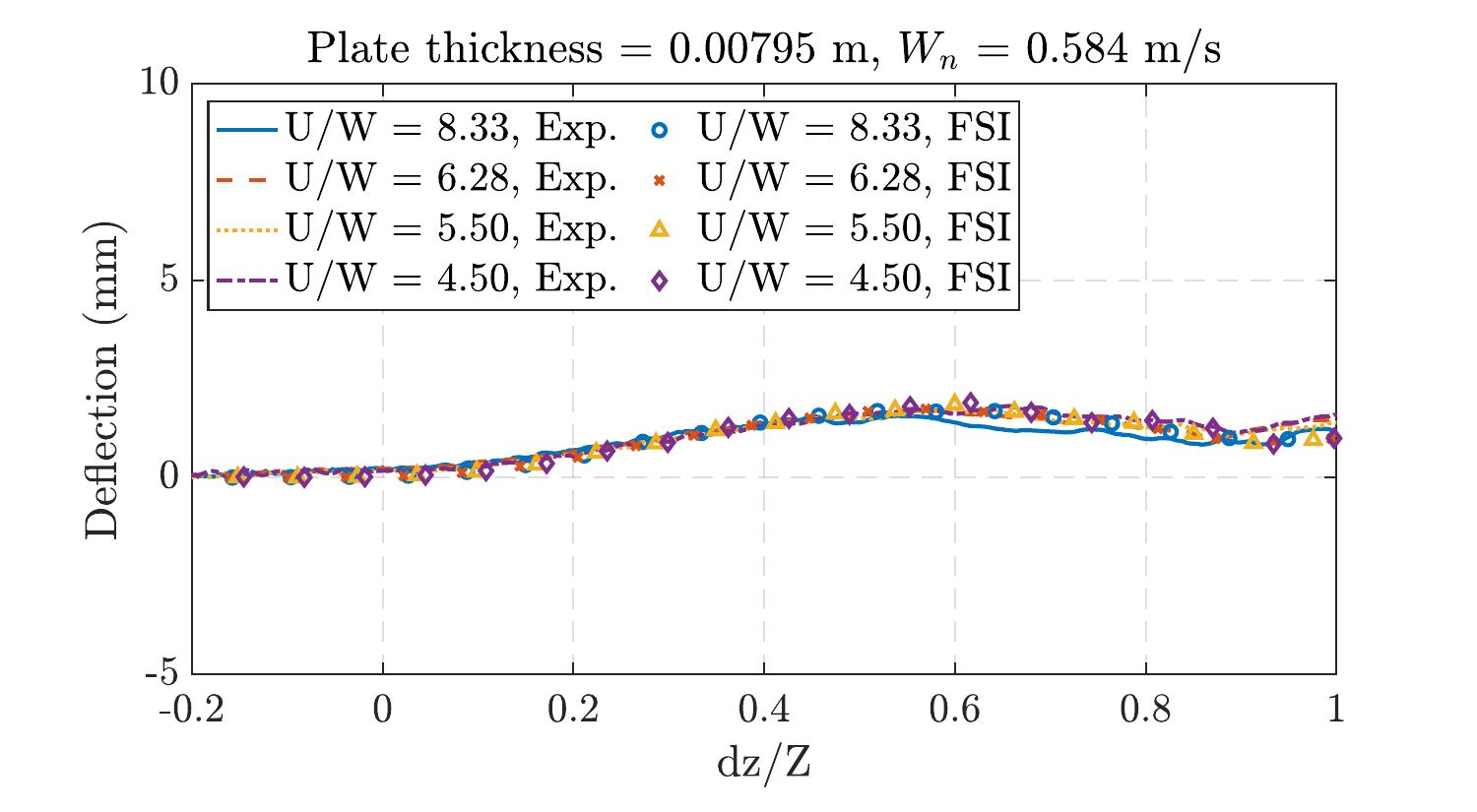}}}
\subfloat[]{\scalebox{0.5}{\includegraphics[width=1\textwidth]{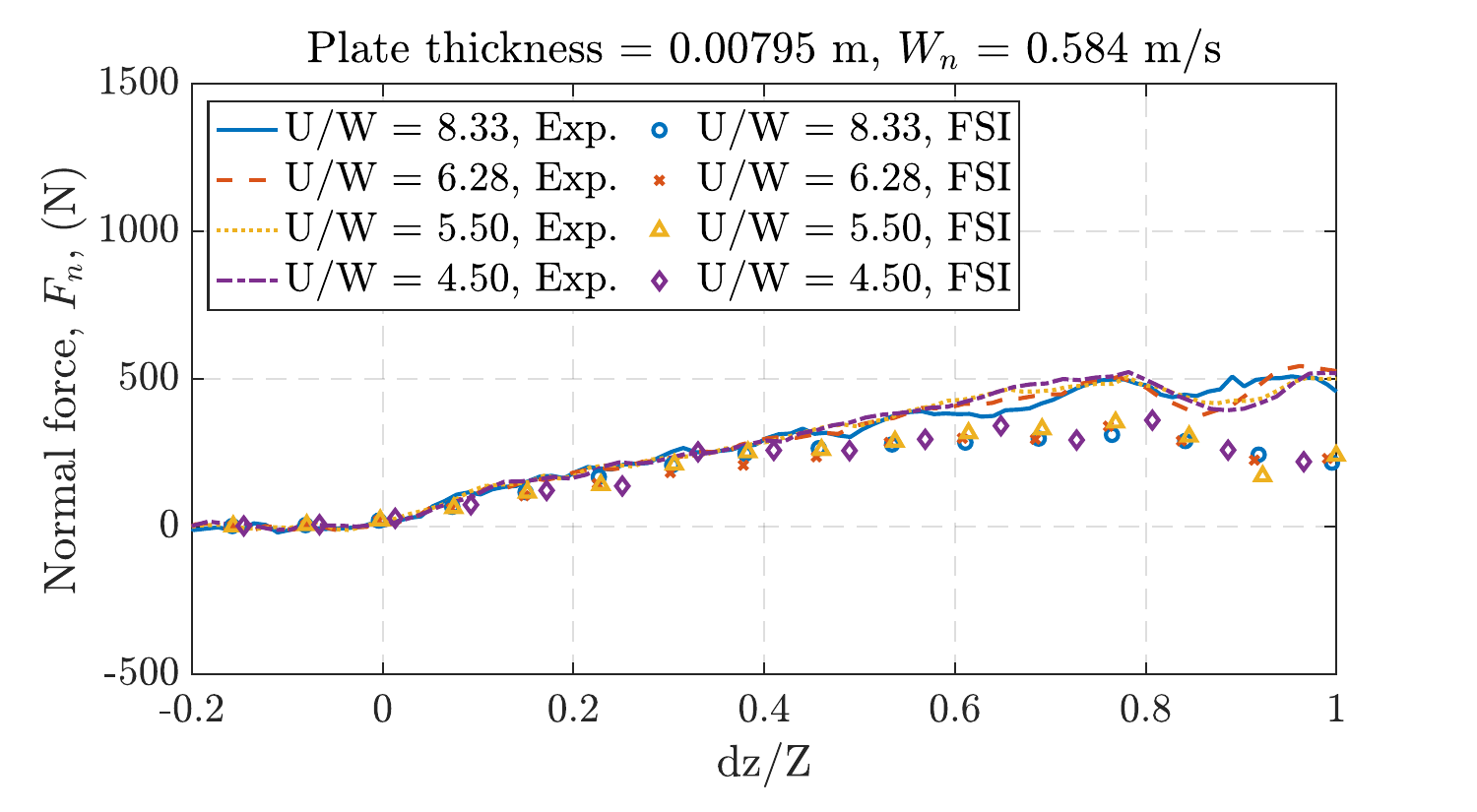}}}\\
\subfloat[]{\scalebox{0.5}{\includegraphics[width=1\textwidth]{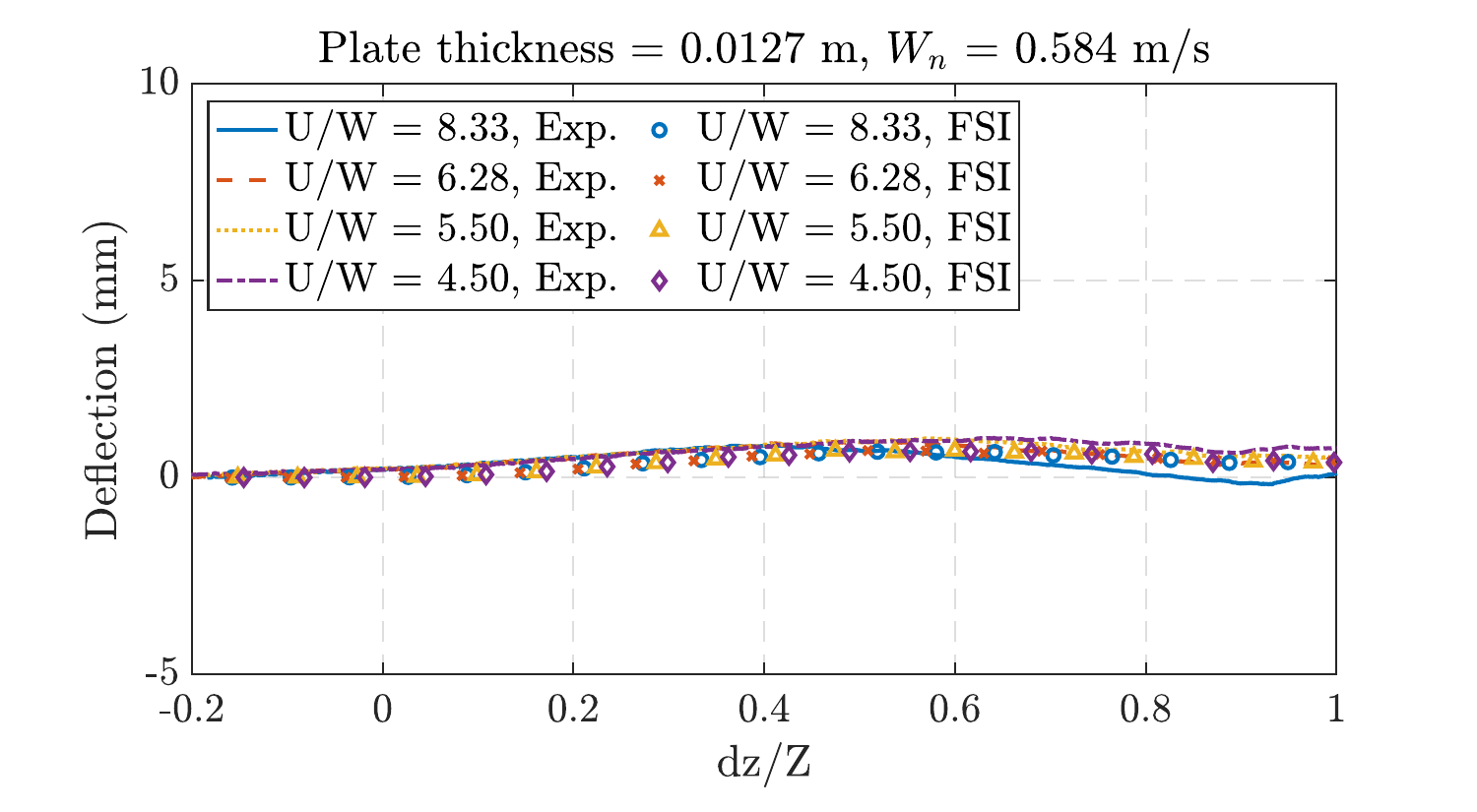}}} 
\subfloat[]{\scalebox{0.5}{\includegraphics[width=1\textwidth]{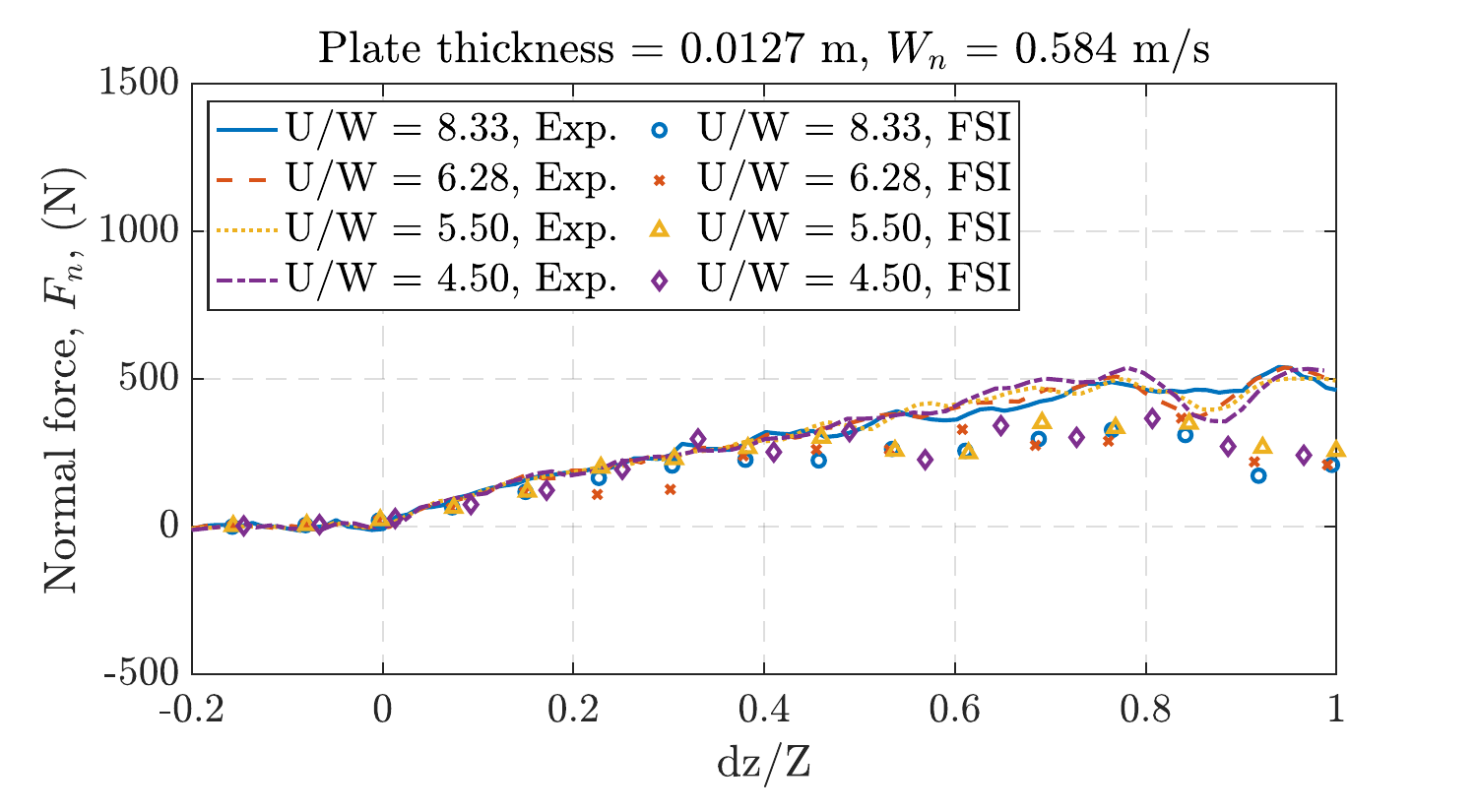}}}
\caption{Plate deflection and normal impact force histories of the highly flexible, moderately deformable, and nearly rigid plates subjected to slamming impacts with $W_n = 0.584$~m/s.}
\end{figure} 
\newpage
 \begin{figure}[h!]
\centering
\captionsetup[subfigure]{justification=centering}
\subfloat[]{\scalebox{0.5}{\includegraphics[width=1\textwidth]{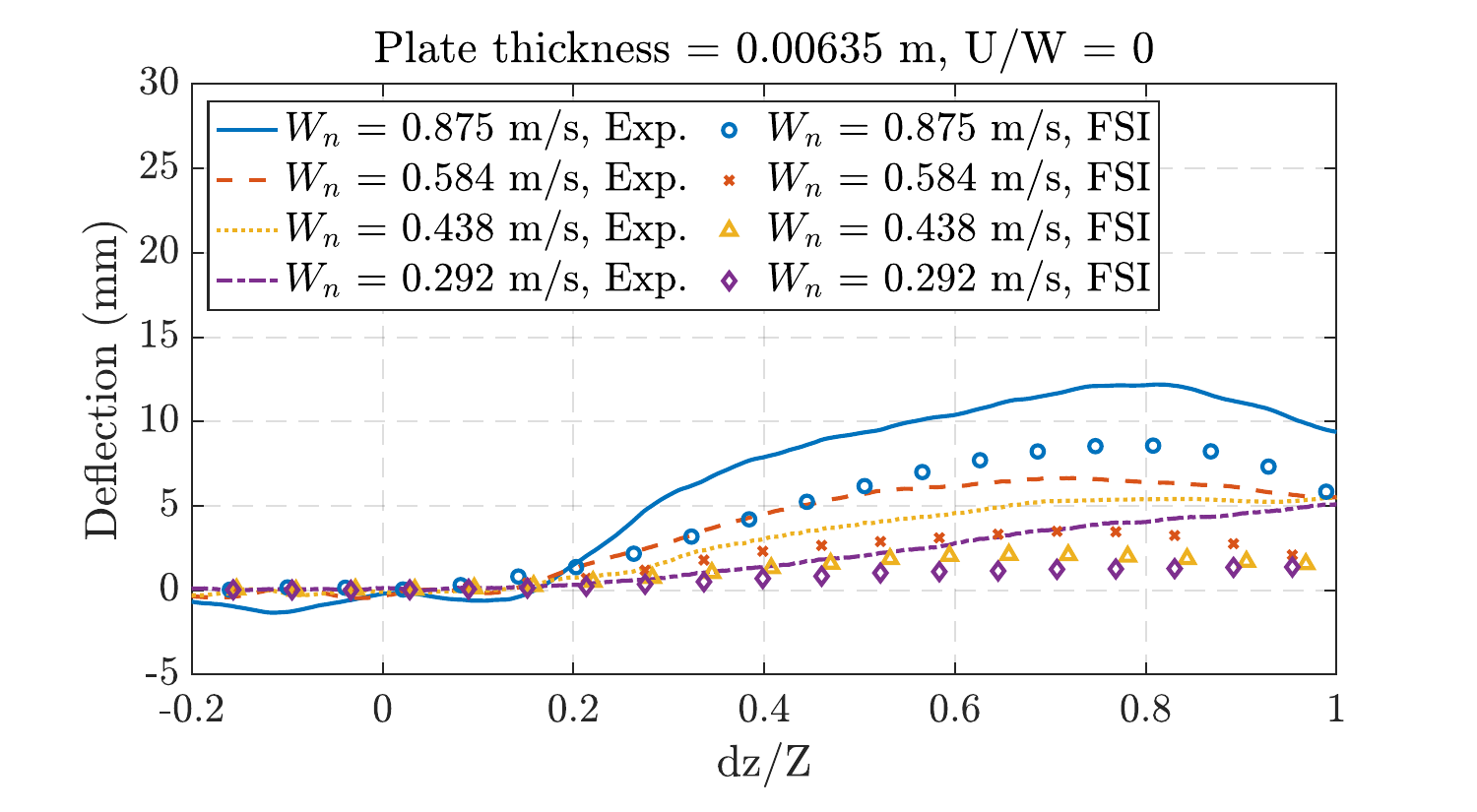}}}
\subfloat[]{\scalebox{0.5}{\includegraphics[width=1\textwidth]{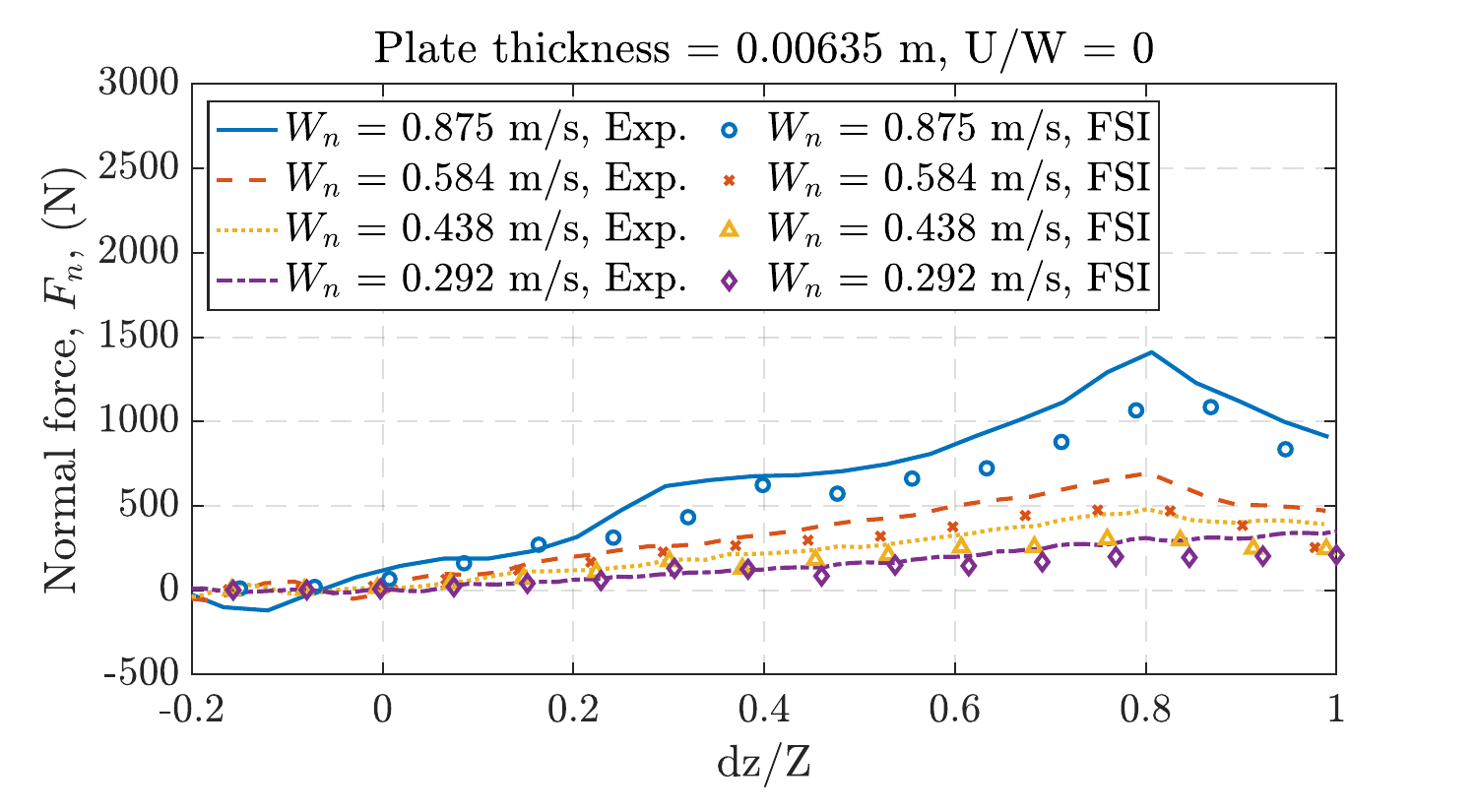}}}\\
\subfloat[]{\scalebox{0.5}{\includegraphics[width=1\textwidth]{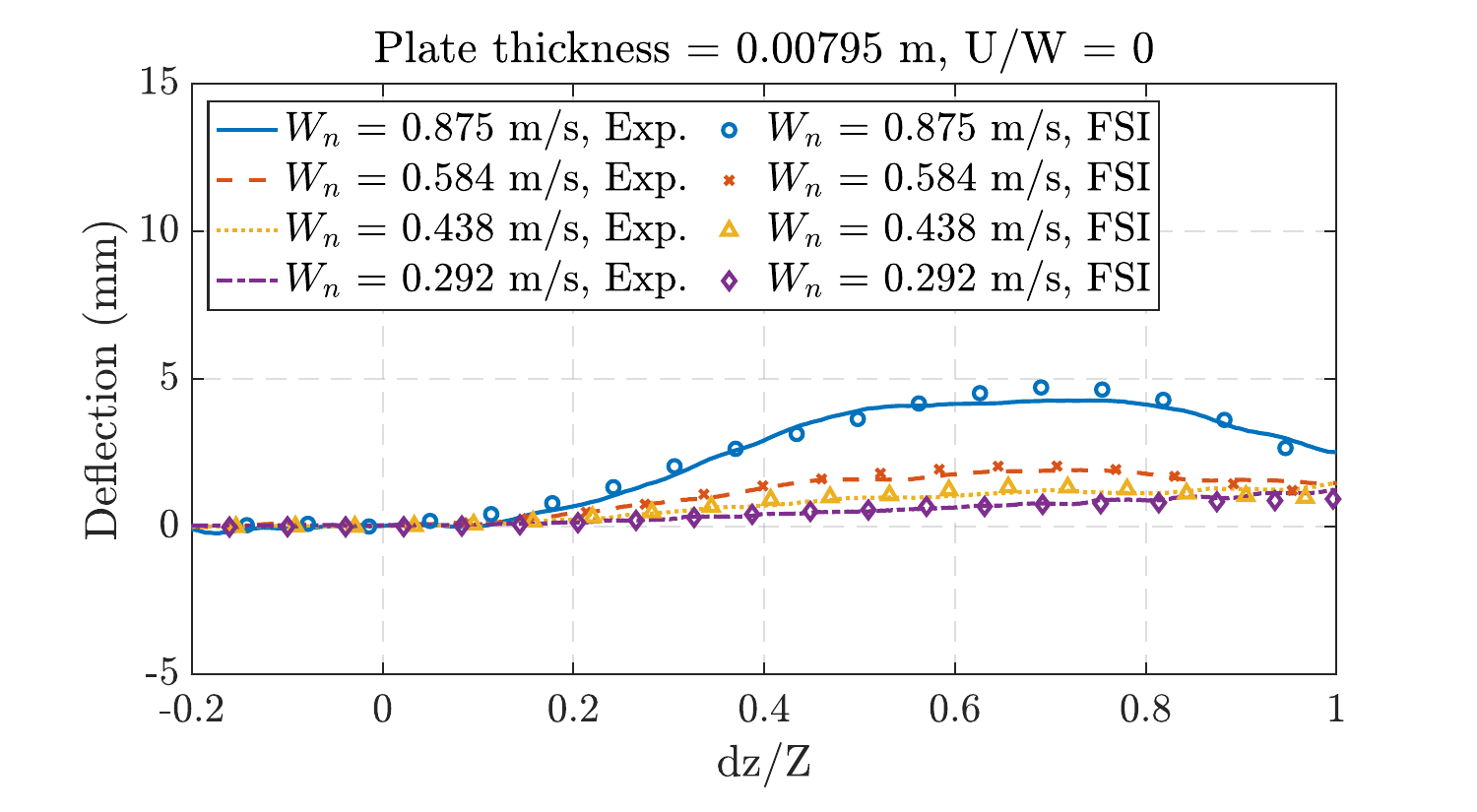}}}
\subfloat[]{\scalebox{0.5}{\includegraphics[width=1\textwidth]{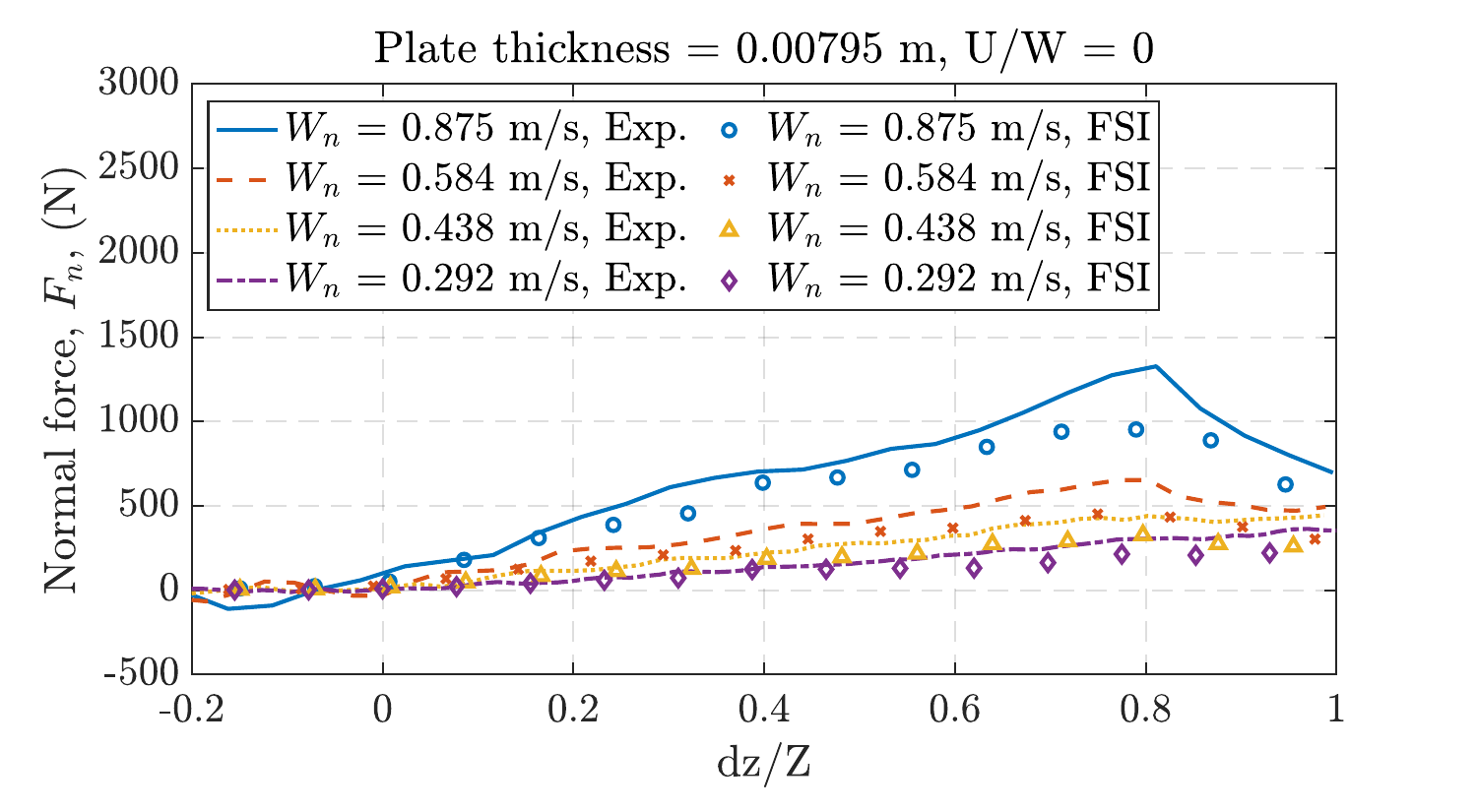}}}\\
\subfloat[]{\scalebox{0.5}{\includegraphics[width=1\textwidth]{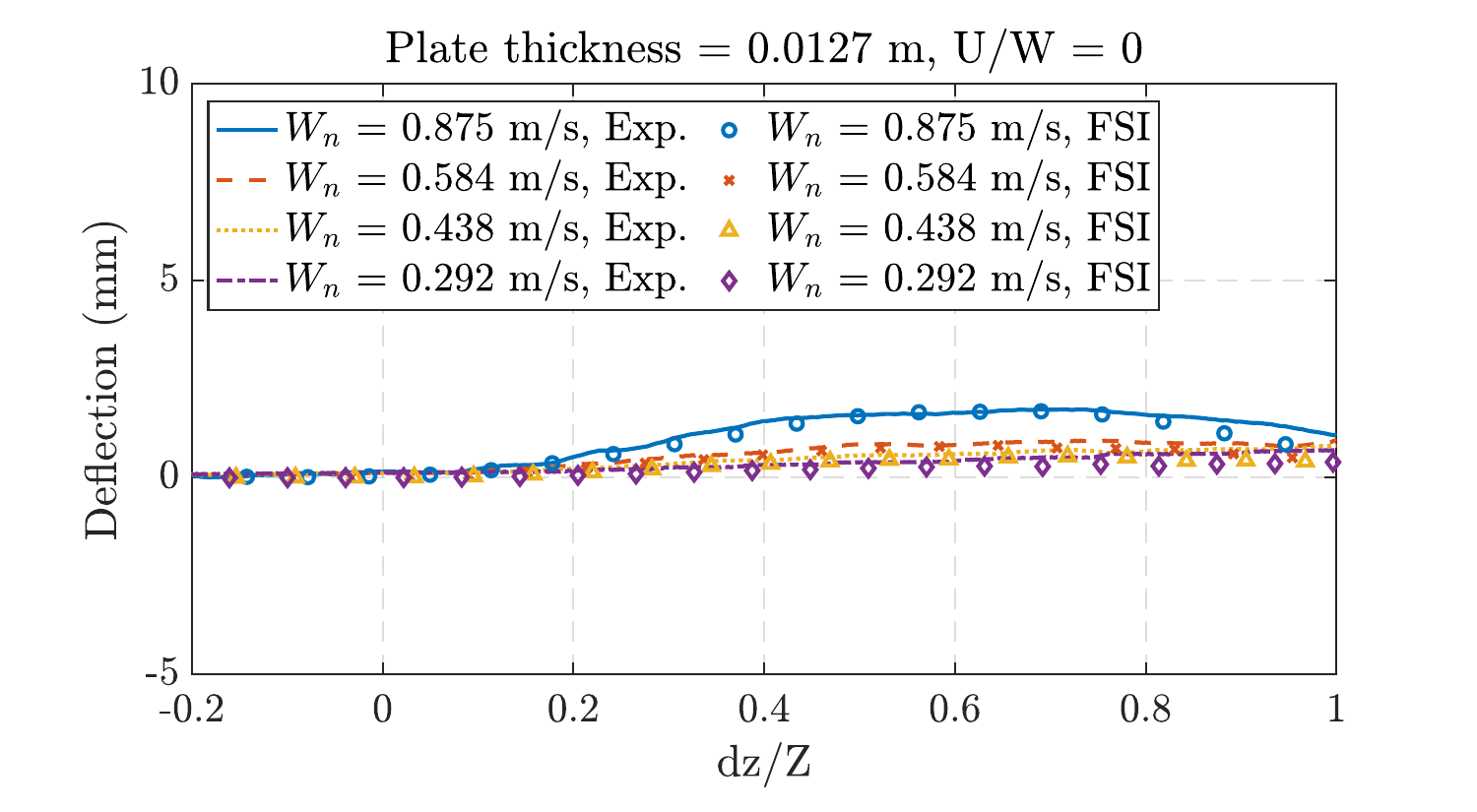}}}
\subfloat[]{\scalebox{0.5}{\includegraphics[width=1\textwidth]{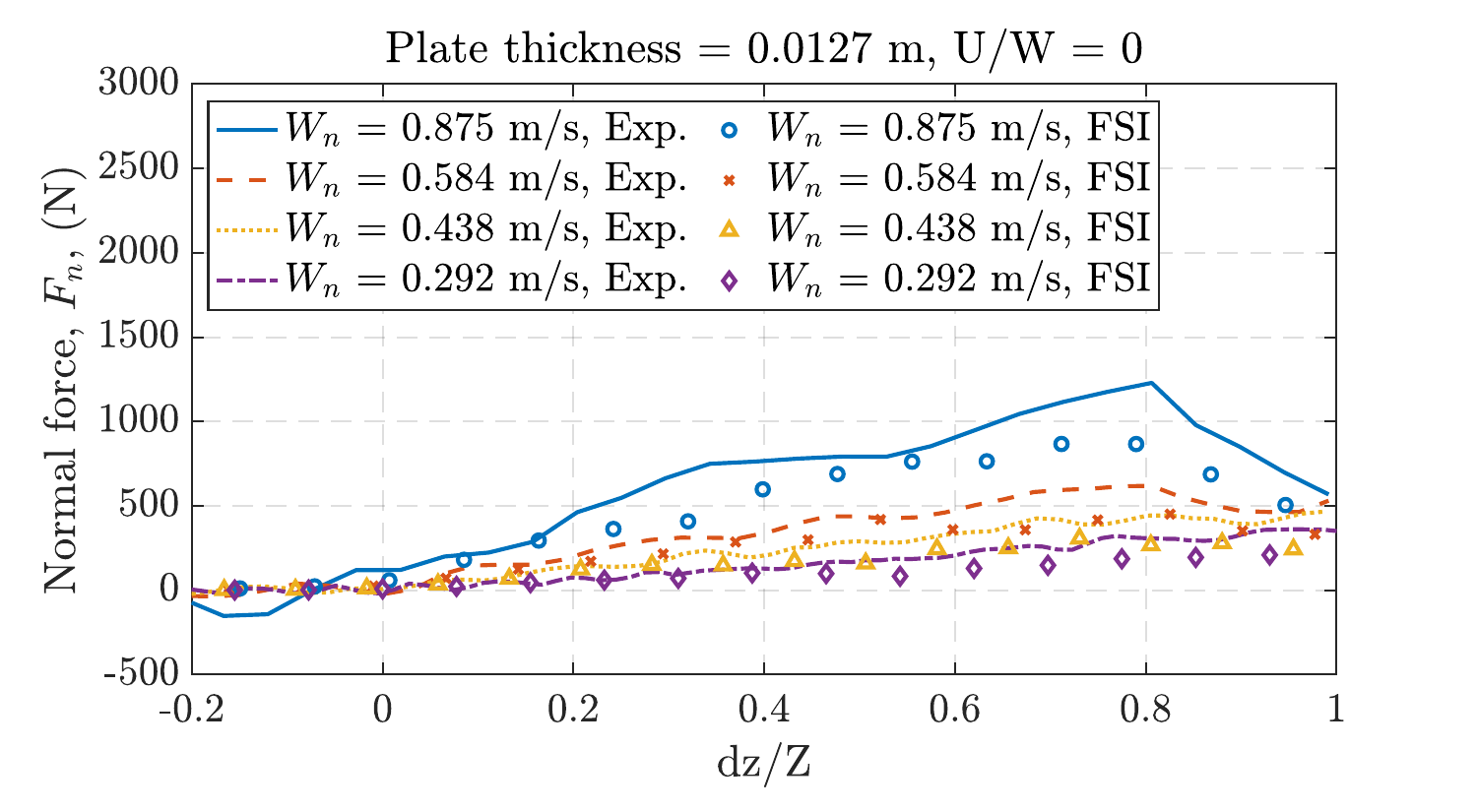}}} 
\caption{Plate deflection and normal impact force histories of the highly flexible, moderately deformable, and nearly rigid plates subjected to slamming impacts with $U/W = 0$.}
\end{figure} 

\newpage
\section{Nonlinear Koopman Quadratic Surface Fitting Method}\label{appendix:b}
Here, we provide detail exposition on the ellipsoidal procedure. Our departure point is the equation of a second degree quadric surface within three dimensional Euclidean space

\begin{equation} \label{surface_fitting}
a_1x^2+a_2y^2+a_3z^2+2a_4yz+2a_5xz+2a_6xy+2a_7x+2a_8y+2a_9z+a_{10} = 0,
\end{equation}
where $x$, $y$, and $z$ are the point cloud data expressed in Cartesian coordinates, and $a_1, a_2, ... a_{10}$ are the least square fitted coefficient parameters. 

Let the invariants of Equation \ref{surface_fitting} under rotation and translation be 

\begin{equation} \label{I}
I = a_1+a_2+a_3,
\end{equation}
\begin{equation} \label{J}
J = a_1a_2+a_2a_3+a_1a_3-a_4^2-a_5^2-a_6^2,
\end{equation}
and 

\begin{equation} \label{K}
K = 
\begin{bmatrix}
a_1 & a_6 & a_5 \\
a_6 & a_2 & a_4 \\
a_5 & a_4 & a_3
\end{bmatrix}.
\end{equation}
The fitting shape of the Equation \ref{surface_fitting} represents an ellipsoid when the constraint $kJ-I^2>0$, where $k$ is a positive real number, is satisfied. Moreover, when $k = 4$, Equation \ref{surface_fitting} is guaranteed to be an ellipsoid \cite{li2004}. The procedure of ellipsoid surface fitting is as follows. Let $p_i(x_i, y_i, z_i)_{i=1}^n$ be the set of points in the added mass point cloud. For each point $p_i(x_i, y_i, z_i)$, let Equation \ref{surface_fitting} be reformulated in the following matrix form

\begin{equation} \label{surface_fitting2}
\mathbf{X}_i^Ta = 0,  
\end{equation}
where

\begin{equation} \label{surface_fitting3}
\mathbf{X}_i = 
\begin{bmatrix}
x_i^2 & y_i^2 & z_i^2 & 2y_iz_i & 2x_iz_i & 2x_iy_i & 2x_i & 2y_i & 2z_i & 1
\end{bmatrix}^T,
\end{equation}
and 

\begin{equation} \label{surface_fitting4}
\mathbf{a}= 
\begin{bmatrix}
v_1 & v_2 &v_3 & v_4& v_5 & v_6 & v_7& v_8& v_9 & v_{10}
\end{bmatrix}^T.
\end{equation}
To ensure Equation \ref{surface_fitting2} corresponds to an ellipsoid, the following condition is imposed

\begin{linenomath*}
\begin{equation}\label{min_condition}
\,\,\,\,\,\min \norm{D \mathbf{a}}^2 \,\,\,\,\,\text{s.t.}\,\,\,\,\,
kJ-I^2 = 1. 
\end{equation}
\end{linenomath*}
$D$ is a design matrix in $\mathbb{R}^{10 \times n}$ defined as $D = \begin{bmatrix} \mathbf{X}_1 & \mathbf{X}_2 & ... & \mathbf{X}_n\end{bmatrix}$. Finally, we solve the following system of equations

\begin{equation} \label{LM}
\begin{gathered}
DD^T\mathbf{v} = \lambda C\mathbf{v}, \\ 
\mathbf{v}^TC\mathbf{v} = 1, 
\end{gathered}
\end{equation}
using the Lagrange multiplier method, to minimize the condition in Equation \ref{min_condition}. Herein,

\begin{equation} \label{C}
C= 
\begin{bmatrix}
-1 & \frac{k}{2}-1 & \frac{k}{2} - 1 & 0 &  0 & 0 & 0 & 0 & 0 & 0 \\
\frac{k}{2} - 1 & -1 & \frac{k}{2} - 1 & 0 &  0 & 0 & 0 & 0 & 0 & 0 \\
\frac{k}{2} - 1 & \frac{k}{2} - 1 & -1 & 0 &  0 & 0 & 0 & 0 & 0 & 0  \\
0 & 0 & 0 & -k & 0 & 0 & 0 & 0 & 0 & 0  \\
0 & 0 & 0 & 0 & -k & 0 & 0 & 0 & 0 & 0  \\
0 & 0 & 0 & 0 & 0 & -k  & 0 & 0 & 0 & 0  \\
0 & 0 & 0 & 0 & 0 & 0  & 0 & 0 & 0 & 0  \\
0 & 0 & 0 & 0 & 0 & 0  & 0 & 0 & 0 & 0  \\
0 & 0 & 0 & 0 & 0 & 0  & 0 & 0 & 0 & 0  \\
0 & 0 & 0 & 0 & 0 & 0  & 0 & 0 & 0 & 0  \\
\end{bmatrix}.
\end{equation} 
The system $DD^T\mathbf{v} = \lambda C\mathbf{v}$ recovers the general eigenvector associated with the unique positive eigenvalues when $k>3$ \cite{li2004}. In other words, Equation \ref{LM} has a unique solution when $k>3$. To arrive at a solution for Equation \ref{LM}, we consider the following matrix decomposition

\begin{equation} \label{C_2}
\begin{gathered}
DD^T = 
\begin{bmatrix}
S_{11} & S_{12} \\
S_{12}^T & S_{22}\\
\end{bmatrix} 
\,\,\,\,\, \text{and} \,\,\,\,\,
\mathbf{v}  = 
\begin{bmatrix}
\mathbf{v}_1 \\
\mathbf{v}_2\\
\end{bmatrix}.
\end{gathered}
\end{equation} 
In the equation above, $S_{11} \in \mathbb{R}^{6 \times 6}$, $S_{12} \in \mathbb{R}^{6 \times 4}$, $S_{22} \in \mathbb{R}^{4 \times 4}$, $\mathbf{v}_1 \in \mathbb{R}^{6 \times 1}$, and $\mathbf{v}_2 \in \mathbb{R}^{4 \times 1}$. Equation \ref{LM} becomes

\begin{equation} \label{LM_2}
\begin{gathered}
(S_{11}-\lambda C_1)\mathbf{v}_1+S_{12}\mathbf{v}_2 = 0, \\
S_{12}^T\mathbf{v}_1+S_{22}\mathbf{v}_2 = 0.
\end{gathered}
\end{equation} 
Note that the matrix $C$ has the eigenvalues $\left[k-3, -\frac{k}{2}, -\frac{k}{2}, -k, -k, -k, 0, 0, 0, 0\right]$ and that the element $c_{ij}$ in matrix $C$ is zero when the corresponding $i$ and $j$ are greater than 6. 

Assume the given data set is not coplanar and therefore $S_{22}$ is nonsingular, Equation \ref{LM_2} is subsequently rewritten as

\begin{equation} \label{LM_3}
\begin{gathered}
\mathbf{a}_2 = -S_{22}^{-1}S_{12}^T\mathbf{v}_1,\\
(S_{11}-S_{12}S_{22}^{-1}S_{12}^T)\mathbf{v}_1=\lambda C_1\mathbf{v}_1. \\
\end{gathered}
\end{equation} 
In cases when $(S_{11}-S_{12}S_{22}^{-1}S_{12}^T)$ is positive definite, the general solution to Equation \ref{LM} is $\mathbf{u} = \left[\mathbf{u}_1^T, \mathbf{u}_2^T \right]^T$ where $\mathbf{u}_1$ is the eigenvector associated with the general system in Equation \ref{LM_3} and $\mathbf{u}_2 = -S_{22}^{-1}S_{12}^T\mathbf{u}_1$. On the other hand, in cases when $(S_{11}-S_{12}S_{22}^{-1}S_{12}^T)$ is singular, $\mathbf{u}_1$ is replaced by the eigenvector associated with the largest eigenvalue of the system.

While the above surface fitting method guarantees an ellipsoid when $k = 4$, this condition only represents a subset of an ellipsoid. To ensure an ellipsoid is recovered, the following iterative procedure is carried out. 

\begin{algorithm}[H]
\SetAlgoLined
\KwResult{k}
Initialize k to a sufficiently large number\;
Solve Equation \ref{LM}\;
  \eIf{fitting is an ellipsoid}{
   STOP\;
   }{
        Run a binary search to identify a k that produces an ellipsoid\;
        Perform bisection algorithm to identify a minimum k that produces an ellipsoid\;
      }
 \caption{Iterative ellipsoid surface fitting}
\end{algorithm} \label{algorithm1}

\bibliographystyle{unsrt}  
\bibliography{references}

\end{document}